\documentclass{jpp}
\usepackage{graphicx}
\usepackage{epstopdf, epsfig}
\usepackage{hyperref}

\shorttitle{Radiation whose decay violates the inverse-square law}
\shortauthor{H. Ardavan}

\title{The electromagnetic radiation whose decay violates the inverse-square law: detailed mathematical treatment of an experimentally realized example} 

\author{Houshang Ardavan\aff{}
  \corresp{\email{ardavan@ast.cam.ac.uk}}}
  
\affiliation{\aff{}Institute of Astronomy, University of Cambridge, Madingley Road, Cambridge CB3 0HA, UK}  

\begin{document}

\maketitle

\begin{abstract}
I analyse and numerically evaluate the radiation field generated by an experimentally realized embodiment of an electric polarization current whose rotating distribution pattern moves with linear speeds exceeding the speed of light in vacuum.  I find that the flux density of the resulting emission (i) has a dominant value and is linearly polarized within a sharply delineated radiation beam whose orientation and polar width are determined by the range of values of the linear speeds of the rotating source distribution, and (ii) decays with the distance $d$ from the source as $d^{-\alpha}$ in which the value of $\alpha$ lies between $1$ and $2$ (instead of being equal to $2$ as in a conventional radiation) across the beam.  In that the rate at which boundaries of the retarded distribution of such a source change with time depends on its duration monotonically, this is an intrinsically transient emission process: temporal rate of change of the energy density of the radiation generated by it has a time-averaged value that is negative (instead of being zero as in a conventional radiation) at points where the envelopes of the wave fronts emanating from the constituent volume elements of the source distribution are cusped.  The difference in the fluxes of power across any two spheres centred on the source is in this case balanced by the change with time of the energy contained inside the shell bounded by those spheres.  These results are relevant not only to long-range transmitters in communications technology but also to astrophysical objects containing rapidly rotating neutron stars (such as pulsars) and to the interpretation of the energetics of the multi-wavelength emissions from sources that lie at cosmological distances (such as radio and gamma-ray bursts).  The analysis presented in this paper is self-contained and supersedes my earlier works on this problem.
\end{abstract}

\section{Introduction
\label{sec:introduction}}

Radiation problems in electrodynamics are customarily analysed in the frequency domain with the far-field approximation and under the assumption that retarded solution of Maxwell's equations for the electromagnetic field can be written down in analogy with the classical expression for the retarded potential.  These constraints and presuppositions relinquish the possibility of detecting a host of effects {\it ab initio} when the problem involves constructive interference of the emitted waves and formation of propagating caustics.  Neither can the sudden changes that characterize the solutions to these problems be easily discerned without recourse to an analysis in the time domain, nor can the emitted waves that are described by such solutions be approximated by plane waves (as effected by the far-field approximation) when they have cusped envelopes that propagate into the far zone.  The {\it a priori} assumption that the retarded field like the retarded potential automatically satisfies the boundary conditions at infinity is moreover unfounded as we shall see in this paper.

A case in point is the problem of finding the radiation generated by an extended source whose distribution pattern rigidly rotates with linear speeds exceeding the speed of light in vacuum.  Such a source is not incompatible with the requirements of special relativity because its superluminally moving distribution pattern is created by the correlated motion of aggregates of subluminally moving charged particles~\citep{GinzburgVL:vaveaa, BolotovskiiBM:VaveaD, BolotovskiiBM:Radbcm}.  This and other types of superluminal sources have already been created in the laboratory~\citep{ArdavanA:Exponr,BolotovskiiBM:Radsse}.

In this paper I present a detailed mathematical treatment of this problem in the time domain that is based on first principles.  The results I obtain turn out to be radically different from those of other treatments of this problem that are based on commonly made assumptions and approximations~\citep{Hewish2,HannayJH:ComIGf,McDonald,HannayJH:JMP,HannayJH:Speapc,
Hannay_Morphology,Hannay:09,Contopoulos:2012}.  I will pinpoint the assumptions and approximations responsible for this discrepancy and explain why they fail in the present instance.  I will also devote an appendix to illustrating Hadamard's method for extracting the finite part of a divergent integral~\citep{HadamardJ:lecCau} which seems to be less widely known than the other two pivotal methods used in this analysis: the time-domain version~\citep{BurridgeR:Asyeir} of the uniform asymptotic expansion near a caustic~\citep{ChesterC:Extstd} and the method of steepest descent~\citep[see, e.g.,][]{BenderOrszag}.  To the extent that (i) it is self-contained, (ii) it presents a more exact and thorough analysis of the problem, (iii) it demonstrates how the requirements of the conservation of energy are met in the present case and (iv) it includes, for the first time, numerical results that depict the characteristic features of the generated radiation comprehensively, this paper supersedes my earlier works on this problem~\citep{ArdavanH:Genfnd,ArdavanH:Speapc,ArdavanH:Morph,ArdavanH:Funda,ArdavanH:Inad}.

I start with an analytic expression for a generic electric polarization whose sinusoidal distribution pattern rotates with a constant angular velocity (figure~\ref{F1}).  This expression represents a single Fourier component of any source whose distribution pattern rotates rigidly.  A discretized version of such a polarization can be created in the laboratory by surrounding a dielectric ring with an array of electrode pairs that oscillate with the same frequency but differing phases (figures~\ref{F2} and \ref{F3}).  In \S~\ref{sec:source}, I will specify the accuracy with which the discrete distribution of the moving source created by such a device matches the continuous distribution described by the original analytic expression and will list an experimentally viable set of values for the parameters of this device to emphasize that the propagation speed of the created distribution can easily exceed the speed of light in vacuum.   
 
In \S~\ref{sec:potential}, I show that to satisfy the required boundary conditions at infinity the free-space radiation field of an accelerated superluminal source has to be calculated (in the Lorenz gauge) by means of the retarded solution of the wave equation for the electromagnetic {\it potential}.  There is a fundamental difference between the classical expression for the retarded potential and the corresponding retarded solution of the wave equation that governs the electromagnetic field.  We will see that while the boundary contribution to the retarded solution for the potential can always be rendered equal to zero by means of a gauge transformation that preserves the Lorenz condition, the boundary contribution to the retarded solution of the wave equation for the field cannot be assumed to be zero {\it a priori}.

An integral representation of the radiation field of an extended charge-current with a rigidly rotating distribution pattern is obtained from the retarded solution of the wave equation for the potential in \S~\ref{sec:formulation}.  The field that arises from each constituent volume element of the rotating distribution pattern of such a source (in this paper labelled by its position at time $t=0$) acts as the Green's function for the present problem (\S\S~\ref{subsec:constraint} and \ref{subsec:field}).  I derive an expression for this Green's function in \S~\ref{subsec:Green's function} and show that it is singular on the envelope of the wave fronts that emanate from the superluminally rotating volume element acting as its source (figure~\ref{F5}).  Outside the envelope -- a tube-like surface consisting of two sheets that tangentially meet along a spiralling cusp (figure~\ref{F6}) -- only one wave front passes through the observation point at any given observation time; but inside the envelope three distinct wave fronts, emitted at three distinct values of the retarded time, simultaneously pass through each observation point (figure~\ref{F4}).  It is the coalescence of two of the contributing retarded times on the envelope of wave fronts that gives rise to the constructive interference of the waves and so the divergence of the Green's function on this surface.  At an observation point on the cusp locus of the envelope all three of the contributing retarded times coalesce and the Green's function has a higher-order singularity (figure~\ref{F7}).    
  
In \S~\ref{subsec:bifurcation}, I introduce the notion of {\it bifurcation surface}: a two-sheeted cusped surface reciprocal to the envelope of wave fronts that resides in the space of source points, instead of residing in the space of observation points, and issues from the observation point, instead of issuing from a source point (figure~\ref{F8}).  Intersection of the bifurcation surface of an observation point with the volume of the source divides this volume into two parts.  The source elements inside the bifurcation surface make their contributions toward the observed field at three distinct values of the retarded time, while the source elements outside the bifurcation surface make their contributions at a single value of the retarded time (as a subluminally moving source would).  The source elements inside and close to the bifurcation surface, for which the values of two of the contributing retarded times approach one another, and the source elements inside the bifurcation surface close to its cusp, for which all three values of the contributing retarded times coalesce, are by far the dominant contributors toward the strength of the observed field.  This is reflected in the fact that the phase of the integrand of the integral defining the Green's function (i.e., the space-time distance between the observation point and source points) has two stationary points, occurring on the two sheets of the bifurcation surface, which coalesce for the source elements on the cusp locus of the bifurcation surface (in this paper referred to as $C$).  By applying the time-domain version of the method already developed by~\citet{ChesterC:Extstd} and~\citet{BurridgeR:Asyeir} for this type of integral, I calculate a uniform asymptotic approximation to the value of the Green's function near the cusp locus of the bifurcation surface in \S~\ref{subsec:Expansion}.  
    
The Green's function for the present problem has a complicated singularity structure: it diverges only if one of the sheets of the bifurcation surface is approached from inside this surface but it remains finite (with values that in general differ on opposite sides of the cusp) if either of these sheets is approached from outside the bifurcation surface (figures~\ref{F9} and \ref{F10}).  Consequently, when the expression for the retarded potential in terms of this Green's function is treated as a generalized function, so that it can be differentiated under the integral sign to obtain the field, the result is a divergent integral.  This is the kind of divergence, well understood in the context of generalized functions, that occurs when the orders of two limiting operations (here, integration and differentiation) are interchanged.  It can be handled, as illustrated by the example given in appendix~\ref{appA}, by means of Hadamard's regularization technique~\citep{HadamardJ:lecCau}.  

We will see in \S~\ref{subsec:Hadamard} that Hadamard's finite part of the resulting divergent integral that represents the field of a constituent ring of the source distribution consists of two types of terms: (i) boundary terms extending over the intersections of the two sheets of the bifurcation surface with the source distribution, i.e., the terms that embody the contributions from the discontinuities of the Green's function and (ii) a three-dimensional integral extending over the volume of the source that is equivalent to the classical expression for the radiation field of an extended source in terms of the retarded value of the electric charge-current density.  In this paper I refer to the part of the radiation from a superluminally rotating source that is described by the boundary terms in question as the {\it unconventional component of the radiation}.  

The bifurcation surface of an observation point intersects the rotating distribution pattern of the source at points which approach the observer along the radiation direction with the speed of light at the retarded time.  The source elements that lie on the cusp locus of the bifurcation surface approach the observer along the radiation direction not only with the speed of light but also with zero acceleration at the retarded time (\S~\ref{sec:field}).  Conversely, the cusp loci of the envelopes of wave fronts that emanate from the superluminally rotating volume elements of the source distribution span a radiation beam in the space of observation points that is composed of constructively interfering waves or caustics.  Geometries of the cusp loci in the spaces of source points (figure~\ref{F11}) and observation points (figure~\ref{F12}) and the parts they play in determining the source elements responsible for, and the regions occupied by, the unconventional radiation will be discussed in \S~\ref{subsec:Cusp}.

Section~\ref{sec:Ev} will be devoted to demonstrating that the integral representation of the part of the field that arises from the volume of the source is the same as that for the field of any other time-dependent extended source regardless of whether the volume elements of the source make their contributions toward the observed field at single or multiple values of the retarded time, i.e. regardless of whether the source distribution lies entirely (or partly) inside the bifurcation surface of the observation point (\S~\ref{subsec:Ev1}) or outside it (\S~\ref{subsec:Ev3}).

The part of the radiation field that arises from the discontinuities of the Green's function, i.e., the part describing the unconventional component of the radiation, is given by the difference between two surface integrals each extending over the intersection of the source distribution with one of the sheets of the bifurcation surface (\S~\ref{sec:Eb}).  The phase of the oscillating exponential factor in the integrand of one of these integrals (the one associated with the singular sheet of the bifurcation surface which contains a conical vertex) has a vanishing derivative with respect to the radial coordinate of source points along a two-dimensional curve (in this paper referred to as $S$), while that of the other integral (the one associated with the regular sheet of the bifurcation surface) has no stationary points.  For an observation point in the far zone, the locus $S$ of stationary points lies extremely close to the cusp locus $C$ of the bifurcation surface (figure~\ref{F11}): the separation between these two loci shrinks as $R_P^{-2}$ with the distance $R_P$ of the observer from the source (\S~\ref{subsec:locus}).  

Given that the cusp $C$ constitutes one of the limits of integration in the expression for the unconventional radiation field, its proximity to the locus of stationary points $S$ of the integrand of the integral over the singular sheet of the bifurcation surface means that the contributions of the two neighbouring critical loci $C$ and $S$ toward the value of this integral cannot be taken into account properly without resorting to a technique more discerning than a direct numerical integration.  In \S~\ref{sec:Eb}, I perform the integration with respect to the radial coordinate in the integral in question by the method of steepest descent~\citep[see, e.g.,][]{BenderOrszag}.  I regard the radial coordinate everywhere in the expression for the unconventional radiation field as complex and invoke Cauchy's integral theorem to deform the original paths of integration along the real axis into contours of steepest descent in the complex plane through the critical points of the integral (\S~\ref{subsec:paths}).  The critical points consist in each case of the original boundaries of integration along the real axis and the stationary points (if any) of the phases of the exponential factors in the integrand.  

The range of integration along the real axis, i.e., the radial extent of the portion of the source that contributes toward the value of the unconventional field at the observation point, is determined by the intersection of the bifurcation surface with the source distribution and so changes as the position of the observation point changes (figure~\ref{F11}).  To find the  distribution of this radiation over all angles we therefore have to determine the paths of steepest descent for different ranges of values of the polar coordinate of the observation point separately.  In the case of observation points located inside the region (coloured orange) that is bounded by the two hyperbolas in figure~\ref{F12}, for which the loci $C$ and $S$ both intersect the source distribution (as shown in figure~\ref{F11}), I will analyse the paths of steepest descent through the critical points of the integral over the singular sheet of the bifurcation surface in \S~\ref{subsec:PhiMinus1} and those through the boundary points of the integral over the regular sheet in \S~\ref{subsec:PhiPlus1}.  In the case of observation points located inside the region (coloured yellow) that encompasses the equatorial plane in figure~\ref{F12}, for which the entire source distribution lies within the bifurcation surface, the field receives no contributions from the loci $C$ or $S$ and the integration can be performed accurately along the real axis (\S~\ref{subsec:Resultant2}).  In the case of observation points located in the narrow transition intervals between the above regions, for which only one of the loci $C$ or $S$ intersect the source distribution, one can find the relevant paths of steepest descent as outlined in \S~\ref{sec:transitional}. 
                   
Outcomes of the analyses in \S\S~\ref{subsec:PhiMinus1} and \ref{subsec:PhiPlus1} enable us to express the boundary fields (i.e., the two contributions toward the value of the unconventional field from the singular and regular sheets of the bifurcation surface) each as a sum of the integrals over the steepest-descent paths that pass through their critical points and any paths at infinity that are needed to close the integration contours~\citep{BenderOrszag}.  Phases of the decaying exponential factors in the integrands of the integrals over the steepest-descent paths are all multiplied by an integer designating the ratio of the radiation frequency to the rotation frequency [i.e., the number of  wavelengths of the polarization wave train that fits around the circumference of the dielectric ring hosting the sinusoidal source distribution (figure~\ref{F1})].  Even for moderate values of this integer (of the order of $10$) the main contributions toward the value of each integral come from short segments of the steepest-descent paths next to the critical points from which they issue.  In \S\S~\ref{subsec:AsymptoticForMinusCin} and \ref{subsec:AsymptoticForPlusCin}, I accordingly approximate the values of the boundary fields by ignoring the connecting paths at infinity and by performing the integration along each steepest-descent path only as far as a point beyond which the change in the resulting value of the integral becomes negligible (to within a pre-specified level of accuracy).  

The asymptotic approximations to the values of the two boundary integrals found in \S\S~\ref{subsec:AsymptoticForMinusCin} and \ref{subsec:AsymptoticForPlusCin} will be combined in \S\S~\ref{subsec:Resultant} and \ref{subsec:Resultant2} and their resultant will be added to the contribution from the volume of the source found in \S~\ref{sec:Ev} to obtain the total radiation field in various regions of the space of observation points outside the transitional intervals in \S~\ref{sec:total}.
                   
The results arrived at in \S~\ref{sec:total} yield the electromagnetic field generated by a polarization current density that, while having an azimuthally rotating distribution pattern, flows in an arbitrary direction.  I will determine the flux density of energy and the state of polarization of the radiation described by this field for the following two specific cases corresponding to two differently designed versions of the experimental device sketched in figure~\ref{F2}: for a current that flows axially, i.e., parallel to the rotation axis (\S~\ref{subsec:energy1}) and for a current that flows radially perpendicular to the rotation axis (\S~\ref{subsec:energy2}).    
      
Numerical evaluation of each integral in the expression for the total radiation field for which the integration with respect to the radial coordinate is performed along a steepest-descent path involves solving a transcendental equation -- one that defines the path in question -- at every point of the integration domain.  Moreover, the integrands of such integrals mostly have gradients whose values along their corresponding steepest-descent paths are not only large at the critical points from which the paths issue but also increase as the distance of the observer from the source increases.  To render the time required for evaluating such integrals manageable, therefore, only discrete sets of values of the quantities that characterize the radiation will be plotted in \S~\ref{sec:numerical} instead of continuous curves.  

In \S~\ref{sec:numericalIa}, I discuss the characteristic features of the emission from a polarization current parallel to the rotation axis for which the range of values of the source speed across the dielectric (in figures~\ref{F1} and \ref{F2}) is such that the non-spherically decaying part of the radiation propagates between the polar angles $60^\circ$ and $70^\circ$ (and $110^\circ$ and $120^\circ$).  I first present, in figure~\ref{F21}, the full angular distribution of the time-averaged value of the radial component of normalized Poynting vector (in a logarithmic scale) at a relatively close distance to the source: at ${\hat R}_P=10$, where ${\hat R}_P$ denotes the radial coordinate $R_P$ of the observation point in units of a light-cylinder radius.  (The light-cylinder radius $c/\omega$ is the distance from the rotation axis at which a distribution pattern rigidly rotating with the angular velocity $\omega$ would attain a linear speed equal to the speed of light in vacuum $c$.)  The factor by which the Poynting vector is normalized here, and elsewhere in this paper, is the mean value of the power that propagates across the sphere ${\hat R}_P=10$ per unit solid angle.  Only the radiation distribution in $0\le\theta_P\le90^\circ$ will be shown because this distribution is symmetric both with respect to the equatorial plane and around the rotation axis ($\theta_P$ denotes the polar coordinate of the observation point $P$ measured from the axis of rotation).  The rapid changes in the magnitude of the Poynting vector in figure~\ref{F21} occur when the cusp locus of the bifurcation surface associated with the observation point enters or leaves the source distribution; they reflect the presence or absence of source elements that approach the observation point along the radiation direction with the speed of light and zero acceleration at the retarded time.  

The angular distribution of the radiation at the larger values $10^2$ to $10^6$ of ${\hat R}_P$ will be presented only between the polar angles $60^\circ$ and $70^\circ$ where this distribution changes with distance (figure~\ref{F22}).  The angular distribution of the radiation in the rest of the interval $0\le\theta_P\le90^\circ$ is the same as that shown in figure~\ref{F21} at all distances.  To facilitate the comparison between these distributions, I will vertically shift the plot of each distribution by the number of decibels by which their ordinates would have changed if the magnitude of the Poynting vector for this part of the radiation had diminished as ${\hat R}_P^{-2}$ with distance.  The separation between the shifted distributions in this and the corresponding figures presented in \S\S~\ref{sec:numericalIb} and \ref{sec:numericalII} will be a measure of the degree to which the dependence of the Poynting vector on distance departs from that predicted by the inverse-square law.  I will obtain a quantitative measure of this departure by plotting logarithm of the radial component of the Poynting vector versus logarithm of distance at various polar angles inside the non-spherically decaying radiation beam (figure~\ref{F24}).  From the slope of the curve fitted to these data one will be able to infer the value of the exponent $\alpha$ in the power-law dependence $R_P^{-\alpha}$ of the radial component of the Poynting vector on distance at various polar angles inside the non-spherically decaying radiation beam (figure~\ref{F25}).  In \S~\ref{sec:numericalIa}, I will also (i) plot the angular distribution of the radiation at various distances in polar coordinates (figure~\ref{F23}) and (ii) point out how the requirements of the conservation of energy (discussed in appendix~\ref{AppC}) are met in this case.

Corresponding results for the emission from another polarization current parallel to the rotation axis whose rotating distribution pattern moves with the linear speeds $c$ and $1.2c$ at the inner and outer radii of the dielectric (in figures~\ref{F1} and \ref{F2}) are presented in \S~\ref{sec:numericalIb}.  The new feature of the radiation in this case, where the non-spherically decaying beam encompasses the equatorial plane, is that the magnitude of the radial component of Poynting vector exhibits a prominent maximum within a narrowing solid angle centred on the plane of rotation (figures~\ref{F26}, \ref{F27}, and \ref{F28}).  The narrow equatorial radiation beam in question stems from an additional mechanism of focusing which comes into play whenever the observation point is closer to the equatorial plane than half the width of the source distribution normal to this plane (\S~\ref{subsec:locus}).  Though significantly more intense than the radiation at other angles when observed close to the source, the equatorial beam will be shown to decay faster with distance than the rest of the non-spherically decaying beam (figure~\ref{F29}).  

For comparison, I will also plot the radial component of normalized Poynting vector (using the same normalization factor) for the radiation generated by a source that is the same as the source generating the non-spherically decaying radiation depicted by curve $a$ of figure~\ref{F26} in every respect (has the same dimensions, the same oscillation frequency, the same current density, \ldots) except that its sinusoidal distribution pattern is stationary.  We will see that even at the relatively short distance ${\hat R}_P=10$ from the source the intensity of the radiation generated by the superluminally rotating source exceeds that of the conventional radiation generated by a corresponding stationary source (depicted in curve $s$ of figure~\ref{F26}) by more than a factor of $300$ on the equatorial plane.
                   
In \S~\ref{sec:numericalII}, I will present the numerical results for a polarization current that differs from that analysed in \S~\ref{sec:numericalIa} only in having a direction everywhere perpendicular (rather than parallel) to the rotation axis (figures~\ref{F30}--\ref{F33}).  The only feature in this case that is radically different from its counterpart in the case of an axial current is the state of polarization of the resulting radiation.  The emissions discussed in \S\S~\ref{sec:numericalIa} and \ref{sec:numericalIb} are both linearly polarized everywhere with position angles parallel to the rotation axis.  We will see that the non-spherically decaying part of the radiation described in \S~\ref{sec:numericalII} is also linearly polarized but with a fixed position angle perpendicular to the rotation axis (figure~\ref{F34}).  The part of the unconventional radiation that propagates in the region next to the equatorial plane (coloured yellow in figure~\ref{F12}), on the other hand, turns out to be elliptically polarized with a position angle that changes as the polar angle of the observation point changes (figure~\ref{F35}). 
            
An essential tool for the derivation of the results reported in this paper is the long established but scarcely used technique by Hadamard for extracting the finite part of a divergent integral~\citep{HadamardJ:lecCau}.  As an illustrative example, derivative of a simple double integral is evaluated, with respect to its free parameter, in appendix~\ref{appA}.  Like the integrand in the expression for the Green's function for the present problem, the integrand in this example contains a Dirac delta function whose argument is a cubic function of one of the integration variables.  Depending on the order in which one performs the integration with respect to the two variables of integration, one obtains two different values for the derivative of this integral, one finite and one divergent.  The paradox is resolved (i.e., the value of the derivative of the integral remains unchanged when the order of integration is changed) once we interpret the divergent integral as a generalized function and equate it to its Hadamard's finite part.                                                      
                   
In appendix~\ref{appB}, I will explain why a conventional approach to the problem formulated in \S~\ref{sec:formulation} fails to capture the unusual features of the radiation described in this paper.  The contributions that arise from the differentiation of the limits of integration in the classical form of the retarded potential (i.e., from the boundaries of the retarded distribution of the source) will be shown to be divergent at any observation points for which the value of the potential at the observation time depends on three coalescing values of the retarded time.  We will see that the more familiar treatment of the retarded potential as a classical function merely replaces the singularities of the Green's function for the present problem by corresponding singularities in the limits of integration.  In contrast to the singularities of the Green's function which can be rigorously handled by Hadamard's regularization technique, however, the singularities encountered in the limits of integration vitiate the differentiability of the retarded potential {\it ab initio}.  

Constancy of the width of the solid angle over which the Poynting vector decays non-spherically might seem to contravene the conservation of energy at first sight.  In the case of a conventional radiation field, for which the derivative of the electromagnetic energy density with respect to time vanishes when time averaged, the continuity equation stating the conservation of energy~\citep[see, e.g.,][]{JacksonJD:Classical} requires that the flux of energy into any closed region (e.g., into the volume bounded by two spheres centred on the source) should equal the flux of energy out of that region.  However, because the boundaries of the support of the retarded distribution of the present source change with time at a rate that depends on the time elapsed since the source was switched on monotonically (appendix~\ref{AppC}), the radiation process analysed in this paper never attains a steady state.  I will evaluate the time-averaged value of the temporal rate of change of the energy density carried by the non-spherically decaying part of the radiation in appendix~\ref{AppC} and show that it is negative at points where the envelopes of the wave fronts emanating from the constituent volume elements of the source distribution are cusped.  In the case of the present radiation process, which is intrinsically transient, the flux of energy into a closed region is always smaller than the flux of energy out of it because the electromagnetic energy contained in that region decreases with time (\S~\ref{sec:conclusion} and appendix~\ref{AppC}).  

In the last seven paragraphs of the concluding section of the paper (\S~\ref{sec:conclusion}), I briefly remark on the implications of the present results for a diverse set of disciplines ranging from astrophysics (e.g., the emission mechanism of pulsars and the interpretation of the energetic requirements of the distant sources of radio and gamma-ray bursts) to communications technology (e.g., antenna theory and the design of long-range transmitters).                          
                
\section{An experimentally realized superluminal source distribution}
\label{sec:source}

\begin{figure}
\centerline{\includegraphics[width=9cm]{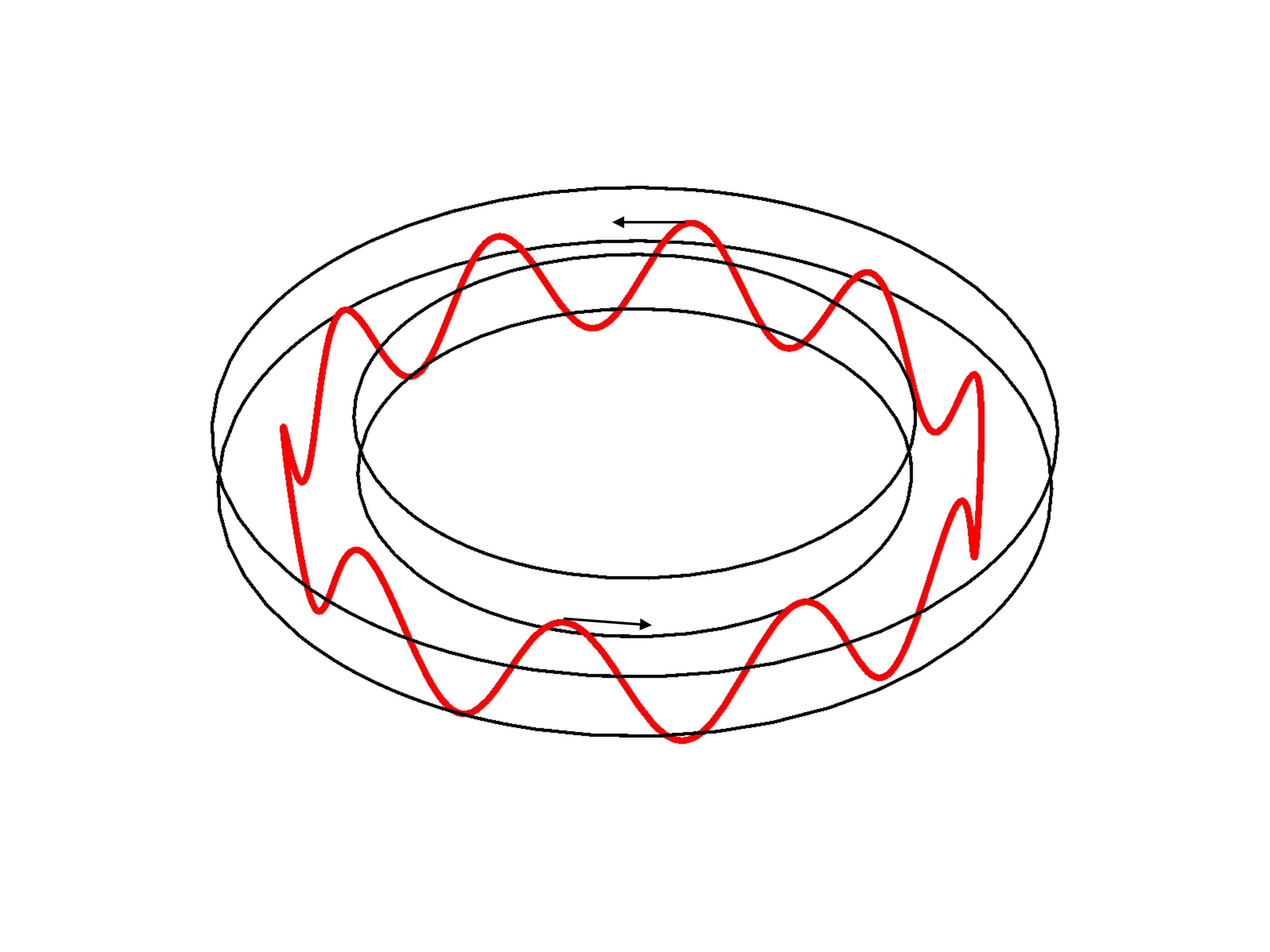}}
\caption{Schematic representation of the distribution pattern of the electric polarization described by~(\ref{E1}) at a given $(r, z)$.  The circles designate the edges of the dielectric ring hosting the polarization and the sinusoidal curve designates the rigidly rotating wave train whose linear speed $r\omega$ (along the shown arrows) exceeds the speed of light in vacuum.} 
\label{F1}
\end{figure}

Consider a distribution of electric polarization ${\bf P}$ whose components in a cylindrical coordinate system $(r,\varphi,z)$ are given by
\begin{equation}
P_{r,\varphi,z}(r,\varphi,z,t)=s_{r,\varphi,z}(r,z)\cos[m(\varphi-\omega t)], 
\label{E1}
\end{equation}
in which $t$ (assumed to be $\ge0$) is time, $\omega$ is a constant angular frequency, ${\mathbf s}(r,z)$ is an arbitrary vector function with a finite support in $r>c/\omega$ and $m$ is a positive integer ($c$ denotes the speed of light in vacuum).  At a given time $t$, the azimuthal dependence of the polarization (\ref{E1}) along each circle of radius $r$ within the source is the same as that of a sinusoidal wave train, of wavelength $2\pi r/m$, whose $m$ cycles fit around the circumference of the circle smoothly.  As time elapses, this wave train propagates around each circle of radius $r$ with a linear speed $r\omega$ that exceeds the speed of light $c$, i.e., rotates about the $z$-axis rigidly (figure~\ref{F1}).  This is a generic source: one can construct the Fourier representation of any distribution with a uniformly rotating pattern, $P_{r,\varphi,z}(r,\varphi-\omega t,z)$, by the superposition over $m$ of terms of the form $s_{r,\varphi,z}(r,z,m)\cos[m(\varphi-\omega t)]$. 

\begin{figure}
\centerline{\includegraphics[width=13cm]{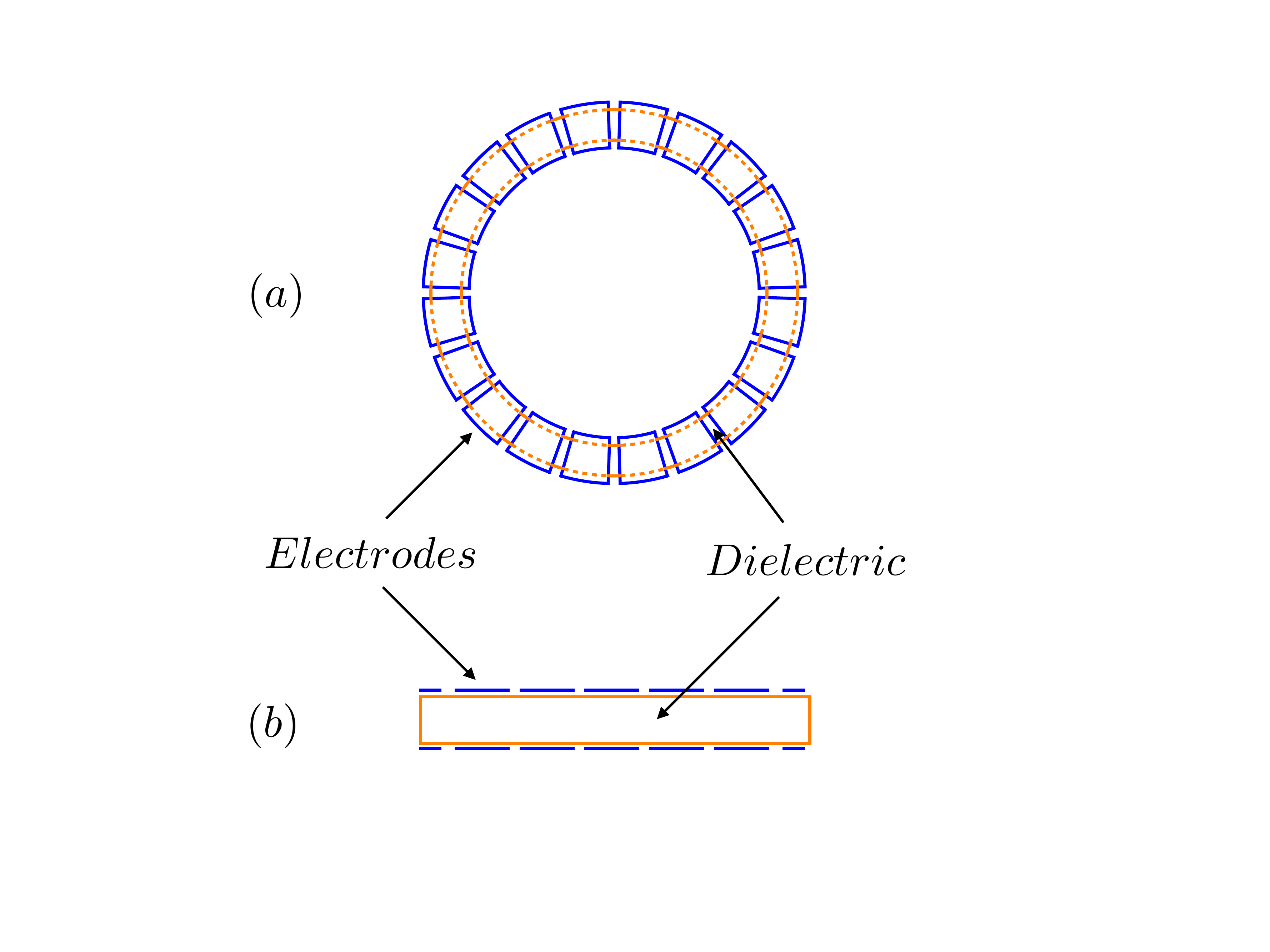}}
\caption{Schematic view of the experimental apparatus (a) from above and (b) from the side, showing the boundaries of the dielectric medium (in orange) and the electrode pairs (in blue).}
\label{F2}
\end{figure}

\begin{figure}
\centerline{\includegraphics[width=6cm]{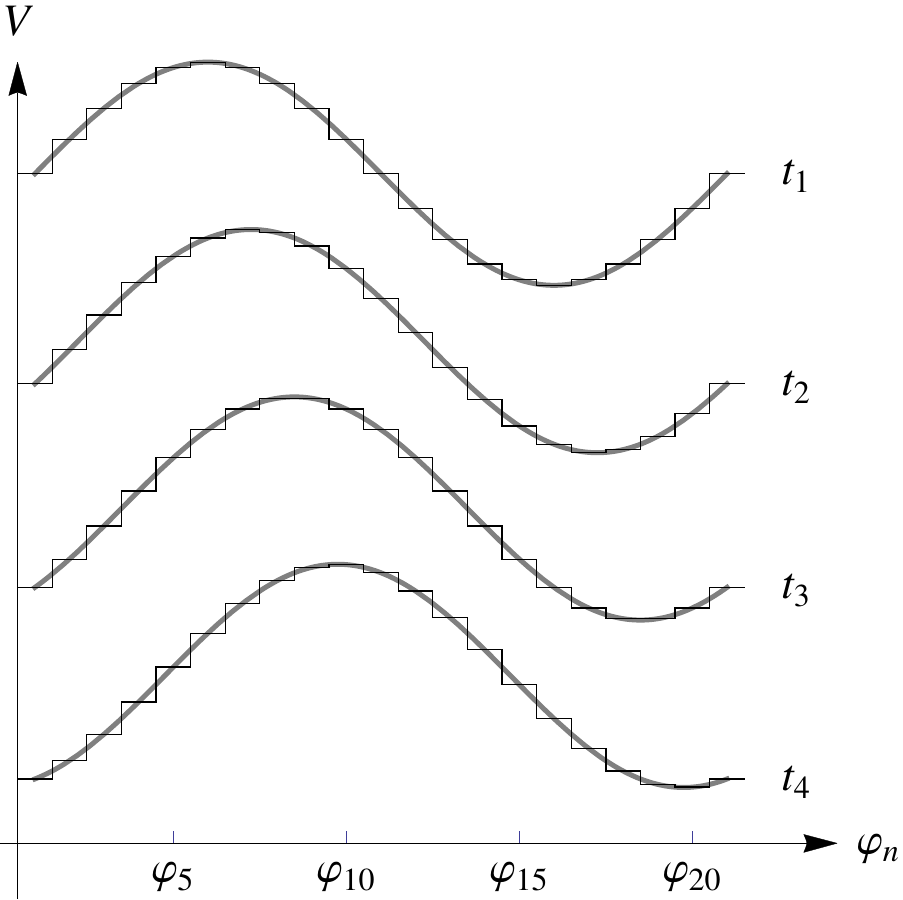}}
\caption{The oscillating voltage $V$ on each electrode pair versus the $\varphi$ coordinate ($\varphi_n=2\pi n/N$ with $n=1,\cdots, 21$) of the centre of that electrode at four equally-spaced consecutive times ($t_1<t_2<t_3<t_4$).  The electrodes oscillate with the same frequency but differing phases.  It can be seen that the phase difference between the oscillations of the adjacent electrode pairs sets this discretized wave train in motion.  The fundamental Fourier component of the resulting discretized polarization, here depicted by a solid sinusoidal curve, thus moves in the azimuthal direction with a speed that can exceed the speed of light in vacuum, even though the charges whose separation creates the polarization move in a different direction with a different speed.}
\label{F3}
\end{figure}

Equation (\ref{E1}) corresponds to a laboratory-based source that has been experimentally implemented~\citep{ArdavanA:Exponr}.  The apparatus in the performed experiments consists of  a circular ring made of a dielectric material, with an array of $N$ electrode pairs that are placed beside each other around its circumference.  With a sufficiently large value of $N$ (to be specified below), a sinusoidal distribution of polarization can be generated along the length of the dielectric by applying a voltage to each pair independently (figure~\ref{F2}).  The distribution pattern of this polarization can then be set in motion by energizing the electrodes with phase-controlled time-varying voltages.  One can synthesize the transverse polarization wave $\cos[m(\varphi-\omega t)]$ moving around the ring by driving each electrode pair with a harmonically oscillating voltage whose frequency is fixed but whose phase depends on the position of the pair around the ring (figure~\ref{F3}).

To estimate the required value of $N$, let us note that the $(\varphi,t)$ dependence of the polarization that is thus generated by the discrete set of electrodes described above has the form
\begin{equation}
P(\varphi,t)=\sum_{k=0}^{N-1}\Pi\left(k-\frac{N\varphi}{2\pi}\right)\cos\left[m\left(\omega t-\frac{2\pi k}{N}\right)\right],
\label{E2}
\end{equation}  
in which $\Pi(x)$ denotes the rectangle function, a function that is unity when $\vert x\vert<\textstyle{\frac{1}{2}}$ and zero when $\vert x\vert>\textstyle{\frac{1}{2}}$.  [For any given $k$, the function $\Pi(k-N\varphi/2\pi)$ is non-zero only over the interval $(2k-1)\pi/N<\varphi<(2k+1)\pi/N$.]  When the electrodes operate over a time interval exceeding $2\pi/\omega$, the generated polarization is a periodic function of $\varphi$ for which the range of values of $\varphi$ correspondingly exceeds the period $2\pi$.

The Fourier-series representation of $\Pi(k-N\varphi/2\pi)$ with the period $2\pi$ is given by
\begin{equation}
\Pi\left(k-\frac{N\varphi}{2\pi}\right)=\frac{1}{N}+\sum_{n=1}^\infty\frac{2}{n\pi}\sin\left(\frac{n\pi}{N}\right)\cos\left[n\left(\varphi-\frac{2\pi k}{N}\right)\right].
\label{E3}
\end{equation}
If we now insert~(\ref{E3}) in~(\ref{E2}) and use formula (4.21.16) of ~\citet{Olver} to rewrite the product of the two cosines in the resulting expression as the sum of two cosines, we obtain two infinite series, each involving a single cosine and extending over $n=1,2,\cdots,\infty$.  These two infinite series can then be combined (by replacing $n$ in one of them by $-n$ everywhere and performing the summation over $n=-1,-2,\cdots,-\infty$) to arrive at
\begin{equation}
P(\varphi,t)=\sum_{n=-\infty}^\infty\frac{1}{n\pi}\sin\left(\frac{n\pi}{N}\right)\sum_{k=0}^{N-1}\cos\left[m\omega t-n\varphi+2\pi(n-m)\frac{k}{N}\right],
\label{E4}
\end{equation} 
 in which the order of summations with respect to $n$ and $k$ has been interchanged and the contribution $N^{-1}$ on the right-hand side of~(\ref{E2}) has been incorporated into the $n=0$ term: the coefficient $(n\pi)^{-1}\sin(n\pi/N)$ has the value $N^{-1}$ when $n=0$.  

The finite sum over $k$ can be evaluated by means of the geometric progression.  The result, according to formula (1.341.3) of ~\citet{Gradshteyn}, is 
\begin{equation}
\sum_{k=0}^{N-1}\cos\left[m\omega t-n\varphi+\frac{2\pi(n-m)k}{N}\right]=\cos\left[m\omega t-n\varphi+\frac{\pi(n-m)(N-1)}{N}\right]\frac{\sin[(n-m)\pi]}{\sin[\frac{(n-m)\pi}{N}]}.
\label{E5}
\end{equation}
The right-hand side of~(\ref{E5}) vanishes when $(n-m)/N$ is different from an integer.  If $n=m+l N$, where $l$ is an integer, on the other hand, the above sum would have the value $N\cos(m\omega t-n\varphi)$, as can be seen by directly inserting $n=m+l N$ in the left-hand side of~(\ref{E5}).  Performing the summation with respect to $k$ in~(\ref{E4}), we therefore obtain
\begin{equation}
P(\varphi,t)=\frac{N}{m\pi}\sin\left(\frac{m\pi}{N}\right)\Big\{\cos[m(\varphi-\omega t)]+\sum_{l\ne0}(-1)^l\left(1+\frac{N l}{m}\right)^{-1}\cos[(N l+m)\varphi-m\omega t]\Big\},
\label{E6}
\end{equation}
since only those terms of the infinite series survive for which $n$ has the value $m+l N$ with an $l$ that ranges over all integers from $-\infty$ to $\infty$.

I have written out the $l=0$ term of the series in~(\ref{E6}) explicitly in order to emphasize the following points.  The parameter $N/m$, which signifies the number of electrode pairs within a wavelength of the polarization wave train, need not be large for the factor $(m\pi/N)^{-1}\sin(m\pi/N)$ to be close to unity: this factor equals $0.9$ even when $N/m$ is only $4$.  Moreover, if the travelling polarization wave $\cos[m(\varphi-\omega t)]$ that is associated with the $l=0$ term has a phase speed $r\omega$ that is only moderately superluminal, the phase speeds $r\omega/\vert1+N l/m\vert$ of the waves described by all the other terms in the series would be subluminal.  Not only would these other polarization waves have amplitudes that are by the factor $\vert1+N l/m\vert^{-1}$ smaller than that of the wave associated with the fundamental Fourier component $l=0$, but also they would generate electromagnetic fields whose characteristics (such as their rate of decay with distance) are different from those generated by the superluminally moving polarization wave.

The fundamental ($l=0$) Fourier component of the discretized polarization current that is created by the present device thus has precisely the same $(\varphi,t)$ dependence as that which is described in~(\ref{E1}) above.  Neither the reduction in its amplitude, which arises from the departure of the value of $(m\pi/N)^{-1}\sin(m\pi/N)$ from unity, nor the presence of the other low-amplitude waves that are superposed on it, makes any difference to the fact that the fundamental Fourier component of the discretized wave created in $r>c/\omega$ rotates uniformly with a superluminal speed (figure~\ref{F3}).  Linearity of the emission process ensures that the radiation that is generated by an individual term of the series in~(\ref{E6}) is not in any way affected by those that are generated by the other terms in this series.

For the distribution pattern of the created polarization current to be moving, it is however essential that the number of electrode pairs per wavelength of this pattern, $N/m$, exceed $2$.  For $N/m=2$, the $l=-1$ term is proportional to $\cos[m(\varphi+\omega t)]$ and so describes a wave that has the same amplitude as, and travels with the same speed in the opposite direction to, the wave described by the $l=0$ term.  The fundamental wave is thus turned into a standing wave when $N/m$ has a value as low as $2$.

Note, finally, that the speed of light is easily attainable.  The adjacent electrode pairs are energized to oscillate out of phase, so that there is a time difference $\Delta t$ between the instants at which the oscillatory applied voltages on adjacent electrodes attain their maximum amplitude.  The variation thus produced in the distribution pattern of the induced polarization current results in the azimuthal propagation of this distribution pattern around the ring with the speed $\Delta\ell/\Delta t$, where $\Delta\ell$ is the distance between the centres of the adjacent electrode pairs.  The phase difference between oscillations of two adjacent electrode pairs, $\Delta\Phi$, and the energizing time delay $\Delta t$ are related by $\Delta\Phi=2\pi\nu\Delta t$, where $\nu$ is the oscillation frequency of the applied voltage.  The generated wave train can retain its shape while rotating around the ring only if it contains an integral number of wavelengths of the sinusoidal distribution pattern of the current, i.e., if the phase difference $\Delta\Phi$ is constrained by $N\Delta\Phi=2\pi m$, where $m$ is an integer [the integer appearing in~(\ref{E1}) which also connects $\nu$ to the angular frequency of rotation of the wave train, $\omega$, via $2\pi\nu=m\omega$].  The propagation speed of the distribution pattern of the polarization current is therefore given by $\Delta\ell/\Delta t=2\pi\nu{\bar r}/m$, in which ${\bar r}=N\Delta\ell/(2\pi)$ denotes the mean radius of the dielectric ring. 

This speed can exceed the speed of light in vacuum, $c$, for a large set of experimentally viable values of the parameters $N$, $\Delta\ell$, $\Delta\Phi$, $\nu$ and $m$. In the case of an apparatus consisting of $N=72$ electrode pairs for which $\Delta\ell=1$ cm, for example, energizing the electrodes with the phase difference $\Delta\Phi=25$ degrees and the frequency $\nu=2.5$ GHz results in a polarization current whose distribution pattern has the form of a sinusoidal wave train, containing $m=5$ wavelengths, and propagates around the ring of mean radius ${\bar r}=11.46$ cm with the speed $\Delta\ell/\Delta t=1.2 c$. 

To be able to calculate the field generated by the polarization current ${\bf j}=\partial{\bf P}/\partial t$, we need an explicit expression also for the amplitude ${\mathbf s}$ of the polarization (\ref{E1}).  A choice that both corresponds to a simple model of the experimentally realized source distribution discussed in~\citet{ArdavanA:Exponr}, and can adequately illustrate the salient features of the resulting radiation, is one in which ${\mathbf s}$ vanishes outside the rectangular region
\begin{equation}
{\mathcal S}^\prime:\qquad{\hat r}_L\leq{\hat r}\leq{\hat r}_U,\quad-{\hat z}_0\leq{\hat z}\leq{\hat z}_0,
\label{E7}
\end{equation}
of the (${\hat r},{\hat z}$)-plane and is constant inside it.  In this expression, the dimensions ${\hat r}_L$, ${\hat r}_U$ and ${\hat z}_0$ of the rectangular cross-section of the annulus bounding the polarization distribution are all constant.

\section{Fundamental role of the retarded potential in electrodynamics of superluminal sources}
\label{sec:potential}

In the classical theory of electromagnetic radiation, Maxwell's equations are most commonly reduced to wave equations by one of the following two methods.
\begin{enumerate}
\item
One of the fields is eliminated between Maxwell's equations by differentiation to obtain a wave equation for the other field: e.g., the electric field ${\bf E}$ is eliminated to obtain the wave equation
\begin{equation}
{\bf\nabla}^2{\bf B}-{1\over c^2}{\partial^2{\bf B}\over\partial t^2}=
-{4\pi\over c}{\bf\nabla\times j}
\label{E8}
\end{equation}
for the magnetic field ${\bf B}$~\citep[see, e.g.,][p. 246]{JacksonJD:Classical}. 
\item
The fields are expressed in terms of potentials.  In the Lorenz gauge, the electromagnetic fields 
\begin{equation}
{\mathbf E}=-\nabla_P \Phi-\frac{1}{c}\frac{\partial{\mathbf A}}{\partial t_P},\quad\quad{\mathbf B}=\nabla_P{\mathbf{\times A}},
\label{E9}
\end{equation}
are expressed in terms of a four-potential $A^\mu$ that satisfies the wave equation
\begin{equation}
\quad{\bf\nabla}^2A^\mu-{1\over c^2}{\partial^2A^\mu\over\partial t^2}=
-{4\pi\over c}j^\mu,\quad\mu=0,\cdots, 3,
\label{E10}
\end{equation}
where $({\bf x},t)$ and $({\bf x}_P,t_P)$ are the space-time coordinates of the source points and the observation point $P$, and $\mu=0$ and $\mu=1,2,3$ respectively designate the temporal and spatial components of $A^\mu=(\Phi,\, {\mathbf A})$ and $j^\mu=(\rho c,\, {\mathbf j})$ in a Cartesian coordinate system~\citep[see, e.g.,][]{JacksonJD:Classical}.  
\end{enumerate}
We shall see below that, in free space, the retarded solutions to the above two wave equations [(\ref{E8}) and (\ref{E10})] do not always have the same form.

The solution to the initial-boundary value problem for (\ref{E10}) inside a closed surface $\partial{\cal D}$ is given by
\begin{eqnarray}
A^\mu({\bf x}_P,t_P)&=&{1\over c}\int_0^{t_P}{\rm d}t\int_{\cal D}{\rm d}^3{\bf x}\,j^\mu G+{1\over4\pi}\int_0^{t_P}{\rm d}t
\int_{\partial{\cal D}}{\rm d}^2{\bf x}\cdot(G\nabla A^\mu-A^\mu\nabla G)\nonumber\\*
&&-{1\over 4\pi c^2}\int_{\cal D}{\rm d}^3{\bf x}
\Big(A^\mu{\partial G\over\partial t}-G{\partial A^\mu\over\partial t}\Big)_{t=0},\qquad
\label{E11}
\end{eqnarray}
in which $G$ is the Green's function and ${\cal D}$ is the volume enclosed by the surface $\partial{\cal D}$~\citep[see, e.g.,][p. 893]{MorsePM:Methods1}.  The potential that arises from a time-dependent localized source in unbounded space decays as ${R_P}^{-1}$ when ${\hat R}_P\gg1$, so that for an arbitrary free-space potential
the second term in~(\ref{E11}) would be of the same order of magnitude ($\sim{R_P}^{-1}$) as the first term in the limit that the boundary $\partial{\cal D}$ tends to infinity.  However,
even potentials that satisfy the Lorenz condition
${\bf\nabla\cdot A}+c^{-1}\partial\Phi/\partial t=0$ are arbitrary to within a solution of the homogeneous wave equation: the gauge transformation
\begin{equation}
{\bf A}\to{\bf A}+\nabla\Lambda,
\qquad\Phi\to\Phi-\partial\Lambda/\partial t,
\label{E12}
\end{equation}
preserves the Lorenz condition if $\nabla^2\Lambda-c^{-2}\partial^2\Lambda/\partial t^2=0$~\citep[see]{JacksonJD:Classical}.  One can always use this gauge freedom in the choice of the potential to render the boundary contribution [the second term in~(\ref{E11})] equal to zero, since this term, too, satisfies the homogenous wave equation.  Under the null initial conditions $A^\mu|_{t=0}=(\partial A^\mu/\partial t)_{t=0}=0$, assumed in this paper, the contribution from the third term in~(\ref{E11}) is, moreover, identically zero.

In the absence of boundaries, i.e., in the limit where $\partial{\cal D}$ lies at infinity, the retarded Green's function for~(\ref{E8}) and (\ref{E10}) has the form
\begin{equation}
G({\bf x}, t;{\bf x}_P, t_P)={\delta(t-t_P+R/c)\over R},
\label{E13}
\end{equation}
where $\delta$ is the Dirac delta function and $R$ is the magnitude of the separation ${\bf R}\equiv{\bf x}_P-{\bf x}$ between the observation point ${\bf x}_P$ and the source point ${\bf x}$.  Irrespective of how the radiation field decays in the limit $R\to\infty$, therefore, the potential $A^\mu$ due to a localized source distribution in an unbounded space which is switched on at $t=0$, can be calculated from the first term in~(\ref{E11})
\begin{equation}
A^\mu({\bf x}_P,t_P)=\frac{1}{c}\int{\rm d}^3 {\bf x}\,{\rm d}t\, j^\mu({\bf x},t)\frac{\delta(t-t_P+R/c)}{R}.
\label{E14}
\end{equation}
Whatever the Green's function for the problem may be in the presence of boundaries, it would approach that in~(\ref{E13}) in the limit where the boundaries tend to infinity.

Next, let us consider the wave equation that governs the magnetic field ${\bf B}$.  One can write the solution to the initial-boundary value problem for~(\ref{E8}) as
\begin{eqnarray}
B_i({\bf x}_P,t_P)&=&{1\over c}\int_0^{t_P}{\rm d}t\int_{\cal D}{\rm d}^3{\bf x}\,({\bf\nabla\times j})_i G+{1\over4\pi}\int_0^{t_P}{\rm d}t\int_{\partial{\cal D}}{\rm d}^2{\bf x}\cdot(G\nabla B_i-B_i\nabla G)\nonumber\\*
&&-{1\over 4\pi c^2}\int_{\cal D}{\rm d}^3{\bf x}\Big(B_i{\partial G\over\partial t}-G{\partial B_i\over\partial t}\Big)_{t=0},\quad
\label{E15}
\end{eqnarray}
where $i=1,2,3$ designate the components of ${\bf B}$ and ${\bf\nabla\times j}$ in a Cartesian coordinate system~\citep{MorsePM:Methods1}.  In contrast to (\ref{E11}), here we no longer have the freedom that was offered by the gauge transformation~(\ref{E12}) to make the boundary term [the second term in~(\ref{E15})] zero.  In other words, the retarded solution to the wave equation for the field cannot be written down in analogy with~(\ref{E14}) as is done in certain textbooks~\citep[see, e.g.,][p. 246]{JacksonJD:Classical}.  

There is a fundamental difference between the classical expression for the retarded potential and the corresponding retarded solution of the wave equation that governs the electromagnetic field: while the boundary contribution to the retarded solution for the potential can always be rendered equal to zero by means of a gauge transformation that preserves the Lorenz condition, the boundary contribution to the retarded solution of the wave equation for the field may be neglected only if it diminishes with distance faster than the contribution of the source density [the first term in~(\ref{E15})] in the far zone.  In the case of a source whose distribution pattern rotates superluminally, where the radiation field decays non-spherically (more slowly than ${\hat R}_P^{-1}$) with distance, the boundary term in the retarded solution (\ref{E15}) for the field is in fact {\em larger} than the source term of this solution, in the limit where the closed surface $\partial{\cal D}$ tends to infinity~\citep{ArdavanH:Funda}.  Given that the distribution of the radiation field of an accelerated superluminal source in the far zone is not known {\em a priori}, to be prescribed as a boundary condition, it follows that the only way one can calculate the free-space radiation field of such sources is via the retarded solution for the potential.

\section{Formulation of the problem}
\label{sec:formulation}
\subsection{Extended polarization currents whose distribution patterns propagate faster than light in vacuum}
\label{subsec:constraint}
The experimentally realized source distribution described in \S~\ref{sec:source} is a generic member of a wide class of rotating source distributions.  Any electric polarization ${\bf P}$ whose distribution pattern rotates uniformly with the constant angular frequency $\omega$ gives rise to a charge density $\rho=-{\mathbf\nabla}\cdot{\mathbf P}$ and a current density ${\mathbf j}=\partial{\mathbf P}/\partial t$ that, like ${\mathbf P}$ itself, depend on the azimuthal angle $\varphi$ in only the combination
\begin{equation}
{\hat\varphi}=\varphi-\omega t,
\label{E16}
\end{equation}
i.e., are of the forms
\begin{eqnarray}
\left[\matrix{ P_{r,\varphi,z}(r,\varphi,z,t)\cr\rho(r,\varphi,z,t)\cr j_{r,\varphi,z}(r,\varphi,z,t)\cr}\right]&=&\left[\matrix{P_{r,{\varphi},z}(r,{\hat\varphi},z,t)\cr\rho(r,{\hat\varphi},z,t)\cr j_{r,{\varphi},z}(r,{\hat\varphi},z,t)\cr}\right],
\label{E17}
\end{eqnarray}
where $(r,\varphi,z)$ are, as in \S~\ref{sec:source}, the cylindrical polar coordinates based on the the axis of rotation, $t$ (assumed to be $\ge0$) is time and $P_{r,\varphi,z}$ and $j_{r,\varphi,z}$ are the cylindrical components of ${\mathbf P}$ and ${\mathbf j}$, respectively.  

In~(\ref{E16}) and (\ref{E17}) the coordinates $t$ and $\varphi$ both range over $(0,\infty)$ but the coordinate ${\hat\varphi}$ has a limited range of length $2\pi$, e.g.,
\begin{equation}
0\le{\hat\varphi}<2\pi.
\label{E18}
\end{equation}
As can be seen from the alternative form $\varphi={\hat\varphi}+\omega t$ of~(\ref{E16}), ${\hat\varphi}$ is a Lagrangian coordinate that labels the rotating volume elements of the current distribution on each circle $r=$const, $z=$const, by their azimuthal positions at the time $t=0$.  This coordinate cannot range over a wider interval because the aggregate of volume elements that constitute a rotating source in its entirety can at most occupy an azimuthal interval of length $2\pi$ at any given time (e.g., at $t=0$).  The polarization distribution $P_{r,\varphi,z}(r,\varphi,z,t)=s_{r,\varphi,z}(r,z)\cos(m{\hat\varphi})$ given in~(\ref{E1}), on which the analysis in the following sections will be based, is an example of this class of sources in which the range of ${\hat\varphi}$ is likewise subject to the constraint~(\ref{E18}).  

Note that beyond $r=c/\omega$ (which I will refer to as the \textit {light cylinder}) the distribution patterns of the above charge-current densities move with linear speeds $r\omega$ exceeding the speed of light in vacuum, $c$.  This is not inconsistent with the requirements of special relativity because the superluminally moving pattern is created by the coordinated motion of aggregates of subluminally moving particles~\citep{GinzburgVL:vaveaa, BolotovskiiBM:VaveaD, BolotovskiiBM:Radbcm}.  Not only is a superluminal current distribution of this type already generated in the laboratory~\citep[see][and \S~\ref{sec:source}]{ArdavanA:Exponr, BolotovskiiBM:Radsse}, but it also occurs in the magnetospheres of astrophysical objects containing rapidly rotating neutron stars such as pulsars~\citep[see][and \S~\ref{sec:conclusion}]{ArdavanH:Nature, SpitkovskyA:Oblique, ArdavanH:Pul, Contopoulos:2012, Tchekhovskoy:etal}. 

\subsection{Radiation field of a superluminally rotating charge-current distribution}
\label{subsec:field}

According to (\ref{E9}), the radiation fields associated with the retarded potential (\ref{E14}) are given by  
\begin{equation}
\left[\matrix{{\mathbf E}\cr{\mathbf B}\cr}\right]=\frac{1}{c^2}\int{\rm d}^3 {\bf x}\,{\rm d}t\,\frac{\delta^\prime(t-t_P+R/c)}{R}\left[\matrix{{\mathbf j}-\rho c\,{\hat{\mathbf n}}\cr{\hat{\mathbf n}}{\mathbf\times}{\mathbf j}\cr}\right],
\label{E19}
\end{equation}
where $\delta^\prime$ denotes the derivative of $\delta$ with respect to its argument and ${\hat{\bf n}}=\nabla_P R={\bf R}/R$.  The terms arising from the differentiation of $R^{-1}$ which describe static fields (terms that are non-zero even when the charge-current distribution is time independent) have been discarded here: in addition to decaying faster with distance, these terms are negligibly smaller than the retained terms in cases where the radiation frequency is appreciably larger than the rotation frequency [i.e., the integer $m$ in~(\ref{E1}) appreciably exceeds unity].  For an observation point that is located at infinity, the unit vector ${\hat{\bf n}}$ is independent of the the integration variables $({\bf x},t)$ and can be taken outside the the above integrals to obtain ${\bf B}={\hat{\bf n}}\times{\bf E}$ in the limit $\vert{\bf x}_P\vert\to\infty$.  Since we will be concerned also with observation points that lie at finite distances from the source, however, I will take the dependence of ${\hat{\bf n}}$ on ${\bf x}$ and ${\bf x}_P$ into account and treat ${\bf E}$ and ${\bf B}$ as two independent vectors in this paper.  

For the purposes of calculating the fields generated by the sources in~(\ref{E17}) and (\ref{E18}), the space-time of source points may be marked either with $({\mathbf x},t)=(r,\varphi,z,t)$ or with the coordinates $(r,{\hat\varphi},z,t)$ that naturally appear in the description of such rotating sources.  Once ${\hat\varphi}$, with the range $(0,2\pi)$, is adopted as one of the coordinates, either $t$ or $\varphi$ (which have unlimited ranges) could be used to track the time evolution of the rotating source point $(r,{\hat\varphi},z)$.

Changing the variables of integration in~(\ref{E19}) from $({\mathbf x},t)=(r,\varphi,z,t)$ to $(r,{\hat\varphi},z,\varphi)$ and introducing the dimensionless coordinates ${\hat r}=r\omega/c$ and ${\hat z}=z\omega/c$, we obtain
\begin{equation}
\left[\matrix{{\mathbf E}\cr{\mathbf B}\cr}\right]=\frac{1}{\omega} \sum_{k=1}^\infty\int_{\mathcal S}{\hat r}{\textrm d}{\hat r}\,{\textrm d}{\hat\varphi}\,{\textrm d}{\hat z}\,\int_{{\hat\varphi}+2(k-1)\pi}^{{\hat\varphi}+2k\pi}{\textrm d}{\varphi}\,\frac{\delta^\prime(g-\phi)}{{\hat R}}\left[\matrix{{\mathbf j}-\rho c\,{\hat{\mathbf n}}\cr{\hat{\mathbf n}}{\mathbf\times}{\mathbf j}\cr}\right],
\label{E20}
\end{equation}
where
\begin{equation}
{\hat R}=[({\hat z}-{\hat z}_P)^2+{{\hat r}_P}^2+{\hat r}^2-2{\hat r}_P{\hat r}\cos(\varphi-\varphi_P)]^{1/2},
\label{E21}
\end{equation}
\begin{equation}
{\hat{\bf n}}=\{[{\hat r}_P-{\hat r}\cos(\varphi-\varphi_P)]{\hat{\bf e}}_{r_P}-{\hat r}\sin(\varphi-\varphi_P){\hat{\bf e}}_{\varphi_P}-({\hat z}-{\hat z}_P){\hat{\bf e}}_{z_P}\}/{\hat R},
\label{E22}
\end{equation}
the function $g({\hat r},\varphi,{\hat z};{\hat r}_P,\varphi_P,{\hat z}_P)$ is defined by
\begin{equation}
g\equiv\varphi-\varphi_P+{\hat R},
\label{E23}
\end{equation}
the variable $\phi$ in the argument of the delta function stands for
\begin{equation}
\phi\equiv{\hat\varphi}-{\hat\varphi}_P\qquad{\rm with}\qquad{\hat\varphi}_P\equiv\varphi_P-\omega t_P,
\label{E24}
\end{equation}
and $({\hat{\mathbf e}}_{r_P},{\hat{\mathbf e}}_{\varphi_P},{\hat{\mathbf e}}_{z_P})$ are the cylindrical base vectors at the observation point $(r_P,\varphi_P,z_P)$.   In~(\ref{E20}), the domain of integration over the $({\hat r},{\hat\varphi},{\hat z})$ space consists of the support ${\mathcal S}$ of the source density $j^\mu$ and the range of integration with respect to $\varphi$ is given by the extended interval of azimuthal angle traversed by the source in the course of its rotations prior to the observation time $t_P$.  

I have expressed the range of $\varphi$ integration as a sum of the intervals of length $2\pi$ that the element initially located at ${\hat\varphi}$ traverses during each of its individual rotations: $k$ is a positive integer enumerating successive rotation periods (the first rotation period being designated by $k=1$) and the summation extends over the set of rotations executed by the source over its lifetime.  Given $({\hat r},{\hat\varphi},{\hat z})$ and $(r_P,\varphi_P,z_P, t_P)$, there are a limited number of values of $k$ for which $g-\phi$ vanishes and so the integral in~(\ref{E20}) is non-zero.  In other words, the contribution received from the source point $({\hat r},{\hat\varphi},{\hat z})$ at the space-time observation point $(r_P,\varphi_P,z_P, t_P)$ is made during a limited number of its (earlier) rotation periods (see appendix~\ref{appB}). 

\subsection{The Green's function for the problem and its loci of singularities}
\label{subsec:Green's function}

To put the current density ${\mathbf j}=j_r{\hat{\mathbf e}}_r+j_\varphi{\hat{\mathbf e}}_\varphi+j_z{\hat{\mathbf e}}_z$ into a form suitable for performing the integration with respect to $\varphi$, we need to express the $\varphi$-dependent base vectors $({\hat{\mathbf e}}_r,{\hat{\bf e}}_\varphi,{\hat{\mathbf e}}_z)$ associated with the source point $(r,\varphi,z)$ in terms of the constant base vectors $({\hat{\mathbf e}}_{r_P},{\hat{\mathbf e}}_{\varphi_P},{\hat{\mathbf e}}_{z_P})$ at the observation point $(r_P,\varphi_P,z_P)$:
\begin{equation}
\left[\matrix{{\hat{\mathbf e}}_r\cr {\hat{\mathbf e}}_\varphi\cr {\hat{\mathbf e}}_z\cr}\right]=\left[\matrix{\cos(\varphi-\varphi_P)&\sin(\varphi-\varphi_P)&0\cr
-\sin(\varphi-\varphi_P)&\cos(\varphi-\varphi_P)&0\cr
0&0&1\cr}\right]\left[\matrix{{\hat{\bf e}}_{r_P}\cr {\hat{\bf e}}_{\varphi_P}\cr {\hat{\bf e}}_{z_P}\cr}\right].
\label{E25}
\end{equation}
Once the resulting expression,
\begin{equation}
{\bf j}=[j_r\cos(\varphi-\varphi_P)-j_\varphi\sin(\varphi-\varphi_P)]{\hat{\mathbf e}}_{r_P}+[j_r\sin(\varphi-\varphi_P)+j_\varphi\cos(\varphi-\varphi_P)]{\hat{\mathbf e}}_{\varphi_P}+j_z{\hat{\mathbf e}}_{z_P},
\label{E26}
\end{equation}
and the expression in (\ref{E22}) for ${\hat{\bf n}}$ are inserted in (\ref{E20}) and $\delta^\prime(g-\phi)$ is written as $- \partial\delta(g-\phi)/\partial{\hat\varphi}$ [see (\ref{E24})], we arrive at
\begin{equation}
\left[\matrix{{\mathbf E}\cr{\mathbf B}\cr}\right]=-\frac{1}{\omega}\sum_{n=1}^2\sum_{j=1}^3\int_{\mathcal S}{\hat r}{\textrm d}{\hat r}\,{\textrm d}{\hat\varphi}\,{\textrm d}{\hat z}\,\frac{\partial G_{nj}}{\partial{\hat\varphi}}\left[\matrix{{\bf u}_{nj}\cr{\bf v}_{nj}\cr}\right],
\label{E27}
\end{equation}
with
\begin{eqnarray}
    &\left[
     \begin{array}{c}
     {\bf u}_{11}\\
     {\bf u}_{12}\\
     {\bf u}_{13}
     \end{array} \right]= \left[
      \begin{array}{c}
     j_r{\hat{\bf e}}_{r_P}+j_\varphi{\hat{\bf e}}_{\varphi_P}\\
      -j_\varphi{\hat{\bf e}}_{r_P}+j_r{\hat{\bf e}}_{\varphi_P} \\
     j_z{\hat{\bf e}}_{z_P}
\end{array} \right],\nonumber\\
\label{E28}
\end{eqnarray}
\begin{eqnarray}
    &\left[
     \begin{array}{c}
     {\bf u}_{21}\\
     {\bf u}_{22}\\
     {\bf u}_{23}
     \end{array} \right]= \rho c\left[
      \begin{array}{c}
     {\hat r}{\hat{\bf e}}_{r_P}\\
      {\hat r}{\hat{\bf e}}_{\varphi_P} \\
     -{\hat r}_P{\hat{\bf e}}_{r_P}+({\hat z}-{\hat z}_P){\hat{\bf e}}_{z_P}
\end{array} \right],\nonumber\\
\label{E29}
\end{eqnarray}
\begin{equation}
\left[\matrix{{\bf v}_{11}&{\bf v}_{12}&{\bf v}_{13}}\right]=\left[\matrix{0&0&0}\right],
\label{E30}
\end{equation}
and 
\begin{eqnarray}
    & \left[
     \begin{array}{c}
     {\bf v}_{21}\\
     {\bf v}_{22}\\
     {\bf v}_{23}
     \end{array} \right]=\left[
      \begin{array}{c}
     -({\hat z}-{\hat z}_P){\bf u}_{12}+{\hat r}j_z{\hat{\bf e}}_{\varphi_P}+{\hat r}_P j_\varphi{\hat{\bf e}}_{z_P}\\  
    {\hat{\bf e}}_{z_P}\times{\bf v}_{21}+{\hat r}_P j_r{\hat{\bf e}}_{z_P} \\
     -{\hat r}_P j_z{\hat{\bf e}}_{\varphi_P}-{\hat r}j_\varphi{\hat{\bf e}}_{z_P}
\end{array} \right],\nonumber\\
\label{E31}
\end{eqnarray}
in which
\begin{equation}
\left[\matrix{G_{n1}\cr G_{n2}\cr G_{n3}\cr}\right]=\sum_{k=1}^\infty\int_{{\hat\varphi}+2(k-1)\pi}^{{\hat\varphi}+2k\pi} {\rm d}\varphi\,{\delta(g-\phi)\over {\hat R}^n}\left[\matrix{\cos(\varphi-\varphi_P)\cr \sin(\varphi-\varphi_P)\cr 1\cr}\right]
\label{E32}
\end{equation}
denotes the outcome of the remaining integration with respect to $\varphi$.  Note that the dependence on ${\hat\varphi}$ of the limits of integration in~(\ref{E32}) does not contribute toward the values of the derivatives of $G_{nj}$ with respect to ${\hat\varphi}$ [see~(\ref{B7})].

The function $G_{nj}({\hat r},{\hat\varphi},{\hat z};{\hat r}_P,{\hat\varphi}_P,{\hat z}_P)$ here acts as the Green's function for the present problem.  It describes the Li\'enard-Wiechert field that arises from an individual volume element of the rotating distribution pattern of the source.  If we specialize the current distribution to a rotating point charge $q$, i.e., let $j_r=j_z=0$ and $j_\varphi=r_s\omega q\delta(r-r_s)\delta({\hat\varphi})\delta(z)$ with a constant $r_s$, then (\ref{E27}) at an observation point in the far zone would describe the familiar field of synchrotron radiation when $r_s<c/\omega$ and a synergic field combining attributes of both synchrotron and \v Cerenkov emissions when $r_s>c/\omega$~\citep[see, e.g.,][]{ArdavanH:Speapc}.  

\begin{figure}
\centerline{\includegraphics[width=11cm]{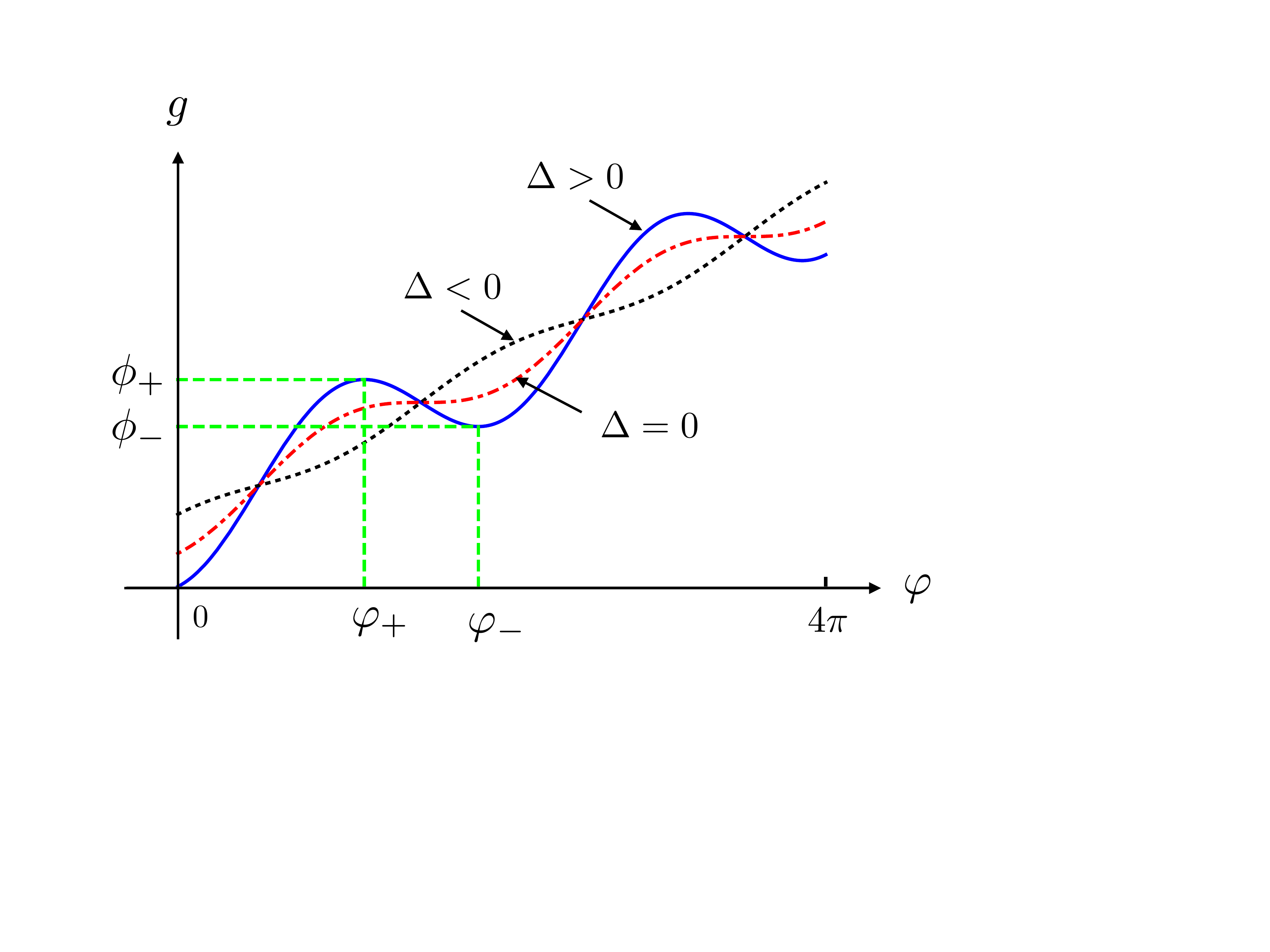}}
\caption{Generic forms of the function $g(\varphi)$ for source points whose $({\hat r},{\hat z})$ coordinates lie across the boundary $\Delta=0$ delineating the projection of the cusp curve of the bifurcation surface onto the $({\hat r},{\hat z})$ plane (see figure~\ref{F11}).  Depending on whether $\phi$ lies outside or inside the interval $(\phi_-,\phi_+)$, contributions are made toward the observed field [i.e., the argument $g(\varphi)-\phi$ of the Dirac delta function in (\ref{E20}) vanishes] at either one or three retarded positions of the source.  For a horizontal line $g=\phi$ that either approaches an extremum of $g(\varphi)$ from inside the interval $(\phi_-,\phi_+)$ or passes through an inflection point of $g(\varphi)$, two or all three of the retarded positions in question coalesce and so their contributions interfere constructively to form caustics.  This figure is for ${\hat r}=3$ and only shows two rotation periods.  At higher speeds, the difference between the values of $\phi_+$ and $\phi_-$ can be large enough for a horizontal line $g=\phi$ to intersect $g(\varphi)$ over more than one rotation period (see figure~\ref{F36}).  Contributions toward the observed field can thus arise, not only from one or three, but from any odd number of retarded positions of the source.  There are contributions from more than three retarded times whenever the rotation period of the source is shorter than the time taken by the collapsing sphere $\vert{\bf x}-{\bf x}_P\vert=c(t-t_P)$, centred on the observation point $P$, to cross the orbit of the source.}
\label{F4}
\end{figure}

\begin{figure}
\centerline{\includegraphics[width=8cm]{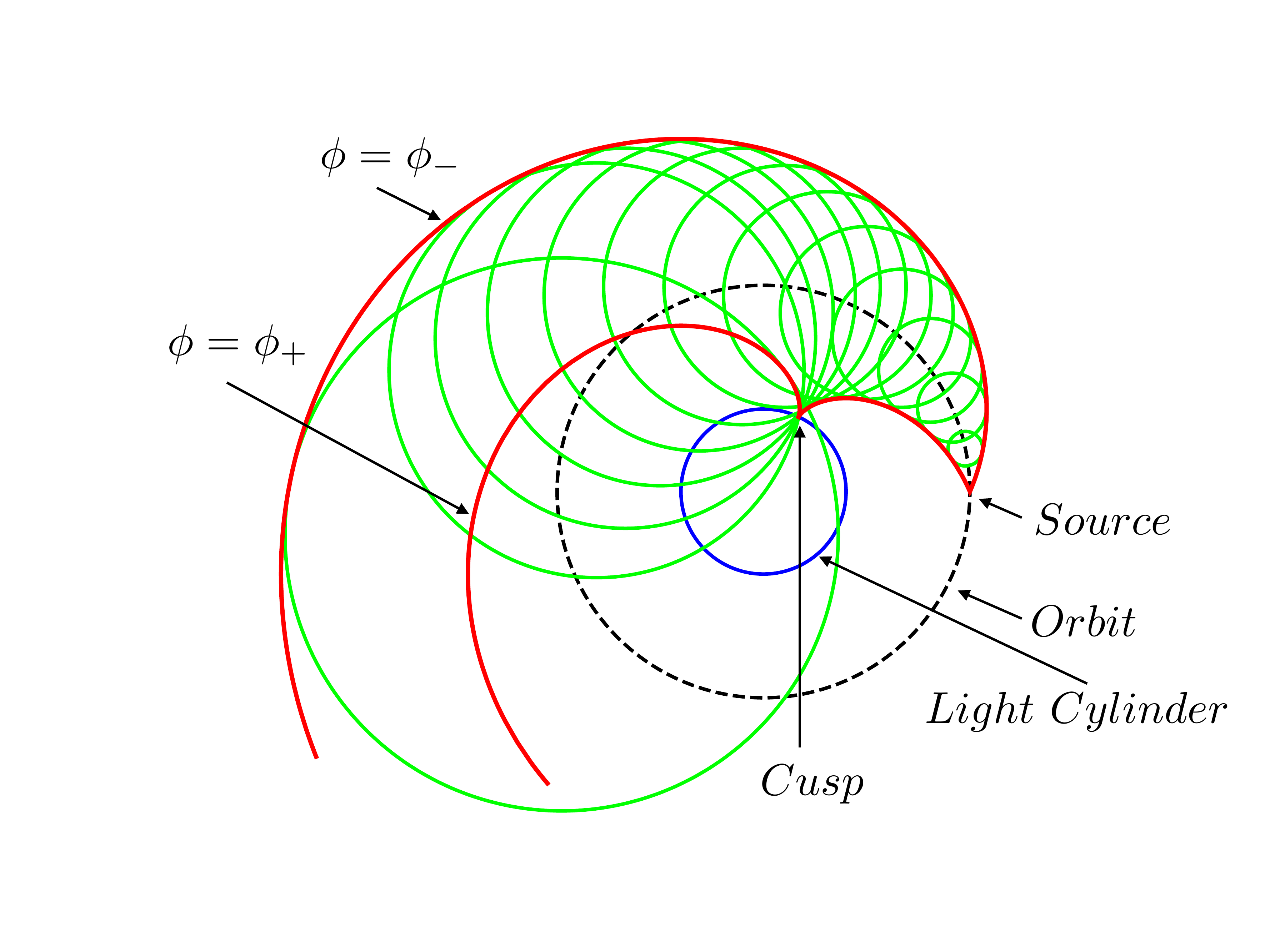}}
\caption{Cross sections with the plane ${\hat z}_P={\hat z}$ of the spherical wave fronts emanating from a rotating source point.  This source has an angular frequency of rotation, $\omega$, that is constant and a speed, $r\omega$, that exceeds the speed of light $c$ in vacuum.  The larger circle depicts the orbit of the source and the smaller circle the light cylinder $r=c/\omega$.  The heavier (red) curves show the intersection of the envelope of these wave fronts (see figure~\ref{F6}) with the plane of rotation.}
\label{F5}
\end{figure}

\begin{figure}
\centerline{\includegraphics[width=10cm]{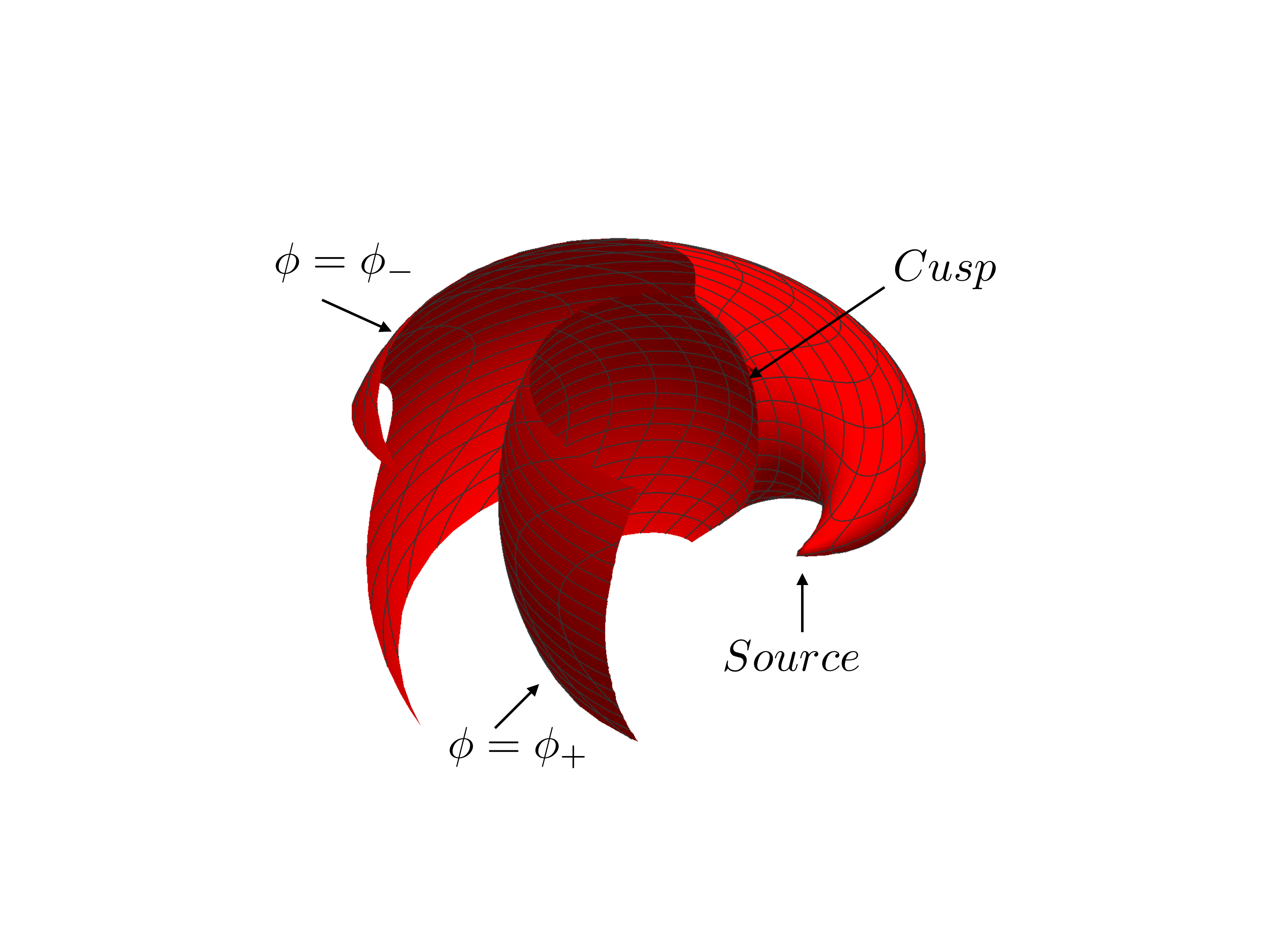}}
\caption{Three-dimensional view [in the space $({\hat r}_P,{\hat\varphi}_P,{\hat z}_P)$ of observation points] of the envelope of wave fronts emanating from the rotating source point $({\hat r},{\hat\varphi},{\hat z}$).  This envelope consists of two sheets that tangentially meet along a cusp (see figure~\ref{F7}).  The singular sheet, i.e., the sheet that issues from the source point with an initial conical shape, is that described by ${\hat\varphi}_P={\hat\varphi}-\phi_-({\hat r}_P,{\hat z}_P;{\hat r}, {\hat z})$.}
\label{F6}
\end{figure}

\begin{figure}
\centerline{\includegraphics[width=9cm]{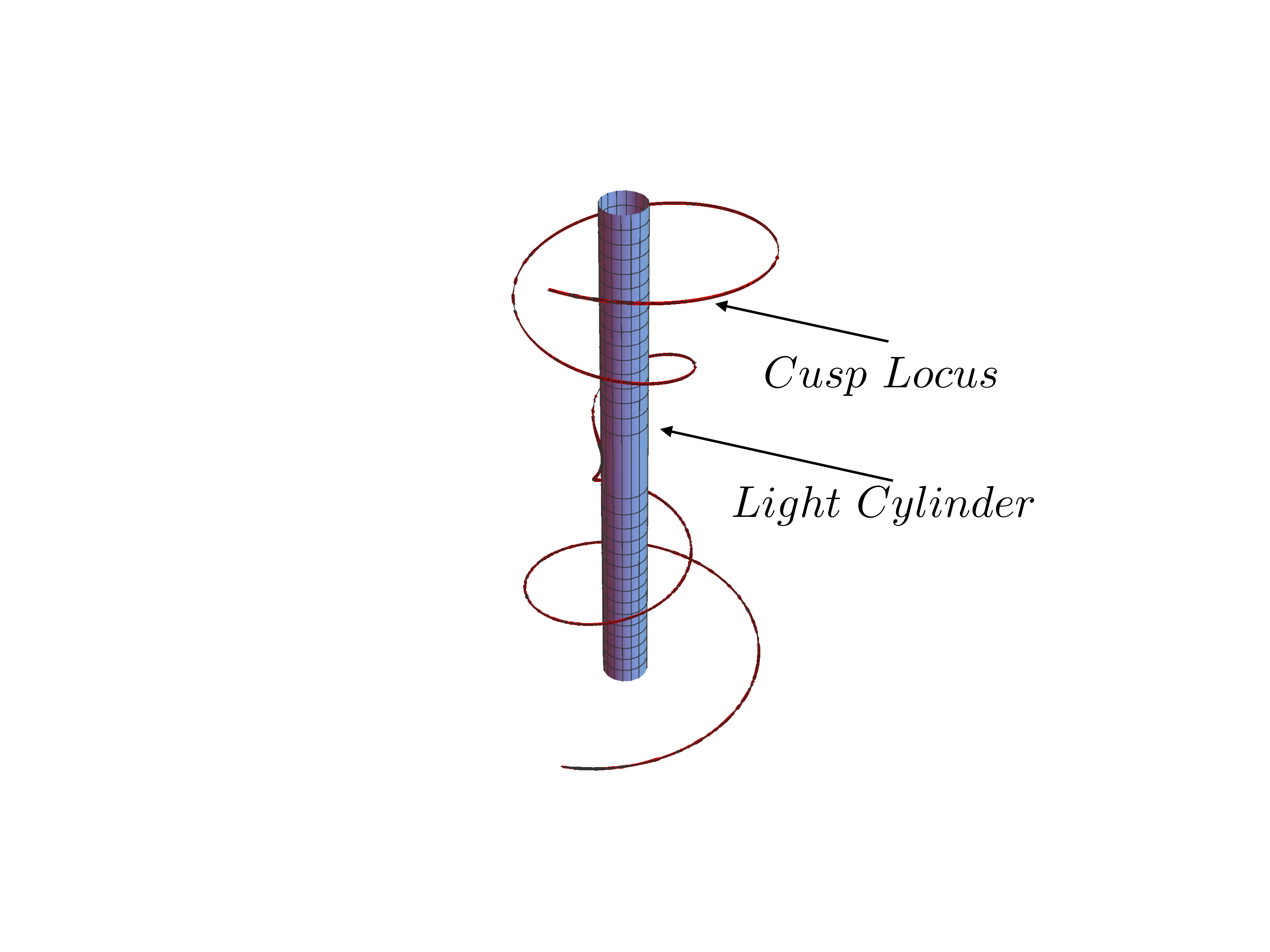}}
\caption{The cusp along which the two sheets of the envelope of wave fronts meet and are tangent to one another.  This cusp touches and is tangent to the light cylinder ${\hat r}_P=1$ on the plane ${\hat z}_P={\hat z}$ and spirals outward into the far field on the hyperbolic surface of revolution $\Delta({\hat r}_P,{\hat z}_P;{\hat r}, {\hat z})=0$ (see figure~\ref{F12}).}
\label{F7}
\end{figure}

Depending on the value of 
\begin{equation}
\Delta=({{\hat r}_P}^2-1)({\hat r}^2-1)-({\hat z}-{\hat z}_P)^2
\label{E33}
\end{equation}
for a given source point $(r,{\hat\varphi},z)$ with $r\omega>c$, the $\varphi$-dependence of the function $g$ that appears in the definition of the Green's function $G_{nj}$ in~(\ref{E32}) has one of the generic forms shown in figure~\ref{F4}.  As can be seen from the curve labelled $\Delta>0$ in this figure, there are values, 
\begin{equation}
\varphi_\pm=\varphi_P+2k\pi-\arccos\left(\frac{1\mp\Delta^{1/2}}{{\hat r}{\hat r}_P}\right),
\label{E34}
\end{equation}
of the retarded position of the source point at which
\begin{equation}
\frac{\partial g}{\partial\varphi}=1+\frac{{\hat r}{\hat r}_P\sin(\varphi-\varphi_P)}{{\hat R}}
\label{E35}
\end{equation}
vanishes and so $G_{nj}$ diverges.  These turning points of $g$ occur at source points for which $\partial(R\vert_{\varphi={\hat\varphi}+\omega t})/\partial t=-c$, i.e., the source points that approach the observer, along the radiation direction ${\hat{\bf n}}$, with the speed of light at the retarded time.  The inflection point of $g$ (see the curve labelled $\Delta=0$ in figure~\ref{F4}), at which
\begin{equation}
\frac{\partial^2g}{\partial\varphi^2}\Big\vert_{\varphi=\varphi_\pm}=\mp\frac{\Delta^{1/2}}{{\hat R}_\pm}
\label{E36}
\end{equation}
in addition vanishes, occurs at source points that approach the observer not only with the wave speed but also with zero acceleration at the retarded time, i.e., for which both $\partial(R\vert_{\varphi={\hat\varphi}+\omega t})/\partial t=-c$ and $\partial^2(R\vert_{\varphi={\hat\varphi}+\omega t})/\partial t^2=0$ at the time when $g\vert_{\varphi={\hat\varphi}+\omega t}=\phi$ and $\partial g/\partial\varphi=\partial^2 g/\partial\varphi^2=0$.  In (\ref{E36}),
\begin{equation}
{\hat R}_\pm= [({\hat z}-{\hat z}_P)^2+{\hat r}^2+{{\hat r}_P}^2-2(1\mp\Delta^{1/2})]^{1/2}
\label{E37}
\end{equation}
is the value of ${\hat R}$ at the extrema $\varphi_\pm$ of $g$.

The envelope of the wave fronts emanating from a given rotating source element $({\hat r},{\hat\varphi},{\hat z})$, on which $\partial g/\partial\varphi$ vanishes, consists of the rigidly rotating two-sheeted surface ${\hat\varphi}-{\hat\varphi}_P=g(\varphi_\pm)$ in the space $({\hat r}_P,{\hat\varphi}_P,{\hat z}_P)$ of observation points. This surface, which is shown in figures.~\ref{F5} and \ref{F6}, is described by
\begin{equation}
\phi_\pm\equiv{\hat\varphi}_\pm-{\hat\varphi}_P=\varphi_\pm-\varphi_P+{\hat R}_\pm
\label{E38}
\end{equation}
[see~(\ref{E23}), (\ref{E24}), (\ref{E34}) and (\ref{E37})].  The two sheets of this surface tangentially meet along a cusp on which $\partial^2g/\partial\varphi^2$ as well as $\partial g/\partial\varphi$ vanishes (see figures~\ref{F6} and \ref{F7}).  Three distinct wave fronts, emitted at three differing values of the retarded time, pass through any given observation point inside the envelope.  At an observation point located on the envelope or its cusp, respectively two or all three of these waves coalesce and interfere constructively (see figure~\ref{F4}).

\subsection{Bifurcation surface of an observation point}
\label{subsec:bifurcation}

\begin{figure}
\centerline{\includegraphics[width=8cm]{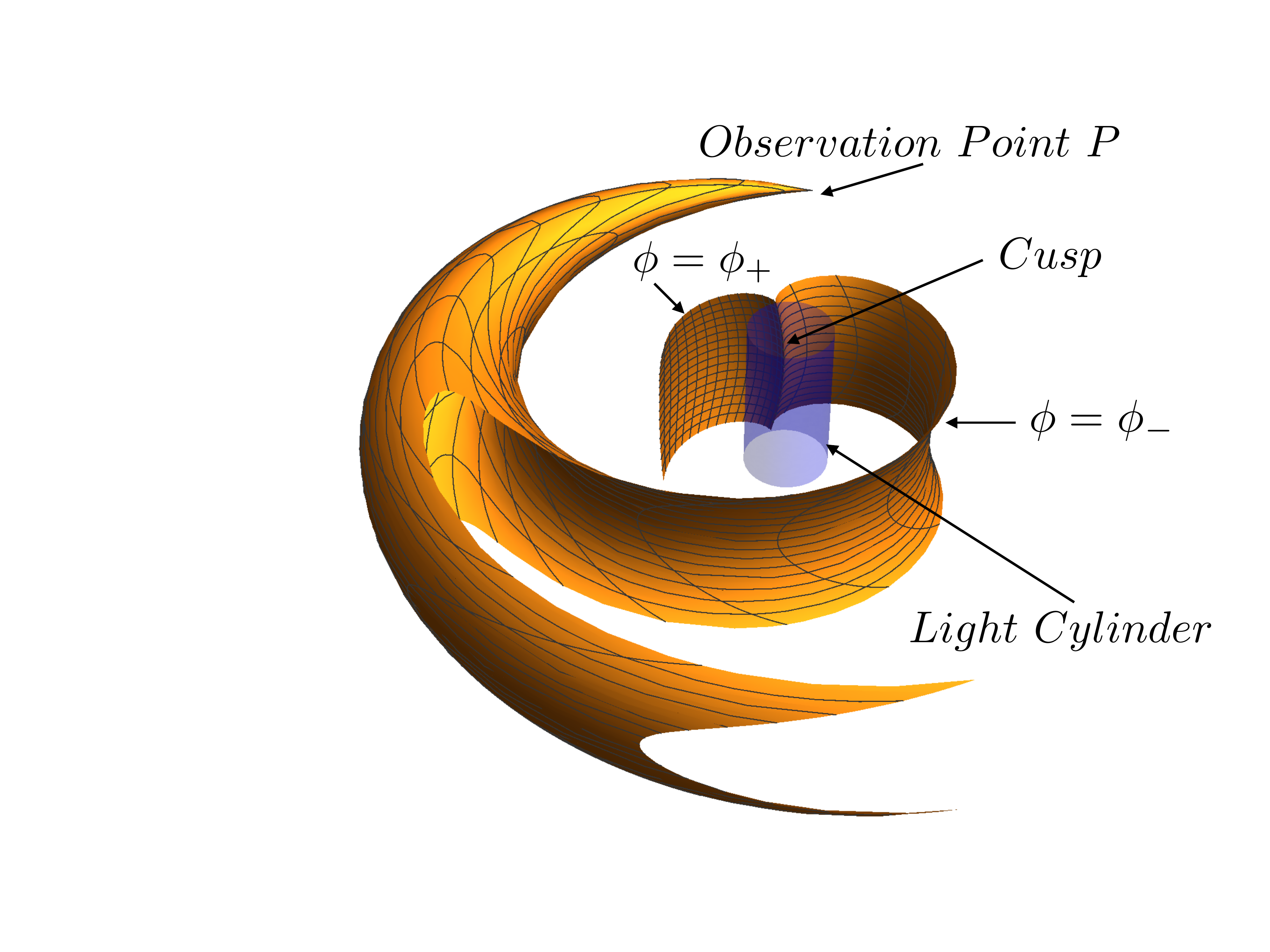}}
\caption{The two sheets $\phi=\phi_\pm$ of the bifurcation surface issuing from the observation point $P$, the cusp $C$ of this surface and the light cylinder ${\hat r}=1$.  In contrast to the envelope of wave fronts which resides in the space of observation points, the surface shown here resides in the space $(r,{\hat\varphi},z)$ of source points: it is the locus of source points that approach $P$, along the radiation direction, with the speed of light at the retarded time.  The two sheets of this surface meet along a cusp that tangentially touches the light cylinder at ${\hat z}={\hat z}_P$ and moves outward spiralling around the rotation axis on the hyperbolic surface of revolution $\Delta({\hat r},{\hat z};{\hat r}_P,{\hat z}_P)=0$ (see figure~\ref{F11}).  The source points on this cusp approach the observer along the radiation direction not only with the speed of light but also with zero acceleration at the retarded time.  The source would normally be distributed over a finite volume close to the light cylinder.  If the position of the observation point is such that the cusp shown here intersects the source distribution, there will be wave fronts with differing emission times that are received simultaneously: while the source points outside the bifurcation surface make their contributions toward the value of the observed field at a single instant of retarded time, the source points inside this surface make their contributions at $3$ (or $5,7,\cdots$) distinct instants of retarded time.}
\label{F8}
\end{figure}

Reciprocally, the locus in the space of source points $({\hat r},{\hat\varphi},{\hat z})$ on which $\partial g/\partial\varphi$ vanishes is a two-sheeted cusped surface issuing from the fixed observation point $P$ (see figure~\ref{F8}).  I refer to this locus, which is described by (\ref{E38}) for fixed values of $({\hat r}_P,{\hat\varphi}_P,{\hat z}_P)$ rather than fixed values of $({\hat r},{\hat\varphi},{\hat z})$, as the \emph{bifurcation surface} of the observation point $P$.  The two sheets $\phi=\phi_+$ and $\phi=\phi_-$ of this surface, respectively referred to as the regular and singular sheets, meet along the following cusp:
\begin{eqnarray}
    &C: \left\{
      \begin{array}{c}
     {\hat r}={\hat r}_C({\hat z})=[1+({\hat z}-{\hat z}_P)^2/({{\hat r}_P}^2-1)]^{1/2},\\
     \varphi=\varphi_C({\hat z})=\varphi_P+2k\pi-\arccos[1/({\hat r}{\hat r}_P)],
\end{array} \right.
\label{E39}
\end{eqnarray}
where $k$ is the same integer as that appearing in (\ref{E20}).   I refer to both $C$ and its projection onto the $(r,z)$ plane as the \emph{cusp locus of the bifurcation surface}; whether it is $C$ itself or its projection that is referred to will be clear from the context.
 
The source points inside the bifurcation surface, close to its cusp, make their contributions toward the observed value of the field at three distinct retarded positions in their trajectory (where a horizontal line $g=\phi$ in figure~\ref{F4} intersects the curve $\Delta>0$ between its extrema), while those outside the bifurcation surface make their contributions at a single retarded position (where the curve $\Delta<0$ is intersected by $g=\phi$ in figure~\ref{F4}).  For the source points on the bifurcation surface (i.e., those for which $g=\phi_\pm$ in figure~\ref{F4}), two of the contributing retarded positions coalesce at the extrema of the curve $\Delta>0$ in figure~\ref{F4} giving rise to a divergent value of the Green's function at $P$.  For the source points located on the cusp locus $C$ of the bifurcation surface (i.e., those for which $\Delta=0$ in figure~\ref{F4}), all three of the contributing retarded positions coalesce at the inflection point of the curve $\Delta=0$ in figure~\ref{F4} giving rise to a higher-order singularity in $G_{nj}$.  In the following section, I use the time-domain version~\citep{BurridgeR:Asyeir} of the method of~\citet{ChesterC:Extstd} to derive a uniform asymptotic approximation to the value of $G_{nj}$ for the source points close to the cusp $C$ of the bifurcation surface. 

\subsection{A uniform asymptotic approximation to the value of the Green's function near the cusp locus of the bifurcation surface}
\label{subsec:Expansion}

As long as the observation point does not coincide with the source point, the function $g(\varphi)$ is analytic and the following transformation of the integration variable in (\ref{E32}) from $\varphi$ to $\nu$ is permissible
\begin{equation}
g(\varphi)=\textstyle{\frac{1}{3}}\nu^3-{c_1}^2\nu+c_2,
\label{E40}
\end{equation}
in which 
\begin{equation}
c_1=[\textstyle{\frac{3}{4}}(\phi_+-\phi_-)]^{1/3}\quad\textrm{and}\quad c_2=\textstyle{\frac{1}{2}}(\phi_++\phi_-),
\label{E41}
\end{equation}
are chosen such that the values of the two functions on opposite sides of (\ref{E40}) coincide at their extrema when $\Delta$ is positive.  Thus an alternative exact expression for $G_{nj}$ is
\begin{equation}
G_{nj}=\sum_{k=1}^\infty {\mathcal H}\int_{-\infty}^\infty{\textrm d}\nu\,f_{nj}\delta(\textstyle{\frac{1}{3}}\nu^3-{c_1}^2\nu+c_2-\phi),
\label{E42}
\end{equation}
where
\begin{equation}
\left[\matrix{f_{n1}\cr f_{n2}\cr f_{n3}\cr}\right]=\frac{1}{{\hat R}^n}\frac{{\textrm d}\varphi}{{\textrm d}\nu}\left[\matrix{\cos(\varphi-\varphi_P)\cr \sin(\varphi-\varphi_P)\cr 1\cr}\right],
\label{E43}
\end{equation}
and the step function ${\mathcal H}$ is non-zero only if the argument of the delta function in (\ref{E32}) vanishes within the original domain of integration ${\hat\varphi}+2(k-1)\pi\le\varphi\le{\hat\varphi}+2k\pi$, i.e., if $g\vert_{\varphi={\hat\varphi}+2k\pi}-\phi\ge0$ but $g\vert_{\varphi={\hat\varphi}+2(k-1)\pi}-\phi\le0$.  Inserting $\varphi={\hat\varphi}+2k\pi$ and $\varphi={\hat\varphi}+2(k-1)\pi$ in the definition of $g$ in (\ref{E23}) and simplifying the resulting expressions by means of (\ref{E24}), we can write this step function as
\begin{equation}
{\mathcal H}={\rm H}\left[{\hat R}\vert_{\varphi={\hat\varphi}}-\omega t_P+2k\pi\right]-{\rm H}\left[{\hat R}\vert_{\varphi={\hat\varphi}}-\omega t_P+2(k-1)\pi\right],
\label{E44}
\end{equation}
in which ${\rm H}(x)$ denotes the Heaviside step function.  

In cases where the distance ${\hat R}_P=({\hat r}_P^2+{\hat z}_P^2)^{1/2}$ of the observation point from the origin of the reference frame is much larger than the coordinates ${\hat r}$ and ${\hat z}$ of the source point, (\ref{E44}) reduces to    
\begin{equation}
{\mathcal H}_\infty={\rm H}[{\hat R}_P-\omega t_P+2k\pi]-{\rm H}[{\hat R}_P-\omega t_P+2(k-1)\pi],\qquad{\hat R}_P\gg1.
\label{E45}
\end{equation}
The step function ${\cal H}$ picks out the particular rotation cycle (or cycles) during which the signal that reaches the observation point $({\hat r}_P,\varphi_P,{\hat z}_P)$ at the observation time $t_P$ is emitted by the source point $({\hat r},{\hat\varphi},{\hat z})$.  

Note that $c_1({\hat r},{\hat z};{\hat r}_P,{\hat z}_P)$ in the expression for $G_{nj}$ vanishes on the cusp locus of the bifurcation surface where $\Delta$ equals zero and $\phi_-=\phi_+$.  The leading term in the asymptotic expansion of the integral in (\ref{E42}) in the vicinity of the cusp locus of the bifurcation surface, i.e., for small $c_1$, can be found by replacing $f_j$ by $p_j+q_j\nu$,
\begin{equation}
G_{nj}\simeq\sum_{k=1}^\infty {\mathcal H}\int_{-\infty}^\infty{\textrm d}\nu\,(p_{nj}+q_{nj}\nu)\delta(\textstyle{\frac{1}{3}}\nu^3-{c_1}^2\nu+c_2-\phi),\qquad c_1\ll 1,
\label{E46}
\end{equation}
where 
\begin{equation}
p_{nj}=\textstyle{\frac{1}{2}}(f_{nj}\vert_{\varphi=\varphi_-}+f_{nj}\vert_{\varphi=\varphi_+}),
\label{E47}
\end{equation}
\begin{equation}
 q_{nj}=\textstyle{\frac{1}{2}}{c_1}^{-1}(f_{nj}\vert_{\varphi=\varphi_-}-f_{nj}\vert_{\varphi=\varphi_+})
\label{E48}
\end{equation}
\cite[see][and note that $\varphi=\varphi_-$ maps onto $\nu=c_1$ and $\varphi=\varphi_+$ onto $\nu=-c_1$]{ChesterC:Extstd}.  To evaluate the integral in (\ref{E46}) we need to know the roots of the cubic function that appears in the argument of the Dirac $\delta$ function in this expression.  Depending on whether the source point is located inside or outside the bifurcation surface, the roots of 
\begin{equation}
\textstyle{\frac{1}{3}}\nu^3-{c_1}^2\nu+c_2-\phi=0
\label{E49}
\end{equation}
for $\Delta>0$ are given, respectively, by 
\begin{equation}
\nu=\nu_\ell=2c_1\cos\left(\textstyle{\frac{2}{3}}\ell\pi+\textstyle{\frac{1}{3}}\arccos\chi\right),\quad\vert\chi\vert<1,
\label{E50}
\end{equation}
with $\ell=0,1,$ and $2$ , or by
\begin{equation}
\nu=\nu_{\rm out}=2c_1{\textrm{sgn}}(\chi)\cosh\left(\textstyle{\frac{1}{3}}{\textrm{arccosh}}\vert\chi\vert\right),\quad\vert\chi\vert>1,
\label{E51}
\end{equation}
where 
\begin{equation}
\chi=\frac{3(\phi-c_2)}{2{c_1}^3}.
\label{E52}
\end{equation}
Note that $\chi$ equals $1$ on the sheet $\phi_+$ of the bifurcation surface and $-1$ on the sheet $\phi_-$.

The integral multiplying $p_j$ in (\ref{E46}) therefore has the following value when the source point lies inside the bifurcation surface ($\vert\chi\vert<1)$ and (\ref{E49}) has the three roots given in (\ref{E50}),
\begin{eqnarray}
\int_{-\infty}^\infty{\textrm d}\nu\delta(\textstyle{\frac{1}{3}}\nu^3-{c_1}^2\nu+c_2-\phi)&=&\sum_{\ell=0}^2\vert\nu_\ell^2-c_1^2\vert^{-1}\nonumber\\*
&=&\sum_{\ell=0}^2  c_1^{-2}\left\vert 4\cos^2\left(\textstyle{\frac{2}{3}}\ell\pi+\textstyle{\frac{1}{3}}\arccos\chi\right)-1\right\vert^{-1},\quad\vert\chi\vert<1.\nonumber\\*
\label{E53}
\end{eqnarray}
Using the trigonometric identity $4\cos^2\alpha-1=\sin 3\alpha/\sin\alpha$, we can write this as
\begin{eqnarray}
\int_{-\infty}^\infty{\textrm d}\nu\delta(\textstyle{\frac{1}{3}}\nu^3-{c_1}^2\nu+c_2-\phi)
&=& c_1^{-2}(1-\chi^2)^{-1/2}\sum_{\ell=0}^2\left\vert\sin\left(\textstyle{\frac{2}{3}}\ell\pi+\textstyle{\frac{1}{3}}\arccos\chi\right)\right\vert\nonumber\\
&=&2 c_1^{-2}(1-\chi^2)^{-1/2}\cos\left({\textstyle\frac{1}{3}}\arcsin\chi\right),\quad\vert\chi\vert<1,\quad
\label{E54}
\end{eqnarray}
in which I have evaluated the sum by adding the sine functions two at a time.  When the source point lies outside the bifurcation surface ($\vert\chi\vert>1$), the above integral receives a contribution only from the single value of $\nu$ given in (\ref{E51}) and we obtain
\begin{equation}
\int_{-\infty}^\infty{\textrm d}\nu\delta(\textstyle{\frac{1}{3}}\nu^3-{c_1}^2\nu+c_2-\phi)= c_1^{-2}(\chi^2-1)^{-1/2}\sinh\left(\textstyle{\frac{1}{3}}{\textrm{arccosh}}\vert\chi\vert\right),\quad\vert\chi\vert>1,\quad
\label{E55}
\end{equation}
where this time I have used the identity $4\cosh^2\alpha-1=\sinh(3\alpha)/\sinh\alpha$.

The second part of the integral in (\ref{E46}) can be evaluated in exactly the same way.  It has the value
\begin{equation}
\int_{-\infty}^\infty{\textrm d}\nu\,\nu\delta(\textstyle{\frac{1}{3}}\nu^3-{c_1}^2\nu+c_2-\phi)=-2 c_1^{-1}(1-\chi^2)^{-1/2}\sin\left(\textstyle{\frac{2}{3}}\arcsin\chi\right),\quad\vert\chi\vert<1,\quad
\label{E56}
\end{equation}
when the source point lies inside the bifurcation surface ($\vert\chi\vert<1$) and the value
\begin{equation}
\int_{-\infty}^\infty{\textrm d}\nu\,\nu\delta(\textstyle{\frac{1}{3}}\nu^3-{c_1}^2\nu+c_2-\phi)= c_1^{-1}(\chi^2-1)^{-1/2}{\textrm{sgn}}(\chi)\sinh\left(\textstyle{\frac{2}{3}}{\textrm{arccosh}}\vert\chi\vert\right),\,\,\vert\chi\vert>1,
\label{E57}
\end{equation}
when the source point lies outside the bifurcation surface ($\vert\chi\vert>1$).  Inserting (\ref{E54})--(\ref{E57}) in (\ref{E46}), we obtain
\begin{equation}
G_{nj}^{\rm{in}}\simeq\sum_{k=1}^\infty 2{\mathcal H} c_1^{-2}(1-\chi^2)^{-1/2}[p_{nj}\cos(\textstyle{\frac{1}{3}}\arcsin\chi)-c_1q_{nj}\sin(\textstyle{\frac{2}{3}}\arcsin\chi)],\quad\vert\chi\vert<1,\quad
\label{E58}
\end{equation}
and
\begin{eqnarray}
G_{nj}^{\rm{out}}\simeq\sum_{k=1}^\infty {\mathcal H} c_1^{-2}(\chi^2-1)^{-1/2}[p_{nj}\sinh(\textstyle{\frac{1}{3}}{\textrm{arccosh}}\vert\chi\vert)+c_1q_{nj}{\textrm{sgn}}(\chi)\sinh(\textstyle{\frac{2}{3}}{\textrm{arccosh}}\vert\chi\vert)],\nonumber\\*
\vert\chi\vert>1,\qquad\qquad
\label{E59}
\end{eqnarray}
where $G_{nj}^{\rm{in}}$ and $G_{nj}^{\rm{out}}$ denote the values of $G_{nj}$ over $\Delta\geq0$ inside and outside the bifurcation surface, respectively.  

For the source points in $\Delta<0$, the functions $\phi_-$ and $\phi_+$ are complex conjugate of one another so that the coefficient $c_2$ is still real but $c_1$ is pure imaginary: the relevant cube root of $\phi_+-\phi_-$ in the expression for $c_1$ is in this case given by $-{\rm i}\vert\phi_+-\phi_-\vert^{1/3}$, which casts the first member of (\ref{E41}) into the form $c_1=-{\rm i}[{\textstyle\frac{3}{4}}\vert\phi_+-\phi_-\vert]^{1/3}$.  Since neither $g(\varphi)$ nor the cubic expression to which $g$ is transformed have any extrema in this case, there is only one real solution to (\ref{E49}) when $\Delta<0$ and $c_1$ is pure imaginary.  This solution, which can be found by writing the coefficient ${c_1}^2$ of $\nu$ (in which $c_1$ is pure imaginary) as $-\vert c_1\vert^2$ prior to solving the cubic, is given by
\begin{equation}
\nu=2 c_1\sinh\left(\textstyle{\frac{1}{3}}{\textrm{arcsinh}}\chi^\prime\right),
\label{E60}
\end{equation}
with 
\begin{equation}
\chi^\prime=\frac{3(\phi-c_2)}{2\vert c_1\vert^3}.
\label{E61}
\end{equation}
Following the same procedure as that employed in deriving (\ref{E55}) and (\ref{E57}), we obtain
\begin{equation}
G_{nj}^{\rm{sub}}\simeq\sum_{k=1}^\infty {\mathcal H} c_1^{-2}({\chi^\prime}^2+1)^{-1/2}[p_{nj}\cosh(\textstyle{\frac{1}{3}}{\textrm{arcsinh}}\chi^\prime)+\vert c_1\vert q_{nj}\sinh(\textstyle{\frac{2}{3}}{\textrm{arcsinh}}\chi^\prime)],\quad\vert\chi^\prime\vert>1,
\label{E62}
\end{equation}
for the value $G_{nj}^{\rm sub}$ of the Green's function in $\Delta<0$ where the source points approach the observer with subluminal speeds.

To complete the derivation of $G_{nj}$, we need to evaluate the coefficients $p_j$ and $q_j$ which are defined by (\ref{E47}), (\ref{E48}) and (\ref{E43}).   The indeterminate quantities ${\textrm d}\varphi/{\textrm d}\nu\vert_{\varphi=\varphi_\pm}$ that appear in these definitions have to be found by repeated differentiation of (\ref{E40}) with respect to $\nu$, and the evaluation of the resulting relations 
\begin{equation}
\frac{\partial g}{\partial\varphi}\frac{{\textrm d}\varphi}{{\textrm d}\nu}=\nu^2-{c_1}^2,
\label{E63}
\end{equation}
and
\begin{equation}
\frac{\partial^2g}{\partial\varphi^2}\left(\frac{{\textrm d}\varphi}{{\textrm d}\nu}\right)^2+\frac{\partial g}{\partial\varphi}\frac{{\textrm d}^2\varphi}{{\textrm d}\nu^2}=2\nu,
\label{E64}
\end{equation}
at $\varphi=\varphi_\pm$.  This procedure, which amounts to applying the l'H\^opital rule, yields
\begin{equation}
\frac{{\textrm d}\varphi}{{\textrm d}\nu}\Big\vert_{\varphi=\varphi_\pm}=\left(\frac{2c_1{\hat R}_\pm}{\Delta^{1/2}}\right)^{1/2}.
\label{E65}
\end{equation}
Equation (\ref{E65}) together with (\ref{E43}), (\ref{E34}) and (\ref{E37}), now yield the following values of
\begin{equation}
\left[\matrix{f_{n1}\cr f_{n2}\cr f_{n3}\cr}\right]_{\varphi=\varphi_\pm}=\frac{1}{{\hat r}{\hat r}_P{\hat R}_\pm^{n-1/2}}\left(\frac{2c_1}{\Delta^{1/2}}\right)^{1/2}\left[\matrix{1\mp\Delta^{1/2}\cr -{\hat R}_\pm\cr {\hat r}{\hat r}_P\cr}
\right],\nonumber\\
\label{E66}
\end{equation}
and hence the following values of 
\begin{equation}
\left[\matrix{p_{n1}\cr p_{n2}\cr p_{n3}\cr}\right]=\frac{1}{{\hat r}{\hat r}_P}\left(\frac{c_1}{2\Delta^{1/2}}\right)^{1/2}
\left[\matrix{{\hat R}_+^{-n+{\textstyle\frac{1}{2}}}+{\hat R}_-^{-n+{\textstyle\frac{1}{2}}}+\Delta^{1/2}({\hat R}_-^{-n+{\textstyle\frac{1}{2}}}-{\hat R}_+^{-n+{\textstyle\frac{1}{2}}})\cr -({\hat R}_-^{-n+{\textstyle\frac{3}{2}}}+{\hat R}_+^{-n+{\textstyle\frac{3}{2}}})\cr {\hat r}{\hat r}_P({\hat R}_-^{-n+{\textstyle\frac{1}{2}}}+{\hat R}_+^{-n+{\textstyle\frac{1}{2}}})\cr}
\right],
\label{E67}
\end{equation}
and
\begin{equation}
\left[\matrix{q_{n1}\cr q_{n2}\cr q_{n3}\cr}\right]=\frac{1}{{\hat r}{\hat r}_P(2c_1\Delta^{1/2})^{1/2}}
\left[\matrix{{\hat R}_-^{-n+{\textstyle\frac{1}{2}}}-{\hat R}_+^{-n+{\textstyle\frac{1}{2}}}+\Delta^{1/2}({\hat R}_-^{-n+{\textstyle\frac{1}{2}}}+{\hat R}_+^{-n+{\textstyle\frac{1}{2}}})\cr {\hat R}_+^{-n+{\textstyle\frac{3}{2}}}-{\hat R}_-^{-n+{\textstyle\frac{3}{2}}}\cr {\hat r}{\hat r}_P({\hat R}_-^{-n+{\textstyle\frac{1}{2}}}-{\hat R}_+^{-n+{\textstyle\frac{1}{2}}})\cr}
\right]
\label{E68}
\end{equation}
[see (\ref{E47}) and (\ref{E48})].  For $\Delta<0$, the functions $c_1$ and $\Delta^{1/2}$ in (\ref{E67}) and (\ref{E68}) are respectively given by $-{\rm i}\vert\phi_+-\phi_-\vert^{1/3}$ and $-{\rm i}\vert\Delta\vert^{1/2}$, so that $p_j$ and $q_j$ are real also in this case: ${\hat R}_-$ is the complex conjugate of ${\hat R}_+$ [see (\ref{E37})].  

\begin{figure}
\centerline{\includegraphics[width=7cm]{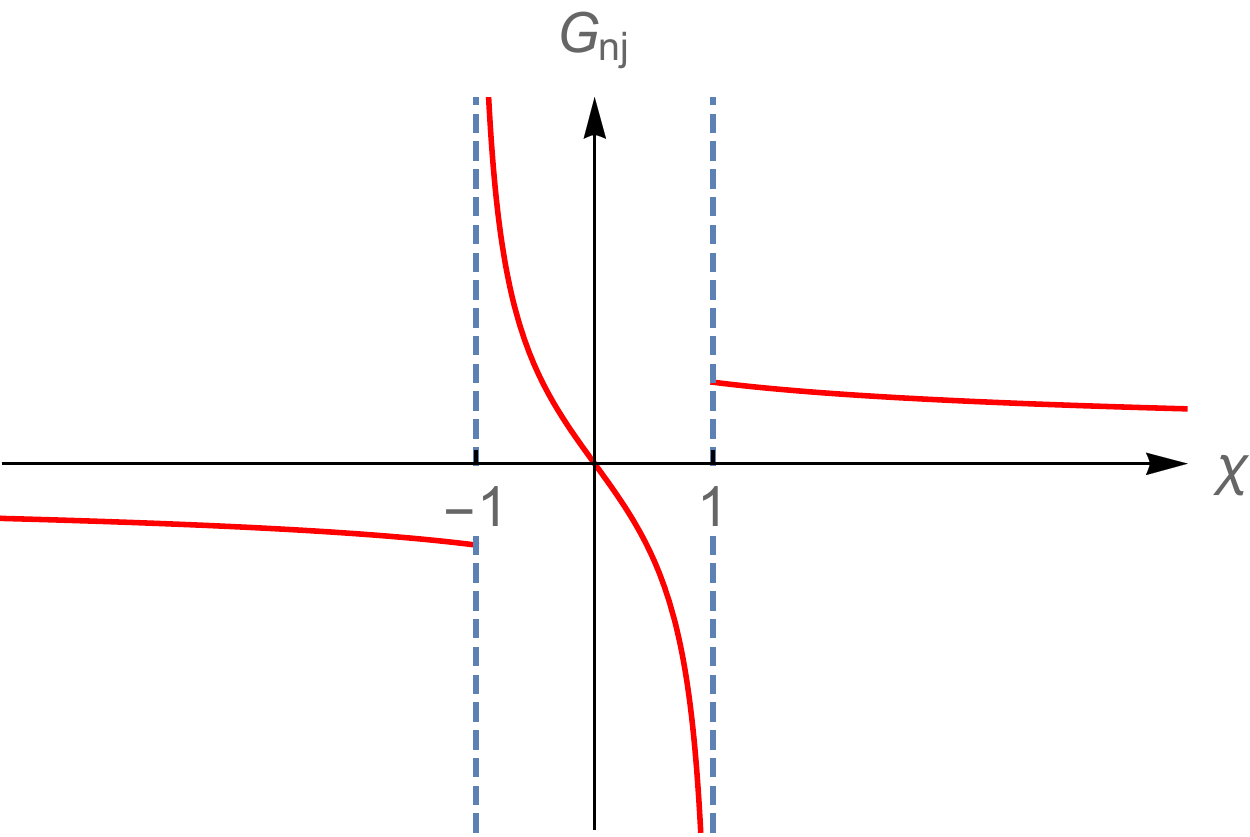}}
\caption{Dependence of the Green's function $G_{nj}$ on $\chi$ in cases where $q_{nj}$ is positive and appreciably greater than $\vert p_{nj}/c_1\vert$.  The two sheets $\phi_+$ and $\phi_-$ of the bifurcation surface map onto the distinct values $\chi=1$ and $\chi=-1$ of $\chi$, respectively, even at the cusp locus of the bifurcation surface where the separation $\phi_+-\phi_-$ of these two sheets vanishes.  The Green's function thus diverges only for source points inside the bifurcation surface whose retarded positions coalesce when they approach this surface or its cusp from $\vert\chi\vert<1$.}
\label{F9}
\end{figure}

\begin{figure}
\centerline{\includegraphics[width=7cm]{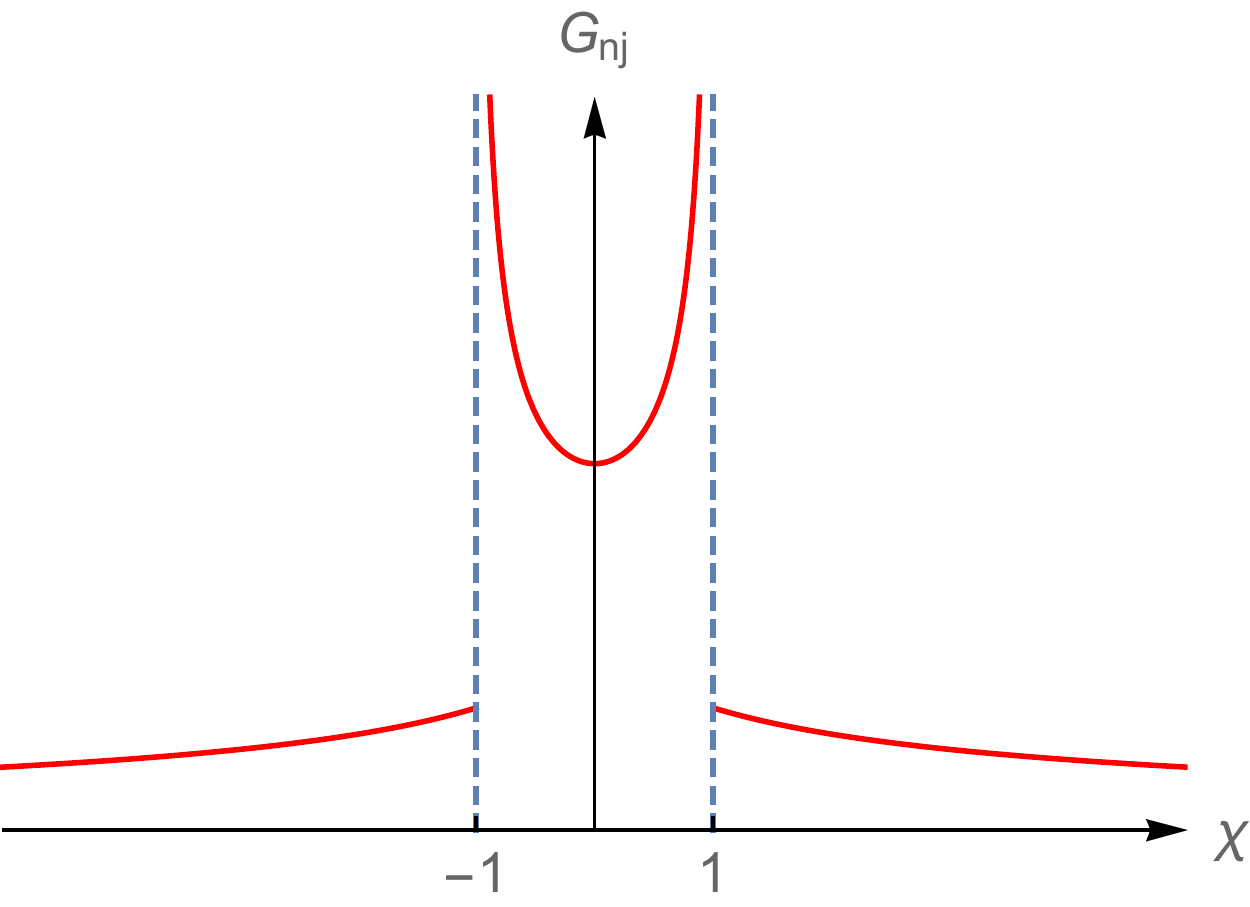}}
\caption{Dependence of the Green's function $G_{nj}$ on $\chi$ in cases where $p_{nj}$ is positive and appreciably greater than $\vert c_1q_{nj}\vert$ (see also figure~\ref{F9}).}
\label{F10}
\end{figure}

The two-dimensional loci $\chi=\pm1$ across which the resulting expression
\begin{equation}
G_{nj}=\left\{\begin{array}{lll}
G_{nj}^{\rm{in}}  &        \Delta>0, \,\,\vert\chi\vert<1\\
G_{nj}^{\rm{out}}  &      \Delta\geq0, \,\,\vert\chi\vert\geq1\\
G_{nj}^{\rm{sub}}  &       \Delta<0, \,\,\vert\chi^\prime\vert>1
\end{array}
\right.
\label{E69}
\end{equation}
for the Green's function changes form correspond to the two sheets $\phi_\pm$ of the bifurcation surface, respectively.  As a source point $(r,{\hat\varphi},z)$ in the vicinity of the cusp $C$ approaches the bifurcation surface from inside, i.e., as $\chi\to1-$ or $\chi\to-1+$, $G_{nj}^{\rm in}$ diverges.  However, as a source point approaches either one of the sheets of the bifurcation surface from outside, the numerator and the denominator in (\ref{E59}) vanish simultaneously and $G_{nj}^{\rm out}$ tends to a finite limit,
\begin{equation}
G_{nj}^{\rm out}\big\vert_{\phi=\phi_\pm}= G_{nj}^{\rm out}\big\vert_{\chi=\pm1}=\textstyle{\frac{1}{3}} c_1^{-2}\left(p_{nj}\pm2c_1 q_{nj}\right).
\label{E70}
\end{equation}
Note that $c_1$, and hence $p_{nj}$ and $q_{nj}$, are independent of $k$ [see (\ref{E34}), (\ref{E38}) and (\ref{E41})]. The only $k$-dependent functions appearing in the expressions for $G_{nj}^{\rm out}\big\vert_{\phi=\phi_\pm}$ are the step functions ${\mathcal H}\vert_{\phi=\phi_\pm}$ which can be summed over $k$ to obtain unity [see (\ref{E44})].  Thus the Green's function $G_{nj}$ is singular only on the inner side of the bifurcation surface (see figures~\ref{F9} and~\ref{F10}).

\subsection{Hadamard's finite part of the divergent integral representing the field}
\label{subsec:Hadamard}

It follows from (\ref{E58}) and (\ref{E69}) that the factor $\partial G_{nj}/\partial{\hat\varphi}$ in the integrand of the integral (\ref{E27}) diverges as $(1-\chi^2)^{-3/2}$ and so has a non-integrable singularity on the bifurcation surface where $\chi^2$ equals 1.  This singularity has arisen because we differentiated the retarded potential (\ref{E14}) under the integral sign when calculating the field.  Had we evaluated the integral in (\ref{E14}) prior to differentiating it we would have found a singularity-free expression.  Interchanging the orders of integration and differentiation is mathematically permissible when the integrand is discontinuous only if one treats the resulting integral as a generalized function and so one handles any non-integrable singularities that consequently arise by means of Hadamard's regularization technique \citep[see][and the illustrative example in appendix~\ref{appA}]{HadamardJ:lecCau, HoskinsRF:GenFun, ArdavanH:JMP99}.  

Hadamard's procedure consists of performing an integration by parts and discarding the divergent (integrated) term in the resulting expression.  The remaining finite part is the value that Hadamard's regularization assigns to the integral; in the present case, it is the value we would have obtained if we had first evaluated the finite integral representing the retarded potential and had differentiated the result $A^\mu({\textbf x}_P,t_P)$ of that evaluation subsequently.  (The more direct approach, in which the potential is first evaluated and then differentiated, cannot of course be carried out for any realistic source distribution analytically.)

The ${\hat\varphi}$ coordinates ${\hat\varphi}_\pm$ of the two sheets of the bifurcation surface depend on the observation time $t_P$ [see (\ref{E38}) and (\ref{E24})], so that these two sheets move across the ${\hat\varphi}$ extent of the source distribution as $t_P$ elapses.  If the position of the observation point is such that the cusp locus of the bifurcation surface intersects the source distribution, the two sheets of this surface (which tangentially meet at the cusp) will divide the volume of the source into a part that lies inside and a part that lies outside the bifurcation surface.  The Lagrangian coordinates ${\hat\varphi}$ designating the initial azimuthal positions of the constituent volume elements of a source that fully occupies an annular region range over the interval $0\le{\hat\varphi}<2\pi$.  The $({\hat r},{\hat z})$ coordinates of these source elements either fall in $\Delta\ge0$ or in $\Delta<0$.  The elements in $\Delta\ge0$ are always divided into two sets: a set inside the bifurcation surface for which ${\hat\varphi}_-\le{\hat\varphi}\le{\hat\varphi}_+$  and so the Green's function $G_{nj}$ has the form $G_{nj}^{\rm in}$ and a set outside for which ${\hat\varphi}$ lies either in $(0,{\hat\varphi}_-)$ or in $({\hat\varphi}_+,2\pi)$ and so $G_{nj}$ has the form $G_{nj}^{\rm out}$ [see (\ref{E69})].  On the other hand, if the position of the observation point is such that $\Delta<0$ for all values of $({\hat r},{\hat z})$ in ${\cal S}^\prime$ [see (\ref{E7})], then the source lies entirely outside the bifurcation surface and $G_{nj}$ has the form $G_{nj}^{\rm sub}$.  Note that, for certain space-time coordinates of the observation point $P$, the values of ${\hat\varphi}_-$ and ${\hat\varphi}_+$ that lie in the interval $(0,2\pi)$ could correspond to different rotation periods, i.e., to different values of $k$ [see (\ref{E34}), (\ref{E37}) and (\ref{E38})].  To simplify the notation, here I adopt an observation time $t_P$ at which the values of ${\hat\varphi}_-$ and ${\hat\varphi}_+$ that lie in the interval $(0,2\pi)$ correspond to the same rotation period $k$. 

Breaking up the volume of integration in the expression for one of the radiation fields, e.g., ${\mathbf E}$, into the domains of validity of $G_{nj}^{\rm in}$, $G_{nj}^{\rm out}$ and $G_{nj}^{\rm sub}$, we can therefore write the ${\hat\varphi}$-integral over ${\bf u}_{nj}$ in (\ref{E27}) as
\begin{eqnarray}
{\mathbf I}_{\hat\varphi}&\equiv&\int_0^{2\pi}{\textrm d}{\hat\varphi}\,{\mathbf u}_{nj}\frac{\partial G_{nj}}{\partial{\hat\varphi}}\nonumber\\*
&=&{\rm H}(\Delta)\left[\left(\int_0^{{\hat\varphi}_-}+\int_{{\hat\varphi}_+}^{2\pi}\right){\textrm d}{\hat\varphi}\,{\mathbf u}_{nj}\frac{\partial G_{nj}^{\rm out}}{\partial{\hat\varphi}}+\int_{{\hat\varphi}_-}^{{\hat\varphi}_+}{\textrm d}{\hat\varphi}\,{\mathbf u}_{nj}\frac{\partial G_{nj}^{\rm in}}{\partial{\hat\varphi}}\right]\nonumber\\*
&&+{\rm H}(-\Delta)\int_0^{2\pi}{\textrm d}{\hat\varphi}\,{\mathbf u}_{nj}\frac{\partial G_{nj}^{\rm sub}}{\partial{\hat\varphi}}.
\label{E71}
\end{eqnarray}
If we now integrate every term of the above expression by parts, recall that ${\hat\varphi}=0$ labels the same source point as does ${\hat\varphi}=2\pi$, and use the fact that the exact version of $G_{nj}$ given in (\ref{E32}) is periodic in ${\hat\varphi}$ as well as in $\varphi$ (with the same period $2\pi$), we arrive at
\begin{eqnarray}
{\mathbf I}_{\hat\varphi}&=&{\rm H}(\Delta)\Bigg\{\left[{\mathbf u}_{nj}\left(G_{nj}^{\rm in}-G_{nj}^{\rm out}\right)\right]_{{\hat\varphi}={\hat\varphi}_-}^{{\hat\varphi}={\hat\varphi_+}}-\left(\int_0^{{\hat\varphi}_-}+\int_{{\hat\varphi}_+}^{2\pi}\right){\textrm d}{\hat\varphi}\,\frac{\partial{\mathbf u}_{nj}}{\partial{\hat\varphi}}G_{nj}^{\rm out}\nonumber\\*
&&-\int_{{\hat\varphi}_-}^{{\hat\varphi}_+}{\textrm d}{\hat\varphi}\,\frac{\partial{\mathbf u}_{nj}}{\partial{\hat\varphi}}G_{nj}^{\rm in}\Bigg\}-{\rm H}(-\Delta)\int_0^{2\pi}{\textrm d}{\hat\varphi}\,\frac{\partial{\mathbf u}_{nj}}{\partial{\hat\varphi}}G_{nj}^{\rm sub},
\label{E72}
\end{eqnarray}
an expression that reduces to 
\begin{eqnarray}
{\mathbf I}_{\hat\varphi}&=&{\rm H}(\Delta)\left[{\mathbf u}_{nj}\left(G_{nj}^{\rm in}-G_{nj}^{\rm out}\right)\right]_{{\hat\varphi}={\hat\varphi}_-}^{{\hat\varphi}={\hat\varphi_+}}
-\int_0^{2\pi}{\textrm d}{\hat\varphi}\,\frac{\partial{\mathbf u}_{nj}}{\partial{\hat\varphi}}G_{nj},\nonumber\\
\label{E73}
\end{eqnarray}
once the integrals over $G_{nj}^{\rm in}$, $G_{nj}^{\rm out}$ and $G_{nj}^{\rm sub}$ are combined in the light of (\ref{E69}).  

We have seen in the last paragraph of \S~\ref{subsec:Expansion} that the value of $G_{nj}^{\rm in}$ at ${\hat\varphi}={\hat\varphi}_\pm$ diverges (figures~\ref{F9} and~\ref{F10}).  The Hadamard finite part of ${\mathbf I}_{\hat\varphi}$ is therefore given by the right-hand side of (\ref{E73}) without the divergent terms involving $G_{nj}^{\rm in}\vert_{{\hat\varphi}={\hat\varphi}_-}$ and $G_{nj}^{\rm in}\vert_{{\hat\varphi}={\hat\varphi}_+}$,  
\begin{equation}
{\rm  Fp}\{{\mathbf I}_{\hat\varphi}\}=-{\rm H}(\Delta){\mathbf u}_jG_{nj}^{\rm out}\big\vert_{{\hat\varphi}={\hat\varphi}_-}^{{\hat\varphi}={\hat\varphi_+}}-\int_0^{2\pi}{\textrm d}{\hat\varphi}\,\frac{\partial{\mathbf u}_{nj}}{\partial{\hat\varphi}}G_{nj},
\label{E74}
\end{equation}
where ${\rm Fp}\{{\mathbf I}_{\hat\varphi}\}$ denotes the Hadamard finite part of the divergent integral ${\mathbf I}_{\hat\varphi}$ \citep[see][]{HadamardJ:lecCau,HoskinsRF:GenFun}.  This procedure applies also to the expression for the radiation field ${\bf B}$ in (\ref{E27}) except that ${\bf u}_{nj}$ in ~(\ref{E71})-(\ref{E74}) is everywhere replaced by ${\bf v}_{nj}$.  

Once the integrals with respect to ${\hat\varphi}$ in (\ref{E27}) are equated to the expression on the right-hand side of (\ref{E74}) and its counterpart for ${\bf B}$, we find that
\begin{equation}
\left[\matrix{{\mathbf E}\cr{\mathbf B}\cr}\right]=\left[\matrix{{\mathbf E}^{\rm v}\cr{\mathbf B}^{\rm v}\cr}\right]+\left[\matrix{{\mathbf E}_+^{\rm b}\cr{\mathbf B}_+^{\rm b}\cr}\right]-\left[\matrix{{\mathbf E}_-^{\rm b}\cr{\mathbf B}_-^{\rm b}\cr}\right]
\label{E75}
\end{equation}
with
\begin{equation}
\left[\matrix{{\mathbf E}^{\rm v}\cr{\mathbf B}^{\rm v}\cr}\right]=\frac{1}{\omega}\sum_{n=1}^2\sum_{j=1}^3\int_{\mathcal S}{\hat r}{\textrm d}{\hat r}\,{\textrm d}{\hat\varphi}\,{\textrm d}{\hat z}\,G_{nj}\frac{\partial}{\partial{\hat\varphi}}\left[\matrix{{\mathbf u}_{nj}\cr{\mathbf v}_{nj}\cr}\right],
\label{E76}
\end{equation}
and
\begin{equation}
\left[\matrix{{\mathbf E}_\pm^{\rm b}\cr{\mathbf B}_\pm^{\rm b}\cr}\right]=\frac{1}{\omega}\sum_{n=1}^2\sum_{j=1}^3\int_{\mathcal S^\prime}{\hat r}{\textrm d}{\hat r}\,{\textrm d}{\hat z}\,{\rm H}(\Delta)\,G_{nj}^{\rm out}\left[\matrix{{\mathbf u}_{nj}\cr{\mathbf v}_{nj}\cr}\right]\Bigg\vert_{{\hat\varphi}={\hat\varphi}_\pm} ,
\label{E77}
\end{equation}
where ${\mathcal S^\prime}$ is the projection of the support ${\mathcal S}$ of the source distribution onto the $(r,z)$ plane [see (\ref{E7})].  The term $\matrix{[{\bf E}^{\rm v}&{\bf B}^{\rm v}]}$ constitutes the contribution from the entire volume of the source while the terms $\matrix{[{\bf E}_\pm^{\rm b}&{\bf B}_\pm^{\rm b}]}$ denote the contributions from the discontinuities of the Green's function on the two sheets $\phi=\phi_\pm$ of the bifurcation surface, respectively.  We will see that the terms $\matrix{[{\bf E}_\pm^{\rm b}&{\bf B}_\pm^{\rm b}]}$ describe unconventional radiation fields with characteristics that turn out to differ from any previously known radiation fields.

\section{Radiation field of the experimentally realized source distribution}
\label{sec:field} 

For the charge and current densities $\rho=-{\mathbf{\nabla\cdot P}}$ and ${\mathbf j}=\partial{\mathbf P}/\partial t$ associated with the polarization distribution $P_{r,\varphi,z}=\Re[s_{r,\varphi,z}({\hat r},{\hat z})\exp(-{\rm i}m{\hat\varphi})]$ in (\ref{E1}) the source terms $[{\mathbf u} _{nj}\,\,{\mathbf v} _{nj}]$ defined by (\ref{E28})--(\ref{E31}) reduce to the real part of
\begin{eqnarray}
\left[\matrix{{\mathbf u}_{nj}({\hat r},{\hat z},{\hat\varphi})\cr{\mathbf v}_{nj}({\hat r},{\hat z},{\hat\varphi})\cr}\right]&=&{\textrm i}m\omega\exp(-{\textrm i}m{\hat\varphi})\left[\matrix{{\tilde{\mathbf u}}_{nj}({\hat r},{\hat z})\cr{\tilde{\mathbf v}}_{nj}({\hat r},{\hat z})\cr}\right],\qquad 0\le{\hat\varphi}<2\pi,
\label{E78}
\end{eqnarray}
where
\begin{eqnarray}
    &\left[
     \begin{array}{c}
     {\tilde{\bf u}}_{11}\\
     {\tilde{\bf u}}_{12}\\
     {\tilde{\bf u}}_{13}
     \end{array} \right]= \left[
      \begin{array}{c}
     s_r{\hat{\bf e}}_{r_P}+s_\varphi{\hat{\bf e}}_{\varphi_P}\\
      -s_\varphi{\hat{\bf e}}_{r_P}+s_r{\hat{\bf e}}_{\varphi_P} \\
     s_z{\hat{\bf e}}_{z_P}
\end{array} \right],\nonumber\\
\label{E79}
\end{eqnarray}
\begin{eqnarray}
    &\left[
     \begin{array}{c}
     {\tilde{\bf u}}_{21}\\
     {\tilde{\bf u}}_{22}\\
     {\tilde{\bf u}}_{23}
     \end{array} \right]= s_0\left[
      \begin{array}{c}
     {\hat r}{\hat{\bf e}}_{r_P}\\
      {\hat r}{\hat{\bf e}}_{\varphi_P} \\
     -{\hat r}_P{\hat{\bf e}}_{r_P}+({\hat z}-{\hat z}_P){\hat{\bf e}}_{z_P}
\end{array} \right],\nonumber\\
\label{E80}
\end{eqnarray}
\begin{equation}
\left[\matrix{ {\tilde{\bf v}}_{11}& {\tilde{\bf v}}_{12}& {\tilde{\bf v}}_{13}}\right]=\left[\matrix{0&0&0}\right],
\label{E81}
\end{equation}
and 
\begin{eqnarray}
    & \left[
     \begin{array}{c}
      {\tilde{\bf v}}_{21}\\
      {\tilde{\bf v}}_{22}\\
      {\tilde{\bf v}}_{23}
     \end{array} \right]=\left[
      \begin{array}{c}
     -({\hat z}-{\hat z}_P){\tilde{\bf u}}_{12}+{\hat r}s_z{\hat{\bf e}}_{\varphi_P}+{\hat r}_P s_\varphi{\hat{\bf e}}_{z_P}\\  
    {\hat{\bf e}}_{z_P}\times {\tilde{\bf v}}_{21}+{\hat r}_P s_r{\hat{\bf e}}_{z_P} \\
     -{\hat r}_P s_z{\hat{\bf e}}_{\varphi_P}-{\hat r}s_\varphi{\hat{\bf e}}_{z_P}
\end{array} \right],\nonumber\\
\label{E82}
\end{eqnarray}
with 
\begin{equation}
s_0=\frac{s_\varphi}{{\hat r}}+\frac{{\rm i}}{m}{\hat{\bf\nabla}}\cdot{\bf s}.
\label{E83}
\end{equation}
Here, ${\hat{\bf\nabla}}\cdot{\bf s}$ denotes the divergence of ${\bf s}$ with respect to the dimensionless coordinates $({\hat r},\varphi,{\hat z})$.  

Insertion of (\ref{E78}) in (\ref{E76}) and (\ref{E77}) results in the following expressions,
\begin{equation}
\left[\matrix{{\mathbf E}^{\rm v}\cr{\mathbf B}^{\rm v}\cr}\right]=m^2\sum_{n=1}^2\sum_{j=1}^3\int_{\mathcal S}{\hat r}{\textrm d}{\hat r}\,{\textrm d}{\hat\varphi}\,{\textrm d}{\hat z}\,\exp(-{\textrm i}m{\hat\varphi})G_{nj}\left[\matrix{{\tilde{\bf u}}_{nj}\cr{\tilde{\bf v}}_{nj}\cr}\right],
\label{E84}
\end{equation}
and
\begin{equation} 
\left[\matrix{{\mathbf E}_\pm^{\rm b}\cr{\mathbf B}_\pm^{\rm b}\cr}\right]={\textrm i}m\sum_{n=1}^2\sum_{j=1}^3\int_{\mathcal S^\prime}{\hat r}{\textrm d}{\hat r}\,{\textrm d}{\hat z}\,{\rm H}(\Delta)\exp(-{\textrm i}m{\hat\varphi}_\pm) G_{nj}^{\rm out}\big\vert_{{\hat\varphi}={\hat\varphi}_\pm}\left[\matrix{{\tilde{\bf u}}_{nj}\cr{\tilde{\bf v}}_{nj}\cr}\right],
\label{E85}
\end{equation}
whose real parts describe the contributions from the volume of the source and from the two sheets of the bifurcation surface toward the total radiation field $\matrix{[{\bf E}&{\bf B}]}$, respectively [see (\ref{E75})].  The values $G_{nj}^{\rm out}\vert_{{\hat\varphi}={\hat\varphi}_\pm}$ of the Green's function that appear in (\ref{E85}) are given by (\ref{E70}), (\ref{E67}) and (\ref{E68}). 

\begin{figure}
\centerline{\includegraphics[width=11cm]{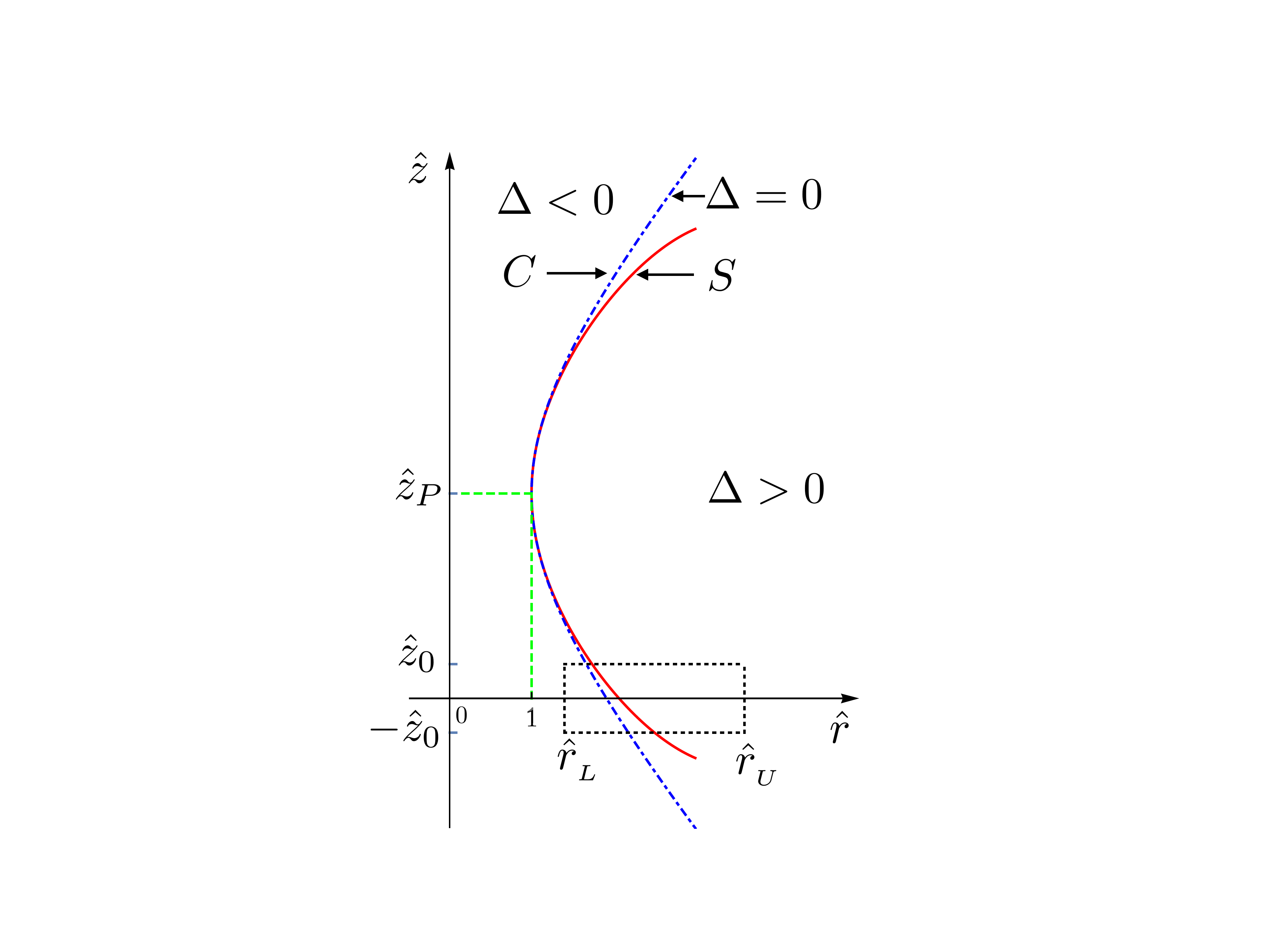}}
\caption{The dash-dotted curve is the projection of the cusp locus of the bifurcation surface, $C$, onto the $({\hat r},{\hat z})$ plane, i.e., the projection of the locus of source points that approach the observer along the radiation direction with the speed of light and zero acceleration at the retarded time [see (\ref{E39})].  The solid curve (in red) is the locus $S$ of the stationary points of the function $\phi_-$, i.e., the stationary points of the phase of the exponential factor that appears in the integrand of the expression for the field $[{\bf E}^{\rm b}_-\quad{\bf B}^{\rm b}_-]$ [see (\ref{E85}) and (\ref{E109})].  The dotted rectangle represents the boundary of the support ${\cal S}^\prime$ of the source term ${\bf s}$ defined in (\ref{E7}), i.e., the boundary of the projection of the source distribution described in \S~\ref{sec:source} onto the $({\hat r},{\hat z})$ plane. The part of the source distribution whose projection lies to the left of curve $C$, for which $\Delta<0$, only generates a spherically-decaying conventional field.  Whether the cusp locus $C$ intersects the source distribution (as shown here) or lies to the left or right of the domain ${\cal S}^\prime$ is dictated by the polar coordinate $\theta_P$ of the observation point $P$ [see (\ref{E89})].  In plotting this figure, I have placed the observation point close to the source (at ${\hat r}_P={\hat z}_P=3$) in order to render the separation between $C$ and $S$ visible.  As ${\hat R}_P$ increases, these two curves overlap and tend toward the vertical.  For ${\hat R}_P\gg1$, the radial distance between $C$ and $S$ at an arbitrary ${\hat z}$ diminishes as ${\hat R}_P^{-2}$ [see (\ref{E110})].}
\label{F11}
\end{figure} 

\subsection{Cusp locus $C$ and its dual role in the spaces of source points and observation points}
\label{subsec:Cusp}

The boundary terms $\matrix{[{\bf E}^{\rm b}_\pm&{\bf B}^{\rm b}_\pm]}$ receive contributions only from those source elements whose $({\hat r},{\hat z})$ coordinates fall within the region $\Delta\geq0$ shown in figure~\ref{F11}, i.e., for which ${\hat r}\geq{\hat r}_C({\hat z})$ [see (\ref{E39})].  In other words, $\matrix{[{\bf E}^{\rm b}_\pm&{\bf B}^{\rm b}_\pm]}$ are non-zero either when the projection of the cusp locus of the bifurcation surface $C$ onto the $({\hat r},{\hat z})$ plane intersects the domain ${\mathcal S}^\prime$ described by (\ref{E7}), in which case ${\hat r}_L\le{\hat r}_C\le{\hat r}_U$ for $-{\hat z}_0\le{\hat z}\le{\hat z}_0$ as shown in figure~\ref{F11}, or when ${\hat r}_C\le{\hat r}_L$ for these values of ${\hat z}$ and the entire radial extent of the source lies inside the bifurcation surface.  The intersection of ${\cal S}^\prime$ and $\Delta\ge0$ which constitutes the domain of integration in (\ref{E85}) thus changes as the location of the cusp locus $C$ (which depends on the position of the observation point) changes.

The parametric equation ${\hat r}={\hat r}({\hat z})$, $\varphi=\varphi({\hat z})$, of the cusp locus of the bifurcation surface associated with a given observation point $({\hat r}_P,{\hat\varphi}_P,{\hat z}_P)$ at the observation time $t_P$ was derived in (\ref{E39}).  If we rewrite the two members of (\ref{E39}) in terms of the dimensionless polar coordinates ${\hat R}_P=({\hat r}_P^2+{\hat z}_P^2)^{1/2}$, $\theta_P=\arccos({\hat z}_P/{\hat R}_P)$, of the observation point $P$ and solve them for $\theta_P$ and $\varphi_P$ as functions of $({\hat r},\varphi,{\hat z})$ and ${\hat R}_P$, we obtain
  
\begin{eqnarray}
&C: \left\{
      \begin{array}{ll}
\theta_P=\theta_P^{\rm c}({\hat r},{\hat z})\equiv\arccos\Big\{\frac{1}{{\hat R}_P{\hat r}}\Big[\frac{{\hat z}}{{\hat r}}\pm\left({{\hat r}}^2-1\right)^{1/2}\left({{\hat R}_P}^2-1-\frac{{{\hat z}}^2}{{{\hat r}}^2}\right)^{1/2}\Big]\Big\},\\
\varphi_P=\varphi_P^{\rm c}({\hat r},\varphi,{\hat z})\equiv\varphi-2k\pi+\arccos\left(\frac{1}{{\hat r}{\hat R}_P\sin\theta_P}\right),
\end{array}\right.
\label{E86}
\end{eqnarray}
where the $\pm$ correspond to the two halves of the cusp curve below and above the plane ${\hat z}={\hat z}_P$, respectively,  and $k$ is the positive integer enumerating successive rotation periods [see (\ref{E20}) and (\ref{E34})].  

The angle between the asymptotes to the hyperbola representing the projection of the cusp locus $C$ onto the $({\hat r},{\hat z})$-plane, as well as the radial coordinate of the point of intersection of this hyperbola with the plane ${\hat z}=0$, depend on the coordinate $\theta_P$ of the observation point (see figure~\ref{F11}).  As the observation point $P$ at a given distance ${\hat R}_P$ moves from the upper half of the rotation axis ($\theta_P=0$) towards the plane of rotation ($\theta_P=\pi/2$), the point of intersection of the cusp locus $C$ with the plane ${\hat z}=0$ gradually shifts across this plane from a large value ($\lim_{{\hat R}_P\to\infty}{\hat r}_C=\csc\theta_P$) of the radial coordinate ${\hat r}$ towards the upper boundary ${\hat r}_U$ of the source distribution, across the radial extent of the source distribution towards its lower boundary ${\hat r}_L$ and eventually towards the light cylinder ${\hat r}=1$.  The cusp $C$ will thus lie to the right of the source distribution shown in figure~\ref{F11} when $0<\theta_P\le\theta_L^{\rm c}$, where
\begin{equation}
\theta_L^{\rm c}=\theta_P^{\rm c}\big\vert_{{\hat r}={\hat r}_U,{\hat z}={\hat z}_0},
\label{E87}
\end{equation}
intersects this source distribution while $\theta_L^{\rm c}\le\theta_P\le\theta_U^{\rm c}$, where
\begin{equation}
\theta_U^{\rm c}=\theta_P^{\rm c}\big\vert_{{\hat r}={\hat r}_L,{\hat z}=-{\hat z}_0},
\label{E88}
\end{equation}
and lies in $1<{\hat r}<{\hat r}_L$ (to the left of this source distribution) when $\theta_U^{\rm c}\le\theta_P\le\pi/2$.  [Recall that ${\hat r}_L$ and ${\hat r}_U$ designate the inner and outer radial boundaries of the source distribution (see figure~\ref{F11}).] 

\begin{figure}
\centerline{\includegraphics[width=10cm]{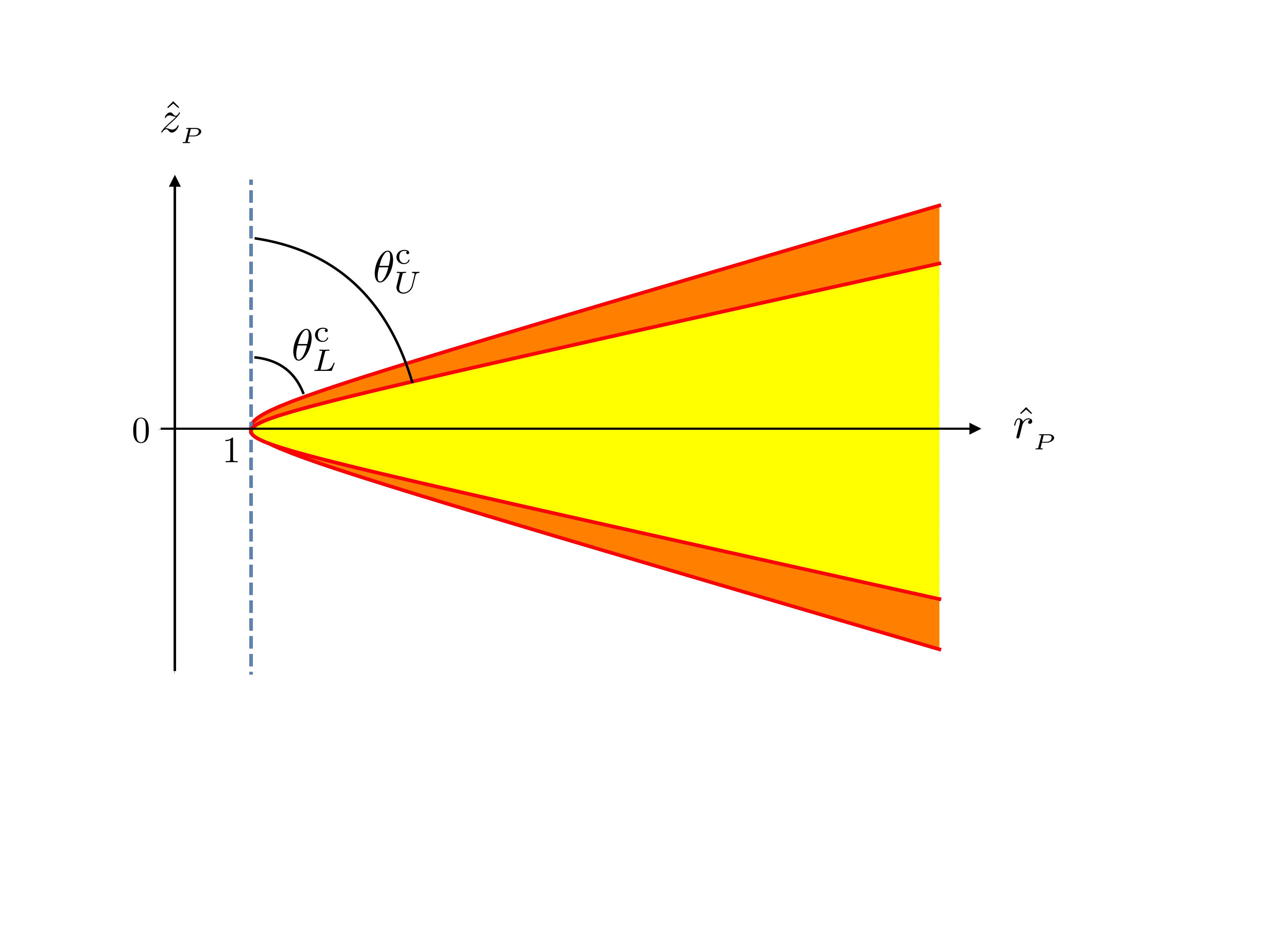}}
\caption{Counterpart of figure~\ref{F11} in the $({\hat r}_P,\varphi_P, {\hat z}_P)$-space of observation points.  While the cusp locus $C$ in figure~\ref{F11} is described by $\Delta=0$ for fixed values of $({\hat r}_P,{\hat z}_P)$, the hyperbolas shown here are described by $\Delta=0$ for fixed values of the source coordinates $({\hat r},{\hat z})$: the values $({\hat r}_U,{\hat z}_0)$ and $({\hat r}_L,-{\hat z}_0)$.  If the observation point $P$ lies in the space (coloured orange) between the hyperbolas, then the cusp locus $C$ of the bifurcation surface intersects the source distribution shown in figure~\ref{F11}.  But if the observation point $P$ lies in the space (coloured yellow) that is bounded by the inner hyperbola, then $\Delta$ is positive throughout the source distribution and the cusp locus $C$ lies to the left of the source distribution shown in figure~\ref{F11}.  On the other hand, at observation points in $0\le\theta_P\le\theta_L^{\rm c}$ and $\pi-\theta_L^{\rm c}\le\theta_P\le\pi$ (outside the coloured regions), $\Delta$ is negative throughout the source distribution and the cusp locus $C$ lies to the right of the source distribution shown in figure~\ref{F11}.  In cases where the lower boundary of the source distribution shown in figure~\ref{F11} falls on or within the light cylinder, i.e., ${\hat r} _L\le1$ but ${\hat r}_U>1$, the two arms of the inner hyperbola shown here coalesce onto the ${\hat r}_P$-axis and the cusp locus of the bifurcation surface intersects the source distribution for all points of the (expanded orange) space inside the outer hyperbola.}
\label{F12}
\end{figure}

Thus the cusp locus $C$ would intersect the support (\ref{E7}) of the source distribution only if the observation point lies within one of the following conical shells, 
\begin{eqnarray}
\left\{
      \begin{array}{ll}
\qquad\theta_L^{\rm c}\leq\theta_P\leq\theta_U^{\rm c},\\
\pi-\theta_U^{\rm c}\leq\theta_P\leq\pi-\theta_L^{\rm c}.
\end{array}\right.
\label{E89}
\end{eqnarray}
The first member of (\ref{E86}), on which the definitions in (\ref{E87}) and (\ref{E88}) are based, reduces to 
\begin{equation}
\theta_P^{\rm c}=\arcsin\left(\frac{1}{{\hat r}}\right)-\left(\frac{{\hat z}}{{\hat r}}\right){{\hat R}_P}^{-1}\pm\frac{1}{2}\left({\hat r}^2-1\right)^{1/2}{\hat R}_P^{-2}+\cdots
\label{E90}
\end{equation}
in the far zone where ${\hat R}_P\gg1$.  The leading term in this expansion (in powers of ${\hat R}_P^{-1}$) together with (\ref{E87}) and (\ref{E88}) shows that, in the limit ${\hat R}_P\to\infty$, the angles $\theta_L^{\rm c}$ and $\theta_U^{\rm c}$ reduce to $\arcsin(1/{\hat r}_U)$ and $\arcsin(1/{\hat r}_L)$, respectively.  We shall see below that the radiation field $\matrix{[{\bf E}&{\bf B}]}$ has radically differing characteristics in each of the three disjoint regions of space separated by these two cones (see figure~\ref{F12}).

\section{ The part of the field arising from the volume of the source}
\label{sec:Ev}
\subsection{Evaluation of $[{\bf E}^{\rm v}\quad{\bf B}^{\rm v}]$ at observation points for which $\theta_L^{\rm c}\le\theta_P\le\pi-\theta_L^{\rm c}$}
\label{subsec:Ev1}

When the polar coordinate $\theta_P$ of the observation point lies in $\theta_L^{\rm c}\le\theta_P\le\pi-\theta_L^{\rm c}$, there are volume elements within the source distribution ${\cal S}$ that approach $P$ along the radiation direction with a speed exceeding $c$ at the retarded time.  For such source elements $\Delta$ is positive.  Depending on whether the ${\hat\varphi}$ coordinates of these elements lie inside or outside the bifurcation surface associated with the observation point $P$, the Green's function $G_{nj}$ that appears in (\ref{E84}) has either the value $G_{nj}^{\rm in}$ or the value $G_{nj}^{\rm out}$ [see (\ref{E69})].  There are also source elements lying in $\Delta<0$ for which $G_{nj}$ has the value $G^{\rm sub}_{nj}$ (see figure~\ref{F11}).  The expression in (\ref{E84}) can therefore be written as
\begin{eqnarray}
\left[\matrix{{\mathbf E}^{\rm v}\cr{\mathbf B}^{\rm v}\cr}\right]&=&m^2\sum_{n=1}^2\sum_{j=1}^3\int_{\cal S^\prime}{\hat r}{\rm d}{\hat r}{\rm d}{\hat z}\,\left[\matrix{{\tilde{\bf u}}_{nj}\cr{\tilde{\bf v}}_{nj}\cr}\right]\Bigg\{{\rm H}(\Delta)\Bigg[\int_{{\hat\varphi}_-}^{{\hat\varphi}_+}{\rm d}{\hat\varphi}\exp(-{\rm i}m{\hat\varphi})G_{nj}^{\rm in}\nonumber\\*
&&+\left(\int_0^{{\hat\varphi}_-}+\int_{{\hat\varphi}_+}^{2\pi}\right){\rm d}{\hat\varphi}\exp(-{\rm i}m{\hat\varphi})G_{nj}^{\rm out}\Bigg]+{\rm H}(-\Delta)\int_0^{2\pi}{\rm d}{\hat\varphi}\exp(-{\rm i}m{\hat\varphi})G_{nj}^{\rm sub}\Bigg\},\nonumber\\*
\label{E91}
\end{eqnarray}
in which $G_{nj}^{\rm in}$, $G_{nj}^{\rm out}$ and $G_{nj}^{\rm sub}$ are given by (\ref{E58}), (\ref{E59}) and (\ref{E62}).

If we change the variable of integration in the integral over ${\hat\varphi}_-\le{\hat\varphi}\le{\hat\varphi}_+$ in (\ref{E91}) from ${\hat\varphi}$ to 
\begin{equation}
\psi=-2c_1\sin\left({\textstyle\frac{1}{3}}\arcsin\chi\right),
\label{E92}
\end{equation}
in which $\chi$ depends on ${\hat\varphi}$ as in (\ref{E52}), and note that the inversion of (\ref{E92}) yields
\begin{equation}
{\hat\varphi}={\textstyle\frac{1}{3}}\psi^3-c_1^2\psi+c_2+{\hat\varphi}_P,
\label{E93}
\end{equation}
we obtain
\begin{equation}
\int_{{\hat\varphi}_-}^{{\hat\varphi}_+}{\rm d}{\hat\varphi}\exp(-{\rm i}m{\hat\varphi})G_{nj}^{\rm in}=2\sum_{k=1}^\infty\int_{-c_1}^{c_1}{\rm d}\psi{\cal H}(p_{nj}+q_{nj}\psi)\exp\left[-{\rm i}m\left({\textstyle\frac{1}{3}}\psi^3-c_1^2\psi+{\hat\varphi}_P+c_2\right)\right]
\label{E94}
\end{equation}
[see (\ref{E58})].  The variables of integration in the remaining two integrals inside the square bracket in (\ref{E91}) can be similarly transformed from ${\hat\varphi}$ to
\begin{equation}
\Psi=2c_1{\rm sgn}(\chi) \cosh\left({\textstyle\frac{1}{3}}{\rm arccosh}\vert\chi\vert\right).
\label{E95}
\end{equation}
This transformation and its inverse
\begin{equation}
{\hat\varphi}={\rm sgn}(\chi)\left({\textstyle\frac{1}{3}}\Psi^3-c_1^2\Psi\right)+c_2+{\hat\varphi}_P,
\label{E96}
\end{equation}
then result in
\begin{eqnarray}
\Bigg(\int_0^{{\hat\varphi}_-}+\int_{{\hat\varphi}_+}^{2\pi}\Bigg){\rm d}{\hat\varphi}G_{nj}^{\rm out}&=&\sum_{k=1}^\infty
\Bigg(\int_{\Psi_L}^{-2c_1}+\int_{2c_1}^{\Psi_U}\Bigg){\rm d}\Psi{\cal H}(p_{nj}+q_{nj}\Psi)\nonumber\\*
&&\times\exp\left[-{\rm i}m\left({\textstyle\frac{1}{3}}\Psi^3-c_1^2\Psi+c_2+{\hat\varphi}_P\right)\right],
\label{E97}
\end{eqnarray}
where 
\begin{equation}
\Psi_L=-2c_1\cosh\left[\frac{1}{3}{\rm arccosh}\left(\frac{3}{2}\frac{{\hat\varphi}_P+c_2}{c_1^3}\right)\right]
\label{E98}
\end{equation}
and 
\begin{equation}
\Psi_U=2c_1\cosh\left[\frac{1}{3}{\rm arccosh}\left(\frac{3}{2}\frac{2\pi-{\hat\varphi}_P-c_2}{c_1^3}\right)\right]
\label{E99}
\end{equation}
are the values of $\Psi$ corresponding to ${\hat\varphi}=0$ and ${\hat\varphi}=2\pi$, respectively [see (\ref{E52}) and (\ref{E59})].  For any given observation point with the space-time coordinates (${\hat R}_P$, $\theta_L^{\rm c}\le\theta_P\le\pi-\theta_L^{\rm c}$, $\varphi_P$, $t_P$) the value of $k$ (in $c_2$) that is selected by the step function ${\cal H}$ (or its far-field version ${\cal H}_\infty$) will automatically render the arguments of the ${\rm arccosh}$ functions in (\ref{E98}) and (\ref{E99}) positive and yield a positive $\Psi_U$ and a negative $\Psi_L$.

Rather than evaluating the remaining integral in (\ref{E91}) by substituting the expression for $G_{nj}^{\rm sub}$ in its integrand, here I replace $G_{nj}$ in (\ref{E84}) by its original representation (\ref{E32}) to write this integral as
\begin{eqnarray}
\sum_{j=1}^3\int_{\mathcal S^\prime}{\hat r}{\textrm d}{\hat r}\,{\textrm d}{\hat z}&{\rm H}(-\Delta)&\left[\matrix{{\tilde{\bf u}}_{nj}\cr{\tilde{\bf v}}_{nj}\cr}\right]\int_0^{2\pi}{\rm d}{\hat\varphi}\exp(-{\rm i}m{\hat\varphi})G_{nj}^{\rm sub}
=\sum_{k=1}^\infty\int_{\mathcal S^\prime}{\hat r}{\textrm d}{\hat r}\,{\textrm d}{\hat z}\,{\rm H}(-\Delta)\nonumber\\*
&&\times\int_0^{2\pi}{\textrm d}{\hat\varphi}\,\exp(-{\textrm i}m{\hat\varphi})
\int_{{\hat\varphi}+2(k-1)\pi}^{{\hat\varphi}+2k\pi}{\textrm d}\varphi\frac{\delta(g-\phi)}{{\hat R}^n}\nonumber\\*
&&\times\left\{\cos(\varphi-\varphi_P)\left[\matrix{{\tilde{\bf u}}_{n1}\cr{\tilde{\bf v}}_{n1}\cr}\right] + \sin(\varphi-\varphi_P)\left[\matrix{{\tilde{\bf u}}_{n2}\cr{\tilde{\bf v}}_{n2}\cr}\right]+\left[\matrix{{\tilde{\bf u}}_{n3}\cr{\tilde{\bf v}}_{n3}\cr}\right]\right\}.\qquad
\label{E100}
\end{eqnarray}
Given that $g$ is a monotonic function of $\varphi$ in $\Delta<0$ and that the integrand in (\ref{E100}) is periodic in $\varphi$ with the period $2\pi$, it makes no difference which period, i.e., which value of $k$, makes the contribution received at the observation time.  We can therefore replace the range of the $\varphi$-integral by $(0,2\pi)$ (omitting the summation over $k$) and perform the trivial integration with respect to ${\hat\varphi}$ to obtain 
\begin{eqnarray}
\lefteqn{\sum_{j=1}^3\int_{\mathcal S^\prime}{\hat r}{\textrm d}{\hat r}\,{\textrm d}{\hat z}\,{\rm H}(-\Delta)\left[\matrix{{\tilde{\bf u}}_{nj}\cr{\tilde{\bf v}}_{nj}\cr}\right]\int_0^{2\pi}{\rm d}{\hat\varphi}\exp(-{\rm i}m{\hat\varphi})G_{nj}^{\rm sub}=\exp(-{\textrm i}m{\hat\varphi}_P)\int_{\mathcal S^\prime}{\hat r}{\textrm d}{\hat r}\,{\textrm d}{\hat z}\,{\rm H}(-\Delta)}\qquad\nonumber\\*
&&\times\int_0^{2\pi}{\textrm d}\varphi\frac{\exp(- {\textrm i}mg)}{{\hat R}^n}\Bigg\{\cos(\varphi-\varphi_P)\left[\matrix{{\tilde{\bf u}}_{n1}\cr{\tilde{\bf v}}_{n1}\cr}\right] + \sin(\varphi-\varphi_P)\left[\matrix{{\tilde{\bf u}}_{n2}\cr{\tilde{\bf v}}_{n2}\cr}\right]+\left[\matrix{{\tilde{\bf u}}_{n3}\cr{\tilde{\bf v}}_{n3}\cr}\right]\Bigg\}\qquad\,\,\,
\label{E101}
\end{eqnarray}
for the value of the last integral in (\ref{E91}).

Inserting (\ref{E94}), (\ref{E97}) and (\ref{E101}) in (\ref{E91}), we arrive at
\begin{eqnarray}
&\left[\matrix{{\mathbf E}^{\rm v}\cr{\mathbf B}^{\rm v}\cr}\right]&=m^2\exp(-{\rm i}m{\hat\varphi}_P)\sum_{n=1}^2\int_{\cal S^\prime}{\hat r}{\rm d}{\hat r}{\rm d}{\hat z}\,\Bigg\{{\rm H}(\Delta)\sum_{j=1}^3\sum_{k=1}^\infty\exp(-{\rm i}m c_2)\left[\matrix{{\tilde{\bf u}}_{nj}\cr{\tilde{\bf v}}_{nj}\cr}\right]\nonumber\\*
&&\times\Bigg(\int_{\Psi_L}^{\Psi_U}-\int_{-2c_1}^{2c_1}+2\int_{-c_1}^{c_1}\Bigg){\rm d}\Psi(p_{nj}+q_{nj}\Psi){\cal H}\exp\left[-{\rm i}m\left({\textstyle\frac{1}{3}}\Psi^3-c_1^2\Psi\right)\right]+{\rm H}(-\Delta)\nonumber\\* 
&&\times\int_0^{2\pi}{\textrm d}\varphi\frac{\exp(- {\textrm i}mg)}{{\hat R}^n}\Bigg(\cos(\varphi-\varphi_P)\left[\matrix{{\tilde{\bf u}}_{n1}\cr{\tilde{\bf v}}_{n1}\cr}\right]+ \sin(\varphi-\varphi_P)\left[\matrix{{\tilde{\bf u}}_{n2}\cr{\tilde{\bf v}}_{n2}\cr}\right]+\left[\matrix{{\tilde{\bf u}}_{n3}\cr{\tilde{\bf v}}_{n3}\cr}\right]\Bigg)\Bigg\},\nonumber\\*
\label{E102}
\end{eqnarray}
where the integration variable $\psi$ in (\ref{E94}) has been renamed $\Psi$ and the limits of integration in (\ref{E97}) have been placed in alternative positions.  For small $c_1$, the difference between the values of the two integrals over $-2c_1\le\Psi\le2c_1$ and $-c_1\le\Psi\le c_1$ is negligibly small compared to that of the first integral inside the parentheses.  Once these two integrals are ignored, the remaining $k$-dependent function in the resulting expression can be summed,
\begin{equation}
\sum_{k=1}^\infty{\cal H}=1,
\label{E103}
\end{equation}
since the $k$-depndence of $c_2$ does not influence the value of $\exp(-{\rm i}m c_2)$  [see (\ref{E34}), (\ref{E38}) and (\ref{E41})].  This enables us to obtain the asymptotic value of the integral over $\Psi_L\le\Psi\le\Psi_U$ by simply extending its range,
\begin{eqnarray}
\int_{\Psi_L}^{\Psi_U}&&{\rm d}\Psi(p_{nj}+q_{nj}\Psi)\exp\left[-{\rm i}m\left({\textstyle\frac{1}{3}}\Psi^3-c_1^2\Psi\right)\right]\nonumber\\*
&&\simeq2\int_0^{\infty}{\rm d}\Psi\,\left\{p_{nj}\cos\left[m\left({\textstyle\frac{1}{3}}\Psi^3-c_1^2\Psi\right)\right]-{\rm i}q_{nj}\Psi\sin\left[m\left({\textstyle\frac{1}{3}}\Psi^3-c_1^2\Psi\right)\right]\right\}, m\gg1,\qquad\,\,\,
\label{E104}
\end{eqnarray}
because the phase of the exponential factor in its integrand is in a canonical form as it stands.

The imaginary part of the $\Psi$-integral in (\ref{E104}) can be obtained by differentiating the real part of this integral with respect to $c_1^2$ and dividing the resulting expression by $m$.  From (9.5.1) of ~\citet{Olver} and (\ref{E103}) and (\ref{E104}) it follows, therefore, that
\begin{eqnarray}
\left[\matrix{{\mathbf E}^{\rm v}\cr{\mathbf B}^{\rm v}\cr}\right]&&\simeq m^2\exp(-{\rm i}m{\hat\varphi}_P)\sum_{n=1}^2\int_{\cal S^\prime}{\hat r}{\rm d}{\hat r}{\rm d}{\hat z}\,\Bigg\{{\rm H}(\Delta)\sum_{j=1}^3 2\pi m^{-1/3}\exp(-{\rm i}m c_2)\left[\matrix{{\tilde{\bf u}}_{nj}\cr{\tilde{\bf v}}_{nj}\cr}\right]\nonumber\\*
&&\times\left[p_{nj}{\textrm {Ai}}\left(-m^{2/3}c_1^2\right)+{\textrm i}m^{-1/3}q_{nj}{\textrm {Ai}}^\prime\left(-m^{2/3}c_1^2\right)\right]+{\rm H}(-\Delta)\nonumber\\*
&&\times\int_0^{2\pi}{\textrm d}\varphi\frac{\exp(- {\textrm i}mg)}{{\hat R}^n}\Bigg(\cos(\varphi-\varphi_P)\left[\matrix{{\tilde{\bf u}}_{n1}\cr{\tilde{\bf v}}_{n1}\cr}\right] + \sin(\varphi-\varphi_P)\left[\matrix{{\tilde{\bf u}}_{n2}\cr{\tilde{\bf v}}_{n2}\cr}\right]+\left[\matrix{{\tilde{\bf u}}_{n3}\cr{\tilde{\bf v}}_{n3}\cr}\right]\Bigg)\Bigg\},\nonumber\\*
&&\nonumber\\*
&&\qquad\qquad\qquad\qquad\qquad\qquad\qquad\qquad\theta_L^{\rm c}\le\theta_P\le\pi-\theta_L^{\rm c},\quad m\gg1,
\label{E105}
\end{eqnarray}
in which ${\textrm {Ai}}$ and ${\textrm {Ai}}^\prime$ are the Airy function and the derivative of the Airy function with respect to its argument, respectively.  

On the other hand, evaluation of the leading term in the asymptotic expansion of the $\varphi$-integral in 
\begin{eqnarray}
\left[\matrix{{\mathbf E}^{\rm v}\cr{\mathbf B}^{\rm v}\cr}\right]&=&m^2\sum_{n=1}^2\int_{\cal S^\prime}{\hat r}\,{\rm d}{\hat r}\,{\rm d}{\hat z}\int_0^{2\pi}{\textrm d}\varphi\frac{\exp[- {\textrm i}m(g+{\hat\varphi}_P)]}{{\hat R}^n}\nonumber\\*
&&\times\Bigg(\cos(\varphi-\varphi_P)\left[\matrix{{\tilde{\bf u}}_{n1}\cr{\tilde{\bf v}}_{n1}\cr}\right] + \sin(\varphi-\varphi_P)\left[\matrix{{\tilde{\bf u}}_{n2}\cr{\tilde{\bf v}}_{n2}\cr}\right]+\left[\matrix{{\tilde{\bf u}}_{n3}\cr{\tilde{\bf v}}_{n3}\cr}\right]\Bigg)
\label{E106}
\end{eqnarray}
by the method of \citet{ChesterC:Extstd} results in exactly the same expression as that which multiplies ${\rm H}(\Delta)$ in (\ref{E105}).  Hence, the expression in (\ref{E106}) gives the combined contributions of the source elements in both $\Delta<0$ and $\Delta>0$.  It turns out that this expression could have been directly obtained in the present case by performing the integration with respect to $t$ in (\ref{E19}), even though such a procedure is not generally applicable to cases in which the retarded time is multi-valued (see appendix~\ref{appB}).

\subsection{Evaluation of $[{\bf E}^{\rm v}\quad{\bf B}^{\rm v}]$ at observation points for which $0<\theta_P\le\theta_L^{\rm c}$ or $\pi-\theta_L^{\rm c}\le\theta_P<\pi$}
\label{subsec:Ev3}

Equation (\ref{E106}) applies also to an observation point for which the entire source lies outside the bifurcation surface, i.e., for which ${\hat r}_U<{\hat r}_C$ (see figure~\ref{F11}).  None of the source elements in ${\cal S}^\prime$ can approach observers that are located in $0<\theta_P\le\theta_L^{\rm c}$ or $\pi-\theta_L^{\rm c}\le\theta_P<\pi$ with a superluminal speed along the radiation direction.  As a result, $\Delta$ is negative throughout the source distribution (\ref{E7}) and the field $\matrix{[{\bf E}^{\rm v}&{\bf B}^{\rm v}]}$ that is generated outside these two cones (i.e., outside the coloured regions in figure~\ref{F12}) is the same as any other conventional radiation field.  

It is customary, when deriving (\ref{E106}) from (\ref{E19}), to replace the term $\partial\rho/\partial t$ that results from the integration with respect to $t$ by $-c\,{\bf{\nabla\cdot j}}$ (from the equation of continuity) and to apply a subsequent integration by parts with respect to ${\bf x}$ (by means of the divergence theorem) to write the conventional radiation field as
\begin{equation}
\left[\matrix{{\mathbf E}^{\rm v}\cr{\mathbf B}^{\rm v}\cr}\right]=\frac{1}{c^2}\int{\rm d}^3{\bf x}\,{\rm d}t\,\frac{\delta(t-t_P+R/c)}{R}\,{\hat{\bf n}}\times\frac{\partial}{\partial t}\left[\matrix{{\hat{\bf n}}\times{\bf j} \cr-{\bf j}\cr}\right].
\label{E107}
\end{equation}
However, of the two equivalent formulations given by (\ref{E106}) and (\ref{E107}), I will be using the former which can be more easily combined with the expressions I will derive for $\matrix{[{\bf E}^{\rm b}_\pm&{\bf E}^{\rm b}_\pm]}$.  

\section{The part of the field arising from the discontinuities of the Green's function}
\label{sec:Eb}

\subsection{Locus of stationary points, $S$, of the phase of the exponential factor in the expression for $[{\bf E}^{\rm b}_-\quad{\bf B}^{\rm b}_-]$}             
\label{subsec:locus}

Despite the apparent symmetry between the two terms $G_{nj}^{\rm out}\vert_{{\hat\varphi}_\pm}\exp(-{\rm i}m{\hat\varphi}_\pm)$ in the expressions for $\matrix{[{\bf E}^{\rm b}_\pm&{\bf B}^{\rm b}_\pm]}$ in (\ref{E85}), these two contributions toward the value of the unconventional radiation field differ radically: the phase ${\hat\varphi}_-$ of the first exponential factor is stationary, as a function of ${\hat r}$, along a curve in the $(r,\varphi,z)$ space while the phase ${\hat\varphi}_+$ of the second exponential has no stationary points.

Differentiation of the functions $\phi_\pm={\hat\varphi}_\pm-{\hat\varphi}_P=g(\varphi_\pm)$ with respect to ${\hat r}$ yields  
\begin{equation}
\frac{\partial\phi_\pm}{\partial{\hat r}}=\frac{{\hat r}^2-1\pm\Delta^{1/2}}{{\hat r}{\hat R}_\pm}
\label{E108}
\end{equation}
[see (\ref{E38})].  Hence $\partial\phi_-/\partial{\hat r}$ vanishes for all ${\hat z}$, along the projection ${\hat r}={\hat r}_S({\hat z})$ of the following three-dimensional curve onto the $(r,z)$ plane,
\begin{eqnarray}
    &S: \left\{
      \begin{array}{c}
     {\hat r}={\hat r}_S({\hat z})=\{\textstyle{\frac{1}{2}}({{\hat r}_P}^2+1)-[ \textstyle{\frac{1}{4}}({{\hat r}_P}^2-1)^2-({\hat z}-{\hat z}_P)^2]^{1/2}\}^{1/2}, \\
      \varphi=\varphi_S({\hat z})=\varphi_P+2k\pi-\arccos({\hat r}_S/{\hat r}_P).
\end{array} \right.
\label{E109}
\end{eqnarray}
Along this curve which is depicted in figure~\ref{F11} the two derivatives $\partial g/\partial\varphi$ and $\partial g/\partial{\hat r}$ of the argument of the Dirac delta function in (\ref{E32}) vanish simultaneously and $\Delta^{1/2}={\hat r}^2-1$ [see (\ref{E34}) and (\ref{E35})].  At the point ${\hat z}={\hat z}_P$ on $S$ the derivative $\partial g/\partial{\hat z}$ also vanishes [since $g$ depends on ${\hat z}$ only through $({\hat z}-{\hat z}_P)^2$], so that the derivatives of the phase of the Green's function (\ref{E32}) with respect to all three of the integration variables in the expression for the field $\matrix{[{\bf E}&{\bf B}]}$ (i.e., $\partial g/\partial\varphi$, $\partial g/\partial{\hat r}$ and $\partial g/\partial{\hat z}$) vanish simultaneously.  I refer to both $S$ and its projection onto the $(r,z)$ plane as the \emph{locus of stationary points} of $\phi_-$; whether it is $S$ itself or its projection that is referred to will be clear from the context.

Note that the coordinates of a far-field observation point need to satisfy ${\hat R}_P\ge2\cot\theta_P\csc\theta_P$ for ${\hat r}_S$ to be real; otherwise, the expression inside the curly brackets in (\ref{E109}) would be negative.  This constraint reflects the fact that the projection of locus $S$ onto the $({\hat r},{\hat z})$ plane curves away from the rotation axis (see figure~\ref{F11}).  The locus $S$ becomes more parallel to the rotation axis as the observation point moves into the far zone.  But, no matter how large ${\hat R}_P$ may be, there are always ranges of values of the polar angle $\theta_P$ (close to $0$ and to $\pi$) for which the function $\phi_-$ has no stationary points.

By comparing (\ref{E39}) and (\ref{E109}), we can see that the radial separation between the curves $C$ and $S$ in figure~\ref{F11} is exceedingly small when the observation point lies either in the far zone, ${\hat R}_P\gg1$, or close to the plane of rotation $\theta_P=\pi/2$, so that $\vert{\hat z}-{\hat z}_P\vert\ll1$ throughout the localized source distribution (\ref{E7}),
\begin{eqnarray}
&{\hat r}_S-{\hat r}_C\simeq\left\{
      \begin{array}{c}
\cot^4\theta_P/(2{\hat R}_P^2\sin\theta_P), \qquad{\hat R}_P\gg1\\
      {\textstyle\frac{1}{2}}({\hat z}-{\hat z}_P)^4/({\hat r}_P^2-1)^3,\qquad\vert{\hat z}-{\hat z}_P\vert\ll1. 
      \end{array} \right.
\label{E110}
\end{eqnarray}
In other words, the locus $S$ of the stationary points of $\phi_-$ is essentially coincident with the cusp locus of the bifurcation surface (i.e.\ with the locus $C$ of the source elements that approach the observation point with the speed of light and zero acceleration at the retarded time) in these cases.  This notwithstanding, (\ref{E33}) and (\ref{E108}) show that the value of the function $\Delta$ undergoes a large change over a small interval in ${\hat r}$: it vanishes on $C$, equals $({\hat r}_S^2-1)^2$ on $S$ and rapidly rises to as large a value as that of ${\hat R}_P^2$ (for ${\hat R}_P\gg1$) a short distance away from $C$.  We will see that the sharp change in the value of $G^{\rm out}_j\vert_{\phi_-}$ resulting from the proximity of the loci $C$ and $S$ renders the numerical evaluation of $\matrix{[{\bf E}^{\rm b}_-&{\bf B}^{\rm b}_-]}$ in the far zone particularly challenging (\S~\ref{sec:numerical}).

If, in analogy with (\ref{E86}), we rewrite the two members of (\ref{E109}) in terms of the dimensionless polar coordinates $({\hat R}_P,\theta_P)$ of the observation point $P$ and solve them for $\theta_P$ and $\varphi_P$ as functions of $({\hat r},{\hat z})$ and ${\hat R}_P$, we obtain
\begin{eqnarray}
&S: \left\{
      \begin{array}{ll}
\theta_P=\theta_P^{\rm s}({\hat r},{\hat z})\equiv\arccos\Big\{\frac{1}{{\hat R}_P{\hat r}}\Big[\frac{{\hat z}}{{\hat r}}+\left({{\hat r}}^2-1\right)^{1/2}\left({{\hat R}_P}^2-{\hat r}^2-\frac{{{\hat z}}^2}{{{\hat r}}^2}\right)^{1/2}\Big]\Big\},\\
\varphi_P=\varphi_P^{\rm s}({\hat r},{\hat z})\equiv\varphi-2k\pi+\arccos\left(\frac{{\hat r}}{{\hat R}_P\sin\theta_P}\right),
\end{array} \right.
\label{E111}
\end{eqnarray}
where $k$ is the positive integer enumerating successive rotation periods.  Hence, the projection of the segment ${\hat z}_P\ge{\hat z}$ of the locus $S$ onto the $({\hat r},{\hat z})$ plane intersects the source distribution (\ref{E7}) at a given $-{\hat z}_0\le{\hat z}\le{\hat z}_0$ if the colatitude of the observation point $P$ lies in the interval $\theta_L^{\rm s}\le\theta_P\le\theta_U^{\rm s}$, where 
\begin{equation}
\theta_L^{\rm s}=\theta_P^{\rm s}\big\vert_{{\hat r}={\hat r}_U,{\hat z}={\hat z}_0}
\label{E112}
\end{equation}
and
\begin{equation}
\theta_U^{\rm s}=\theta_P^{\rm s}\big\vert_{{\hat r}={\hat r}_L,{\hat z}=-{\hat z}_0}.  
\label{E113}
\end{equation}
The radial coordinates of all source elements would exceed ${\hat r}_S$, on the other hand, if $\theta_U^{\rm s}\le\theta_P\le\pi-\theta_U^{\rm s}$ [see (\ref{E7}) and figure~\ref{F11}].  

The loci $C$ and $S$ would both lie within the source distribution (\ref{E7}), at every value of ${\hat z}$ in $-{\hat z}_0\le{\hat z}\le{\hat z}_0$, if $\theta_L\le\theta_P\le\theta_U$, where
\begin{equation}
\theta_L=\theta_P^{\rm s}\big\vert_{{\hat r}={\hat r}_U,{\hat z}=-{\hat z}_0}
\label{E114}
\end{equation}
and
\begin{equation}
\theta_U=\theta_P^{\rm c}\big\vert_{{\hat r}={\hat r}_L,{\hat z}={\hat z}_0}.
\label{E115}
\end{equation}
This can be seen by noting that as the polar coordinate of the observation point increases (at a given ${\hat R}_P$) away from the rotation axis $\theta_P=0$ toward the equatorial plane $\theta_P=\pi/2$, the curve $S$ intersects the entire ${\hat z}$-extent of the source once it passes the corner ${\hat r}={\hat r}_U$, ${\hat z}=-{\hat z}_0$ of the rectangular support ${\cal S}^\prime$ of the source distribution (\ref{E7}) shown in figure~\ref{F11}.  It is curve $C$, on the other hand, that starts leaving the source at the corner ${\hat r}={\hat r}_L$, ${\hat z}={\hat z}_0$ of ${\cal S}^\prime$ after $S$ and $C$ have swept across the ${\hat r}$-extent of the source.  If $\theta_P$ continues to increase past the value $\pi/2$, then $S$ and $C$ (in that order) start entering ${\cal S}^\prime$ from the corner ${\hat r}={\hat r}_L$, ${\hat z}={\hat z}_0$ when $\theta_P=\pi-\theta_U$ and will both intersect the entire ${\hat z}$-extent of the source again when $\pi-\theta_U\le\theta_P\le\pi -\theta_L$.

There are intervals of $\theta_P$ near $\theta_L$ or $\theta_U$ for which only one of the curves $C$ and $S$ intersects the source distribution, mostly over a limited section of $-{\hat z}_0\le{\hat z}\le{\hat z}_0$ (see figure~\ref{F11}).  Evaluation of the radiation field in these transitional intervals whose widths rapidly shrink with increasing distance -- as ${\hat R}_P^{-2}$ for ${\hat R}_P\gg1$ [see (\ref{E110})] -- will be dealt with separately in \S~\ref{sec:transitional}. 

Note that the stationary point ${\hat r}=1$, $\varphi=\varphi_P+2\pi k-\arccos(1/{\hat r})$, ${\hat z}={\hat z}_P$, at which all three derivatives of $g$ vanish and the curves $C$ and $S$ meet tangentially (see figure~\ref{F11}) does not fall within the range of integration in (\ref{E85}) unless (i) there are source elements whose speeds equal the speed of light $c$, and (ii) the observation point lies sufficiently close to the plane of rotation $\theta_P=\pi/2$ for its coordinate ${\hat z}_P={\hat R}_P\cos\theta_P$ to match the coordinate ${\hat z}$ of some source elements.  For the source distribution described in \S~\ref{sec:source}, these requirements are met only if ${\hat r}_L\le1$, i.e., the source elements at the inner radius of the dielectric move with a speed that is smaller than or equal to $c$ and the observation point lies within the following angular interval
\begin{equation}
\frac{\pi}{2}-\arcsin\left(\frac{{\hat z}_0}{{\hat R}_P}\right)\leq\theta_P\leq\frac{\pi}{2}+\arcsin\left(\frac{{\hat z}_0}{{\hat R}_P}\right)
\label{E116}
\end{equation}
encompassing the equatorial plane.  Note also that the width of this interval decreases with increasing distance as ${\hat R}_P^{-1}$ in the far zone.

\subsection{Paths of steepest descent for the exponential kernels $\exp(-{\rm i}m{\phi}_\pm)$}
\label{subsec:paths}

Owing to the proximity of the loci $C$ and $S$ and the rapid variation of $G^{\rm out}_{nj}\vert_{\phi_-}$ in their vicinity, the contributions of the critical points discussed in the preceding section toward the value of the integral in (\ref{E85}) cannot be taken into account properly without resorting to a technique more discerning than a direct numerical integration.  Here I evaluate the ${\hat r}$-integral in (\ref{E85}) by the method of steepest descent \citep[see, e.g.,][]{BenderOrszag}.  I regard the variable of integration ${\hat r}$ as complex, i.e., write 
 \begin{equation}
 {\hat r}=u+{\rm i}v,
 \label{E117}
 \end{equation}
in which $u$ and $v$ are real, and invoke Cauchy's integral theorem to deform the original path of integration into the contours of steepest descent in the complex $(u,v)$-plane that pass through the critical points of the phases ${\phi_\pm}({\hat r},{\hat z})$ at a given ${\hat z}$, i.e., through the stationary point ${\hat r}={\hat r}_S$ and the boundary points ${\hat r}={\hat r}_C$ or ${\hat r}_L$ and ${\hat r}={\hat r}_U$.  We will see that the constant $m$ in the argument of the exponential factor in (\ref{E85}) need not be particularly large for the main contributions to the values of $\matrix{[{\bf E}_\pm^{\rm b}&{\bf B}_\pm^{\rm b}]}$ to come from a limited segment of each path next to the critical point from which it issues. 
 
The first step is to write $\phi_\pm$ as functions of $(u,v,{\hat z};{\hat r}_P,{\hat z}_P)$.  Inserting (\ref{E117}) in (\ref{E33}) and adopting the square root of the resulting complex expression which is positive on the real axis $v=0$, we obtain
 \begin{equation}
 \Delta^{1/2}=d\,\exp({\rm i}\delta),
 \label{E118}
 \end{equation}
 with
 \begin{equation}
 d=\{[({\hat r}_P^2-1)(u^2-v^2-1)-({\hat z}-{\hat z}_P)^2]^2+4({\hat r}_P^2-1)^2u^2v^2\}^{1/4},
 \label{E119}
 \end{equation}
 and
 \begin{equation}
 \delta=\frac{1}{2}\arctan\frac{2({\hat r}_P^2-1)uv}{({\hat r}_P^2-1)(u^2-v^2-1)-({\hat z}-{\hat z}_P)^2}.
 \label{E120}
 \end{equation}
  Equation~(\ref{E37}) together with (\ref{E117}) and (\ref{E118}) then yields
  \begin{equation}
  {\hat R}_\pm=L_\pm\exp({\rm i}\sigma_\pm),
  \label{E121}
  \end{equation}
  in which
  \begin{equation}
  L_\pm=\{[({\hat z}-{\hat z}_P)^2+{\hat r}_P^2+u^2-v^2-2(1\mp d\cos\delta)]^2+4(uv\pm d\sin\delta)^2\}^{1/4},
  \label{E122}
  \end{equation}
  and 
  \begin{equation}
  \sigma_\pm=\frac{1}{2}\arctan\frac{2(uv\pm d\sin\delta)}{({\hat z}-{\hat z}_P)^2+{\hat r}_P^2+u^2-v^2-2(1\mp d\cos\delta)}.
  \label{E123}
  \end{equation}
From Euler's formula $\exp({\rm i}x)=\cos x+{\rm i}\sin x$ and (\ref{E34}), we have
\begin{equation}
\exp[{\rm i}(\varphi_\pm-\varphi_P)]=\frac{1\mp\Delta^{1/2}-{\rm i}{\hat R}_\pm}{{\hat r}{\hat r}_P},
\label{E124}
\end{equation}
so that the corresponding expression for $\varphi_\pm$ can be written as
\begin{eqnarray}
\varphi_\pm&=&\varphi_P+2k\pi+\xi_\pm-\eta+{\rm i}\ln[{\hat r}_P(u^2+v^2)^{1/2}]\nonumber\\*
&&-({\rm i}/2)\ln\{1+d^2+L_\pm^2+2L_\pm\sin\sigma_\pm\mp2d[L_\pm\sin(\sigma_\pm-\delta)+\cos\delta]\},
\label{E125}
\end{eqnarray}
with
\begin{equation}
\xi_\pm=\arctan\frac{-L_\pm\cos\sigma_\pm\mp d\sin\delta}{1\mp d\cos\delta+L_\pm\sin\sigma_\pm},
\label{E126}
\end{equation}
and
\begin{equation}
\eta=\arctan(v/u),
\label{E127}
\end{equation}
in the light of (\ref{E117}), (\ref{E118}), ({\ref{E121}) and (\ref{E124}).

Inserting the above expressions for $\Delta^{1/2}$, ${\hat R}_\pm$ and $\varphi_\pm$ in (\ref{E38}), we finally arrive at the following expressions for the real and imaginary parts of $\phi_\pm={\hat\varphi}_\pm-{\hat\varphi}_P$ as functions of $(u,v,{\hat z},{\hat r}_P,{\hat z}_P)$,
\begin{equation}
\Re[\phi_\pm(u,v)]=L_\pm\cos\sigma_\pm+\xi_\pm-\eta+2k\pi,
\label{E128}
\end{equation}
and
\begin{eqnarray}
\Im[\phi_\pm(u,v)]&=&L_\pm\sin\sigma_\pm+\ln[{\hat r}_P(u^2+v^2)^{1/2}]\nonumber\\*
&&-(1/2)\ln\{1+d^2+L_\pm^2+2L_\pm\sin\sigma_\pm\mp2d[L_\pm\sin(\sigma_\pm-\delta)+\cos\delta]\}.\qquad
\label{E129}
\end{eqnarray}
The path of steepest descent through a given critical point of $\phi_-$ or $\phi_+$ is the curve in the complex $(u,v)$-plane along which the corresponding phase $-{\rm i}m{\hat\varphi}_-$ or $-{\rm i}m{\hat\varphi}_+$ of the relevant exponential factor in (\ref{E85}) has a constant imaginary part and a negative real part~\citep{BenderOrszag}.

To use Cauchy's theorem to express the integrals over ${\hat r}$ in (\ref{E85}) as integrals over such steepest-descent paths, we also need the Jacobians of the transformations that map the real axis onto these paths.  Along each path of steepest descent through a critical point of either $\exp(-{\rm i}m\phi_-)$ or $\exp(-{\rm i}m\phi_+)$, the real part of the relevant $\phi_\pm$ is constant.  So, setting the total derivative of $\Re[\phi_\pm(u,v)]$ equal to zero, we find that the slope of a steepest-descent path is given by
\begin{equation}
\frac{{\rm d}v}{{\rm d}u}=-\frac{\frac{\partial}{\partial u}[\Re(\phi_\pm)]}{\frac{\partial}{\partial v}[\Re(\phi_\pm)]}
=\frac{\frac{\partial}{\partial u}[\Re(\phi_\pm)]}{\frac{\partial}{\partial u}[\Im(\phi_\pm)]}
=\frac{\Re[\frac{\partial\phi_\pm}{\partial{\hat r}}]}{\Im[\frac{\partial\phi_\pm}{\partial{\hat r}}]},
\label{E130}
\end{equation}
where I have used the Cauchy-Riemann relation
\begin{equation}
\frac{\partial}{\partial v}\{\Re[\phi_\pm(u,v)]\}=-\frac{\partial}{\partial u}\{\Im[\phi_\pm(u,v)]\},
\label{E131}
\end{equation}
and the following expression for the derivatives of the complex functions $\phi_\pm(u,v)$ with respect to the complex variable ${\hat r}=u+{\rm i}v$: 
\begin{equation}
\frac{\partial\phi_\pm}{\partial{\hat r}}=\frac{\partial}{\partial u}\{\Re[\phi_\pm(u,v)]\}+{\rm i} \frac{\partial}{\partial u}\{\Im[\phi_\pm(u,v)]\}.
\label{E132}
\end{equation}
Equation~(\ref{E130}) applies to a steepest-descent path through any critical point of either $\phi_-$ or $\phi_+$, irrespective of whether it is parametrized by $u$ or $v$ [i.e., whether its slope is given by the last expression on the right-hand side of (\ref{E130}) or by the inverse of this expression]. 

The functions $\partial\phi_\pm/\partial{\hat r}$ have the values derived in (\ref{E108}).  Writing ${\hat r}$ everywhere in the right-hand side of (\ref{E108}) as $u+{\rm i}v$ and making use of (\ref{E118}) and (\ref{E121}), we arrive at
\begin{equation}
\frac{\partial\phi_\pm}{\partial{\hat r}}=\frac{u^2-v^2-1\pm d\cos\delta+{\rm i}(2uv\pm d\sin\delta)}{(u^2+v^2)^{1/2}L_\pm\exp[{\rm i}(\sigma_\pm+\eta)]},
\label{E133}
\end{equation}
so that insertion of the real and imaginary parts of the above expression in the last member of (\ref{E130}) yields
\begin{equation}
\frac{{\rm d}v}{{\rm d}u}=\cot(\zeta_\pm-\sigma_\pm-\eta),
\label{E134}
\end{equation}
where
\begin{equation}
\zeta_\pm=\arctan\frac{2uv\pm d\sin\delta}{u^2-v^2-1\pm d\cos\delta}.
\label{E135}
\end{equation}
The Jacobian of the transformation from ${\hat r}$ along the real axis to $u$ or $v$ along a steepest-descent path through a critical point of $\phi_\pm$ is therefore given by
\begin{equation}
J_u^\pm=\frac{{\rm d}{\hat r}}{{\rm d}u}=1+{\rm i}\cot(\zeta_\pm-\sigma_\pm-\eta),
\label{E136}
\end{equation}
or
\begin{equation}
 J_v^\pm=\frac{{\rm d}{\hat r}}{{\rm d}v}=\tan(\zeta_\pm-\sigma_\pm-\eta)+{\rm i},
 \label{E137}
 \end{equation}
depending, respectively, on whether the path in question is parametrized by $u$ or by $v$.

We shall see below that a variable, more suitable than either $u$ or $v$, for parametrizing the path of steepest descent through the critical point $u={\hat r}_C$, $v=0$, is the radial distance $w=[(u-{\hat r}_C)^2+v^2]^{1/2}$ from this point.  If we mark the complex ${\hat r}$ plane by a set of polar coordinates centred on the point $u={\hat r}_C$, $v=0$, i.e., write 
\begin{equation}
{\hat r}=u+{\rm i}v={\hat r}_C+w\exp({\rm i}\lambda),
\label{E138}
\end{equation}
and express the Cartesian coordinates $u$ and $v$ in (\ref{E134}) in terms of the polar coordinates $w$ and $\lambda$, we find that
\begin{equation}
\frac{{\rm d}\lambda}{{\rm d}w}=\frac{1}{w}\cot(\zeta_\pm-\sigma_\pm-\eta+\lambda).
\label{E139}
\end{equation}
This together with (\ref{E138}) yields
\begin{equation}
J_w^\pm=\frac{{\rm d}{\hat r}}{{\rm d}w}=\exp({\rm i}\lambda)[1+{\rm i}\cot(\zeta_\pm-\sigma_\pm-\eta+\lambda)]
\label{E140}
\end{equation}
for the Jacobian of the transformation from ${\hat r}$ along the real axis to $w$ along a steepest-descent path through the point ${\hat r}={\hat r}_C$ (i.e., through $w=0$).

It should be noted (for purposes of numerical evaluation of the above expressions) that the real functions that are defined in terms of an $\arctan$, i.e., $\delta$, $\sigma_\pm$, $\xi_\pm$, $\zeta_\pm$ and $\lambda$, are continuous along each steepest-descent path.  Any discontinuities arising from the multi-valuedness of $\arctan$ should be removed by adopting an appropriate branch of $\arctan$.  The multi-valued complex function $c_1(u+{\rm i}v,{\hat z},{\hat r}_P,{\hat z}_P)$ is rendered continuous along every one of the paths described in this section by choosing the cubic root of $\phi_+-\phi_-$ in (\ref{E41}) as follows: along the lower branch of the path ${\cal L}_S$ in figures~\ref{F16} and \ref{F19}, for which $u<0$,
\begin{eqnarray}
&c_1\Big\vert_{{\cal L}_S,u<0}=\left\{
      \begin{array}{c}
[{\textstyle\frac{3}{4}}(\phi_+-\phi_-)]^{1/3}, \qquad{\rm arg}(\phi_+-\phi_-)>0\\
     \exp(2{\rm i}\pi/3){[{\textstyle\frac{3}{4}}(\phi_+-\phi_-)]}^{1/3} ,\qquad{\rm arg}(\phi_+-\phi_-)\le0, 
      \end{array} \right.
\label{E140a}
\end{eqnarray}
along the paths ${\cal L}_C$ in figures~\ref{F16} and \ref{F18},
\begin{eqnarray}
&c_1\Big\vert_{{\cal L}_C}=\left\{
      \begin{array}{c}
 \exp(-2{\rm i}\pi/3)[{\textstyle\frac{3}{4}}(\phi_+-\phi_-)]^{1/3}, \qquad{\rm arg}(\phi_+-\phi_-)>0\\
    {[{\textstyle\frac{3}{4}}(\phi_+-\phi_-)]}^{1/3} ,\qquad{\rm arg}(\phi_+-\phi_-)\le0, 
      \end{array} \right.
\label{E140b}
\end{eqnarray}
and along all other paths $c_1$ is given by $[{\textstyle\frac{3}{4}}(\phi_+-\phi_-)]^{1/3}$ irrespective of the sign of ${\rm arg}(\phi_+-\phi_-)$.  Along the path ${\cal L}_S$, the phase of $c_1(u+{\rm i}v,{\hat z},{\hat r}_P,{\hat z}_P)$ discontinuously changes by $2\pi/3$ across the saddle point at ${\hat r}={\hat r}_S$.  This correspondingly shifts the phase of the Green's function $G_{nj}^{\rm out}\vert_{{\cal L}_S}$ by $\pi$ across ${\hat r}={\hat r}_S$, a phase shift similar to that encountered across a focal point in optics.  The choice in (\ref{E140b}) is dictated by the analytic expression for the approximate value of the phase of $c_1\vert_{{\cal L}_C}$ in the vicinity of ${\hat r}={\hat r}_C$ [see (\ref{E151}) below].  It is understood that the branch cuts for any complex multi-valued functions are selected to lie outside the closed areas in the $(u,v)$ plane around which the contour integrations are performed.

\subsection{Paths of steepest descent through the critical points of the phase ${\phi}_-$ for\\ observation points in $\theta_L\le\theta_P\le\theta_U$ or $\pi-\theta_U\le\theta_P\le\pi-\theta_L$}
\label{subsec:PhiMinus1}

At observation points for which the cusp locus $C$ of the bifurcation surface intersects the source distribution ${\cal S}^\prime$, the ${\hat r}$-integral in the expression for $\matrix{[{\bf E}^{\rm b}_-\,\,\,{\bf B}^{\rm b}_-]}$ in (\ref{E85}) extends over ${\hat r}_C\le{\hat r}\le{\hat r}_U$, since ${\rm H}(\Delta)$ in its integrand would vanish over the remaining segment ${\hat r}_L\le{\hat r}<{\hat r}_C$ of the range of integration with respect to ${\hat r}$ (see \S~\ref{subsec:Cusp} and figure~\ref{F11}).  The critical points of $\phi_-$ for the asymptotic evaluation of the ${\hat r}$-integral for large $m$ would consist, therefore, of the locus ${\hat r}={\hat r}_S$ of stationary points of the phase ${\hat\varphi}_-$ of the exponential and the boundaries ${\hat r}={\hat r}_C$ and ${\hat r}={\hat r}_U$ of the domain of integration. 

The path of steepest descent for the exponential factor $\exp(-{\rm i}m{\hat\varphi}_-)$, in the complex $(u,v)$ plane, through the point $u={\hat r}_S$, $v=0$, at which $\phi_-(u,v)$ has a saddle point is parametrically described by the solution $v=v_S(u)$ of the transcendental equation 
\begin{equation}
\Re[\phi_-(u,v)]=\phi_-({\hat r}_S,0)\equiv\phi_S
\label{E141}
\end{equation}
that satisfies $v_S({\hat r}_S)=0$ and the condition 
\begin{equation}
\gamma_S\equiv\Im[\phi_-(u,v)]\big\vert_{v=v_S(u)}\leq0
\label{E142}
\end{equation}
at all relevant values of the curve parameter $u$ and the fixed parameters $({\hat z},{\hat r}_P,{\hat z}_P)$.  I denote this path which consists of two segments, one in $v<0$ and one in $v>0$, by ${\cal L}_S$ (see figures~\ref{F13}--\ref{F15}).  From the plots of $\gamma_S$ versus $u$, it can be seen that for $v_S(u)\ge0$, the condition in (\ref{E142}) is satisfied by the segment $u\ge{\hat r}_S$ of the solution $v=v_S(u)$ (see figure~\ref{F14}), while for $v_S(u)\le0$, the condition in (\ref{E142}) is satisfied by the segment $u\le{\hat r}_S$ of this solution (see figure~\ref{F15}).

\begin{figure}
\centerline{\includegraphics[width=10cm]{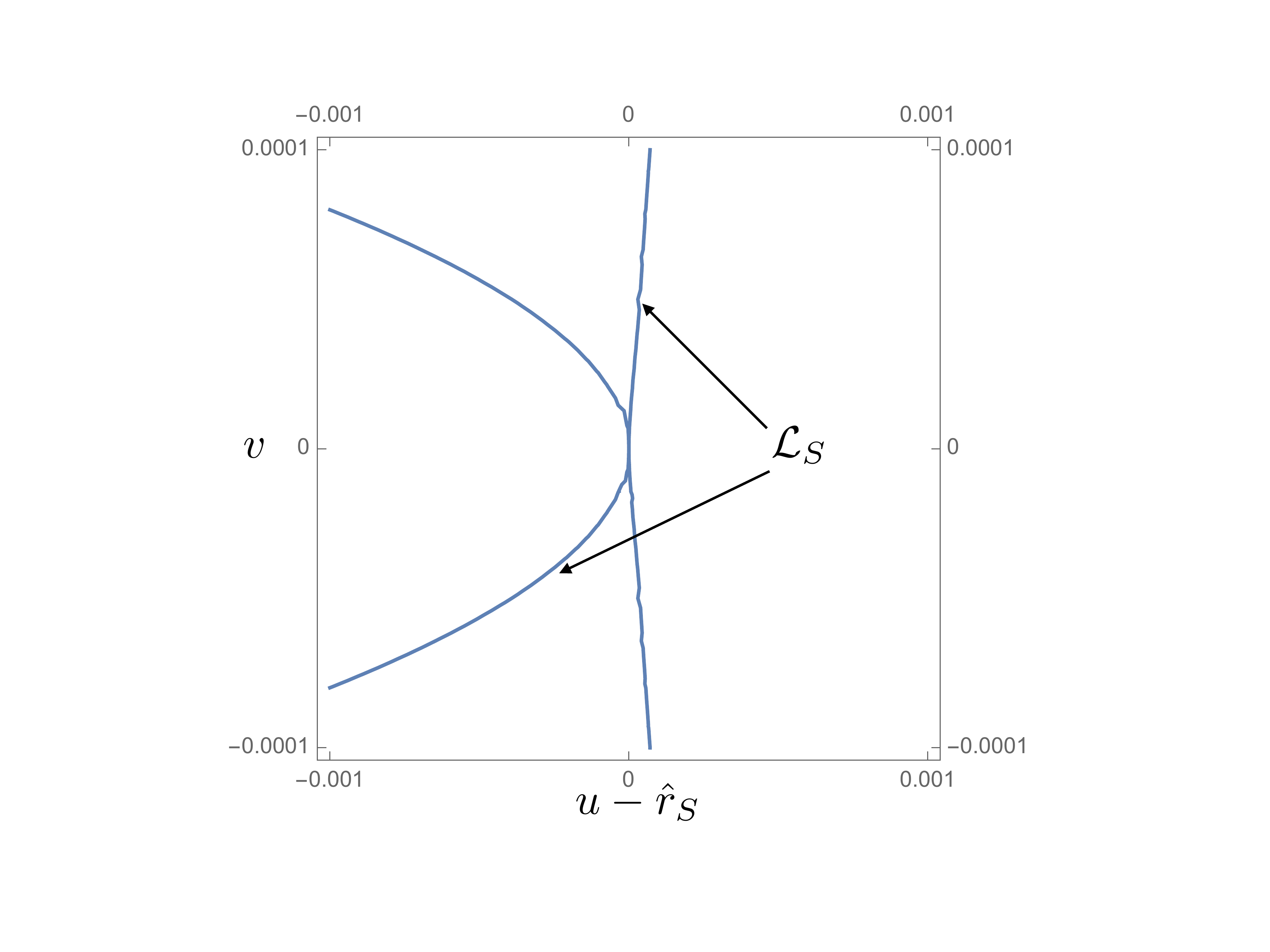}}
\caption{The solutions $v=v_S(u)$ of (\ref{E141}) for ${\hat z}=0$, ${\hat R}_P=100$ and $\theta_P=\pi/3$ in the vicinity of the saddle point $({\hat r}_S,0)$ of the function $\Re[\phi_-(u,v)]$.  As shown by figures~\ref{F14} and \ref{F15}, the segments here designated by ${\cal L}_S$ satisfy the condition in (\ref{E142}) and so constitute the paths of steepest descent through the saddle point $({\hat r}_S,0)$.}
\label{F13}
\end{figure} 

\begin{figure}
\centerline{\includegraphics[width=8.5cm]{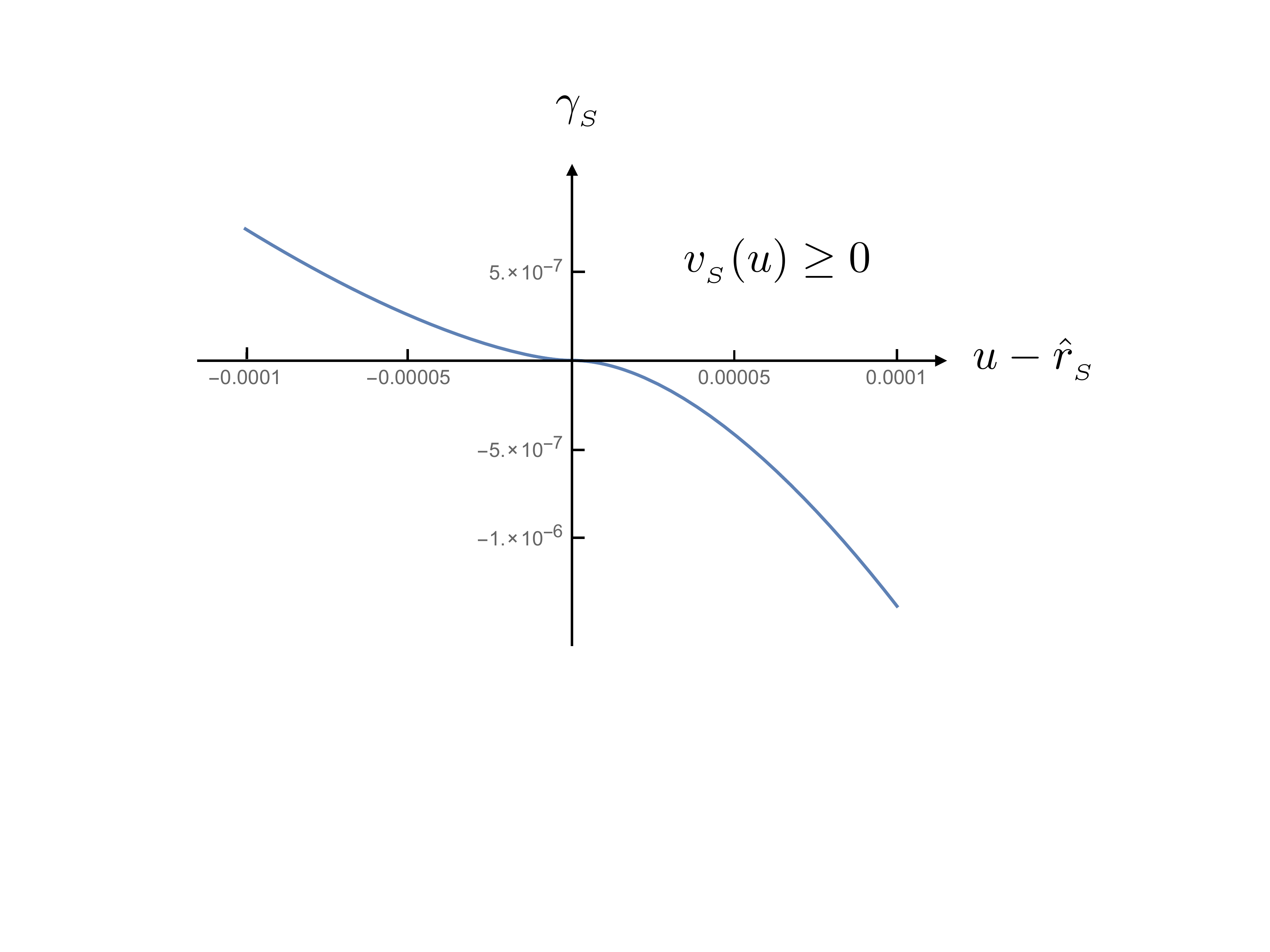}}
\caption{The function $\gamma_S(u)$, here plotted for ${\hat z}=0$, ${\hat R}_P=100$ and $\theta_P=\pi/3$, shows that of the two segments of the solution to (\ref{E141}) for which $v_S(u)\ge0$ (the upper segments in figure~\ref{F13}), only the segment $u\ge{\hat r}_S$ (on the right) satisfies the requirement in (\ref{E142}).}
\label{F14}
\end{figure} 

\begin{figure}
\centerline{\includegraphics[width=8.5cm]{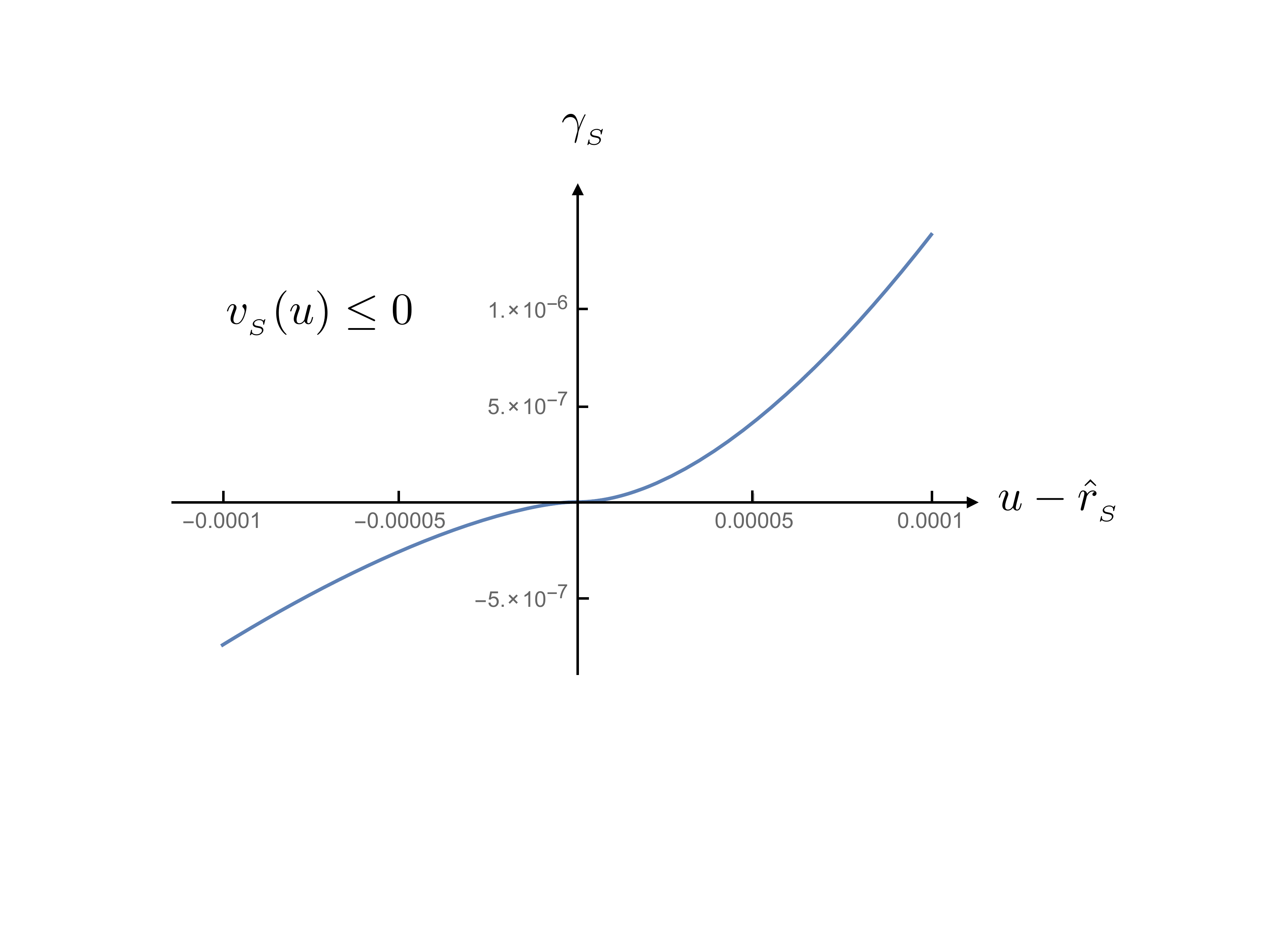}}
\caption{The function $\gamma_S(u)$, here plotted for ${\hat z}=0$, ${\hat R}_P=100$ and $\theta_P=\pi/3$, shows that of the two segments of the solution to (\ref{E141}) for which $v_S(u)\le0$ (the lower segments in figure~\ref{F13}), only the segment $u\le{\hat r}_S$ (on the left) satisfies the requirement in (\ref{E142}).}
\label{F15}
\end{figure} 

When the cusp curve of the bifurcation surface intersects the source distribution, as in figure~\ref{F11}, the point ${\hat r}={\hat r}_C$ (rather than ${\hat r}={\hat r}_L$) constitutes the lower boundary of the domain of integration with respect to ${\hat r}$ in (\ref{E85}).  In terms of the polar coordinates introduced in (\ref{E138}), the path of steepest descent for $\exp[-{\rm i} m\phi_-(w,\lambda)]$ through the boundary point ${\hat r}={\hat r}_C$ is given by the solution $\lambda=\lambda^-_C(w)$ of the transcendental equation  
\begin{equation}
\Re[\phi_-(w,\lambda)]=\phi_-\vert_{{\hat r}={\hat r}_C}\equiv\phi_C
\label{E143}
\end{equation}
that satisfies $\lambda_C^-(0)=-\pi/2$ and the condition
\begin{equation}
\gamma^-_C\equiv\Im[\phi_-(w,\lambda)]\big\vert_{\lambda=\lambda^-_C(w)}\leq0,
\label{E144}
\end{equation}
for all relevant values of $({\hat z}, {\hat r}_P, {\hat z}_P)$ and of the curve parameter $w$.  This path is designated as ${\cal L}_C$ in figure~\ref{F16}.  Note that the requirement $\lambda_C^-(0)=-\pi/2$ on the path issuing from $w=0$ is dictated by the fact that, of the two solutions $\lambda=\lambda(w)$ of (\ref{E143}) through ${\hat r}={\hat r}_C$, in $-\pi/2\le\lambda\le0$ and in $0\le\lambda\le\pi/2$, only the one reducing to $-\pi/2$ at $w=0$ for which ${\rm d}\lambda/{\rm d}w$ is negative can satisfy (\ref{E144}) [see (\ref{E140})].

In contrast to ${\cal L}_S$ that passes through the point ${\hat r}={\hat r}_S$ itself, the path ${\cal L}_C$ cannot be used as a contour of integration that includes the point ${\hat r}={\hat r}_C$ because the functions $G^{\rm out}_{nj}\vert_{\phi=\phi_\pm}$ which appear in the integrand of the integral in (\ref{E85}) are both divergent at ${\hat r}={\hat r}_C$.  To be able to apply Cauchy's integral theorem, we need to confine the domain of integration in the complex plane to one in which the integrand is analytic.  This may be done in the present case by determining the nature of the singularities of $G^{\rm out}_{nj}\vert_{\phi=\phi_\pm}$ at ${\hat r}={\hat r}_C$ and accordingly indenting the paths of steepest descent through this point to excise the singularities of $G^{\rm out}_{nj}\vert_{\phi=\phi_\pm}$ from the domain of integration.   

To approximate $G^{\rm out}_{nj}\vert_{\phi=\phi_\pm}$ in the neighbourhood of ${\hat r}={\hat r}_C$ (in order to determine the nature of their singularities at  this point), we may note that $\phi_\pm$ can be expanded into a Taylor series in powers of $({\hat r}-{\hat r}_C)^{1/2}$ to obtain
\begin{eqnarray}
\phi_\pm&=&\phi_C+\frac{{\hat r}_C^2-1}{{\hat r}_C{\hat R}_C}\left({\hat r}-{\hat r}_C\right)\pm\frac{[2{\hat r}_C({\hat r}_P^2-1)]^{3/2}}{3{\hat R}_C^3}\left({\hat r}-{\hat r}_C\right)^{3/2}\nonumber\\*
&&+\frac{({\hat r}_C^2-1)[{\hat r}_C^2({\hat r}_P^2-1)({\hat R}_C^2+4)+{\hat r}_C^2-1]}{2{\hat r}_C^2{\hat R}_C^5}({\hat r}-{\hat r}_C)^2\nonumber\\*
&&\mp\frac{{\hat r}_C^{1/2}({\hat r}_P^2-1)^{3/2}[3{\hat R}_C^4+4({\hat r}_C^2-1)(3{\hat R}_C^2+5)]}{5\sqrt2{\hat R}_C^7}({\hat r}-{\hat r}_C)^{5/2}+\cdots,
\label{E145}
\end{eqnarray}
for fixed $({\hat z}, {\hat r}_P, {\hat z}_P)$, where ${\hat R}_C=({\hat r}_C^2{\hat r}_P^2-1)^{1/2}$.  Insertion of this in (\ref{E41}) shows that the function $c_1$ appearing in the expression for $G^{\rm out}_{nj}\vert_{\phi=\phi_\pm}$ in (\ref{E70}) has the value
\begin{equation}
c_1\simeq\frac{2^{1/6}}{{\hat R}_C}\left[{\hat r}_C({\hat r}_P^2-1)\left({\hat r}-{\hat r}_C\right)\right]^{1/2} \left[1-\frac{3{\hat R}_C^4+4({\hat r}_C^2-1)(3{\hat R}_C^2+5)}{20{\hat r}_C{\hat R}_C^4}\left({\hat r}-{\hat r}_C\right)\right]
\label{E146}
\end{equation}
near ${\hat r}={\hat r}_C$.  Evaluating $c_1$, $p_j$ and $q_j$ near ${\hat r}={\hat r}_C$ from (\ref{E146}), (\ref{E67}) and (\ref{E68}) and inserting the results in (\ref{E70}), we arrive at
\begin{eqnarray}
G^{\rm out}_{nj}\Big\vert_{\phi=\phi_\pm}&\simeq&\frac{{\hat R}_C^{2-n}} {3{\hat r}_C^2{\hat r}_P({\hat r}_P^2-1)({\hat r}-{\hat r}_C)}\left[\left(\matrix{1\cr -{\hat R}_C\cr {\hat r}_C{\hat r}_P\cr}\right)\right.\nonumber\\*
&&\left.\pm\frac{2^{1/2}{\hat r}_C^{1/2}({\hat r}_P^2-1)^{1/2}({\hat r}-{\hat r}_C)^{1/2}}{{\hat R}_C^2}\left(\matrix{2{\hat R}_C^2+3^{n-1}\cr(-1)^{n-1}{\hat R}_C\cr 3^{n-1}{\hat r}_C{\hat r}_P\cr}\right)\right],\nonumber\\*
&&\qquad\qquad\qquad\qquad 0\le{\hat r}-{\hat r}_C\ll1.
\label{E147}
\end{eqnarray}
The integrand in (\ref{E85}) therefore has both a simple pole and a branch point at ${\hat r}={\hat r}_C$ which should be circumvented by an indentation of the integration contour.  
    
The semi-circular indentation designated as ${\cal L}_\epsilon$ in figure~\ref{F16} is described by
 \begin{equation}
  {\hat r}={\hat r}_C+\epsilon({\hat r}_S-{\hat r}_C)\exp({\rm i}\lambda),\qquad\lambda_0^-\le\lambda\le-\pi/2,
  \label{E148}
  \end{equation}   
where $\epsilon\ll1$ is a real constant and $\lambda_0^-$ is the value of $\lambda$ at which this circular arc intersects the path ${\cal L}_C$.  Since $\epsilon$ is small, the value of $\lambda_0^-$ can be determined from the approximate solution to $\Re(\phi_--\phi_C)=0$ for $w\ll1$.  

To derive approximate solutions to (\ref{E143}) in the vicinity of $w=0$, i.e., to find the paths ${\cal L}_C$ and ${\cal K}_C$ that are shown in figures~\ref{F16} and \ref{F17} when $w\ll1$, we can insert (\ref{E145}) in (\ref{E143}) and set the first two terms of the resulting Taylor expansions of $\Re(\phi_\pm-\phi_C)$ in powers of $w^{1/2}$ equal to zero.  This leads to the following two equations for the dependences $\lambda_C^\pm(w)$ of $\lambda$ on $w$ along ${\cal K}_C$ and ${\cal L}_C$ respectively:
\begin{equation}
\kappa\cos\left({\textstyle\frac{3}{2}}\lambda_C^\pm\right)\pm3\cos\lambda_C^\pm\simeq0,\qquad w\ll1,
\label{E149}
\end{equation}
where  
\begin{equation}
\kappa=\frac{2^{3/2}{\hat r}_C^{5/2}({\hat r}_P^2-1)^{3/2}w^{1/2}}{({\hat r}_C^2-1)({\hat r}_C^2{\hat r}_P^2-1)}.
\label{E150}
\end{equation}
The function $\cos({\textstyle\frac{3}{2}}\lambda_\pm)$ in (\ref{E149}) can be written in terms of $\cos({\textstyle\frac{1}{2}}\lambda_\pm)$ to obtain a cubic equation for $\cos({\textstyle\frac{1}{2}}\lambda_\pm)$ whose relevant roots, i.e., the roots satisfying $\lim_{w\to0}\lambda_C^\pm=-\pi/2$ and the constraints (\ref{E144}) and (\ref{E159}) below, are given by
\begin{equation} 
\lambda_C^\pm\simeq-2\arccos\left\{\mp\kappa^{-1}\left[\left(1+\kappa^2\right)^{1/2}\cos\left(\mu\pm{\textstyle\frac{2}{3}}\pi\right)+\textstyle{\frac{1}{2}}\right]\right\},\qquad w\ll1,
\label{E151}
\end{equation}
with
\begin{equation}
\mu=\frac{1}{3}\arccos\frac{1-{\textstyle\frac{3}{2}}\kappa^2}{(1+\kappa^2)^{3/2}}.
\label{E152}
\end{equation}
Note that, in contrast to their Taylor expansions in powers of $w^{1/2}$, the above expressions for $\lambda_C^\pm$ are valid also at ${\hat z}={\hat z}_P$ where $\kappa$ diverges. 

The angle $\lambda_0^-$ in the description of the indentation ${\cal L}_\epsilon$ in (\ref{E148}) is obtained by evaluating the above expression for $\lambda_C^-$ at $w=\epsilon({\hat r}_S-{\hat r}_C)$,
\begin{equation}
\lambda_0^-=\lambda_C^-\big\vert_{w=\epsilon({\hat r}_S-{\hat r}_C)}.
\label{E153}
\end{equation}
Note that when $w=\epsilon({\hat r}_S-{\hat r}_C)$ and the observation point is sufficiently close to the equatorial plane $\theta_P=\pi/2$ for $\vert{\hat z}-{\hat z}_P\vert$ to be small throughout the source distribution (\ref{E7}), $\kappa$ assumes a small value,
\begin{equation}
\kappa\big\vert_{w=\epsilon({\hat r}_S-{\hat r}_C)}\simeq2\epsilon^{1/2},\qquad\quad\vert{\hat z}-{\hat z}_P\vert\ll1, 
\label{E154}
\end{equation}
[see (\ref{E39}) and (\ref{E110})].  From the following Taylor expansion of (\ref{E151}) in powers of $\kappa$
\begin{equation}
\lambda_C^\pm=-\pi/2\pm\kappa/(3\sqrt2)+\cdots,\quad w\ll1,\quad {\hat z}\ne{\hat z}_P,
\label{E155}
\end{equation} 
it follows, therefore, that $\lambda_C^\pm+\pi/2$ approach zero like $\epsilon^{1/2}$ as $\epsilon$ tends to zero.  That this holds true also when ${\hat z}={\hat z}_P$ follows from a corresponding numerical analysis based on the exact expression for $\Re(\phi_\pm-\phi_C)$. 

Finally, the path of steepest descent of $\exp[-{\rm i}m\phi_-(u,v)]$ through the boundary point ${\hat r}={\hat r}_U$ is given by the solution $u=u^-_U(v)$ of the transcendental equation
\begin{equation}
\Re[\phi_-(u,v)]=\phi_-({\hat r}_U,0)\equiv\phi^-_U
\label{E156}
\end{equation}
that satisfies $u_U^-(0)={\hat r}_U$ and the condition
\begin{equation}
\gamma^-_U\equiv\Im[\phi_-(u,v)]\big\vert_{u=u^-_U(v)}\leq0.
\label{E157}
\end{equation}
In contrast to (\ref{E141}) which was solved for $v$ as a function of $u$, (\ref{E156}) has to be solved for $u$ as a function of $v$ because, otherwise, the Jacobian of the transformation from ${\hat r}$ to $u+{\rm i}v(u)$ would diverge at the point ${\hat r}={\hat r}_U$ [see (\ref{E136})].  From the plots of $\gamma_U^-$ as a function of $v$ for $u=u^-_U(v)\le0$ and $u=u^-_U(v)\ge0$, similar to those shown in figures~\ref{F14} and \ref{F15}, it follows that the requirement expressed in (\ref{E157}) is met only by the segment of $u^-_U(v)$ that lies in $v\ge0$.  This segment which constitutes the path of steepest descent through ${\hat r}={\hat r}_U$ is designated as ${\cal L}_U$ in figure~\ref{F16}. 

\begin{figure}
\centerline{\includegraphics[width=11cm]{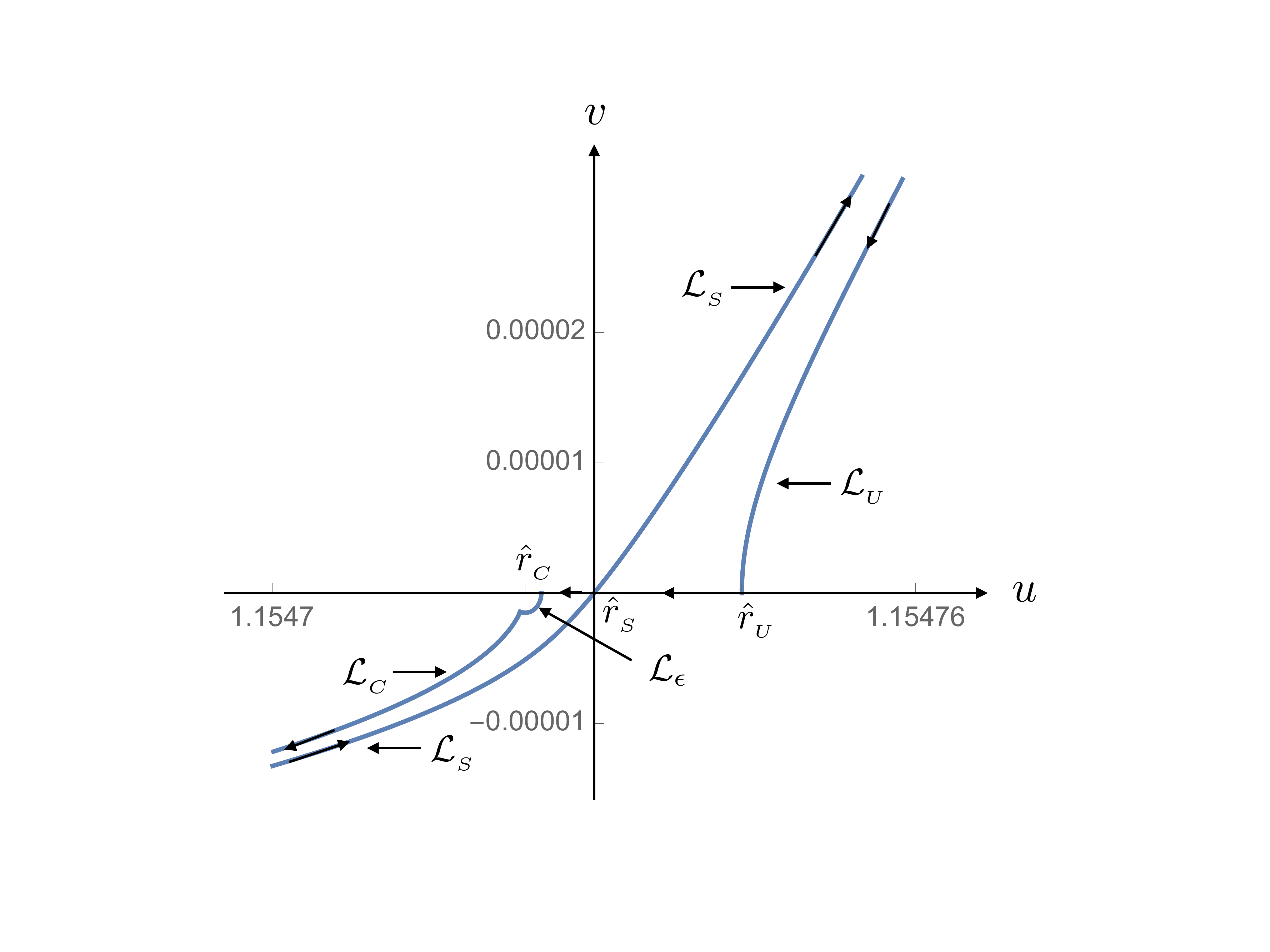}}
\caption{The complex ${\hat r}=u+{\rm i}v$ plane with a shift in the position of the imaginary axis which places the saddle point $({\hat r}_S,0)$ of $\phi_-(u,v,{\hat z},{\hat R}_P,\theta_P)$ at the origin.  The curves ${\cal L}_S$, ${\cal L}_C$ and ${\cal L}_U$ delineate the paths of steepest descent of $\exp(-{\rm i}m\phi_-)$ through the following critical points, respectively: the saddle point $({\hat r}_S,0)$, the cusp point $({\hat r}_C,0)$ and the boundary point $({\hat r}_U,0)$.  Here, the cusp point lies between the lower and upper boundaries $({\hat r}_L,0)$ and $({\hat r}_U,0)$ of the source distribution (see figure~\ref{F11}).  The segment ${\hat r}_C+\epsilon\le u\le{\hat r}_U$ of the real axis, together with ${\cal L}_S$, ${\cal L}_C$, ${\cal L}_U$ and the indentation ${\cal L}_\epsilon$, surrounding the singularity of $G^{\rm out}_j\vert_{\phi=\phi_-}$ at the cusp point, constitute the contours of integration for the evaluation of the part $[{\bf E}^{\rm b}_-\quad{\bf B}^{\rm b}_-]$ of the field given by (\ref{E164}).  The arrows show the adopted directions of integration along the various contours.  This figure is plotted for the following set of values of the parameters: ${\hat R}_P=10^2$, $\theta_P=\pi/3$, ${\hat z}=0$, $m=10$ and ${\hat r}_U=1.15474$.}
\label{F16}
\end{figure}

\subsection{Paths of steepest descent through the critical points of the phase ${\phi}_+$ for observation points in $\theta_L\le\theta_P\le\theta_U$ or $\pi-\theta_U\le\theta_P\le\pi-\theta_L$}
\label{subsec:PhiPlus1}

In contrast to ${\hat\varphi}_-$, the function ${\hat\varphi}_+$ that appears in the expression for $\matrix{[{\bf E}^{\rm b}_+&{\bf B}^{\rm b}_+]}$ in (\ref{E85}) has no extrema.  So, at observation points for which ${\hat r}_L<{\hat r}_C$, the kernel $\exp(-{\rm i}m{\hat\varphi}_+)$ of the ${\hat r}$-integral in (\ref{E85}) has only two critical points: the point ${\hat r}={\hat r}_C$ at which the cusp intersects the source distribution, at a given ${\hat z}$, and the boundary point ${\hat r}={\hat r}_U$.  Since $c_1$ vanishes at ${\hat r}={\hat r}_C$, the value of the Green's function $G_{nj}^{\rm out}$ on $\phi=\phi_+$, too, diverges at this point [see (\ref{E147})].   As in \S~\ref{subsec:PhiMinus1}, therefore, we need to excise this singularity from the domain of integration in the complex plane by introducing an indentation in the path of steepest descent through ${\hat r}={\hat r}_C$.

The path of steepest descent for $\exp[-{\rm i} m\phi_+(w,\lambda)]$ through the boundary point ${\hat r}={\hat r}_C$ is given by the solution $\lambda=\lambda^+_C(w)$ of the transcendental equation  
\begin{equation}
\Re[\phi_+(w,\lambda)]=\phi_+\vert_{{\hat r}={\hat r}_C}=\phi_C
\label{E158}
\end{equation}
that satisfies $\lambda^+_C(0)=-\pi/2$ and the condition
\begin{equation}
\gamma^+_C\equiv\Im[\phi_+(w,\lambda)]\big\vert_{\lambda=\lambda^+_C(w)}\leq0,
\label{E159}
\end{equation}
for all relevant values of $({\hat z}, {\hat r}_P, {\hat z}_P)$ and of the curve parameter $w$.  From the plots of $\gamma^+_C$ as a function of $w$ for $-\pi/2\le\lambda^+_C\le0$ and $0\le\lambda^+_C\le\pi/2$, similar to those shown in figure~\ref{F14} and \ref{F15}, it follows that the requirement expressed in (\ref{E159}) is met only by the segment of $\lambda^+_C(w)$ that lies in $-\pi/2\le\lambda\le-\pi/3$.  This path is designated as ${\cal K}_C$ in figure~\ref{F17}.   

The semi-circular indentation designated as ${\cal K}_\epsilon$ in figure~\ref{F17} is described by
\begin{equation}
{\hat r}={\hat r}_C+\epsilon({\hat r}_S-{\hat r}_C)\exp({\rm i}\lambda),\qquad-\pi/2\le\lambda\le\lambda^+_0,
\label{E160}
\end{equation}
where $\epsilon$ is the small parameter appearing in the description of ${\cal L}_\epsilon$, (\ref{E148}), and
\begin{equation}
\lambda^+_0=\lambda_C^+\Big\vert_{w=\epsilon({\hat r}_S-{\hat r}_C)}
\label{E161}
\end{equation}
is the angle at which this circular arc intersects ${\cal K}_C$ [see (\ref{E151})].  

The path of steepest descent for $\exp(-{\rm i}m{\hat\varphi}_+)$ through the boundary point ${\hat r}={\hat r}_U$ is given by the solution $u=u^+_U(v)$ of the transcendental equation  
\begin{equation}
\Re[\phi_+(u,v)]=\phi_+({\hat r}_U,0)\equiv\phi_U^+,
\label{E162}
\end{equation}
which satisfies the condition
\begin{equation}
\gamma^+_U\equiv\Im[\phi_+(u,v)]\big\vert_{u=u^+_U(v)}\leq0.
\label{E163}
\end{equation}
Invoking, this time, the plots of $\gamma^+_U$ as a function of $v$ for $u_U^+(v)\le0$ and $u_U^+(v)\ge0$, we find that the requirement expressed in (\ref{E163}) is met only by the segment of $u^+_U(v)$ that lies in $v\le0$.  This path is designated as ${\cal K}_U$ in figure~\ref{F17}.   
 
\begin{figure}
\centerline{\includegraphics[width=10cm]{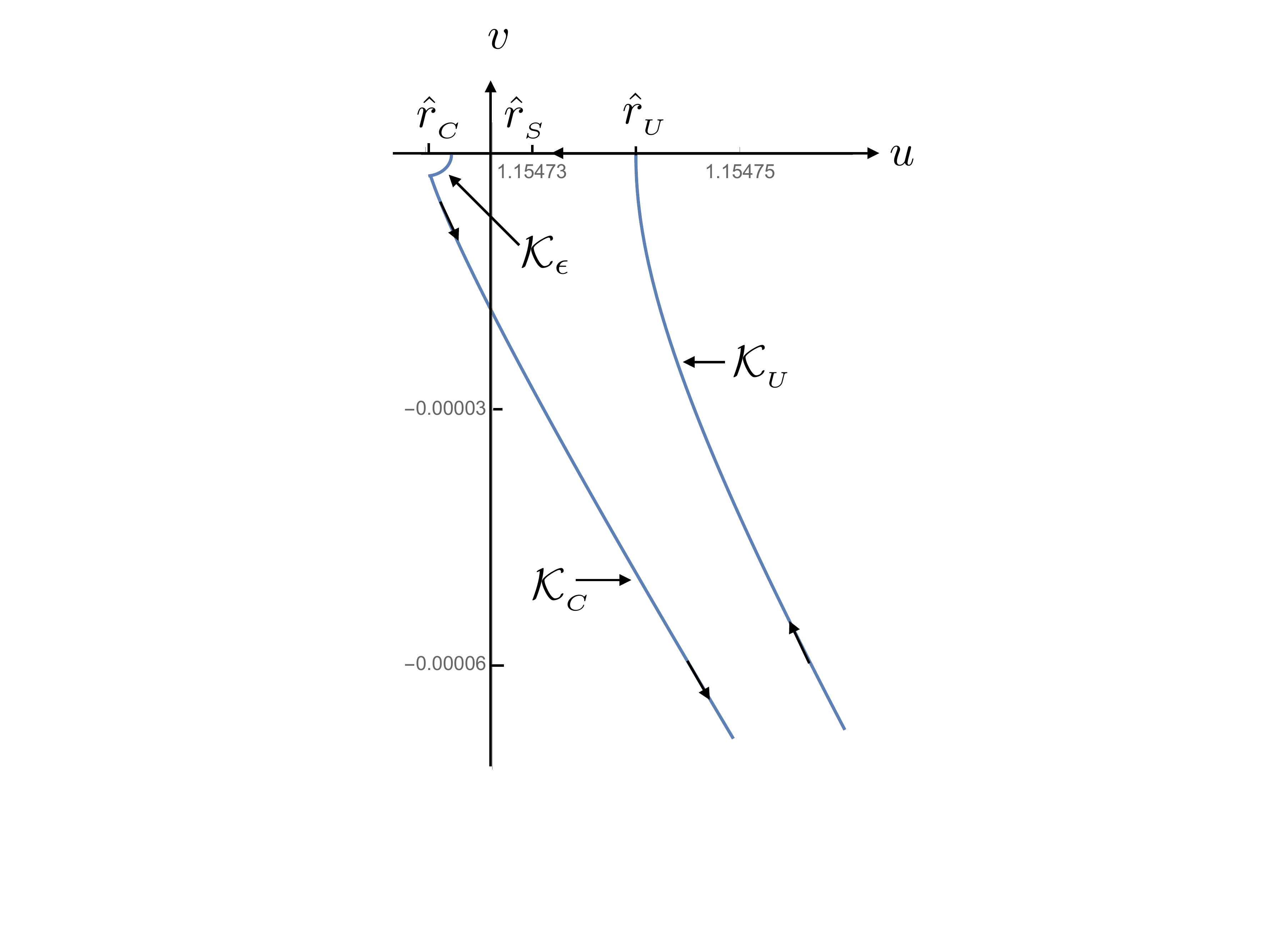}}
\caption{The complex ${\hat r}=u+{\rm i}v$ plane with a shift in the position of the imaginary axis which places the saddle point $({\hat r}_S,0)$ of $\phi_-(u,v,{\hat z},{\hat R}_P,\theta_P)$ at the origin.  The curves ${\cal K}_C$ and ${\cal K}_U$ delineate the paths of steepest descent of $\exp(-{\rm i}m\phi_+)$ through the cusp point $({\hat r}_C,0)$ and the boundary point $({\hat r}_U,0)$, respectively.  Here, the cusp point lies between the lower and upper boundaries $({\hat r}_L,0)$ and $({\hat r}_U,0)$ of the source distribution (see figure~\ref{F11}).  The segment ${\hat r}_C+\epsilon\le u\le{\hat r}_U$ of the real axis, together with ${\cal K}_S$, ${\cal K}_U$ and the indentation ${\cal K}_\epsilon$, surrounding the singularity of $G^{\rm out}_j\vert_{\phi=\phi_+}$ at the cusp point, constitute the contours of integration for the evaluation of the part $[{\bf E}^{\rm b}_+\quad{\bf B}^{\rm b}_+]$ of the field given by (\ref{E166}).  The arrows show the adopted directions of integration along the various contours.  This figure is plotted for the same set of values of the parameters as those for figure~\ref{F16}.}
\label{F17}
\end{figure}

\subsection{Asymptotic value of $[{\bf E}^{\rm b}_-\quad{\bf B}^{\rm b}_-]$ for large $m$ in $\theta_L\le\theta_P\le\theta_U$ or $\pi-\theta_U\le\theta_P\le\pi-\theta_L$}
\label{subsec:AsymptoticForMinusCin}

Having gone into the complex plane ${\hat r}=u+{\rm i}v$, and having delineated the paths of steepest descent for the factor $\exp(-{\rm i}m{\hat\varphi}_-)$ in the integrand of the expression for the field $\matrix{[{\bf E}^{\rm b}_-&{\bf B}^{\rm b}_-]}$ in (\ref{E85}), we are now in a position to use Cauchy's theorem to replace the ${\hat r}$-integral over the segment ${\hat r}_C\le{\hat r}\le{\hat r}_U$ of the real axis in this expression by the sum of a set of integrals over the steepest-descent paths ${\cal L}_C$, ${\cal L}_S$ and ${\cal L}_U$ passing through the critical points of this integral and over the semi-circular path ${\cal L}_\epsilon$ bypassing the singularity of its integrand (see figure~\ref{F16}).  Since the integrals over the steepest-descent paths each have a kernel that exponentially decays away from the critical points ${\hat r}={\hat r}_C$, ${\hat r}={\hat r}_S$ and ${\hat r}={\hat r}_U$, the requirement (set by Cauchy's theorem) that these paths should form a closed contour together with the segment ${\hat r}_C\le u\le{\hat r}_U$ of the real axis is not essential for obtaining an asymptotic approximation to the value of $\matrix{[{\bf E}^{\rm b}_-&{\bf B}^{\rm b}_-]}$.  Even for moderate values ($\sim 10$) of the integer $m$ that appears in the arguments of the exponential factors $\exp(m\gamma^-_C)$, $\exp(m\gamma^-_S)$ and $\exp(m\gamma^-_U)$ in (\ref{E164}) below, accurate values of the integrals over the steepest-descent paths can be obtained by performing each integration over only a limited segment of the corresponding path adjacent to the critical point from which it issues.  The length of the segment over which each integral needs to be evaluated is dictated by the value of $m$ and the degree of required accuracy.  For a given level of accuracy, the larger the value of the harmonic number $m$, the shorter is the required segment.  

Disregarding the negligible contributions from any connecting paths away from the critical points ${\hat r}={\hat r}_C$, ${\hat r}={\hat r}_S$, and ${\hat r}={\hat r}_U$ that may be needed to construct a closed contour out of ${\cal L}_C$, ${\cal L}_S$, ${\cal L}_U$, ${\cal L}_\epsilon$ and the segment of the real axis between ${\hat r}_C$ and ${\hat r}_U$ in figure~\ref{F16}, we can write the ${\hat r}$-integral in the expression for $\matrix{[{\bf E}^{\rm b}_-&{\bf B}^{\rm b}_-]}$ (which extends over ${\hat r}_C\le{\hat r}\le{\hat r}_U$) as the sum of the integrals along the steepest-descent paths ${\cal L}_C$, ${\cal L}_S$, ${\cal L}_U$ and ${\cal L}_\epsilon$ since the path along the real axis is traversed in the direction of decreasing ${\hat r}$ \citep[see, e.g.,][]{BenderOrszag}.  The exponent $-{\rm i}m{\hat\varphi}_-$ of the exponential factor in (\ref{E85}) has the values $m[\gamma_C^--{\rm i}(\phi_C+{\hat\varphi}_P)]$, $m[\gamma_S-{\rm i}(\phi_S+{\hat\varphi}_P)]$ and $m[\gamma_U^--{\rm i}(\phi_U^-+{\hat\varphi}_P)]$ along ${\cal L}_C$, ${\cal L}_S$ and ${\cal L}_U$, respectively [see (\ref{E38}), (\ref{E141})-- (\ref{E144}), (\ref{E156}) and (\ref{E157})].  Hence, in cases where the cusp locus $C$ of the bifurcation surface associated with the observation point $P$ intersects the source distribution (\ref{E7}), the asymptotic value of $\matrix{[{\bf E}^{\rm b}_-&{\bf B}^{\rm b}_-]}$ for large $m$ is given by 
\begin{eqnarray} 
\left[\matrix{{\bf E}^{\rm b}_-\cr{\bf B}^{\rm b}_-\cr}\right]&\simeq&{\textrm i}m\exp(-{\rm i}m{\hat\varphi}_P)\int_{-{\hat z}_0}^{{\hat z}_0}{\textrm d}{\hat z}\Bigg\{\exp(-{\rm i}m\phi_S)\int_{{\cal L}_S}{\rm d}u\,\exp(m\gamma_S) J_u^- \left[\matrix{{\bf \Lambda}_-\cr{\bf \Gamma}_-\cr}\right]\Bigg\vert_{{\hat r}=u+{\rm i}v_S}\nonumber\\*
&&+\exp(-{\rm i}m\phi_U^-)\int_{{\cal L}_U}{\rm d}v\,\exp(m\gamma^-_U)J_v^-\left[\matrix{{\bf \Lambda}_-\cr{\bf \Gamma}_-\cr}\right]\Bigg\vert_{{\hat r}=u_U^-+{\rm i}v}\nonumber\\*
&&+\exp(-{\rm i}m\phi_C)\int_{{\cal L}_C}{\rm d}w\,\exp(m\gamma^-_C) J_w^-\left[\matrix{{\bf \Lambda}_-\cr{\bf \Gamma}_-\cr}\right]\Bigg\vert_{{\hat r}={\hat r}_C+w\exp({\rm i}\lambda_C^-)}\nonumber\\*
&&+{\rm i}\epsilon({\hat r}_S-{\hat r}_C)\int_{-\pi/2}^{\lambda^-_0}{\rm d}\lambda\,\exp[{\rm i}(\lambda-m\phi_-)]\left[\matrix{{\bf \Lambda}_-\cr{\bf \Gamma}_-\cr}\right]\Bigg\vert_{{\hat r}={\hat r}_C+\epsilon({\hat r}_S-{\hat r}_C)\exp({\rm i}\lambda)}\Bigg\},\nonumber\\*
&&\nonumber\\*
&&\qquad\qquad\qquad m\gg1,\qquad\theta_L\le\theta_P\le\theta_U,\qquad\pi-\theta_U\le\theta_P\le\pi-\theta_L,
\label{E164}
\end{eqnarray}
where
\begin{equation}
\left[\matrix{{\bf \Lambda}_\pm\cr{\bf \Gamma}_\pm\cr}\right]
=\sum_{n=1}^2\sum_{j=1}^3\frac{{\hat r}(p_{nj}\pm2c_1q_{nj})}{3 c_1^2}\left[\matrix{{\tilde{\bf u}}_{nj}\cr{\tilde{\bf v}}_{nj}\cr}\right]
\label{E165}
\end{equation}
[see (\ref{E70}) and (\ref{E85})].  In this expression, the functions $\theta_L$ and $\theta_U$ and the Jacobians $J_u^-$, $J_v^-$ and $J_w^-$ are defined in (\ref{E114}), (\ref{E115}), (\ref{E136}), (\ref{E137}) and (\ref{E140}), respectively, and the value of $\lambda^-_0$ is given by (\ref{E153}).  

Integrand of the integral over ${\cal L}_S$ in (\ref{E164}) is singular at $u=0$, ${\hat z}={\hat z}_P$.  From the approximate expression for the integrand of the original integral in (\ref{E85}) in the vicinity of ${\hat r}={\hat r}_S$ at a given value of ${\hat z}\ne{\hat z}_P$, we find, however, that this singularity is integrable: outcome of the integration with respect to ${\hat r}$ of the approximate expression in question turns out to have a logarithmic singularity at ${\hat z}={\hat z}_P$.

The numerical computations described in \S~\ref{sec:numerical} show that the combined contributions of the paths ${\cal L}_C$, ${\cal K}_C$, ${\cal L}_U$ and ${\cal K}_U$ (in figures~\ref{F16} and \ref{F17}) toward the value of the field decays spherically with distance.  The non-spherically decaying contribution -- which turns out to be more dominant and less steeply diminishing with distance the larger the value of $m$ -- is that arising from the path ${\cal L}_S$ which goes through the saddle point at ${\hat r}={\hat r}_S$.

\subsection{Asymptotic value of $[{\bf E}^{\rm b}_+\quad{\bf B}^{\rm b}_+]$ for large $m$ in $\theta_L\le\theta_P\le\theta_U$ or $\pi-\theta_U\le\theta_P\le\pi-\theta_L$}
\label{subsec:AsymptoticForPlusCin}

The phase ${\hat\varphi}_+$ of the exponential in the expression for $\matrix{[{\bf E}^{\rm b}_+&{\bf B}^{\rm b}_+]}$ in (\ref{E85}) has no extrema but the function $G_{nj}^{\rm out}\vert_{\phi=\phi_+}$ that multiplies this exponential is singular [see (\ref{E147})].  Once the singularity of its integrand at ${\hat r}={\hat r}_C$ is circumvented by means of the indentation ${\cal K}_\epsilon$ shown in figure\ref{F17}, the ${\hat r}$-integral over ${\hat r}_C\le{\hat r}\le{\hat r}_U$ in (\ref{E85}) can be approximated by the sum of the integrals over the steepest-descent paths ${\cal K}_C$ and ${\cal K}_U$ and the indentation ${\cal K}_\epsilon$ to obtain 
\begin{eqnarray}
\left[\matrix{{\bf E}^{\rm b}_+\cr{\bf B}^{\rm b}_+\cr}\right]&\simeq&{\textrm i}m\exp(-{\rm i}m{\hat\varphi}_P)\int_{-{\hat z}_0}^{{\hat z}_0}{\textrm d}{\hat z}\Bigg\{\exp(-{\rm i}m\phi_U^+)\int_{{\cal K}_U}{\rm d}v\,\exp(m\gamma^+_U)J_v^+\left[\matrix{{\bf \Lambda}_+\cr{\bf \Gamma}_+\cr}\right]\Bigg\vert_{{\hat r}=u_U^++{\rm i}v}\nonumber\\*
&&+\exp(-{\rm i}m\phi_C)\int_{{\cal K}_C}{\rm d}w\,\exp(m\gamma^+_C)J_w^+\left[\matrix{{\bf \Lambda}_+\cr{\bf \Gamma}_+\cr}\right]\Bigg\vert_{{\hat r}={\hat r}_C+w\exp({\rm i}\lambda_C^+)}\nonumber\\*
&&+{\rm i}\epsilon({\hat r}_S-{\hat r}_C)\int_{-\pi/2}^{\lambda^+_0}{\rm d}\lambda\,\exp[{\rm i}(\lambda-m\phi_+)]\left[\matrix{{\bf \Lambda}_+\cr{\bf \Gamma}_+\cr}\right]\Bigg\vert_{{\hat r}={\hat r}_C+\epsilon({\hat r}_S-{\hat r}_C)\exp({\rm i}\lambda)}\Bigg\},\nonumber\\*
&&\nonumber\\*
&&\qquad\qquad\qquad m\gg1,\qquad\theta_L\le\theta_P\le\theta_U,\qquad\pi-\theta_U\le\theta_P\le\pi-\theta_L,\quad
\label{E166}
\end{eqnarray}
where the functions ${\bf\Lambda}_+$, ${\bf \Gamma}_+$, $J_v^+$ and $\lambda^+_0$ are given by (\ref{E165}),  (\ref{E137}) and (\ref{E161}), respectively.  The domain of validity of this expression in the space of observation points, which is the same as that of (\ref{E164}), is shown in figure~\ref{F12}.

\subsection{Resultant of the boundary fields in $\theta_L\le\theta_P\le\theta_U$ or $\pi-\theta_U\le\theta_P\le\pi-\theta_L$}
\label{subsec:Resultant}

The difference between the expressions in (\ref{E166}) and (\ref{E164}), which constitutes the part of the radiation field $\matrix{[{\bf E}&{\bf B}]}$ denoted as $\matrix{[{\bf E}^{\rm b}_+-{\bf E}^{\rm b}_-&{\bf B}^{\rm b}_+-{\bf B}^{\rm b}_-]}$ [see (\ref{E75})-(\ref{E77})], can be simplified by noting that not only do the leading terms in the Laurent expansions of $G_{nj}^{\rm out}\vert_{\phi=\phi_-}$ and $G_{nj}^{\rm out}\vert_{\phi=\phi_+}$ about the point ${\hat r}={\hat r}_C$ equal one another [see (\ref{E147})], but also $\lambda_0^-$ and $\lambda_0^+$ both approach the value $-\pi/2$ in the limit $\epsilon\to0$ [see (\ref{E153}) and (\ref{E161})].  This means that, as $\epsilon$ tends to zero, the values both of the integrands and of the integration limits in the integrals over $\lambda$ in (\ref{E164}) and (\ref{E166}) approach one another thus rendering the divergent parts of these two integrals equal.
 
From (\ref{E147}) and the corresponding expansions of $\exp(m\gamma_C^\pm)J_w^\pm$ in powers of $w^{1/2}$ it follows that, in the vicinity of the singular point $w=0$, the difference between the integrands of the integrals over ${\cal L}_C$ and ${\cal K}_C$ in (\ref{E164}) and (\ref{E166}) is given by
\begin{eqnarray}
\lefteqn{\exp(m\gamma^+_C)J_w^+\left[\matrix{{\bf \Lambda}_+\cr{\bf \Gamma}_+\cr}\right]\Bigg\vert_{{\hat r}={\hat r}_C+w\exp({\rm i}\lambda_C^+)}-\exp(m\gamma^-_C) J_w^- \left[\matrix{{\bf \Lambda}_-\cr{\bf \Gamma}_-\cr}\right]\Bigg\vert_{{\hat r}={\hat r}_C+w\exp({\rm i}\lambda_C^-)}}\nonumber\\*
&\qquad\simeq&\sum_{n=1}^2\frac{{\hat R}_C^{2-n}}{3{\hat r}_C{\hat r}_P ({\hat r}_P^2-1)w^{1/2}}\Bigg\{\frac{1}{w^{1/2}}[J_w^+\exp(m\gamma_C^+-{\rm i}\lambda_C^+)-J_w^-\exp(m\gamma_C^--{\rm i}\lambda_C^-)]\nonumber\\*
&&\times\Bigg(\left[\matrix{{\tilde{\bf u}}_{n1}\cr{\tilde{\bf v}}_{n1}\cr}\right]-{\hat R}_C\left[\matrix{{\tilde{\bf u}}_{n2}\cr{\tilde{\bf v}}_{n2}\cr}\right]+{\hat r}_C{\hat r}_P\left[\matrix{{\tilde{\bf u}}_{n3}\cr{\tilde{\bf v}}_{n3}\cr}\right]\Bigg)\nonumber\\*
&&+\frac{[2{\hat r}_C({\hat r}_P^2-1)]^{1/2}}{{\hat R}_C^2}[J_w^+\exp(m\gamma_C^+-{\rm i}\lambda_C^+/2)+J_w^-\exp(m\gamma_C^--{\rm i}\lambda_C^-/2)]\nonumber\\*
&&\times\Bigg((2{\hat R}_C^2+3^{n-1})\left[\matrix{{\tilde{\bf u}}_{n1}\cr{\tilde{\bf v}}_{n1}\cr}\right]+(-1)^{n-1}{\hat R}_C\left[\matrix{{\tilde{\bf u}}_{n2}\cr{\tilde{\bf v}}_{n2}\cr}\right]+3^{n-1}{\hat r}_C{\hat r}_P\left[\matrix{{\tilde{\bf u}}_{n3}\cr{\tilde{\bf v}}_{n3}\cr}\right]\Bigg)\Bigg\},\nonumber\\*
&&\nonumber\\*
&& \qquad\qquad\qquad\qquad\qquad\qquad\qquad\qquad\qquad\qquad\qquad\qquad w\ll1,
\label{E167}
\end{eqnarray}
a function whose singularity at $w=0$ is integrable: it can be seen from (\ref{E155}) and the corresponding expansions
\begin{equation}
J_w^\pm=-{\rm i}\pm\kappa/(2\sqrt{2})+\cdots,\qquad w\ll1,\quad {\hat z}\ne{\hat z}_P,
\label{E168}
\end{equation}
\begin{equation}
\gamma_C^\pm=-\frac{{\hat r}_C^2-1}{{\hat r}_C{\hat R}_C}w\pm\frac{2{\hat r}_C^{3/2}({\hat r}_P^2-1)^{3/2}}{3{\hat R}_C^3}w^{3/2}+\cdots,\qquad\quad w\ll1,
\label{E169}
\end{equation}
that the factor $J_w^+\exp(m\gamma_C^+)-J_w^-\exp(m\gamma_C^-)$ vanishes like $w^{1/2}$ in the limit $w\to0$, so that the right-hand side of (\ref{E167}) diverges as $w^{-1/2}$ in this limit.  This can be shown to hold true also for ${\hat z}={\hat z}_P$ by means of a numerical computation.    

In other words, the non-integrable singularities of the two integrands in the integrals over ${\cal L}_C$ and ${\cal K}_C$ partially cancel to yield an integrable singularity.  That this makes the integrals over the indentations ${\cal L}_\epsilon$ and ${\cal K}_\epsilon$ (which were introduced to circumvent the non-integrable singularities of $[\matrix{{\bf\Lambda}_\pm&{\bf\Gamma}_\pm}]$ at $w=0$) superfluous is, at the same time, confirmed by the fact that the integrals over $\lambda$ cancel out of the expression for $\matrix{[{\bf E}^{\rm b}_+-{\bf E}^{\rm b}_-&{\bf B}^{\rm b}_+-{\bf B}^{\rm b}_-]}$ in the limit $\epsilon\to0$. 

Combining (\ref{E164}) and (\ref{E166}), we therefore find that
\begin{eqnarray}
&&\left[\matrix{{\bf E}^{\rm b}_+-{\bf E}^{\rm b}_-\cr{\bf B}^{\rm b}_+-{\bf B}^{\rm b}_-\cr}\right]\simeq{\textrm i}m\exp(-{\rm i}m{\hat\varphi}_P)\int_{-{\hat z}_0}^{{\hat z}_0}{\textrm d}{\hat z}\nonumber\\*
&&\quad\times\Bigg\{
\exp(-{\rm i}m\phi_U^+)\int_{{\cal K}_U}{\rm d}v\,\exp(m\gamma^+_U)J^+_v
\left[\matrix{{\bf \Lambda}_+\cr{\bf \Gamma}_+\cr}\right]\Bigg\vert_{{\hat r}=u_U^++{\rm i}v}\nonumber\\*
&&\qquad\,\,\,-\exp(-{\rm i}m\phi_U^-)\int_{{\cal L}_U}{\rm d}v\,\exp(m\gamma^-_U)J^-_v \left[\matrix{{\bf \Lambda}_-\cr{\bf \Gamma}_-\cr}\right]\Bigg\vert_{{\hat r}=u_U^-+{\rm i}v}\nonumber\\*
&&\qquad\,\,\,-\exp(-{\rm i}m\phi_S)\int_{{\cal L}_S}{\rm d}u\,\exp(m\gamma_S) J^-_u \left[\matrix{{\bf \Lambda}_-\cr{\bf \Gamma}_-\cr}\right]\Bigg\vert_{{\hat r}=u+{\rm i}v_S}\nonumber\\*
&&\qquad\,\,\,+\exp(-{\rm i}m\phi_C)\lim_{\epsilon\to0}\int_{\epsilon({\hat r}_S-{\hat r}_C)}^{w_0}{\rm d}w\,\sum_{\iota=\pm}\iota\exp(m\gamma^\iota_C)J_w^\iota\left[\matrix{{\bf \Lambda}_\iota\cr{\bf \Gamma}_\iota\cr}\right]\Bigg\vert_{{\hat r}={\hat r}_C+w\exp({\rm i}\lambda_C^\iota)}\Bigg\},\nonumber\\*
&&\nonumber\\*
&&\qquad\qquad\qquad m\gg1,\qquad\theta_L\le\theta_P\le\theta_U,\quad\quad\pi-\theta_U\le\theta_P\le\pi-\theta_L,
\label{E170}
\end{eqnarray}
where $w_0$ is a constant of the order of unity denoting the value of $w$ beyond which any contributions from the points along the paths ${\cal L}_C$ and ${\cal K}_C$ are negligible.  This expression applies to the case where the cusp locus $C$ intersects the source distribution.

The fact that the singularities of the integrals over ${\cal L}_C+{\cal K}_C$ and over ${\cal L}_S$ are integrable even when the ranges of these integrals include values of ${\hat z}$ that match that of ${\hat z}_P$ implies that the coincidence of the loci $C$ and $S$ at ${\hat z}={\hat z}_P$ (in figure~\ref{F11}) does not vitiate the applicability of the steepest-descent method used to evaluate the ${\hat r}$-integral in (\ref{E85}).  The length of the path connecting $C$ and $S$ along the real axis of the complex $(u,v)$-plane (in figure~\ref{F16}) shrinks to zero for those source elements whose ${\hat z}$-coordinate equals the ${\hat z}_P$-coordinate of the observation point [see (\ref{E110})].  There is nevertheless a non-zero contribution toward the value of the ${\hat r}$-integral in question from this path in the limit ${\hat z}\to{\hat z}_P$ because the coalescence of $C$ and $S$ results, at the same time, in a higher-order singularity of the integrand in (\ref{E85}): all three derivatives ($\partial g/\partial{\hat r}$, $\partial g/\partial\varphi$ and $\partial g/\partial{\hat z}$) of the argument of the Dirac delta function in (\ref{E32}) with respect to the source coordinates, as well as the second derivative $\partial^2 g/\partial\varphi^2$ simultaneously vanish at the site (${\hat r}=1$, $\varphi=\varphi_P-3\pi/2$, ${\hat z}={\hat z}_P$) of this coalescence (see \S~\ref{subsec:locus}).

\subsection{Resultant of the boundary fields in $\theta_U\le\theta_P\le\pi-\theta_U$}
\label{subsec:Resultant2}

When the source distribution lies in $\Delta>0$ in its entirety and neither of the loci $C$ and $S$ intersect it (see figure~\ref{F11}), there is no need to invoke the method of steepest descent for evaluating the ${\hat r}$-integrals in the expressions for $\matrix{[{\bf E}^{\rm b}_\pm&{\bf B}^{\rm b}_\pm]}$.  Given that the phases $\phi_\pm$ are similar functions of ${\hat r}$ in this case, it is simpler and so more convenient to evaluate the following combination of the two integrals in (\ref{E85}) directly,
\begin{eqnarray}
\left[\matrix{{\bf E}^{\rm b}_+-{\bf E}^{\rm b}_-\cr{\bf B}^{\rm b}_+-{\bf B}^{\rm b}_-\cr}\right]&\simeq&{\textstyle\frac{2}{3}}m\exp(-{\rm i}m{\hat\varphi}_P)\sum_{n=1}^2\sum_{j=1}^3\int_{\cal S^\prime}{\hat r}\,{\rm d}{\hat r}\,{\rm d}{\hat z}\,c_1^{-1}\exp(-{\rm i}m c_2)\left[\matrix{{\tilde{\bf u}}_{nj}\cr{\tilde{\bf v}}_{nj}\cr}\right]\nonumber\\*
&&\times\left[c_1^{-1}p_{nj}\sin\left({\textstyle\frac{2}{3}}m c_1^3\right)
+2{\rm i}q_{nj}\cos\left({\textstyle\frac{2}{3}}m c_1^3\right)\right],\,\, m\gg1,\,\theta_U\le\theta_P\le\pi-\theta_U.\nonumber\\*
\label{E171}
\end{eqnarray}
Here I have used (\ref{E70}) for the asymptotic values of $G^{\rm out}_{nj}\vert_{{\hat\varphi}={\hat\varphi}_\pm}$ and have written $\phi_\pm$ in terms of $c_1$ and $c_2$ with the aid of (\ref{E41}).  

\section{Total radiation field outside the transitional intervals}
\label{sec:total}
In this section I assemble and combine the expressions derived in the preceding sections for the parts $\matrix{[{\bf E}^{\rm v}&{\bf B}^{\rm v}]}$ and $\matrix{[{\bf E}^{\rm b}_+-{\bf E}^{\rm b}_-&{\bf B}^{\rm b}_+-{\bf B}^{\rm b}_-]}$ of the radiation field $\matrix{[{\bf E}&{\bf B}]}$ in various regions of space [see (\ref{E75})].

At observation points for which $0<\theta_P\le\theta_L$ or $\pi-\theta_L\le\theta_P<\pi$ (see figure~\ref{F12}), the field consists entirely of the part arising from the volume of the source which is given by 
\begin{eqnarray}
\left[\matrix{{\bf E}\cr{\bf B}\cr}\right]&=&m^2\exp(-{\textrm i}m{\hat\varphi}_P)\sum_{n=1}^2\int_{-{\hat z}_0}^{{\hat z}_0}{\textrm d}{\hat z}\int_{{\hat r}_L}^{{\hat r}_U}{\textrm d}{\hat r}\,{\hat r}\int_0^{2\pi}{\textrm d}\varphi\frac{\exp(- {\textrm i}mg)}{{\hat R}^n}\Bigg(\cos(\varphi-\varphi_P)\left[\matrix{{\tilde{\bf u}}_{n1}\cr{\tilde{\bf v}}_{n1}\cr}\right]\nonumber\\*
&&+ \sin(\varphi-\varphi_P)\left[\matrix{{\tilde{\bf u}}_{n2}\cr{\tilde{\bf v}}_{n2}\cr}\right]+\left[\matrix{{\tilde{\bf u}}_{n3}\cr{\tilde{\bf v}}_{n3}\cr}\right]\Bigg),\quad0<\theta_P\le\theta_L\quad{\rm or}\quad \pi-\theta_L\le\theta_P<\pi
\label{E172}
\end{eqnarray}
(see \S~\ref{subsec:Ev3}).  Note that at such observation points $\Delta$ is negative for the coordinates $({\hat r},{\hat z})$ of all volume elements of the source.  This field has the same characteristics as a conventional radiation field.

At observation points for which $\theta_L\le\theta_P\le\theta_U$ or $\pi-\theta_U\le\theta_P\le\pi-\theta_L$, (\ref{E75}), (\ref{E106}), and (\ref{E170}) jointly yield
\begin{eqnarray}
&&\left[\matrix{{\bf E}\cr{\bf B}\cr}\right]\simeq\exp(-{\rm i}m{\hat\varphi}_P)\Bigg\{m^2\sum_{n=1}^2\int_{-{\hat z}_0}^{{\hat z}_0}{\rm d}{\hat z}\int_{{\hat r}_L}^{{\hat r}_U}{\rm d}{\hat r}\,{\hat r}
\int_0^{2\pi}{\textrm d}\varphi\frac{\exp(- {\textrm i}mg)}{{\hat R}^n}\nonumber\\*
&&\quad\times\Bigg(\cos(\varphi-\varphi_P)\left[\matrix{{\tilde{\bf u}}_{n1}\cr{\tilde{\bf v}}_{n1}\cr}\right]+\sin(\varphi-\varphi_P)\left[\matrix{{\tilde{\bf u}}_{n2}\cr{\tilde{\bf v}}_{n2}\cr}\right]+\left[\matrix{{\tilde{\bf u}}_{n3}\cr{\tilde{\bf v}}_{n3}\cr}\right]\Bigg)+{\rm i}m\nonumber\\*
&&\quad\times\int_{-{\hat z}_0}^{{\hat z}_0}{\textrm d}{\hat z}\Bigg[\exp(-{\rm i}m\phi_C)\lim_{\epsilon\to0}\int_{\epsilon({\hat r}_S-{\hat r}_C)}^{w_0}{\rm d}w\sum_{\iota=\pm}\iota\exp(m\gamma^\iota_C)J_w^\iota \left[\matrix{{\bf \Lambda}_\iota\cr{\bf \Gamma}_\iota\cr}\right]\Bigg\vert_{{\hat r}={\hat r}_C+w\exp({\rm i}\lambda_C^\iota)}\nonumber\\*
&&\quad+\exp(-{\rm i}m\phi_U^+)\int_{{\cal K}_U}{\rm d}v\,\exp(m\gamma^+_U)J^+_v\left[\matrix{{\bf \Lambda}_+\cr{\bf \Gamma}_+\cr}\right]\Bigg\vert_{{\hat r}=u_U^++{\rm i}v}\nonumber\\*
&&\quad-\exp(-{\rm i}m\phi_U^-)\int_{{\cal L}_U}{\rm d}v\,\exp(m\gamma^-_U)J^-_v\left[\matrix{{\bf \Lambda}_-\cr{\bf \Gamma}_-\cr}\right]\Bigg\vert _{{\hat r}=u_U^-+{\rm i}v}\nonumber\\*
&&\quad-\exp(-{\rm i}m\phi_S)\int_{{\cal L}_S}{\rm d}u\,\exp(m\gamma_S) J^-_u\left[\matrix{{\bf \Lambda}_-\cr{\bf \Gamma}_-\cr}\right]\Bigg\vert_{{\hat r}=u+{\rm i}v_S}\Bigg]\Bigg\},\qquad\qquad\nonumber\\*
&&\nonumber\\*
&&\qquad\qquad\qquad m\gg1,\qquad\theta_L\le\theta_P\le\theta_U,\qquad\pi-\theta_U\le\theta_P\le\pi-\theta_L.
\label{E173}
\end{eqnarray}
When the superluminally moving part of the source distribution extends as far as the light cylinder, i.e., when ${\hat r}_L=1$, the angle $\theta_U$ equals $\pi/2$ and (\ref{E172}) and (\ref{E173}) jointly describe the field throughout space (see figure~\ref{F12}). 

But when ${\hat r}_L\ge1$, as in figure~\ref{F11}, there is a third region, $\theta_U\le\theta_P\le\pi-\theta_U$ (coloured yellow in figure~\ref{F12}), in which the field is described by 
\begin{eqnarray}
\left[\matrix{{\bf E}\cr{\bf B}\cr}\right]&=&2m\exp(-{\rm i}m{\hat\varphi_P})\sum_{n=1}^2\sum_{j=1}^3\int_{-{\hat z}_0}^{{\hat z}_0}{\rm d}{\hat z}\int_{{\hat r}_L}^{{\hat r}_U}{\rm d}{\hat r}\,{\hat r}\left\{p_{nj}\left[\pi m^{2/3}{\rm Ai}\left(-m^{2/3}c_1^2\right)\right.\right.\nonumber\\*
&&\left.\left.
+{\textstyle\frac{1}{3}}c_1^{-2}\sin\left({\textstyle\frac{2}{3}}m c_1^3\right)\right]+{\rm i}q_{nj}\left[\pi m^{1/3}{\rm Ai}^\prime\left(-m^{2/3}c_1^2\right)+{\textstyle\frac{2}{3}}c_1^{-1}\cos\left({\textstyle\frac{2}{3}}m c_1^3\right)\right]\right\}\nonumber\\*
&&\times\exp(-{\rm i}m c_2)\left[\matrix{{\tilde{\bf u}}_{nj}\cr{\tilde{\bf v}}_{nj}\cr}\right],\qquad\qquad m\gg1,\qquad\theta_U\le\theta_P\le\pi-\theta_U,
\label{E174}
\end{eqnarray}
as can be seen from (\ref{E75}), (\ref{E105}) and (\ref{E171}).  The step function ${\rm H}(\Delta)$ in (\ref{E105}) is omitted here because, at observation points for which $\theta_U\le\theta_P\le\pi-\theta_U$, the values of the radial coordinates ${\hat r}$ of all volume elements of the source exceed that of ${\hat r}_C$. 

In the case of a radiation problem involving caustics, such as the present one, it makes a difference whether the generated field is calculated prior to proceeding to the far-field limit or vice versa.  This is because the far-field approximation replaces spherical wave fronts by planar ones thereby vitiating the formation of caustics.  Equations~(\ref{E172})-(\ref{E174}), which hold true irrespective of whether the observer is located in the near or the far zone, can now be numerically evaluated in the radiation field, where ${\hat R}_P\gg1$.  For an informed interpretation of the numerical results (reported in \S~\ref{sec:numerical}) it would be helpful to inspect the far-field versions of the quantities $\Delta$, $c_1$, $c_2$, $p_{nj}$ and $q_{nj}$ that appear in the above equations by replacing them with the following leading terms in their Taylor expansions in powers of ${\hat R}_P^{-1}$:
\begin{equation}
\Delta\simeq({\hat r}^2\sin^2\theta_P-1){\hat R}_P^2,
\label{E175}
\end{equation}
\begin{equation}
c_1\simeq\left(3\tau/2\right)^{1/3},
\label{E176}
\end{equation}
\begin{equation}
c_2\simeq{\hat R}_P-{\hat z}\cos\theta_P+3\pi/2,
\label{E177}
\end{equation}
\begin{equation}
p_{n1}\simeq
({\hat r}\sin\theta_P)^{-1}({\hat r}^2\sin^2\theta_P-1)^{-1/4}(12\tau)^{1/6}{\hat R}_P^{-n-1},
\label{E178}
\end{equation}
\begin{equation}
p_{n2}\simeq-{\hat R}_Pp_{n1},\qquad p_{n3}\simeq{\hat r}\sin\theta_P{\hat R}_P p_{n1},
\label{E179}
\end{equation}
\begin{equation}
q_{n1}\simeq
(16/3)^{1/6}({\hat r}\sin\theta_P)^{-1}({\hat r}^2\sin^2\theta_P-1)^{1/4}\tau^{-1/6}{\hat R}_P^{-n},
\label{E180}
\end{equation}
and
\begin{equation}
q_{n2}\simeq{\textstyle\frac{1}{2}}(-1)^{n-1}{\hat R}_P^{-1}q_{n1},\qquad q_{n3}\simeq{\textstyle\frac{3^{n-1}}{2}}{\hat r}\sin\theta_P{\hat R}_P^{-1}q_{n1},
\label{E181}
\end{equation}
where
\begin{equation}
\tau=({\hat r}^2\sin^2\theta_P-1)^{1/2}-\arctan({\hat r}^2\sin^2\theta_P-1)^{1/2}
\label{E182}
\end{equation}
[see (\ref{E33}), (\ref{E41}), (\ref{E67}) and (\ref{E68})].  Note that the expression in (\ref{E174}), for instance, consists of two types of terms for each value of $j$:  the ones involving $p_{n2}$, $p_{n3}$ and $q_{n1}$ which diminish as ${\hat R}_P^{-n}$ with increasing ${\hat R}_P$ and the ones involving $p_{n1}$, $q_{n2}$ and $q_{n3}$ which diminish as ${\hat R}_P^{-n-1}$ with distance.  Both types of terms need to be kept because there are isolated observation points at which the terms that decay as ${\hat R}_P^{-n}$ cancel out in the expressions for ${\bf E}$ or ${\bf B}$ and the corresponding field decays as ${\hat R}_P^{-n-1}$.  

\section{Evaluation of the field in transitional intervals}
\label{sec:transitional}

The radiation field $\matrix{[{\bf E}&{\bf B}]}$ changes rapidly over the narrow angular intervals $\theta_L^{\rm c}\le\theta_P\le\theta_L$ and $\theta_U\le\theta_P\le\theta_U^{\rm s}$ [see (\ref{E87}), and (\ref{E113})--(\ref{E115})].  In these transitional intervals, at least one of the loci $C$ and $S$ intersects the source distribution across either the entire or a portion of its ${\hat z}$-extent (see figure~\ref{F11}) but $C$ and $S$ do not both intersect the source at every value of ${\hat z}$ in $-{\hat z}_0\le{\hat z}\le{\hat z}_0$ as they do in $\theta_L\le\theta_P\le\theta_U$.  The angular widths of such intervals rapidly decrease with increasing distance: as ${\hat R}_P^{-2}$ for ${\hat R}_P\gg1$ [see (\ref{E110})].  Nevertheless, the total flux of energy close to the source (a quantity which I will use to normalize the Poynting flux in \S~\ref{sec:numerical}) cannot be accurately evaluated without including the contributions from these intervals.

At observation points for which $\theta_L^{\rm c}<\theta_P<\theta_L^{\rm s}$ and ${\hat z}_P>0$, the cusp locus $C$ intersects the source distribution~(\ref{E7}) over ${\hat z}_U^{\rm c}\le{\hat z}\le{\hat z}_0$, where
 \begin{equation}
 {\hat z}_U^{\rm c}={\hat z}_P-({\hat r}_P^2-1)^{1/2}({\hat r}_U^2-1)^{1/2}
 \label{E183}
 \end{equation}
[see (\ref{E39})], while the locus $S$ lies outside the source.  The critical points contributing toward the asymptotic value of the ${\hat r}$-integral in the expression for $\matrix{[{\bf E}^{\rm b}_-& {\bf B}^{\rm b}_-]}$ in (\ref{E85}) are therefore only the ones at ${\hat r}={\hat r}_C$ and ${\hat r}={\hat r}_U$.  The paths of steepest descent that issue from these critical points are found (in the same way as in \S~\ref{subsec:paths}) to be those shown in figure~\ref{F18}.  The critical points contributing toward the asymptotic value of the ${\hat r}$-integral in the expression for $\matrix{[{\bf E}^{\rm b}_+&{\bf B}^{\rm b}_+]}$ are the same as those shown in figure~\ref{F17} except that ${\hat r}_S$ is here greater than ${\hat r}_U$ and so $S$ falls outside the range of integration, instead of inside it.  The total radiation field can be evaluated in this case from a version of (\ref{E173}) in which the integral over ${\cal L}_S$ is absent and the second integration with respect to ${\hat z}$ runs from ${\hat z}_U^{\rm c}$ to ${\hat z}_0$.

\begin{figure}
\centerline{\includegraphics[width=11cm]{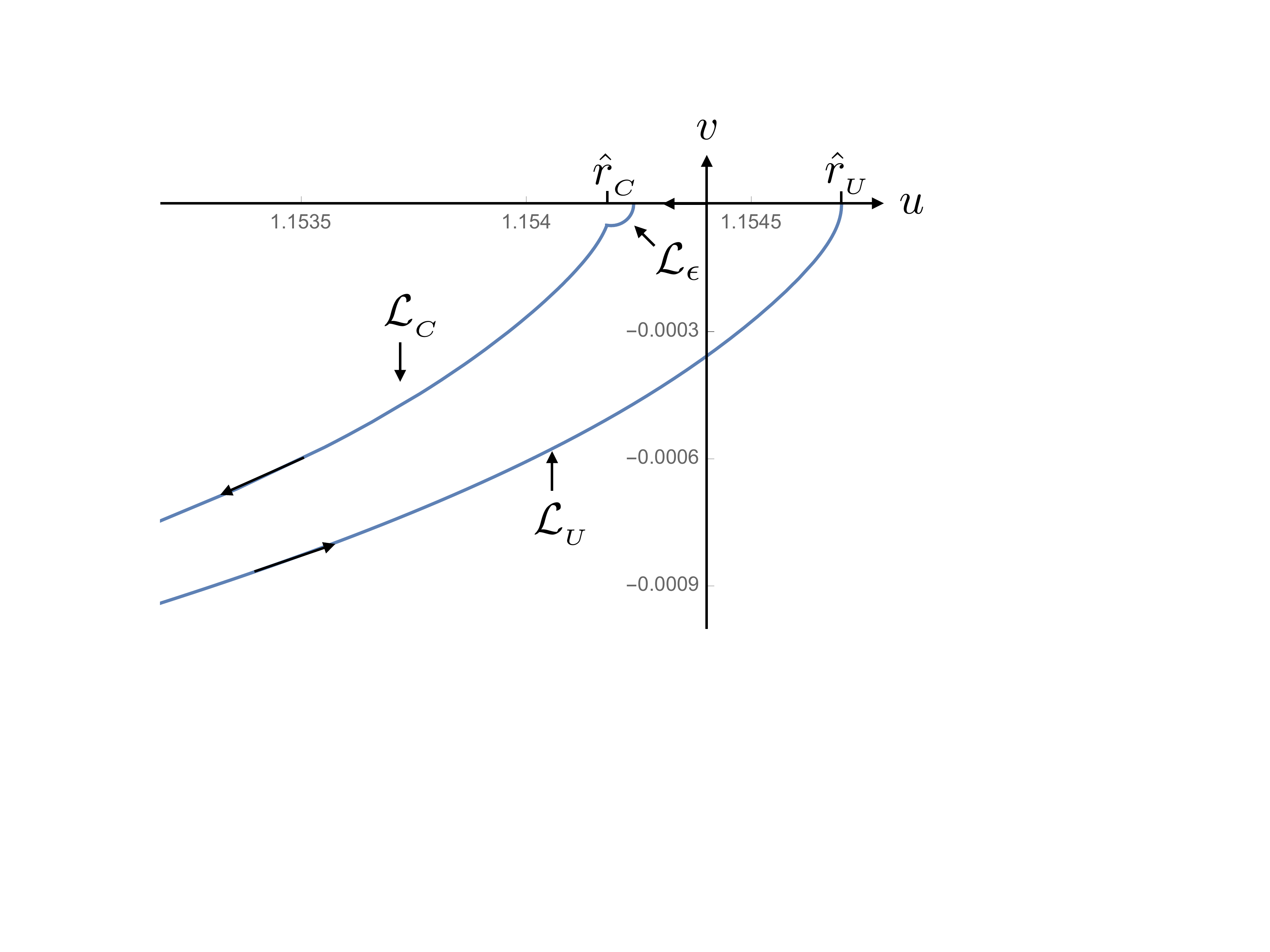}}
\caption{Paths of steepest descent of the exponential kernel $\exp(-{\rm i}m\phi_-)$ for an observation point in the transitional interval $\theta_L^{\rm c}<\theta_P<\theta_L^{\rm s}$ and a source element within ${\hat z}_U^{\rm c}\le{\hat z}\le{\hat z}_0$.   This figure is plotted for the following set of values of the parameters: ${\hat R}_P=10$, $\theta_P=181\pi/540$, $m=10$, ${\hat z}=-0.025$, ${\hat r}_U=1.1547$.  Radial coordinate ${\hat r}_S=1.1548$ of the stationary point of the phase $\phi_-$ here exceeds the outer radius ${\hat r}_U$ of the source distribution.}
\label{F18}
\end{figure} 

At observation points for which $\theta_L^{\rm s}<\theta_P<\theta_U$, ${\hat z}_P>0$, curves $S$ and $C$ both intersect the source distribution (\ref{E7}) but over the limited ranges ${\hat z}_U^{\rm s}\le{\hat z}\le{\hat z}_0$ and $\max({\hat z}_U^{\rm c},-{\hat z}_0)\le{\hat z}\le{\hat z}_0$, respectively, where
 \begin{equation}
 {\hat z}_U^{\rm s}={\hat z}_P-({\hat r}_P^2-{\hat r}_U^2)^{1/2}({\hat r}_U^2-1)^{1/2}
 \label{E184}
 \end{equation}
[see (\ref{E109})].  For ${\hat z}_U^{\rm s}\le{\hat z}\le{\hat z}_0$ the critical points of the ${\hat r}$-integrals in (\ref{E85}) and their corresponding paths of steepest descent are the same as those shown in figures~\ref{F16} and \ref{F17}.  For $\max({\hat z}_U^{\rm c},-{\hat z}_0)\le{\hat z}\le{\hat z}_0$, on the other hand, the critical points are only ${\hat r}={\hat r}_C$ and ${\hat r}={\hat r}_U$ for which the paths of steepest descent are as shown in figure~\ref{F18}.  The total radiation field is in this case given by a version of (\ref{E173}) in which the second integration with respect to ${\hat z}$ is split into two integrals with differing ranges and the contribution from ${\cal L}_S$ is omitted from the integrand of the integral over $\max({\hat z}_U^{\rm c},-{\hat z}_0)\le{\hat z}\le{\hat z}_0$.

Another set of transitional intervals occurs when the value of the polar coordinate $\theta_P$ of the observation point exceeds that of $\theta_U$ [see (\ref{E115})] but is smaller than both $\theta_P^{\rm s}\vert_{{\hat r}={\hat r}_L,{\hat z}={\hat z}_0}$ and $\theta_U^{\rm c}$ [see (\ref{E111}) and (\ref{E88})].  In this case, there are contributions toward the value of the ${\hat r}$-integral in (\ref{E85}) from the critical points ${\hat r}_L$, ${\hat r}_S$ and ${\hat r}_U$ of the exponential kernel $\exp(-{\rm i}m\phi_-)$ if ${\hat z}_L^{\rm c}\le{\hat z}\le{\hat z}_0$ and from the critical points ${\hat r}_C$, ${\hat r}_S$ and ${\hat r}_U$ of this kernel if $-{\hat z}_0\le{\hat z}\le{\hat z}_L^{\rm c}$, where
 \begin{equation}
 {\hat z}_L^{\rm c}={\hat z}_P-({\hat r}_P^2-1)^{1/2}({\hat r}_L^2-1)^{1/2}
 \label{E185}
 \end{equation}
[see (\ref{E39}) and figure~\ref{F11}].  The paths of steepest descent of $\exp(-{\rm i}m\phi_-)$ for ${\hat z}_L^{\rm c}\le{\hat z}\le{\hat z}_0$ can be determined in the same way as in \S~\ref{subsec:paths} and are shown in figure~\ref{F19}.  The corresponding paths of steepest descent issuing from the critical points ${\hat r}_L$ and ${\hat r}_U$ of the ${\hat r}$-integral entailing the exponential kernel $\exp(-{\rm i}m\phi_+)$ in (\ref{E85}) are shown in figure~\ref{F20}.  The contributions made by the source elements in $-{\hat z}_0\le{\hat z}\le{\hat z}_L^{\rm c}$ stem from the same set of critical points (${\hat r}_C$, ${\hat r}_S$ and ${\hat r}_U$) as those encountered in \S\S~\ref{subsec:PhiMinus1} and \ref{subsec:PhiPlus1} and so can be calculated from an appropriate version of (\ref{E173}) for which the steepest-descent paths resemble the ones in figures~\ref{F16} and \ref{F17}. 

At observation points for which $\theta_P^{\rm s}\vert_{{\hat r}={\hat r}_L,{\hat z}={\hat z}_0}\le\theta_P\le\theta_U^{\rm c}$, locus $S$ intersects the source distribution over $-{\hat z}_0\le{\hat z}\le{\hat z}_L^{\rm s}$ and locus $C$ over $-{\hat z}_0\le{\hat z}\le{\hat z}_L^{\rm c}$.  There are contributions in this case toward the value of $\matrix{[{\bf E}_-^{\rm b}&{\bf B}_-^{\rm b}]}$ from the critical points ${\hat r}_C$, ${\hat r}_S$ and ${\hat r}_U$ in $-{\hat z}_0\le{\hat z}\le{\hat z}_L^{\rm c}$, from ${\hat r}_L$, ${\hat r}_S$ and ${\hat r}_U$ in ${\hat z}_L^{\rm c}\le{\hat z}\le{\hat z}_L^{\rm s}$, and from ${\hat r}_L$ and ${\hat r}_U$ in ${\hat z}_L^{\rm s}\le{\hat z}\le{\hat z}_0$.  The corresponding set of steepest-descent paths of $\exp(-{\rm i}m\phi_-)$ and $\exp(-{\rm i}m\phi_+)$ that issue from these points are similar to those in figures~\ref{F16} and \ref{F17} when $-{\hat z}_0\le{\hat z}\le{\hat z}_L^{\rm c}$, similar to those in figures~\ref{F19} and \ref{F20} when ${\hat z}_L^{\rm c}\le{\hat z}\le{\hat z}_L^{\rm s}$ and both similar to that in figure~\ref{F20} when ${\hat z}_L^{\rm s}\le{\hat z}\le{\hat z}_0$.  The contributions in question can be evaluated by means of a version of (\ref{E173}) in which the integration with respect to ${\hat z}$ is split into the listed sub-intervals and the integrations over $u$, $v$ or $w$ are performed along the steepest-descent paths that issue from the critical points appropriate to each sub-interval. 

Finally, at observation points for which $\theta_U^{\rm c}\le\theta_P\le\theta_U^{\rm s}$ the cusp locus $C$ lies entirely in ${\hat r}<{\hat r}_L$ and the intersection of $S$ with the source distribution only occurs in $-{\hat z}_0\le{\hat z}\le{\hat z}_L^{\rm s}$. The relevant critical points for the evaluation of $\matrix{[{\bf E}_-^{\rm b}&{\bf B}_-^{\rm b}]}$ over $-{\hat z}_0\le\theta_P\le{\hat z}_L^{\rm s}$ are ${\hat r}_L$, ${\hat r}_S$ and ${\hat r}_U$ as in figure~\ref{F19}.  The relevant critical points for the evaluations of $\matrix{[{\bf E}_-^{\rm b}&{\bf B}_-^{\rm b}]}$ over ${\hat z}_L^{\rm s}\le\theta_P\le{\hat z}_0$, and of $\matrix{[{\bf E}_+^{\rm b}&{\bf B}_+^{\rm b}]}$ over $-{\hat z}_0\le{\hat z}\le{\hat z}_0$, are only ${\hat r}_L$ and ${\hat r}_U$ as in figure~\ref{F20}.  

\begin{figure}
\centerline{\includegraphics[width=11cm]{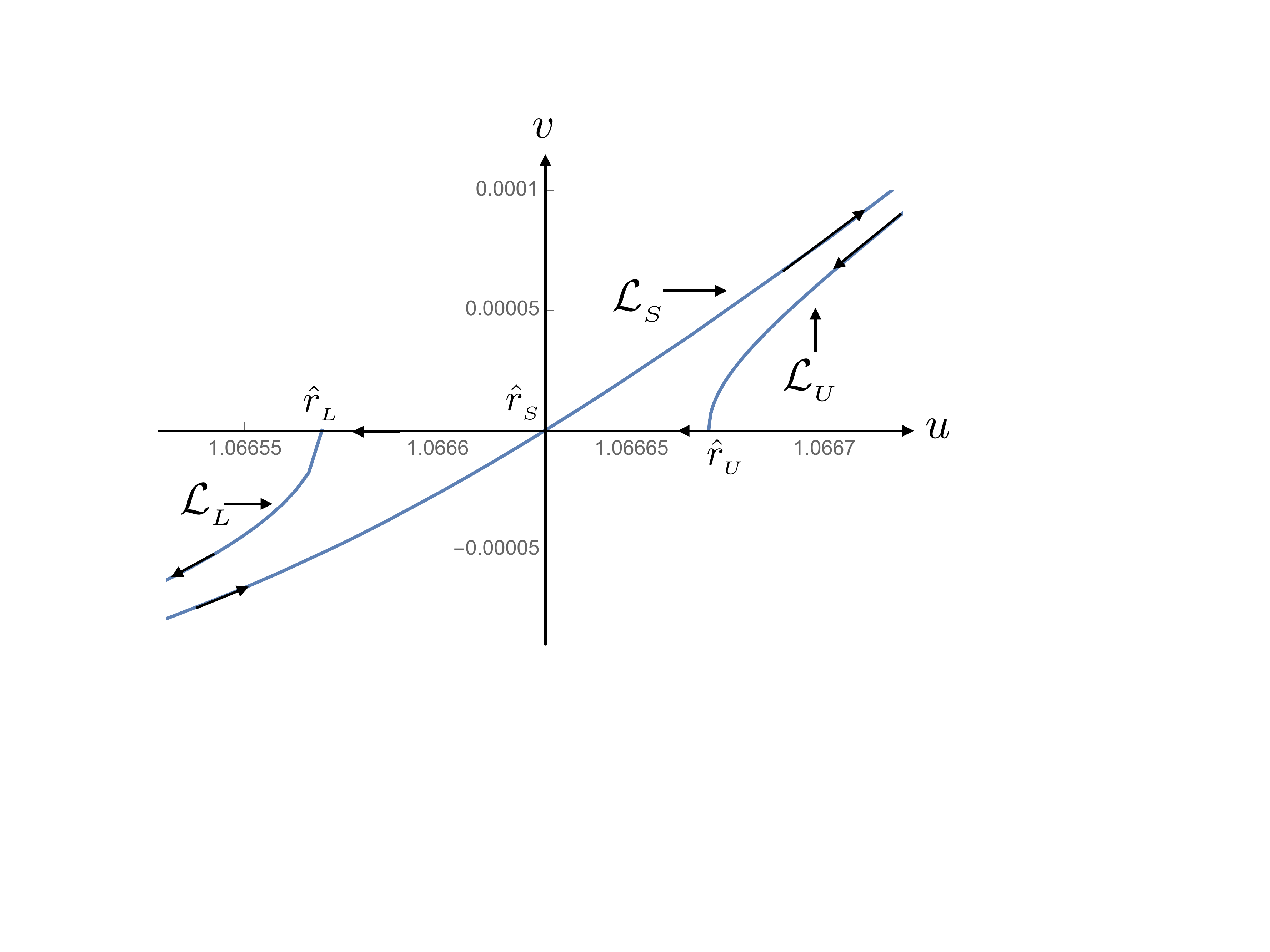}}
\caption{Paths of steepest descent of the exponential kernel $\exp(-{\rm i}m\phi_-)$ for an observation point in the transitional interval $\theta_U\le\theta_P\le\theta_P^{\rm s}\vert_{{\hat r}={\hat r}_L,{\hat z}={\hat z}_0}$ and a source element within ${\hat z}_L^{\rm c}\le{\hat z}\le{\hat z}_0$.  This figure is plotted for the following set of values of the parameters: ${\hat R}_P=10$, $\theta_P=52\pi/135$, $m=10$, ${\hat z}_0=0.1$, ${\hat z}=0.08$, ${\hat r}_L=1.06657$, and ${\hat r}_U=1.06667$.  Radial coordinate of the lower boundary of the source distribution here exceeds the location ${\hat r}_C=1.06652$ of the cusp but the stationary point ${\hat r}_S=1.06662$ of the phase $\phi_-$ falls within the source distribution.}
\label{F19}
\end{figure} 

\begin{figure}
\centerline{\includegraphics[width=11cm]{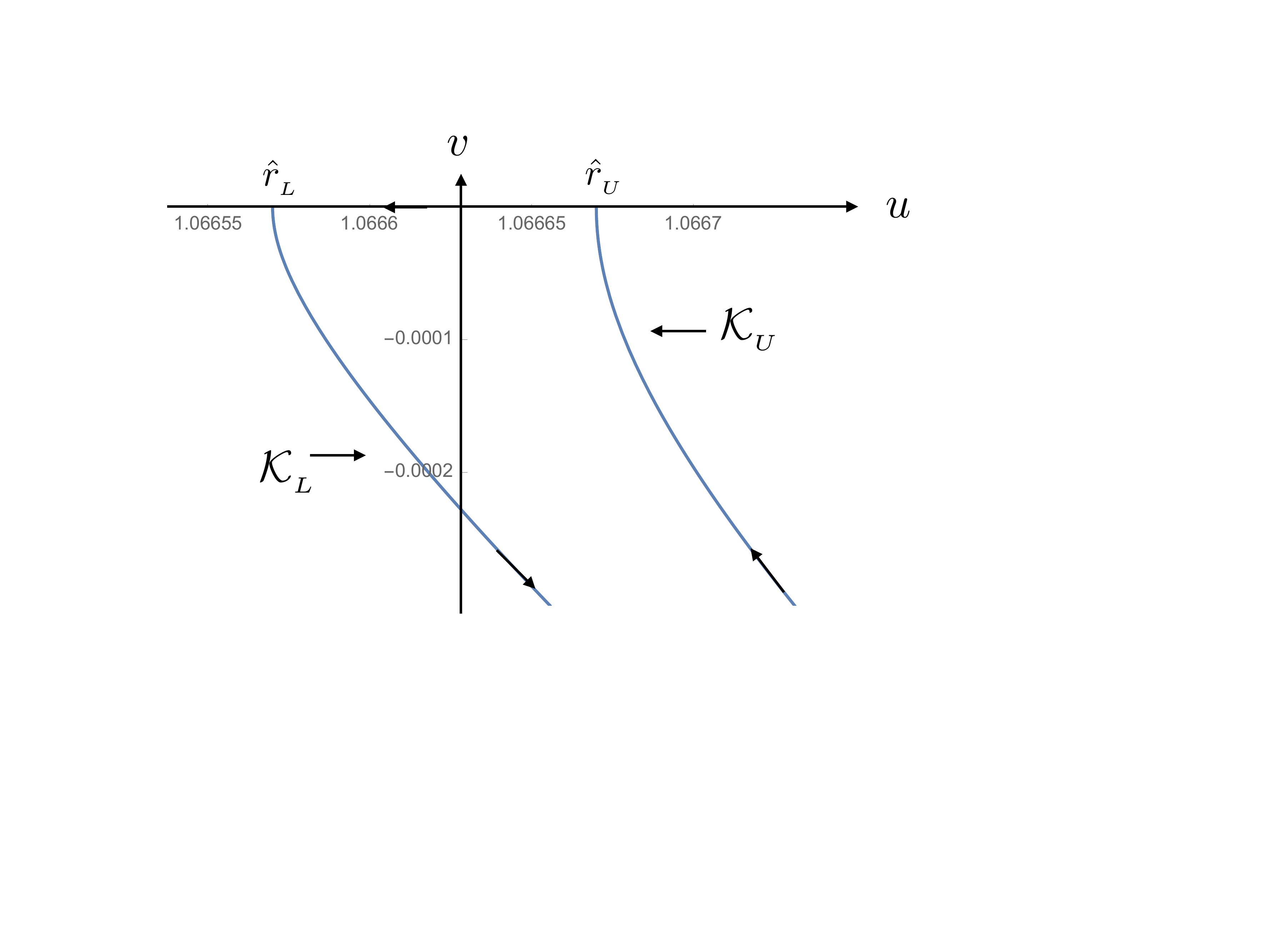}}
\caption{Paths of steepest descent of the exponential kernel $\exp(-{\rm i}m\phi_+)$ for an observation point in the transitional interval $\theta_U\le\theta_P\le\theta_P^{\rm s}\vert_{{\hat r}={\hat r}_L,{\hat z}={\hat z}_0}$ and a source element within ${\hat z}_L^{\rm c}\le{\hat z}\le{\hat z}_0$.  The parameters for this figure have the same values as those for figure~\ref{F19}.}
\label{F20}
\end{figure} 

\section{Flux of energy and state of polarization of the radiation}
\label{sec:energy}

Asymptotic value of the total radiation field $\matrix{[{\bf E}&{\bf B}]}$ derived in \S~\ref{sec:total} depends on the observation time $t_P$ through the oscillating factor $\exp(-{\rm i}m{\hat\varphi}_P)$ [see (\ref{E24})] which multiplies all three of the expressions in (\ref{E172})-(\ref{E174}).  When the cross-product of the real parts of ${\bf E}$ and ${\bf B}$ is averaged over an oscillation period $m\omega t_P$ one finds that
\begin{equation}
\langle\Re({\bf E}){\bf\times}\Re({\bf B})\rangle=\textstyle{\frac{1}{2}}\Re({\bf E\times B^*}),
\label{E186}
\end{equation}
in which ${\bf B}^*$ is the complex conjugate of ${\bf B}$ and the angular brackets denote averaging with respect to ${\hat\varphi}_P$ or $t_P$.  (Note that in the present case ${\bf E\times B^*}$ is not necessarily real.)  The Poynting vector therefore has the time-averaged value 
\begin{equation}
{\bf S}=\frac{c}{8\pi}\Re({\bf E\times B^*}).
\label{E187}
\end{equation}
Mean value of the radial component of the time-averaged Poynting vector over a sphere of radius ${\hat R}_P$ centred at the origin is given by the integral of ${\hat{\bf n}}_\infty\cdot{\bf S}$ over all values of $\theta_P$ and $\varphi_P$ divided by the solid angle $4\pi$ covering the entire sphere,  
\begin{equation}
{\bar S}_n=\frac{1}{4\pi}\int_0^{2\pi}{\rm d}\varphi_P\int_0^\pi{\rm d}\theta_P\,\sin\theta_P\,{\hat{\bf n}}_\infty\cdot{\bf S},
\label{E188}
\end{equation}
where
\begin{equation}
{\hat{\bf n}}_\infty=\sin\theta_P{\hat{\bf e}}_{r_P}+\cos\theta_P{\hat{\bf e}}_{z_P}
\label{E189}
\end{equation}
denotes the unit vector normal to the sphere, i.e., the unit vector along the line joining the origin of the coordinates to the observation point.  Note that the symmetries of the present source render ${\bf S}$ independent of $\varphi_P$ and make the $\theta_P$-integrals over $(0,\pi/2)$ and $(\pi/2,\pi)$ equal to one another.

In contrast to the case of a conventional radiation, in which both ${\bf S}$ and ${\bar S}_n$ have the dependence ${\hat R}_P^{-2}$ on the radius ${\hat R}_P$, the ratio ${\hat{\bf n}}_\infty\cdot{\bf S}/{\bar S}_n$ \citep[defining directive gain in conventional antenna theory,][]{IEEE} is not independent of ${\hat R}_P$ in the present case.  To present the results of the numerical computations in \S~\ref{sec:numerical} in terms of a dimensionless quantity most closely resembling directive gain, here I introduce
\begin{equation}
{\hat{\bf n}}_\infty\cdot{\hat{\bf S}}=\frac{{\hat{\bf n}}_\infty\cdot{\bf S}}{{\bar S}_n\vert_{{\hat R}_P=10}},
\label{E190}
\end{equation}
in which the radial component of time-averaged Poynting vector is normalized by the mean value of the power that propagates across the sphere ${\hat R}_P=10$ per unit solid angle.  

To determine the state of polarization of the radiation, I evaluate the Stokes parameters
\begin{equation}
I=\vert E_\parallel\vert^2+ \vert E_\perp\vert^2,\quad Q=\vert E_\parallel\vert^2- \vert E_\perp\vert^2, 
\label{E191}
\end{equation}
\begin{equation}
U=2\Re\left(E_\parallel E_\perp^*\right),\quad V=-2\Im\left(E_\parallel E_\perp^*\right),
\label{E192}
\end{equation}
\begin{equation}
L=(Q^2+U^2)^{1/2},\quad \psi_S=\frac{1}{2}\arctan\frac{U}{Q},
\label{E193}
\end{equation}
in which $E_\parallel={\hat{\bf e}}_\parallel\cdot{\bf E}$ and $E_\perp={\hat{\bf e}}_\perp\cdot{\bf E}$ are the components of the electric field along the unit vectors ${\hat{\bf e}}_\parallel={\hat{\bf e}}_{\varphi_P}$ and ${\hat{\bf e}}_\perp={\hat{\bf n}}_\infty\times{\hat{\bf e}}_\parallel$.  Together with ${\hat{\bf n}}_\infty$ (the radiation direction), ${\hat{\bf e}}_\parallel$ (which is parallel to the plane of rotation) and ${\hat{\bf e}}_\perp$ (which is perpendicular to both ${\hat{\bf n}}_\infty$ and ${\hat{\bf e}}_\parallel$) constitute the base vectors of a Cartesian reference frame.  

The Poynting fluxes and Stokes parameters I will numerically evaluate in \S~\ref{sec:numerical} are for the emissions that are generated in the cases of the following two differently designed versions of the experimental apparatus described in \S\S~\ref{sec:source} and \ref{subsec:constraint}.

\subsection{Case I: The emission from a polarization parallel to the rotation axis}
\label{subsec:energy1}

In the case where the faces of the electrode pairs shown in figure~\ref{F2} are normal to ${\hat{\bf e}}_z$ and the vector ${\bf s}$ in (\ref{E1}) is spatially uniform in ${\cal S}^\prime$ and zero outside it [see (\ref{E7})], the only non-zero components of the charge-current density are $j_z$ and $\rho$, i.e., the source terms $s_z$ and $s_0$.  Once $s_r$ and $s_\varphi$ in (\ref{E78})-(\ref{E83}) are set equal to zero and $s_0$ is evaluated with the aid of (\ref{E83}) and (\ref{E7}), the vectors ${\tilde{\bf u}}_{nj}$ and ${\tilde{\bf v}}_{nj}$ in these equations reduce to
\begin{equation} 
\left[\matrix{ {\tilde{\bf u}}_{11}& {\tilde{\bf u}}_{12}& {\tilde{\bf u}}_{13}}\right]=s_z\left[\matrix{0&0&{\hat{\bf e}}_{z_P}}\right],
\label{E194}
\end{equation}
\begin{eqnarray}
    \left[
     \begin{array}{c}
     {\tilde{\bf u}}_{21}\\
     {\tilde{\bf u}}_{22}\\
     {\tilde{\bf u}}_{23}
     \end{array} \right]&=& \frac{{\rm i}s_z}{m}\left[\delta({\hat z}+{\hat z}_0)-\delta({\hat z}-{\hat z}_0)\right]\left[
      \begin{array}{c}
     {\hat r}{\hat{\bf e}}_{r_P}\\
      {\hat r}{\hat{\bf e}}_{\varphi_P} \\
     -{\hat r}_P{\hat{\bf e}}_{r_P}+({\hat z}-{\hat z}_P){\hat{\bf e}}_{z_P}
\end{array} \right],
\label{E195}
\end{eqnarray}
\begin{equation} 
\left[\matrix{ {\tilde{\bf v}}_{21}& {\tilde{\bf v}}_{22}& {\tilde{\bf v}}_{23}}\right]=s_z\left[\matrix{{\hat r}{\hat{\bf e}}_{\varphi_P}&-{\hat r}{\hat{\bf e}}_{r_P}&-{\hat r}_P{\hat{\bf e}}_{\varphi_P}}\right],
\label{E196}
\end{equation}
and ${\tilde{\bf v}}_{1j}=0$.  The polarization charges are here confined to the surfaces ${\hat z}=\pm{\hat z}_0$ because ${\bf P}$ is assumed to be spatially constant inside the dielectric and zero outside it.  The delta functions in (\ref{E195}) stem from the fact that ${\hat\nabla}\cdot(s_z{\hat{\bf e}}_z)=\partial s_z/\partial{\hat z}$ and the dependence of $s_z$ on ${\hat z}$ is, according to (\ref{E7}), given by the combination ${\rm H}({\hat z}+{\hat z}_0)- {\rm H}({\hat z}-{\hat z}_0)$ of Heaviside step functions.

To elicit the orientations of the vectors ${\bf E}$ and ${\bf B}$ from the full expressions for these vectors in (\ref{E172})--(\ref{E174}), let ${\bf E}_{nj}$ be the part of the field ${\bf E}$ arising from the source term ${\tilde{\bf u}_{nj}}$ and ${\bf B}_{nj}$ be the part of the field ${\bf B}$ arising from the source term ${\tilde{\bf v}_{nj}}$, i.e., let
\begin{eqnarray}
\left[\matrix{{\bf E}\cr{\bf B}\cr}\right]=\sum_{n=1}^2\sum_{j=1}^3\left[\matrix{{\bf E}_{nj}\cr{\bf B}_{nj}\cr}\right].
\label{E197}
\end{eqnarray}
Then it follows from (\ref{E194})--(\ref{E197}) that in the present case
\begin{equation}
\left[\matrix{{\bf E}\cr{\bf B}\cr}\right]=\left[\matrix{(E_{21}+E_{23}^{r}){\hat{\bf e}}_{r_P}+E_{22}{\hat{\bf e}}_{\varphi_P}+(E_{13}+E_{23}^{z}){\hat{\bf e}}_{z_P}\cr B_{22}{\hat{\bf e}}_{r_P}+(B_{23}+B_{21}){\hat{\bf e}}_{\varphi_P}\cr}\right],
\label{E198}
\end{equation}
where $E_{13}={\hat{\bf e}}_{z_P}\cdot{\bf E}_{13}$, $E_{21}={\hat{\bf e}}_{r_P}\cdot{\bf E}_{21}$, $E_{22}={\hat{\bf e}}_{\varphi_P}\cdot{\bf E}_{22}$, $E_{23}^r={\hat{\bf e}}_{r_P}\cdot{\bf E}_{23}$, $E_{23}^z={\hat{\bf e}}_{z_P}\cdot{\bf E}_{23}$, $B_{21}={\hat{\bf e}}_{\varphi_P}\cdot{\bf B}_{21}$, $B_{22}={\hat{\bf e}}_{r_P}\cdot{\bf B}_{22}$ and $B_{23}={\hat{\bf e}}_{\varphi_P}\cdot{\bf B}_{23}$.  The values of the fields ${\bf E}_{nj}$ and ${\bf B}_{nj}$ in these expressions can be computed by means of (\ref{E172})-(\ref{E174}).  

Hence the component of the time-averaged Poynting vector along the radial direction ${\hat{\bf n}}_\infty$ is in the present case given by
\begin{equation}
{\hat{\bf n}}_\infty\cdot{\bf S}=-\frac{c}{8\pi}\Re[E_\perp(B_{21}^*+B_{23}^*)+\cos\theta_PB_{22}^*E_\parallel],
\label{E199}
\end{equation}
with
\begin{eqnarray}
\left[\matrix{E_\parallel\cr E_\perp\cr}\right]=\left[\matrix{E_{22}\cr  \sin\theta_P(E_{13}+E_{23}^z)-\cos\theta_P(E_{21}+E_{23}^r)\cr}\right]\nonumber\\*
\label{E200}
\end{eqnarray}
[see (\ref{E187}), (\ref{E189}) and (\ref{E198})].  The Stokes parameters for this radiation are given by (\ref{E191})--(\ref{E193}) and (\ref{E200}).

In the limit ${\hat R}_P\to\infty$, it follows from (\ref{E32}), (\ref{E85}) and (\ref{E194})--(\ref{E196}) that (i) the values of $E_{21}$, $E_{22}$, $B_{21}$ and $B_{22}$ are by a factor of the order of ${\hat R}_P^{-1}$ smaller than that of $E_{13}$, (ii) the value of $B_{23}$ approximately equals $-\sin\theta_P E_{13}$ and (iii) the value of ${\tilde{\bf u}}_{23}$ is given by
\begin{equation}
{\tilde{\bf u}}_{23}\simeq-\frac{{\rm i} s_z}{m}{\hat R}\left[\delta({\hat z}+{\hat z}_0)-\delta({\hat z}+{\hat z}_0)\right]{\hat{\bf n}}_\infty,\quad{\hat R}_P\gg1.
\label{E201}
\end{equation}
Hence, (\ref{E198}) and (\ref{E199}) reduce to 
\begin{equation}
{\bf E}\simeq E_{13}{\hat{\bf e}}_{z_P}+\csc\theta_P E^r_{23}{\hat{\bf n}}_\infty,\qquad {\bf B}\simeq{\hat{\bf n}}_\infty\times{\bf E},
\label{E202}
\end{equation}
and ${\hat{\bf n}}_\infty\cdot{\bf S}\simeq c\sin^2\theta_P\vert E_{13}\vert^2/(8\pi)$ at an observation point sufficiently distant from the source for which ${\hat R}\simeq{\hat R}_P\gg~1$.

\subsection{Case II: The emission from a radial polarization perpendicular to the rotation axis}
\label{subsec:energy2}

If the normals to the faces of the electrode pairs shown in figure~\ref{F2} lie along ${\hat{\bf e}}_r$, the only non-zero components of the charge-current density would be $j_r$ and $\rho$, i.e., the source terms $s_r$ and $s_0$.   Once $s_\varphi$ and $s_z$ in (\ref{E78})-(\ref{E83}) are set equal to zero and $s_0$ is evaluated with the aid of (\ref{E83}) and (\ref{E7}), the vectors ${\tilde{\bf u}}_{nj}$ and ${\tilde{\bf v}}_{nj}$ in these equations reduce to
\begin{equation} 
\left[\matrix{ {\tilde{\bf u}}_{11}& {\tilde{\bf u}}_{12}& {\tilde{\bf u}}_{13}}\right]=s_r\left[\matrix{{\hat{\bf e}}_{r_P}&{\hat{\bf e}}_{\varphi_P}&0}\right],
\label{E203}
\end{equation}
\begin{eqnarray}
    \left[
     \begin{array}{c}
     {\tilde{\bf u}}_{21}\\
     {\tilde{\bf u}}_{22}\\
     {\tilde{\bf u}}_{23}
     \end{array} \right]&=& \frac{{\rm i}s_r}{m}\left[\frac{1}{{\hat r}}+\delta({\hat r}-{\hat r}_L)-\delta({\hat r}-{\hat r}_U)\right]\left[
      \begin{array}{c}
     {\hat r}{\hat{\bf e}}_{r_P}\\
      {\hat r}{\hat{\bf e}}_{\varphi_P} \\
     -{\hat r}_P{\hat{\bf e}}_{r_P}+({\hat z}-{\hat z}_P){\hat{\bf e}}_{z_P}
\end{array} \right],
\label{E204}
\end{eqnarray}
\begin{equation} 
\left[\matrix{ {\tilde{\bf v}}_{21}\cr{\tilde{\bf v}}_{22}\cr{\tilde{\bf v}}_{23}}\right]=s_r\left[\matrix{-({\hat z}-{\hat z}_P){\hat{\bf e}}_{\varphi_P}\cr({\hat z}-{\hat z}_P){\hat{\bf e}}_{r_P}+{\hat r}_P{\hat{\bf e}}_{z_P}\cr0}\right],
\label{E205}
\end{equation}
and ${\tilde{\bf v}}_{1j}=0$.  The first factor in (\ref{E204}) stems from the fact that ${\hat\nabla}\cdot(s_r{\hat{\bf e}}_r)=(s_r/{\hat r})+\partial s_r/\partial{\hat r}$ and the ${\hat r}$-dependence of $s_r$ is, according to (\ref{E7}), given by the combination ${\rm H}({\hat r}-{\hat r}_L)-{\rm H}({\hat r}-{\hat r}_U)$ of Heaviside step functions.  In addition to the surface charges on ${\hat r}={\hat r}_L$ and ${\hat r}={\hat r}_U$, there is also a volume distribution of polarization charge in this case.

If, as in the preceding section, we let ${\bf E}_{nj}$ be the part of the field ${\bf E}$ arising from the source term ${\tilde{\bf u}_{nj}}$ and ${\bf B}_{nj}$ be the part of the field ${\bf B}$ arising from the source term ${\tilde{\bf v}_{nj}}$, then (\ref{E197}) together with (\ref{E203})-(\ref{E205}) yield
\begin{equation}
{\bf E}=(E_{11}+E_{21}+E_{23}^r){\hat{\bf e}}_{r_P}+(E_{12}+E_{22}){\hat{\bf e}}_{\varphi_P}+E_{23}^z{\hat{\bf e}}_{z_P}
\label{E206}
\end{equation}
and
\begin{equation}
{\bf B}=B_{22}^r{\hat{\bf e}}_{r_P}+B_{21}{\hat{\bf e}}_{\varphi_P}+B_{22}^z{\hat{\bf e}}_{z_P},
\label{E207}
\end{equation}
where in this case $E_{11}={\hat{\bf e}}_{r_P}\cdot{\bf E}_{11}$, $E_{21}={\hat{\bf e}}_{r_P}\cdot{\bf E}_{21}$, $E_{23}^r={\hat{\bf e}}_{r_P}\cdot{\bf E}_{23}$, $E_{12}={\hat{\bf e}}_{\varphi_P}\cdot{\bf E}_{12}$, $E_{22}={\hat{\bf e}}_{\varphi_P}\cdot{\bf E}_{22}$, $E_{23}^z={\hat{\bf e}}_{z_P}\cdot{\bf E}_{23}$, $B_{22}^r={\hat{\bf e}}_{r_P}\cdot{\bf B}_{22}$, $B_{21}={\hat{\bf e}}_{\varphi_P}\cdot{\bf B}_{21}$ and $B_{22}^z={\hat{\bf e}}_{z_P}\cdot{\bf B}_{22}$ are the components of the fields ${\bf E}_{nj}$ and ${\bf B}_{nj}$ given in (\ref{E172})-(\ref{E174}).  (Note that, because ${\bf E}$ and ${\bf B}$ have different orientations in Cases I and II, the expressions that define $E_{nj}$ and $B_{nj}$ in \S~\ref{subsec:energy1} are not the same as those that define these scalars in this section}.)

From (\ref{E187}), (\ref{E206}) and (\ref{E207}) it follows that the radial component of time-averaged Poynting vector has the value
\begin{equation}
{\hat{\bf n}}_\infty\cdot{\bf S}=\frac{c}{8\pi}\Re[E_\parallel(\sin\theta_P B_{22}^{z*}-\cos\theta_P B_{22}^{r*})-B_{21}^*E_\perp],
\label{E208}
\end{equation}
with
\begin{eqnarray}
\left[\matrix{E_\parallel\cr E_\perp\cr}\right]=\left[\matrix{E_{12}+E_{22}\cr -\cos\theta_P(E_{11}+E_{21}+E_{23}^r)+\sin\theta_P E_{23}^z\cr}\right],\nonumber\\*
\label{E209}
\end{eqnarray}
for the case where the current flows perpendicular to the rotation axis.  The Stokes parameters for this radiation are given by (\ref{E191})--(\ref{E193}) and (\ref{E209}).

In the limit ${\hat R}_P\to\infty$, it follows from (\ref{E32}), (\ref{E85}) and (\ref{E203})--(\ref{E205}) that the values of $E_{21}$ and $E_{22}$ are by a factor of the order of ${\hat R}_P^{-1}$ smaller than that of $E_{23}$, and the following limiting relationships hold: $E^z_{23}\simeq\cot\theta_P E^r_{23}$, $B_{21}\simeq\cos\theta_P E_{11}$, $B_{22}^r\simeq-\cos\theta_P E_{12}$, $B_{22}^z\simeq\sin\theta_P E_{12}$ and
\begin{equation}
{\tilde{\bf u}}_{23}\simeq-\frac{{\rm i} s_r}{m}{\hat R}\left[\frac{1}{{\hat r}}+\delta({\hat r}-{\hat r}_L)-\delta({\hat r}-{\hat r}_U)\right]{\hat{\bf n}}_\infty.
\label{E210}
\end{equation}
Hence, (\ref{E206})--(\ref{E208}) reduce to 
\begin{equation}
{\bf E}\simeq E_{11}{\hat{\bf e}}_{r_P}+E_{12}{\hat{\bf e}}_{\varphi_P}+\csc\theta_P E^r_{23}{\hat{\bf n}}_\infty,\quad {\bf B}\simeq{\hat{\bf n}}_\infty\times{\bf E},
\label{E211}
\end{equation}
and 
\begin{equation}
{\hat{\bf n}}_\infty\cdot{\bf S}\simeq\frac{c}{8\pi}\left[\cos^2\theta_P\vert E_{11}\vert^2+\vert E_{12}\vert^2\right]
\label{E212}
\end{equation}
at an observation point sufficiently distant from the source for which ${\hat R}\simeq{\hat R}_P\gg~1$.

\section{Numerical evaluation of characteristics of the radiation field}
\label{sec:numerical}
\subsection{Case Ia: A polarization parallel to the rotation axis for which the non-spherically decaying radiation beam spans $60^\circ\le\theta_P\le70^\circ$ and $110^\circ\le\theta_P\le120^\circ$}
\label{sec:numericalIa}

I have used Mathematica to compute the total radiation field $\matrix{[{\bf E}&{\bf B}]}$ and the time-averaged value of radial component of its flux density by numerically evaluating the integrals in (\ref{E172})-(\ref{E174}) and inserting the outcome in (\ref{E187}).  The results reported in this section are for the following choice of the dimensionless parameters of the source distribution (described in \S~\ref{sec:source}): ${\hat r}_L=\csc(7\pi/18)$,  ${\hat r}_U=\csc(\pi/3)$, ${\hat z}_0=0.1$, $m=10$ and $s_r=s_\varphi=0$, i.e., for a polarization current density parallel to the rotation axis whose sinusoidal distribution pattern, which consists of $10$ wavelengths of the polarization wave train (figure~\ref{F1}), azimuthally propagates with linear speeds ranging from $r_L\omega=1.0642 c$ to $r_U\omega=1.1547 c$ across the radial extent of the polarized dielectric (see figures~\ref{F1}, \ref{F2} and~\ref{F11}).  The only other parameter entering the expression for the radiation field is $s_z$ which I will assume to be independent of $({\hat r},{\hat z})$, i.e., to be constant over the cross-section of the dielectric.  It is not necessary to specify the value of the constant $s_z$ because we will be normalizing the Poynting vector, which is proportional to $s_z^2$, by dividing it by a quantity that is likewise proportional to $s_z^2$: namely, the average value of the power that propagates across the sphere ${\hat R}_P=10$ per unit solid angle [see (\ref{E188}) and (\ref{E190})].

The above values of the dimensionless parameters can be experimentally realized in a number of different ways.  The number of wavelengths $m$ of the polarization wave train fitting around the circumference of the dielectric ring (figure~\ref{F1}) would be $10$ if there are N=144 electrode pairs and the phases of sinusoidal oscillations of the voltages across adjacent electrodes (figure~\ref{F2}) differ by $\Delta\Phi=360^\circ m/N=25^\circ$.  For a voltage with the oscillation frequency $\nu=2.3$ GHz, the polarization wave train rotates around the dielectric ring with the angular frequency $\omega=2\pi\nu/m=1.45\times 10^9$ radians/sec.  This yields $c/\omega=20.76$ cm for the radius of the light cylinder, i.e., the radius at which the linear speed of the rotating distribution pattern of the source equals $c$.  So, the requirement $\csc(7\pi/18)\le{\hat r}\le\csc(\pi/3)$ on the range of linear speeds $r\omega$ of the polarization wave train in units of $c$ is met if the inner and outer boundaries of the dielectric have the radii $r_L=\csc(7\pi/18)c/\omega=22.09$ cm and $r_U=\csc(\pi/3)c/\omega=23.97$ cm, respectively.  The mean radius ${\bar r}=N\Delta\ell/(2\pi)$ of the dielectric ring would have the value $(r_L+r_U)/2=23.03$ cm if the distance between the centres of adjacent electrodes is $\Delta\ell=1$ cm.  The remaining dimension of the dielectric ring, i.e., its thickness in the direction parallel to the rotation axis, is in this case $2 z_0=4.15$ cm.   

\begin{figure}
\centerline{\includegraphics[width=12cm]{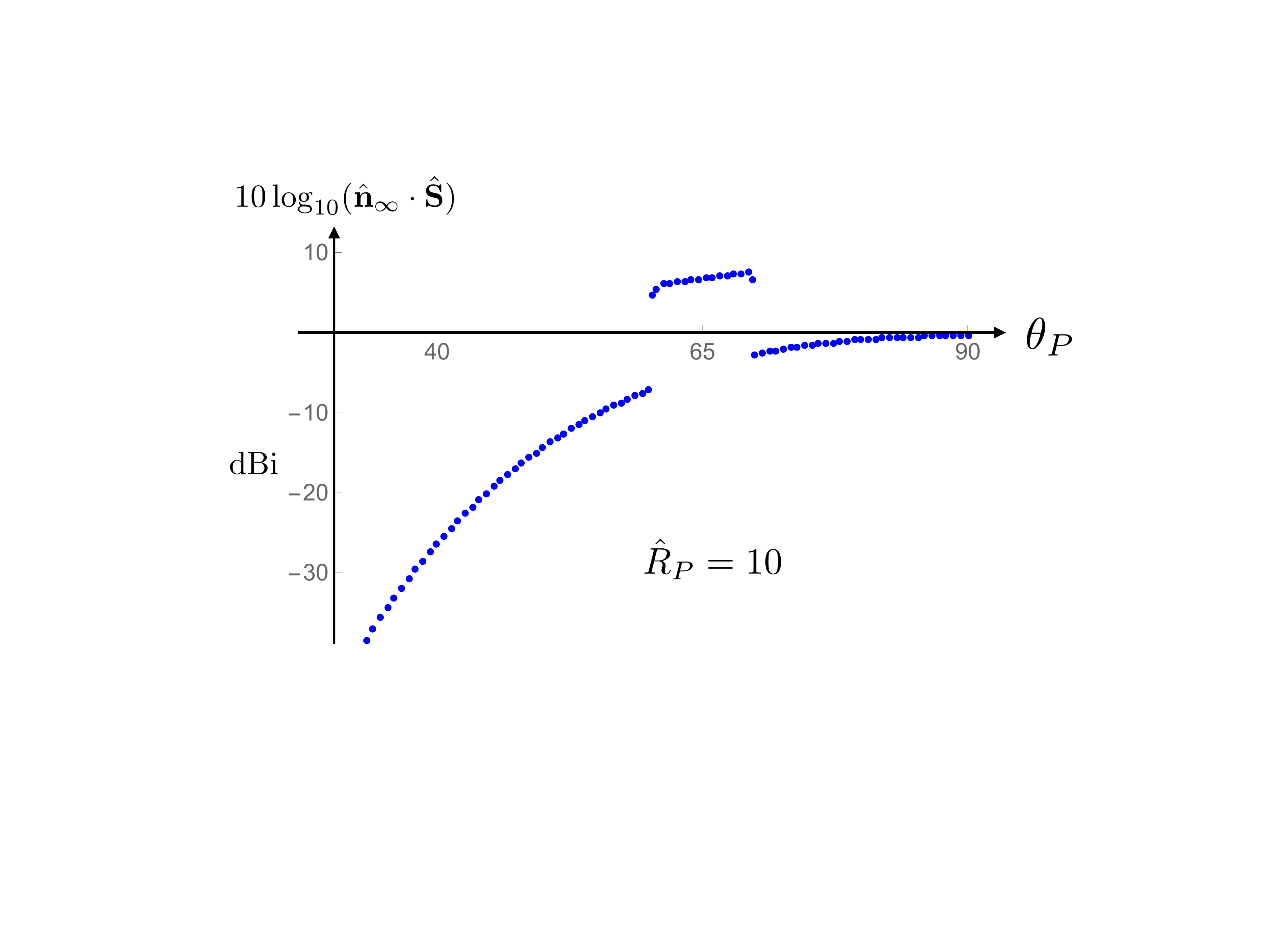}}
\caption{Logarithmic plot of the radial component of normalized Poynting vector ${\hat{\bf S}}$ versus the angle $\theta_P$ between the rotation axis and the radiation direction depicting directive gain of the radiation source at a distance of 10 light-cylinder radii.  Since this distribution is symmetric with respect to the equatorial plane $\theta_P=90^\circ$, its remaining half in $90^\circ<\theta_P<180^\circ$ is not shown here.  Values of the parameters used for plotting this figure are those for Case Ia described in \S~\ref{sec:numericalIa}.  (Only a discrete set of values of ${\hat{\bf n}}_\infty\cdot{\hat{\bf S}}$ are plotted, instead of a continuous curve, to render the required computing time for the points in $60^\circ\le\theta_P\le70^\circ$ manageable.)}
\label{F21}
\end{figure}

Angular distribution of the radiation at a distance of 10 light-cylinder radii is plotted in figure~\ref{F21}.  The vertical axis in this figure shows the normalized component of the Poynting vector ${\bf S}$ along the radiation direction ${\hat{\bf n}}_\infty$ versus the angle $\theta_P$ between the radiation direction and the rotation axis ${\hat{\bf e}}_z$ in degrees [see (\ref{E190}) and (\ref{E199})].  The normalization factor, i.e., the average value of the power that propagates across the sphere ${\hat R}_P=10$ per unit solid angle, has the value ${\bar S}_n\vert_{{\hat R}_P=10}=2.66\times10^{-3}\,\vert j_z\vert^2$ Watt/m${}^2$, in which $\vert j_z\vert=\nu s_z$ stands for the amplitude of the electric current density in units of amp/m${}^2$.  A logarithmic unit of measurement is used along the vertical axis so that changes in the value of ${\hat{\bf n}}_\infty\cdot{\hat{\bf S}}$, called directive gain in antenna theory~\citep{IEEE}, are registered in decibels.  The three-dimensional distribution of the directive gain, by virtue of having an azimuthal symmetry about the rotation axis ($\theta_P=0$) and a reflection symmetry about the equatorial plane ($\theta_P=90^\circ$), can be inferred from the plot shown in figure~\ref{F21}.  

The sharp changes in this distribution occur across the polar angles where the cusp locus $C$ of the bifurcation surface of the observation point either enters (at ${\hat r}={\hat r}_U$ when $\theta_P=60^\circ$) or leaves (at ${\hat r}={\hat r}_L$ when $\theta_P=70^\circ$) the source distribution (see figure~\ref{F11}).  The higher values of the radiation flux at angles for which the cusp $C$ intersects the source distribution reflect the fact that when the observation point is located in $60^\circ\le\theta_P\le70^\circ$ there exist source elements (in ${\hat r}_C\le{\hat r}\le{\hat r}_U$) which approach the  observer with the speed of light and zero acceleration at the retarded time.  On the other hand, characteristics of the emission in $0\le\theta_P\le60^\circ$ (only part of whose distribution is shown here) are the same as those of a conventional radiation: all volume elements of the rotating source approach an observer in $0\le\theta_P\le60^\circ$ with subluminal speeds.  Despite having a similar rate of decay in the present case, the emission in $70^\circ\le\theta_P\le90^\circ$ is however different from a conventional radiation: component of the velocity of each volume element of the source along the radiation direction (i.e., along the line that joins the source element to an observer in $70^\circ\le\theta_P\le90^\circ$) exceeds $c$, while the component of its acceleration along the radiation direction is non-zero.  

\begin{figure}
\centerline{\includegraphics[width=12cm]{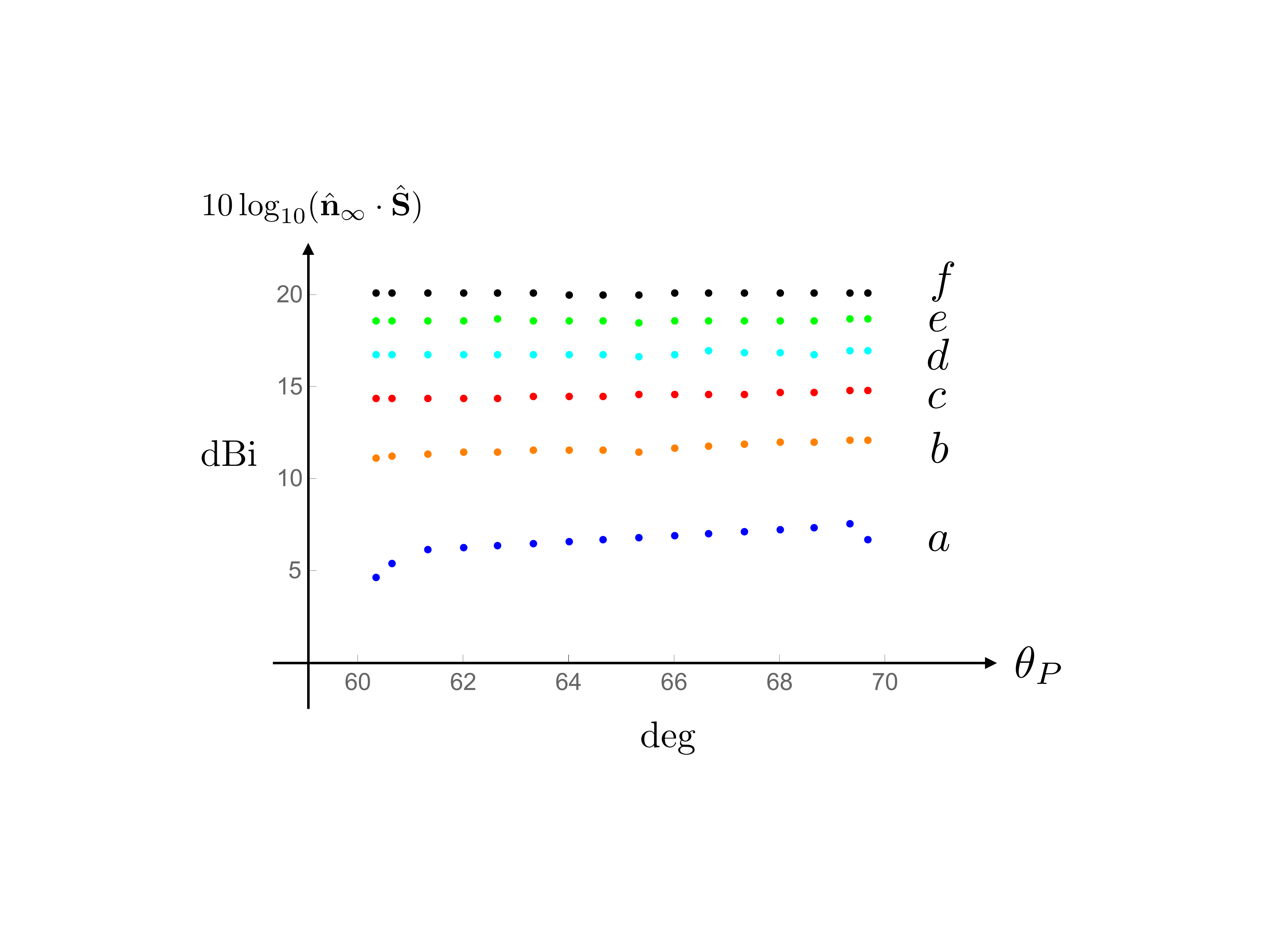}}
\caption{Vertically shifted distributions of the radial component of the normalized Poynting vector ${\hat{\bf S}}$ in $60^\circ\le\theta_P\le70^\circ$ for the following set of values of ${\hat R}_P$: ({\it a}) $10$, ({\it b}) $10^2$, ({\it c}) $10^3$, ({\it d}) $10^4$, ({\it e}) $10^5$ and ({\it f}) $10^6$.  In this figure, the values of the normalizing factors in ${\hat{\bf S}}$ for ${\hat R}_P>10$ have been shifted, relative to that for ${\hat R}_P=10$, by the following amounts: ({\it b}) $20$ dBi, ({\it c}) $40$ dBi, ({\it d}) $60$ dBi, ({\it e}) $80$ dBi and ({\it f}) $100$ dBi.  Note that these shifted curves would have been coincident had the Poynting vector been decaying as ${\hat R}_P^{-2}$.  The parts of these distributions in $0\le\theta_P\le60^\circ$ and $70^\circ\le\theta_P\le 90^\circ$ (which are not plotted here) are coincident with one another and with those for ${\hat R}_P=10$ that are shown in figure~\ref{F21}.  (Values of the parameters used for plotting this figure are those for Case Ia described in \S~\ref{sec:numericalIa}.)}
\label{F22}
\end{figure}

In contrast to the spherically decaying part of the radiation whose angular distribution is independent of distance, the angular distribution of the part of the radiation that propagates into $\theta_L\le\theta_P\le\theta_U$ has a dependence on $\theta_P$ that changes with ${\hat R}_P$.  Figure~\ref{F22} shows the angular distributions of the radial component of the normalized Poynting vector ${\hat{\bf S}}$ (i.e., the component of the Poynting vector along the radiation direction ${\hat{\bf n}}_\infty$ divided by the average value of the power that propagates across the sphere ${\hat R}_P=10$ per unit solid angle) for the following set of values of ${\hat R}_P$ (i.e., distance in units of the light-cylinder radius): ({\it a}) $10$, ({\it b}) $10^2$, ({\it c}) $10^3$, ({\it d}) $10^4$, ({\it e}) $10^5$ and ({\it f}) $10^6$.  To facilitate the comparison between these distributions, I have vertically shifted each of the curves for ${\hat R}_P>10$, relative to that for ${\hat R}_P=10$, by the following amounts in this figure: ({\it b}) $20$ dBi, ({\it c}) $40$ dBi, ({\it d}) $60$ dBi, ({\it e}) $80$ dBi and ({\it f}) $100$ dBi. These are the number of decibels by which $10\log_{10}({\hat{\bf n}}_\infty\cdot{\hat{\bf S}})$ would have changed if the magnitude of the Poynting vector for this part of the radiation had diminished as ${\hat R}_P^{-2}$ with distance.  The remaining parts of these distributions in $0<\theta_P< 60^\circ$ and $70^\circ\le\theta_P\le90^\circ$ are identical in shape to those for ${\hat R}_P=10$ (shown in figure~\ref{F21}) at these angles and coincide when shifted in the same way.

The separation between the shifted distributions in $60^\circ\le\theta_P\le70^\circ$ is a measure of the degree to which the dependence of the Poynting vector on distance departs from the inverse-square law.  Figure~\ref{F22} therefore indicates (i) that the Poynting vector decays more slowly with distance than predicted by the inverse-square law, and (ii) that the rate of decay of the Poynting vector with distance depends on both coordinates, $\theta_P$ and $R_P$, of the observation point. 

\begin{figure}
\centerline{\includegraphics[width=12cm]{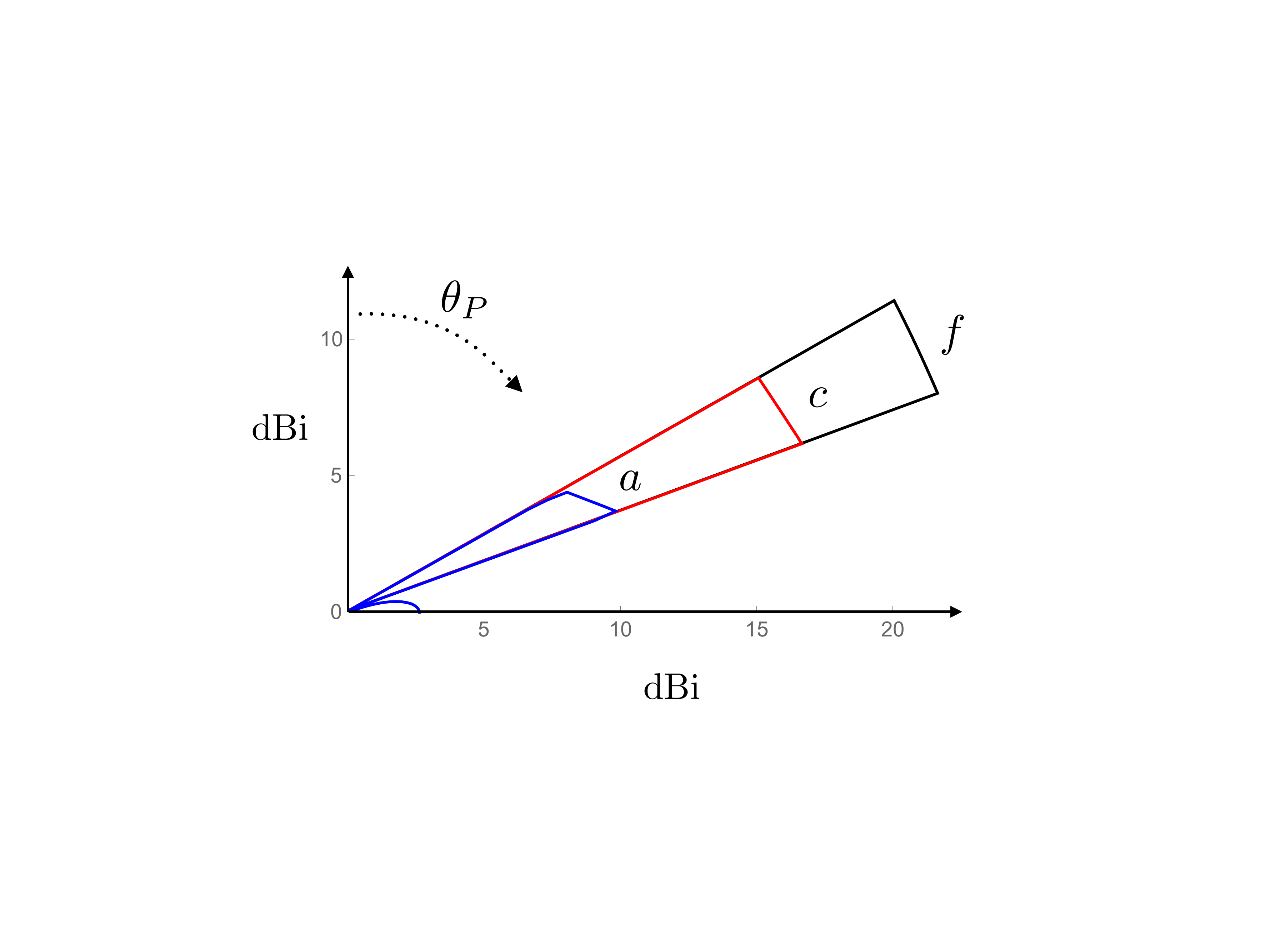}}
\caption{Angular distribution of the radiation in $0\le\theta_P\le90^\circ$ at distances ${\hat R}_P=10$ (curve $a$ of figure~\ref{F22}), ${\hat R}_P=10^3$ (curve $c$ of figure~\ref{F22}), and ${\hat R}_P=10^6$ (curve $f$ of figure~\ref{F22}).  The angle between the radius vector to each point and the vertical axis stands for the polar coordinate $\theta_P$ of the observation point.  The radial coordinate of each point on the curves $a$, $c$ and $f$ stands for the value of $10\log_{10}({\hat{\bf S}})$ that appears in figure~\ref{F22} against its coordinate $\theta_P$ plus $3$ dBi.  The emission in $70^\circ\le\theta_P\le90^\circ$ and the conventional radiation in $0\le\theta_P\le60^\circ$ (which is too weak to show up in this plot) have distance-independent distributions.  Three-dimensional distributions of the radiation patterns at the distances ${\hat R}_P=10, 10^3$ and $10^6$ are given by the surfaces of revolution that result from the reflection of curves $a$, $c$ and $f$ with respect to the horizontal axis followed by their rotation about the vertical axis.   (Values of the parameters used for plotting this figure are those for Case Ia described in \S~\ref{sec:numericalIa}.)}
\label{F23}
\end{figure}

Figure~\ref{F23} shows figure~\ref{F21} (the curve in blue that is marked as $a$) and parts $c$ and $f$ of figure~\ref{F22} in a polar coordinate system.   At each polar angle $\theta_P$, measured from the vertical axis, value of the radial coordinate of a point on curve $a$ shows the directive gain of the radiation source at the observation point (with the radial and polar coordinates ${\hat R}_P=10$ and $\theta_P$) plus $3$ dBi.  The $3$ dBi increase is introduced here to render the value of $10 \log_{10}({\hat{\bf n}}_\infty\cdot{\hat{\bf S}})$ positive, and so representable as a radial coordinate, across a sufficiently wide range of angles.  The three-dimensional angular distribution of the radiation can be obtained by rotating this curve about the vertical axis and reflecting the resulting surface of revolution with respect to the equatorial plane.  The emission in $0<\theta_P< 60^\circ$ is too weak to show up in this figure.  The radial coordinates of the points on curves $c$ and $f$ respectively equal the shifted values of $10 \log_{10}({\hat{\bf n}}_\infty\cdot{\hat{\bf S}})$ plotted in figure~\ref{F22} at the distances ${\hat R}_P=10^3$ and ${\hat R}_P=10^6$ increased by $3$ dBi.  

The non-spherically decaying part of the present radiation is linearly polarized with a fixed position angle: the Stokes parameters essentially have the values $L/I=1$, $V=0$ and $\psi_S=-\pi/2$ in $60^\circ\le\theta_P\le70^\circ$ and $110^\circ\le\theta_P\le120^\circ$ [see (\ref{E191})--(\ref{E193}) and (\ref{E200})]. 

\begin{figure}
\centerline{\includegraphics[width=12cm]{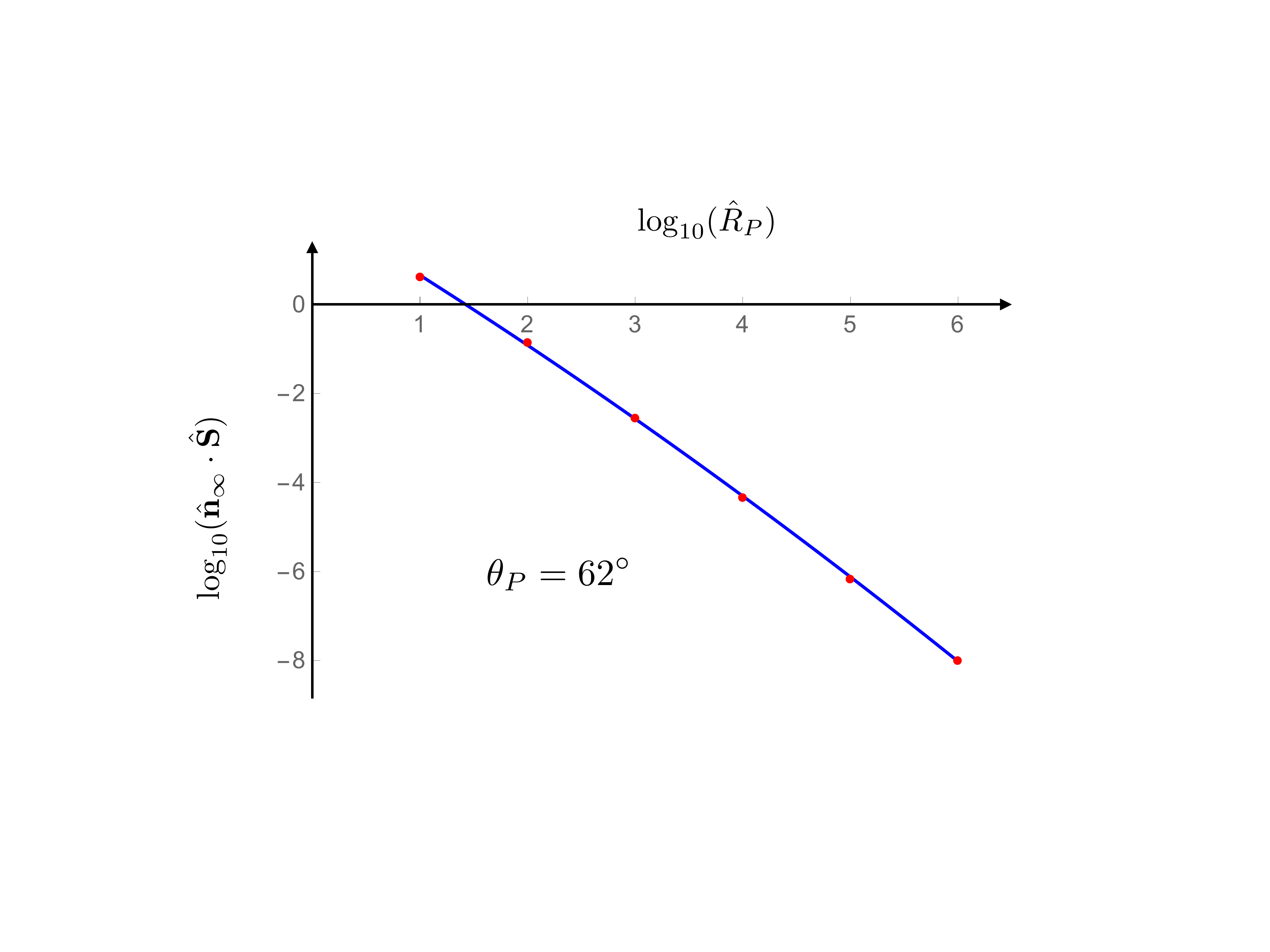}}
\caption{Logarithmic plot of the radial component of normalized Poynting vector versus distance along the generating line of a cone (in this case the cone $\theta_P=62^\circ$) inside the solid angle $60^\circ\le\theta_P\le70^\circ$, $0\le\varphi_P\le360^\circ$, where the radiation depicted in figures~\ref{F21}-\ref{F23} decays non-spherically.  The best fit to the computed points (extracted from figure~\ref{F22}) has the slowly varying slope $-1.45$ (instead of $-2$) in this direction.  (Values of the parameters used for plotting this figure are those for Case Ia described in \S~\ref{sec:numericalIa}.)}
\label{F24}
\end{figure}

In figure~\ref{F24}, I have plotted logarithm of the radial component of normalized Poynting vector versus logarithm of the distance (in units of the light-cylinder radius) at a polar angle ($\theta_P=62^\circ$) within the non-spherically decaying radiation beam depicted in figures~\ref{F21}-\ref{F23}.  The red dots are the data points that appear in figure~\ref{F23} against $\theta_P=62^\circ$ and the blue curve is the best fit to these dots, given by 
$\log({\hat{\bf n}}_\infty\cdot{\hat{\bf S}})=2.15-1.45\log{\hat R}_P-0.04( \log{\hat R}_P)^2$.  This figure shows that the flux of the outward-propagating radiation along $\theta_P=62^\circ$ diminishes with distance as ${\hat R}_P^{-1.45}$ (instead of ${\hat R}_P^{-2}$) over the range of distances ${\hat R}_P=10-10^6$ and that the value of the exponent in this power-law dependence on ${\hat R}_P$ is itself a slowly varying function of distance. 

\begin{figure}
\centerline{\includegraphics[width=12cm]{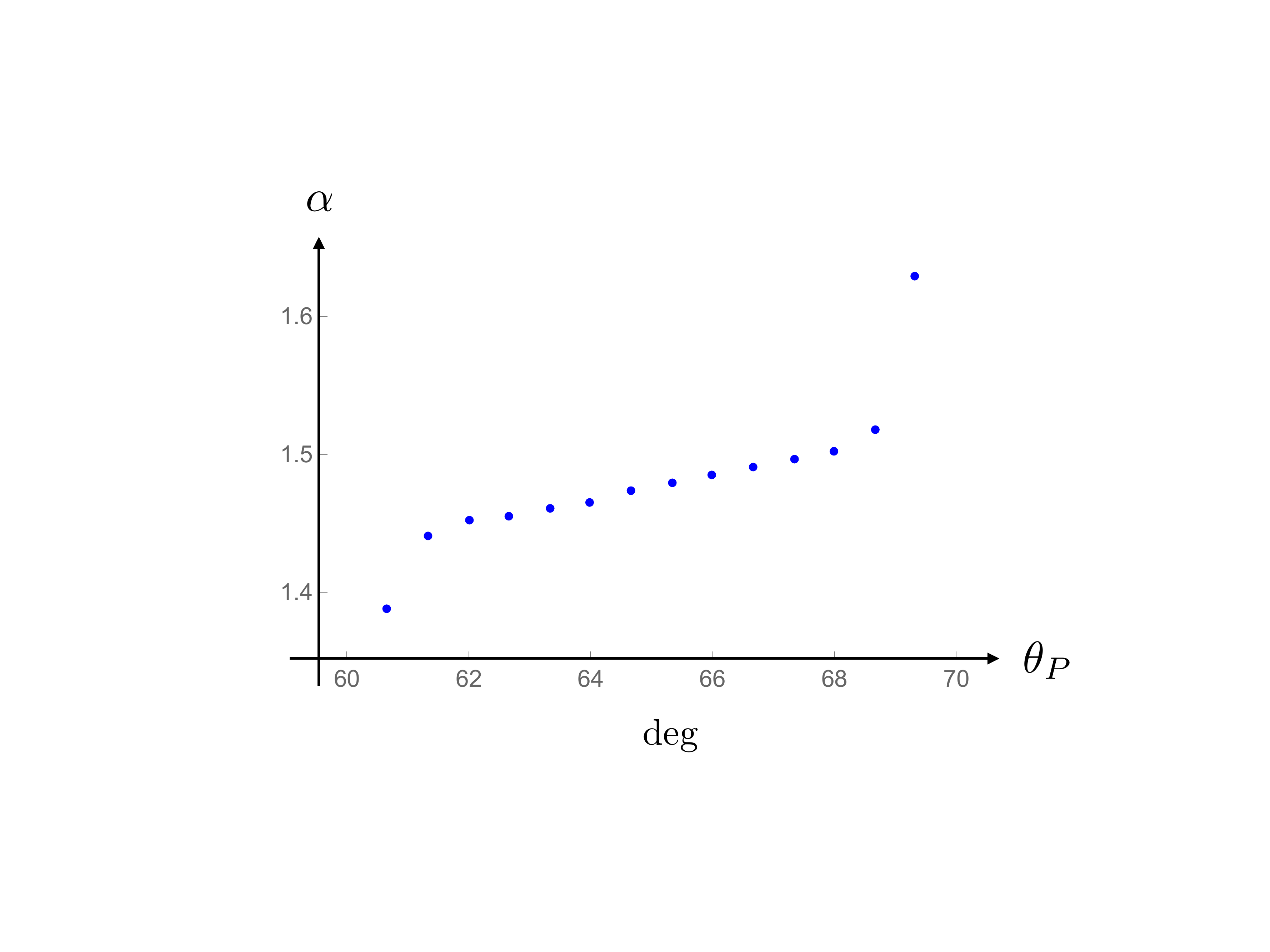}}
\caption{Exponent $\alpha$ in the distance dependence ${\hat R}_P^{-\alpha}$ of the radial component of normalized Poynting vector as a function of the polar angle $\theta_P$ within the solid angle $60^\circ\le\theta_P\le70^\circ$, $0\le\varphi_P\le360^\circ$.  (Values of the parameters used for plotting this figure are those for Case Ia described in \S~\ref{sec:numericalIa}.)}
\label{F25}
\end{figure}

By applying the same procedure to the remaining computed points in figure~\ref{F23}, one can find the exponent $\alpha$ in the distance dependence ${\hat R}_P^{-\alpha}$ of ${\hat{\bf n}}_\infty\cdot{\hat{\bf S}}$ also for other directions within the non-spherically decaying beam.  The result is shown in figure~\ref{F25}.  Thus the departure of the value of $\alpha$ from $2$ occurs over a solid angle whose polar and azimuthal widths are constant.  This departure is less pronounced at the edge $\theta_P=70^\circ$ of the beam where only limited segments of the loci $C$ and $S$ lie within the source distribution (see \S~\ref{sec:transitional}) but increases toward the edge $\theta_P=60^\circ$ as the portion of the source that lies within the bifurcation surface increases in volume (see figure~\ref{F11}). 

Constancy of the width of the solid angle over which the Poynting vector decays as ${\hat R}_P^{-\alpha}$, with $1<\alpha<2$, implies that the power propagating across a sphere of radius ${\hat R}_P$ increases as ${\hat R}_P^{2-\alpha}$ with distance, rather than being independent of ${\hat R}_P$ as in a conventional radiation: a result that at first sight seems to contravene the requirements of the conservation of energy.  However, the constructively interfering waves from the particular set of volume elements of the polarization current that are responsible for the non-spherically decaying signal at a given observation point constitute a radiation beam for which the time-averaged value $\partial{\cal U}/\partial t_P$ of the temporal rate of change of energy density is negative (see appendix~\ref{AppC}) rather than being zero as in a conventional radiation.  This means that the flux of energy out of a closed region (e.g., out of the volume bounded by two large spheres centred on the source) is greater than the flux of energy into it because the amount of energy contained within the region decreases with time.  I have shown in appendix~\ref{AppC} that the computed value of $\partial{\cal U}/\partial t_P$ for the non-spherically decaying radiation described in this section is indeed negative and decays as ${\hat R}_P^{-\beta}$ with a value of $\beta$ whose range and angular dependence are similar to those of $\alpha$ (cf., figures~\ref{fC2} and \ref{E25}).  Since neither the topology of the retarded distribution of the present source nor the temporal rate of change of the energy density of its emission ever attain a steady state, it is not surprising that $\alpha$ should differ from its conventional value $2$.  As pointed out in appendix~\ref{AppC}, the slow decay of the radiation discussed in this paper is in fact required by the conservation of energy given that the time-averaged temporal rate of change of the energy density of this radiation is negative rather than zero.

A final remark is in order: from the values of the mean radius of the dielectric ring (${\bar r}=23.03$ cm) and the wavelength associated with the oscillation frequency of the electrodes ($\lambda=12$ cm) it follows that the distance at which the Fresnel number ${\bar r}^2/(R_P\lambda)$ attains the value unity is $R_P\simeq44$ cm in the present case.  Given that this distance is by many orders of magnitude shorter than the distances over which the radiation from a superluminally rotating source is shown to disobey the inverse-square law (see figure~\ref{F24}), the non-spherical decay discussed in this paper is not in any way related to that which occurs within the Fresnel (or Rayleigh) distance when a conventional radiation beam is focused.

\subsection{Case Ib: A polarization parallel to the rotation axis for which the non-spherically decaying radiation beam encompasses the equatorial plane}
\label{sec:numericalIb}

I have numerically evaluated the integrals in (\ref{E172})-(\ref{E174}), and thereby the time-averaged Poynting vector (\ref{E187}), also for the following choice of dimensionless parameters of the source distribution described in \S~\ref{sec:source}: ${\hat r}_L=1$,  ${\hat r}_U=1.2$, ${\hat z}_0=0.1$, $m=10$ and $s_r=s_\varphi=0$, i.e., for a polarization parallel to the rotation axis whose sinusoidal distribution pattern, consisting of $10$ wavelengths, azimuthally propagates with linear speeds ranging from $c$ to $1.2c$ across the radial extent of the polarized dielectric (see figures~\ref{F1} and \ref{F11}).  The oscillation frequency $\nu$ of the voltages that energize the electrode pairs is here set equal to $2.5$ GHz, so that the angular frequency of rotation of the distribution pattern of the polarization current has the value $\omega=2\pi\nu/m=1.57\times10^9$ radians/sec.  This yields a light cylinder with the radius $c/\omega=19.1$ cm and requires that the dielectric hosting the polarization current should have the mean radius ${\textstyle\frac{1}{2}}(r_L+r_U)=21$ cm and the radial width $ r_U-r_L=3.8$ cm.  These values of the parameters can be experimentally realized by surrounding the dielectric ring with an array of $N=130$ electrode pairs whose centres are a distance $\Delta\ell=1.015$ cm apart and the phases of whose oscillations differ by $\Delta\Phi=27.7^\circ$.  As in \S~\ref{sec:numericalIa}, I have moreover assumed that $s_z$ is independent of $({\hat r},{\hat z})$ and that the axial thickness of the dielectric is $2 z_0=3.8$ cm.  

\begin{figure}
\centerline{\includegraphics[width=11cm]{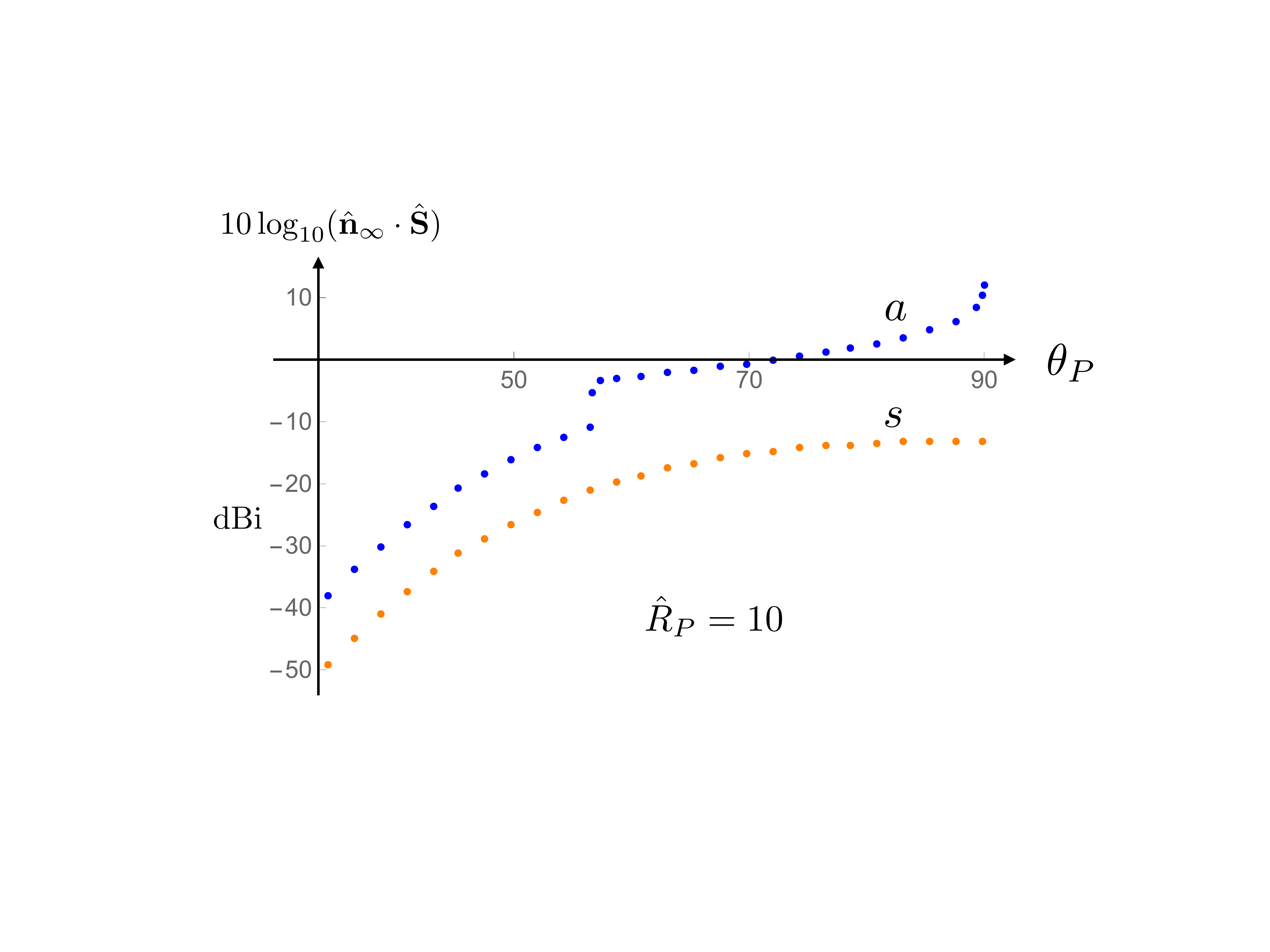}}
\caption{The outward-propagating component of the normalized Poynting vector ${\hat{\bf S}}$ versus the polar coordinate $\theta_P$ of the observation point at the distance ${\hat R}_P=10$ for both a superluminally rotating source (curve $a$) and a corresponding stationary source (curve $s$).  Since these distributions are symmetric with respect to the equatorial plane $\theta_P=90^\circ$, their remaining halves in $90^\circ\le\theta_P\le180^\circ$ are not shown here.  (Values of the parameters used for plotting this figure are those for Case Ib described in \S~\ref{sec:numericalIb}.)}
\label{F26}
\end{figure}

Curve $a$ in figure~\ref{F26} is a plot of the radial component of time-averaged Poynting vector divided by the average power that propagates across the sphere ${\hat R}_P=10$ per unit solid angle (i.e., the directive gain of the present radiation source at a distance of $10$ light-cylinder radii) versus the polar coordinate $\theta_P$ of the observation point [see (\ref{E190}) and (\ref{E199})].  The average power that propagates across the sphere ${\hat R}_P=10$ per unit solid angle is in this case given by ${\bar S}_n\vert_{{\hat R}_P=10}=2.03\times10^{-2}\,\vert j_z\vert^2$ Watt/m${}^2$, where $\vert j_z\vert=\nu s_z$ stands for the amplitude of the electric current density in units of amp/m${}^2$.  Since the source is symmetric with respect to the equatorial plane (see figure~\ref{F11}), so is the distribution of its radiation.  The remaining half of the radiation distribution in $90^\circ\le\theta_P\le180^\circ$ consists therefore of the reflection of the half that is shown in figure~\ref{F26} across the plane $\theta_P=90^\circ$.  The rapid change in the intensity of the radiation at $\theta_P=\theta_L\simeq56.4^\circ$ reflects the penetration of the cusp $C$ of the bifurcation surface associated with the observation point $P$ into the source distribution across its boundary ${\hat r}={\hat r}_U$ (see figure~\ref{F11}).  Once the observation point $P$ is in $\theta_L\le\theta_P\le\theta_U$, certain volume elements of the source (those in ${\hat r}_C\le{\hat r}\le{\hat r}_U$) approach $P$ along the radiation direction with the speed of light and zero acceleration at the retarded time, thus emitting waves that interfere constructively.  The weaker radiation in $0<\theta_P < \theta_L$ consists entirely of the conventional radiation described by (\ref{E172}). 

Curve $s$ in figure~\ref{F26} shows the radial component of normalized Poynting vector for the radiation generated by a source that is the same as the source generating the radiation depicted by curve $a$ in every respect (has the same dimensions, the same oscillation frequency, the same current density, \ldots) except that its sinusoidal distribution pattern is stationary, i.e., is described by
\begin{equation}
P_z(r,\varphi,z,t)=s_z \cos(m\varphi)\cos(m\omega t),
\label{E215}
\end{equation}
instead of (\ref{E1}), and so does not rotate around the dielectric ring.  The normalization factor used for curve $s$ is the same as that for curve $a$: namely the average value of the power arising from the rotating source that propagates across the sphere ${\hat R}_P= 10$ per unit solid angle.  Comparing the two curves we can see that even at the relatively short distance $R_P=10c/\omega=191$ cm from the source, the intensity of the non-spherically decaying radiation generated by the superluminally rotating source exceeds that of the conventional radiation generated by a corresponding stationary source by more than $25$ decibels (i.e., by more than a factor of $300$) on the equatorial plane.

\begin{figure}
\centerline{\includegraphics[width=12cm]{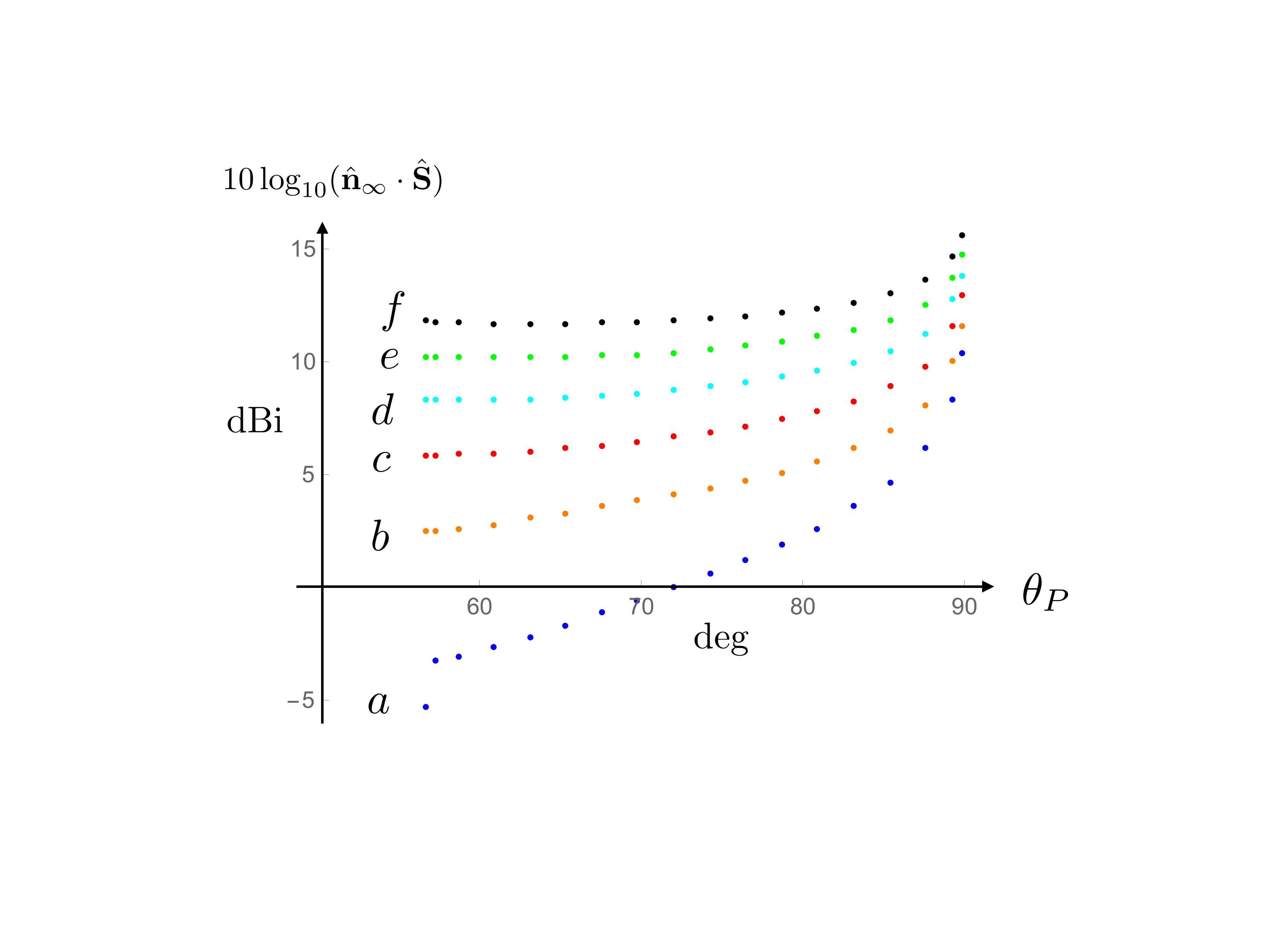}}
\caption{Vertically shifted distributions of the radiation in $\theta_L\le\theta_P\le\pi/2$ at six values of ${\hat R}_P$: ({\it a}) $10$, ({\it b}) $10^2$, ({\it c}) $10^3$, ({\it d}) $10^4$, ({\it e}) $10^5$ and ({\it f}) $10^6$.  As in figure~\ref{F22}, the normalization factor used here is the integral of the Poynting vector over a sphere of radius ${\hat R}_P=10$ divided by $4\pi$.  Vertical coordinates of the points in the distributions at ${\hat R}_P=10^2,\, 10^3,\, 10^4,\, 10^5,\, 10^6$ are respectively raised by $20,\, 40,\,60,\,80,\,100$ dBi relative to those in the distribution at ${\hat R}_P=10$.  The spherically decaying parts of these distributions in $0\le\theta_P\le 56.4^\circ$ are identical in shape to that for ${\hat R}_P=10$ (shown in figure~\ref{F26}) and would coincide if included in this figure.  (Values of the parameters used for plotting this figure are those for Case Ib described in \S~\ref{sec:numericalIb}.)}
\label{F27}
\end{figure}

\begin{figure}
\centerline{\includegraphics[width=11cm]{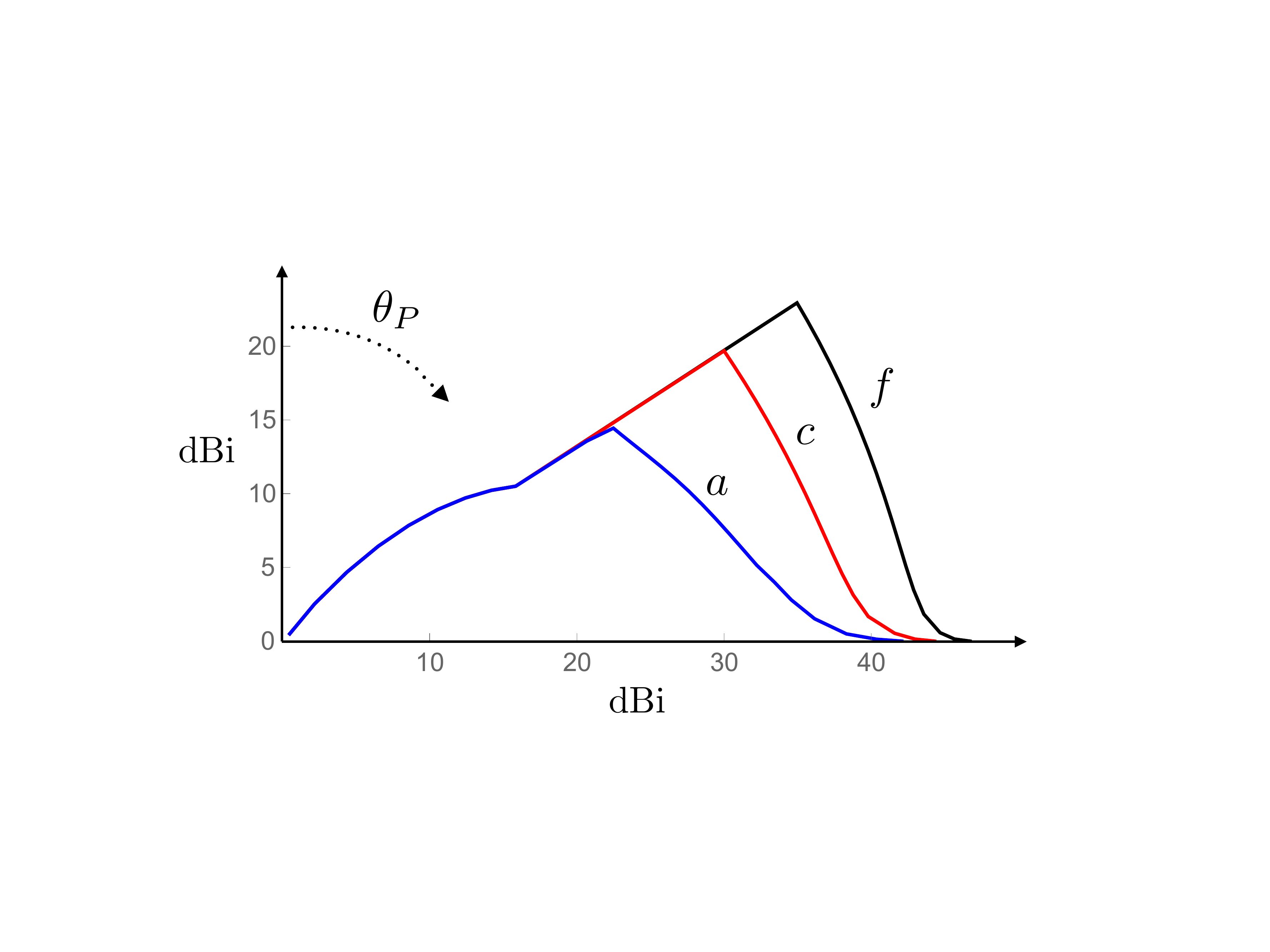}}
\caption{Polar diagrams of the distributions depicted by curve $a$ of figure~\ref{F26} (shown in blue) and curves $c$ and $f$ of figure~\ref{F27} (shown in red and black, respectively).  The angle between the radius vector to each point and the vertical axis stands for the polar coordinate $\theta_P$ of the observation point.  The radial coordinate of each point on the curves $a$, $c$ and $f$ stands for the value of $10\log_{10}({\hat{\bf S}})$ that appears in figure~\ref{F27} against its coordinate $\theta_P$ plus $30$ dBi.    All three distributions coincide in $0\le\theta_P\le56.4^\circ$ where their decay with distance complies with the inverse-square law.  Three-dimensional distributions of the radiation patterns at the distances ${\hat R}_P=10$, ${\hat R}_P=10^3$ and ${\hat R}_P=10^6$ are given by the surfaces of revolution that result from the reflection of curves $a$, $c$ and $f$ with respect to the horizontal axis followed by their rotation about the vertical axis.  (Values of the parameters used for plotting this figure are those for Case Ib described in \S~\ref{sec:numericalIb}.)}
\label{F28}
\end{figure}

\begin{figure}
\centerline{\includegraphics[width=11cm]{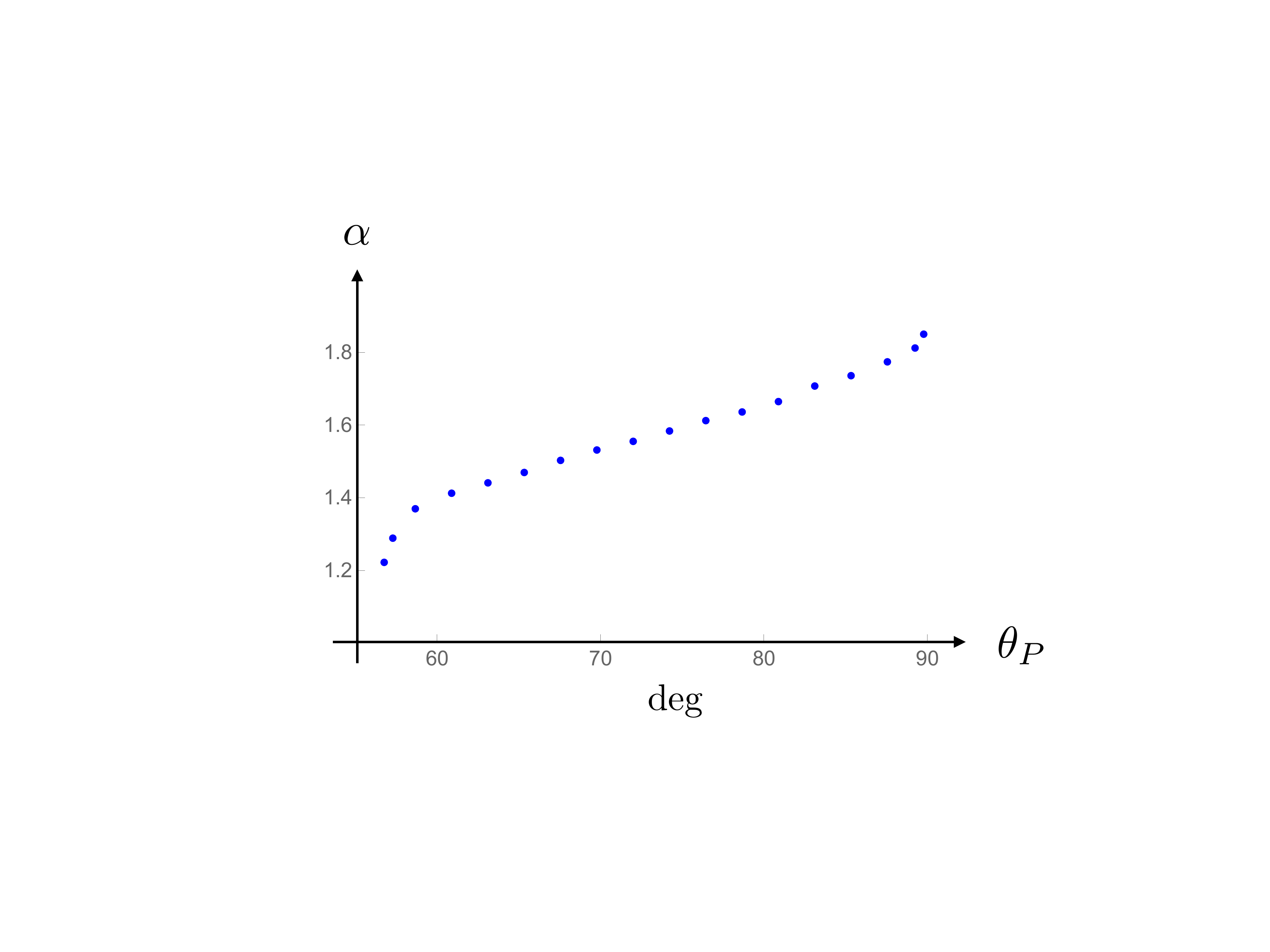}}
\caption{Angular dependence of the exponent $\alpha$ in the power-law ${\hat R}_P^{-\alpha}$ that best describes the decay of the Poynting vector with distance over the range $10\le{\hat R}_P\le10^6$.  Vertical coordinates of the points plotted in this figure were obtained by applying the procedure illustrated in figure~\ref{F24} to the data in figure~\ref{F27}.  (Values of the parameters used for plotting this figure are those for Case Ib described in \S~\ref{sec:numericalIb}.)}
\label{F29}
\end{figure}

Figures \ref{F27}, \ref{F28} and \ref{F29} are the counterparts of figures \ref{F22}, \ref{F23} and \ref{F25} for Case Ib.  Maximum value of the intensity of the radiation depicted in these figures occurs at $\theta_P=\pi/2$ because an additional mechanism of focusing comes into play when the coordinate ${\hat z}_P$ of the observation point falls within the ${\hat z}$-extent $-{\hat z}_0\le{\hat z}\le{\hat z}_0$ of the source distribution, i.e., the stationary point ${\hat z}={\hat z}_P$ of the phases ${\hat\varphi}_\pm$ of the exponential factors appearing in (\ref{E85}) falls within the domain of integration (see \S~\ref{subsec:locus}).  This radiation propagates into a solid angle encompassing the equatorial plane whose polar width decreases as ${\hat R}_P^{-1}$ in the far zone [see (\ref{E116})].  From the fact that the area subtended by the solid angle into which this part of the radiation propagates increases as ${\hat R}_P$ (instead of ${\hat R}_P^2$) while the rate of decay of the Poynting vector with distance for the emission into the equatorial plane is close to ${\hat R}_P^{-1.85}$ (see figure~\ref{F29}), it can be seen that the power carried by the focused equatorial radiation decreases with distance, rather than being constant as in a conventional radiation.  This means that the increase in the flux of energy with distance across surfaces subtending the fixed solid angle within which ${\hat{\bf n}}_\infty\cdot{\hat{\bf S}}$ decays more slowly than ${\hat R}_P^{-2}$ is partly compensated by a corresponding decrease in the flux of energy with distance across surfaces subtending the narrowing solid angle $\pi/2-\arcsin({\hat z}_0/{\hat R}_P)\le\theta_P\le\pi/2+\arcsin({\hat z}_0/{\hat R}_P)$, $0\le\varphi_P<2\pi$ into which the stronger equatorial radiation propagates.  In the case of the present example, therefore, the radiation meets the requirements of the conservation of energy not only by means of the mechanism discussed in appendix~\ref{AppC} but partly by containing a high-intensity beam whose width narrows with distance.

The non-spherically decaying part of the radiation is, as in Case Ia, linearly polarized with a fixed position angle: the Stokes parameters essentially have the values $L/I=1$, $V=0$ and $\psi_S=-\pi/2$ throughout $56.4^\circ\le\theta_P\le123.6^\circ$ [see (\ref{E191})--(\ref{E193}) and (\ref{E200})]. 

\subsection{Case II: A radial polarization for which the non-spherically decaying radiation beam spans $60^\circ\le\theta_P\le70^\circ$ and $110^\circ\le\theta_P\le120^\circ$}
\label{sec:numericalII}

In this section I analyse the emission from an example of the source distribution described in \S~\ref{subsec:energy2} for which the dimensionless parameters appearing in the expressions for the fields [in (\ref{E172})--(\ref{E174})] have the same values as those adopted in \S~\ref{subsec:energy1} except that the direction of the polarization current density is perpendicular (rather than parallel) to the rotation axis.  The polarization current density again has a sinusoidal distribution pattern consisting of $m=10$ wavelengths which azimuthally propagates with linear speeds ranging from ${\hat r}_L=\csc(7\pi/18)$ (in units of $c$), at the inner edge, to ${\hat r}_U=\csc(\pi/3)$ at the outer edge of a dielectric ring with the axial thickness $2{\hat z}_0=0.2$ (in units of the light-cylinder radius $c/\omega$).  The voltages across the electrode pairs have the oscillation frequency $\nu=2.5$ GHz so that the angular frequency of rotation of the polarization current has the value $\omega=2\pi\nu/m=1.57\times10^9$ radians/sec.  This yields a light cylinder with the radius $c/\omega=19.1$ cm and requires that the dielectric hosting the polarization current should have the mean radius ${\textstyle\frac{1}{2}}( r_L+r_U)=21$ cm and the radial width $r_U-r_L=3.8$ cm.   Moreover, the axial thickness of the dielectric is $2 z_0=3.8$ cm and $s_r$, i.e., the non-zero component of ${\bf s}$, is independent of $({\hat r},{\hat z})$.  These values of the parameters can be experimentally realized by surrounding the dielectric ring with an array of $N=130$ electrode pairs whose centres are a distance $\Delta\ell=1.015$ cm apart and the phases of whose oscillations differ by $\Delta\Phi=27.7^\circ$. 

\begin{figure}
\centerline{\includegraphics[width=11cm]{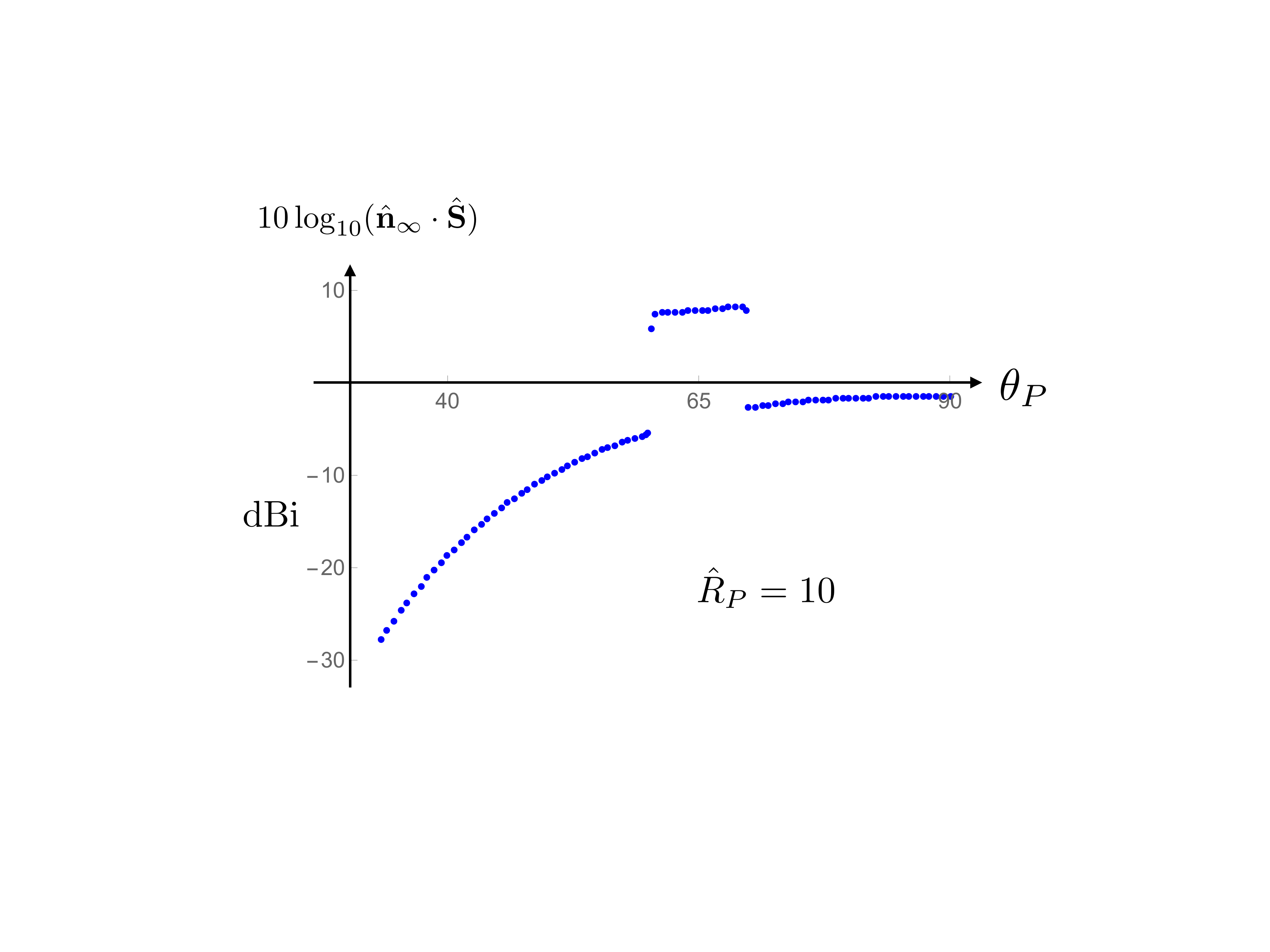}}
\caption{Logarithmic plot of the time-averaged value of the radial component of the normalized Poynting vector versus the polar angle $\theta_P$ for the radiation from the source described in \S~\ref{subsec:energy2} at $10$ light-cylinder radii. The Poynting vector is here divided by the mean value of the flux of power across a sphere of radiaus ${\hat R}_P=10$ (concentric with the ring-shaped source) per unit solid angle.  The vertical axis therefore marks the directivity of the radiation source described in \S~\ref{subsec:energy2} at $10$ light-cylinder radii.  The distribution of this radiation is independent of the azimuthal angle $\varphi_P$ and is symmetric with respect to the equatorial plane $\theta_P=90^\circ$.  The sharp changes across $\theta_P=60^\circ$ and $\theta_P=70^\circ$ reflect the fact that only an observer in $60^\circ<\theta_P<70^\circ$ can receive the cusped radiation generated by the superluminally rotating volume elements of the distribution pattern of the source.  (Values of the parameters used for plotting this figure are those for Case II described in \S~\ref{sec:numericalII}.)}
\label{F30}
\end{figure}

\begin{figure}
\centerline{\includegraphics[width=11cm]{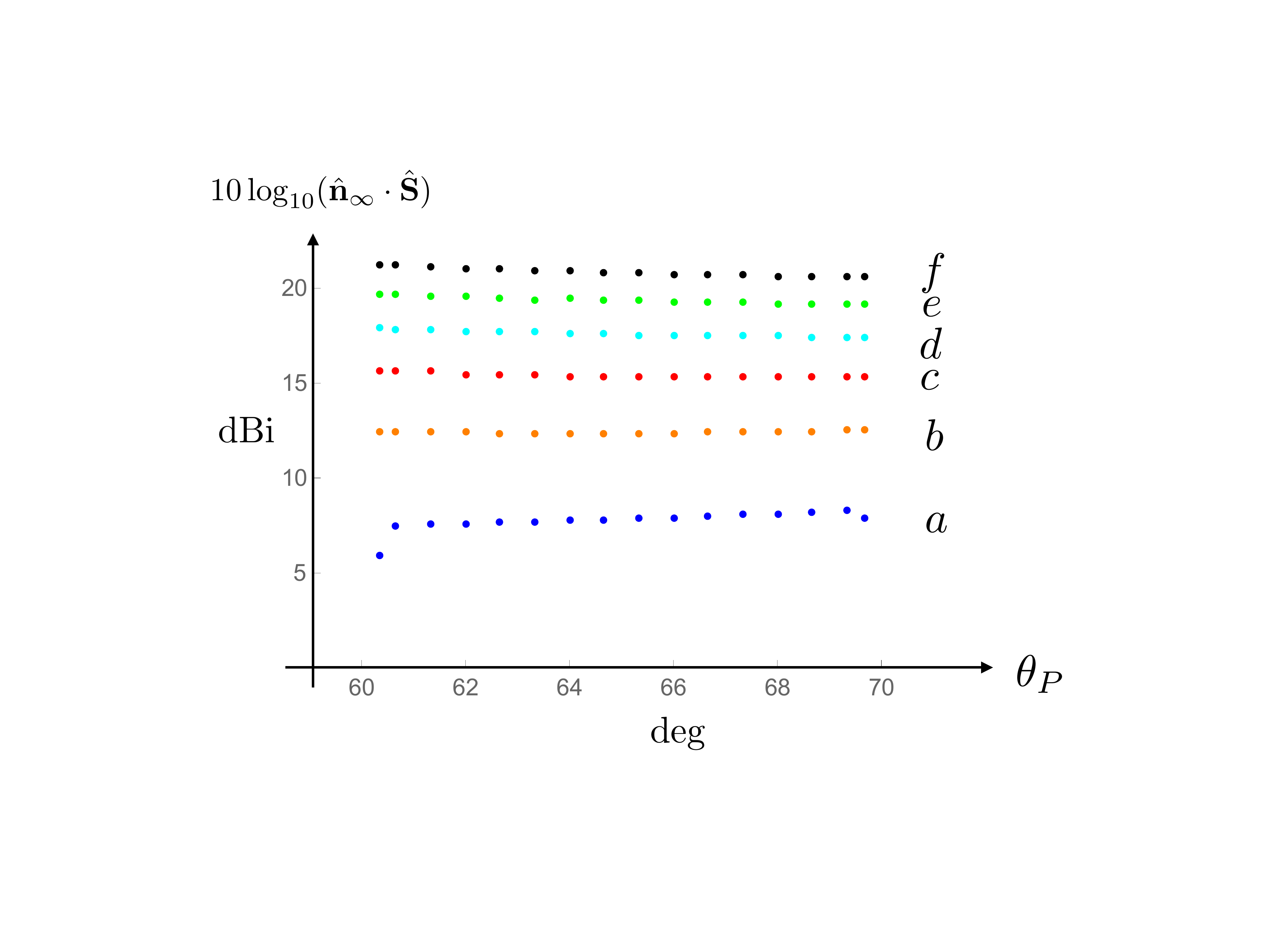}}
\caption{Vertically shifted time-averaged values of the radial component of the normalized Poynting vector over the limited range of polar angles where the cusped radiation from the source described in \S~\ref{subsec:energy2} is observeable. Curves $a$ to $f$ respectively correspond to the values $10$, $10^2$, $\cdots$, $10^6$ of the dimensionless distance ${\hat R}_P$.  The distribution at each ${\hat R}_P$ with a value $\ge10^2$ is here shifted upward relative to the preceding distribution at ${\hat R}_P/10$ by $20$ dBi.  The separation of the curves in this figure is a measure of the degree to which the dependence of the radial component of time-averaged Poynting vector on distance differs from ${\hat R}_P^{-2}$.  Had ${\hat{\bf n}}\cdot{\hat{\bf S}}$ been decaying as ${\hat R}_P^{-2}$, a tenfold increase in the value of distance would have resulted in a $20$ dBi decrease in the value of $10\log_{10}({\hat{\bf n}}\cdot{\hat{\bf S}})$ and so the curves $c$, $d$, $e$ and $f$ would have been coincident with curve $a$.  The parts of the radiation distribution in $0\le\theta_P\le60^\circ$ and $70^\circ\le\theta_P\le90^\circ$ are identical in shape to those for ${\hat R}_P=10$ (shown in figure~\ref{F31}) at all distances and would have coincided had they been included in this figure.  (Values of the parameters used for plotting this figure are those for Case II described in \S~\ref{sec:numericalII}.)}
\label{F31}
\end{figure}

\begin{figure}
\centerline{\includegraphics[width=11cm]{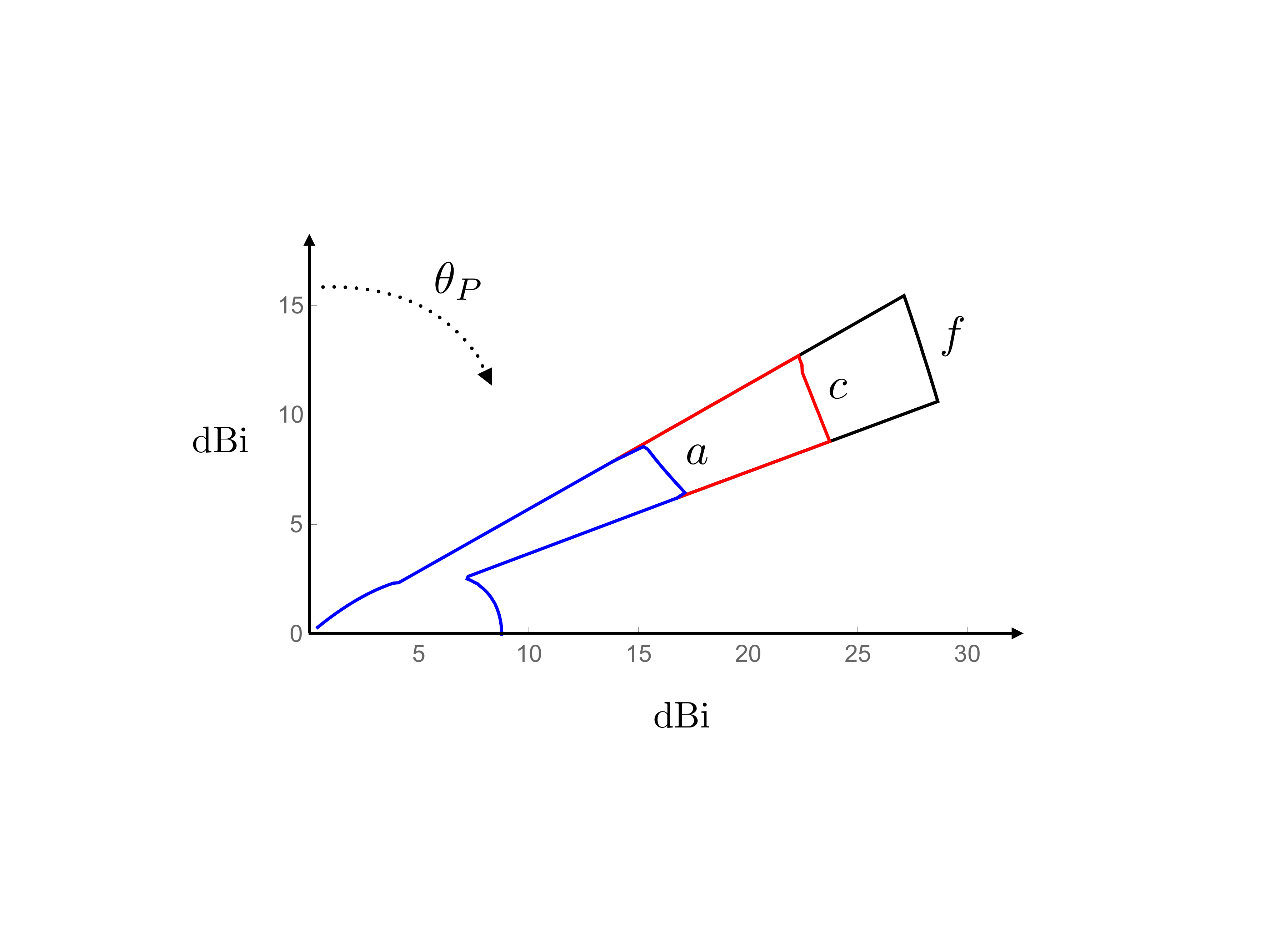}}
\caption{The results shown in figures~\ref{F30} and \ref{F31} are here depicted in polar coordinates.  The value of the radial coordinate of each point on cuve $a$ corresponds to that of the time-averaged radial component of the normalized Poynting vector in logarithmic units (shown on the vertical axis of figure~\ref{F30}) plus $10$ dBi, and the value of the polar angle of each point corresponds to that of $\theta_P$. This holds true also for the points on curves $c$ and $f$ except that their radial coordinates in $60^\circ\le\theta_P\le70^\circ$ respectively correspond to the shifted values of $10\log_{10}({\hat{\bf n}}\cdot{\hat{\bf S}})$ for ${\hat R}_P=10^3$ and $10^6$ shown on the vertical axis of figure~\ref{F31}.  Three-dimensional distributions of the radiation patterns at the distances ${\hat R}_P=10, 10^3$ and $10^6$ are given by the surfaces of revolution that result from the reflection of curves $a$, $c$ and $f$ with respect to the horizontal axis followed by their rotation about the vertical axis.  (Values of the parameters used for plotting this figure are those for Case II described in \S~\ref{sec:numericalII}.)} 
\label{F32}
\end{figure}

\begin{figure}
\centerline{\includegraphics[width=11cm]{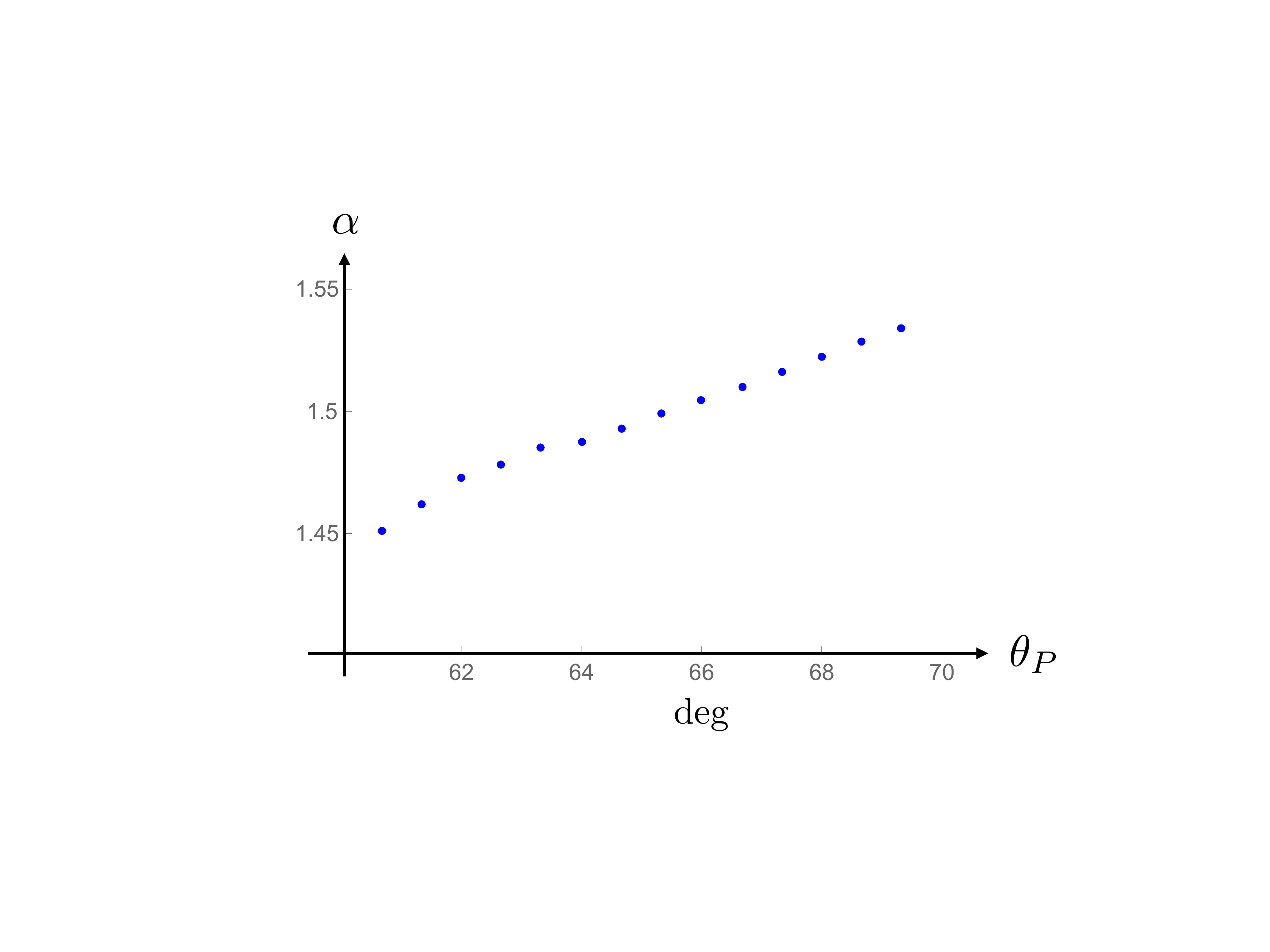}}
\caption{The exponent $\alpha$ in the dependence ${\hat R}_P^{-\alpha}$ of the time-averaged radial component of the Poynting vector on the distance ${\hat R}_P$ at the polar angles $60^\circ\le\theta_P\le70^\circ$ within the cusped radiation beam shown in figure~\ref{F32}.  To derive the value of this exponent I have used the data shown in figure \ref{F31} to plot $\log({\hat{\bf n}}\cdot{\hat{\bf S}})$ versus $\log{\hat R}_P$ at each of the specified $\theta_P$s and to fit the the resulting graph with $\log({\hat{\bf n}}\cdot{\hat{\bf S}})=\alpha^\prime-\alpha\log{\hat R}_P-\alpha^{\prime\prime}(\log{\hat R}_P)^2$ in which $\alpha$, $\alpha^\prime$ and $\alpha^{\prime\prime}$ are constants (as in figure~\ref{F24}).  The values of $\alpha^{\prime\prime}$ in the best fits to the data, though significantly smaller than the corresponding values of $\alpha$ shown here, are also positive. Thus the exponent $\alpha$ is itself a slowly increasing function of ${\hat R}_P$ at any given $\theta_P$.  (Values of the parameters used for plotting this figure are those for Case II described in \S~\ref{sec:numericalII}.)}
\label{F33}
\end{figure}

\begin{figure}
\centerline{\includegraphics[width=11cm]{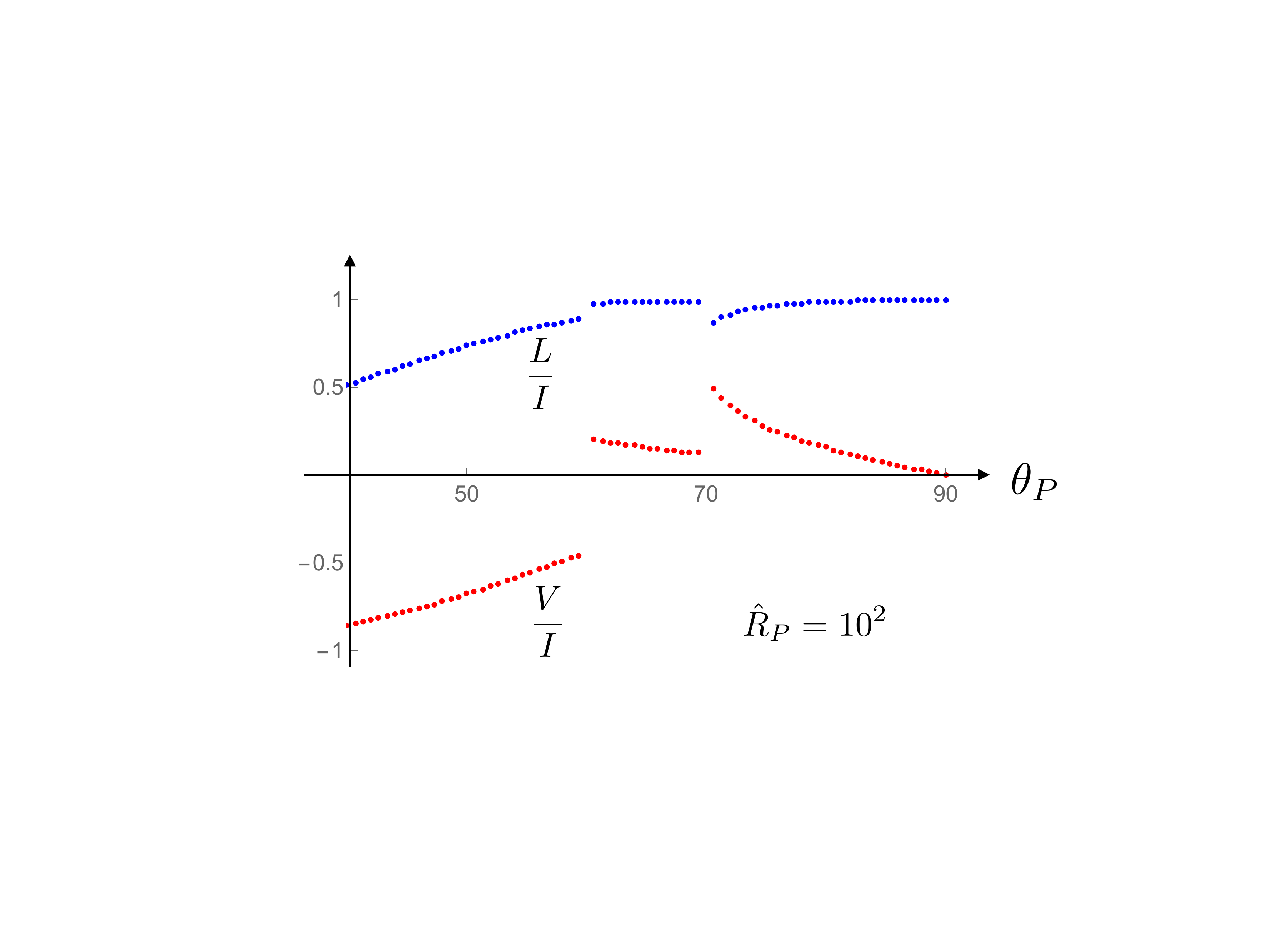}}
\caption{Fractions of linear polarization $L/I$ (the upper blue dots) and circular polarization $V/I$ (the lower red dots) for the radiation generated by an electric current that flows across the radial dimension of the dielectric ring at ${\hat R}_P=10^2$.   (Values of the parameters used for plotting this figure are those for Case II described in \S~\ref{sec:numericalII}.)}
\label{F34}
\end{figure}

\begin{figure}
\centerline{\includegraphics[width=11cm]{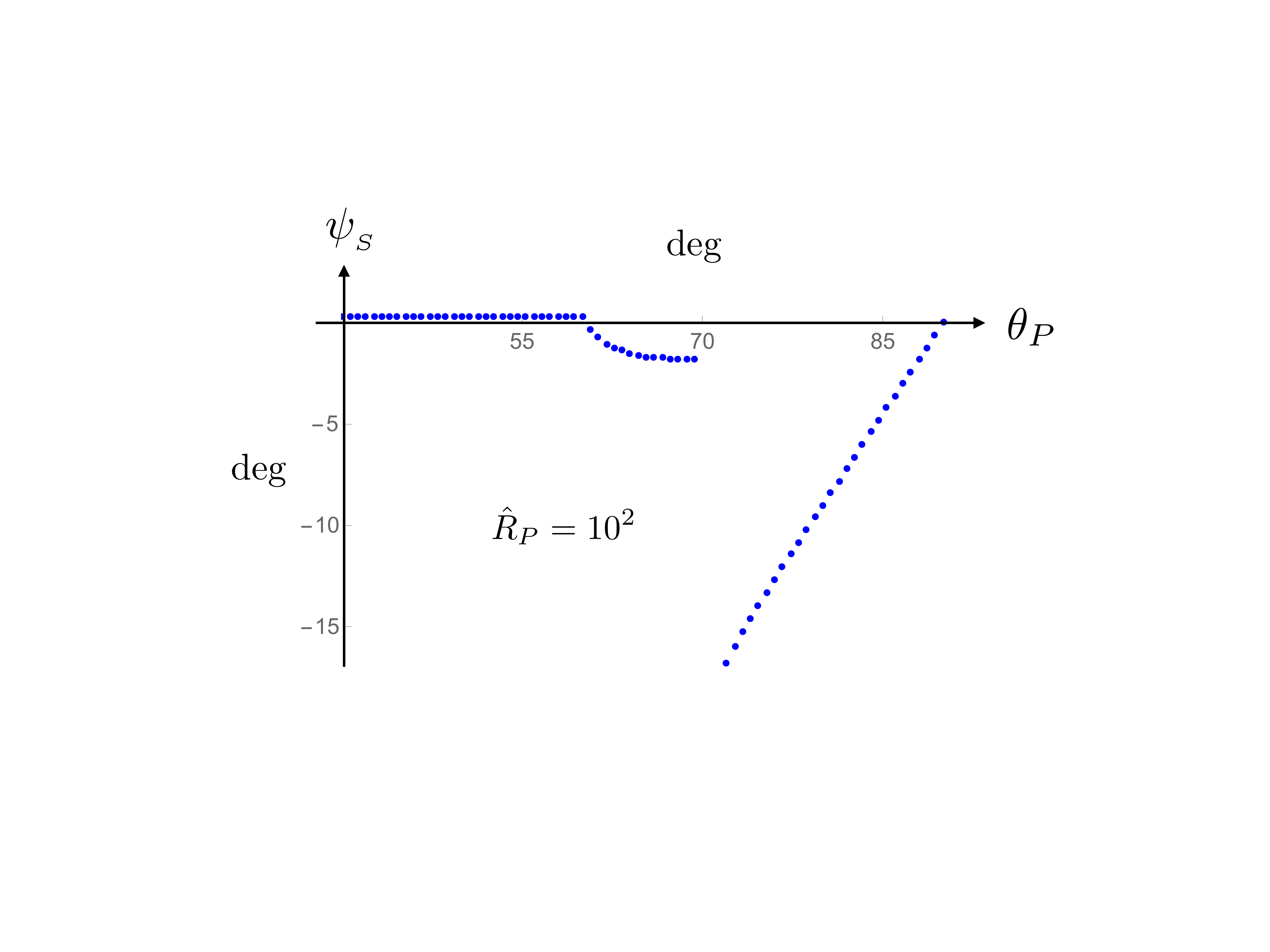}}
\caption{The polarization position angle $\psi_S$ as a function of the polar coordinate $\theta_P$ of the observation point for the radiation generated by an electric current whose direction is everywhere perpendicular to the axis of rotation at ${\hat R}_P=10^2$.   (Values of the parameters used for plotting this figure are those for Case II described in \S~\ref{sec:numericalII}.)  }
\label{F35}
\end{figure}

Figures~\ref{F30}--\ref{F33} are the counterparts of figures~\ref{F21}--\ref{F23} and \ref{F25}, respectively [see (\ref{E190}) and (\ref{E208})].  The normalization factor appearing in (\ref{E190}) has the value ${\bar S}_n\vert_{{\hat R}_P=10}=3.07\times10^{-3}\,\vert j_r\vert^2$ Watt/m${}^2$, where $\vert j_r\vert=\nu s_r$ stands for the amplitude of the electric current density in units of amp/m${}^2$.  The radiation in this case differs from that in Case Ia mainly in its state of polarization.  While essentially linearly polarized with a fixed position angle across the non-spherically decaying beam in $60^\circ\le\theta_P\le70^\circ$ and $110^\circ\le\theta_P\le120^\circ$, this radiation is elliptically polarized with a circular polarization that changes sense across the unconventional beam (figure~\ref{F34}) and has a position angle that sweeps across the radiation distribution in $60^\circ\le\theta_P\le120^\circ$ (figure~\ref{F35}).  Moreover, the direction of polarization of the non-spherically decaying beam is in the present case orthogonal to that of the non-spherically decaying beam in Case Ia, thus reflecting the orthogonality of the directions of the electric current density in these two cases.

\section{Conclusion}
\label{sec:conclusion}

The unconventional properties of the radiation discussed in this paper stem from the collaborative action, at certain observation points, of several focusing mechanisms simultaneously: the space-time distance between the observation point and the constituent volume elements of a superluminally rotating source distribution [i.e., the argument of the delta function in the expression for the retarded potential in (\ref{E14})] is stationary with respect to {\it three} coordinates of certain source elements concurrently.  In addition, these concurring stationary points are not all isolated.  The stationary point with respect to the retarded azimuthal positions of those source elements [which is equivalent to that with respect to their emission times (\S~\ref{subsec:constraint})] results from the coalescence of two other stationary points and so is degenerate.  At each of the original isolated stationary points the rotating source element approaches the observer along the radiation direction with the speed of light at the retarded time.  At the locus of the degenerate stationary points resulting from the coalescence of two of these isolated stationary points (here referred to as the cusp locus $C$) the source elements approach the observer at the retarded time not only with the speed of light but also with zero acceleration along the radiation direction.  

The locus of such degenerate stationary points [which lies at a boundary of the integration domain delineated by the intersection of the cusp $C$ with the source distribution (figure~\ref{F11})] is separated from the locus of source points whose space-time distances from the observer are stationary with respect to the radial coordinate $r$ (here denoted by $S$) by a distance that shrinks to zero when the observation point lies either in the plane of rotation or at infinity.  For an observation point in the plane of rotation, the space-time distance in question is stationary also with respect to the axial coordinate $z$ of those source points that lie on a plane passing through the observation point normal to the $z$-axis.  These critical points result in a Green's function for the problem that is discontinuous on a two-sheeted cusped surface (figure~\ref{F8}) and has non-integrable singularities there (\S~\ref{subsec:Expansion}).  The singularities of this Green's function have been handled in \S\S~\ref{subsec:Hadamard}, \ref{sec:Eb} and \ref{sec:total} analytically.  The complicated integrals representing the regularized values of the fields (\S~\ref{sec:total}) that have had to be evaluated numerically (\S~\ref{sec:numerical}) are free of any singularities. 

The unusual coincidence and proximity of so many critical points, in particular the shrinking (as ${\hat R}_P^{-2}$) of the separation between the cusp locus $C$ and the locus of saddle points $S$ with the distance ${\hat R}_P$ of the observer from the source (figure~\ref{F11}), results in an emission that not only is more intense than a corresponding conventional radiation (figure~\ref{F26}) but in addition decays more slowly with distance than predicted by the inverse-square law: time-averaged value of the radial component of its flux density diminishes with ${\hat R}_P$ as ${\hat R}_P^{-\alpha}$ with an exponent $\alpha$ whose values range between $1$ and $2$ (rather than being equal to $2$, as in a spherically decaying radiation) within the fixed solid angle into which it is beamed (see figures~\ref{F25}, \ref{F29} and \ref{F33}).  

At observation points for which projections of the velocities of all volume elements of the distribution pattern of the source along the radiation direction are subluminal, there are no contributions toward the value of the radiation field from any stationary points.  The radiation in such regions of space (where it may be regarded as a superluminal generalization of synchrotron radiation) obeys the inverse-square law but is still much stronger than that from an identical stationary or subluminally rotating source.  It can be seen from figure~\ref{F26} that even where it has the same characteristics as a conventional radiation (i.e., at polar angles $0\le\theta_P\le56.4^\circ$ in the case of the example plotted in figure~\ref{F26}), the radial component of the time-averaged Poynting vector for the radiation from the rotating source is an order of magnitude larger than that for the radiation from its stationary counterpart.  (Note that the Poynting vector has been normalized in the same way in both cases shown in figure~\ref{F26}: it has been divided by the mean value of the power, emitted by the rotating source, that propagates across the sphere ${\hat R}_P=10$ per unit solid angle.) 

The angular position and extent of the non-spherically decaying component of the radiation (depicted in figures \ref{F23}, \ref{F28} and \ref{F32}) is determined by the values of the linear speeds of the rotating distribution pattern of the polarization current at the inner and outer radii of the dielectric that hosts this current (figure \ref{F1}).  The sudden changes in the value of the flux density across the boundaries of the non-spherically decaying radiation beam reflect the presence or absence of source elements that approach the observer along the radiation direction with the speed of light and zero acceleration at the retarded time.  At observation points within the transition intervals across these boundaries -- transition intervals that become narrower the larger the distance of the observer from the source -- the value of the field does not receive contributions from all the stationary points.  

The exponent $\alpha$ in the power-law decay $R_P^{-\alpha}$ of the flux density of the intense beam itself varies with both the angular position and the distance of the observer.  To show these variations I have presented the plots of the angular distribution of the flux density for the non-spherically decaying component of the radiation (using a logarithmic scale) at the six distances ${\hat R}_P=10$, $10^2$, \ldots , $10^6$ in the same figure (figures~\ref{F22}, \ref{F27}, and \ref{F31}) by moving up the distributions for ${\hat R}_P\ge10^2$ relative to that for ${\hat R}_P=10$ each by the number of decibels ($20$, $40$, \ldots, $100$) that it would have decayed had it been obeying the inverse-square law ${\hat R}_P^{-2}$.  The fact that in figures~\ref{F22}, \ref{F27}, and \ref{F31} the distributions for longer distances lie above those for shorter distances in each of these figures, instead of being coincident, means that the plotted flux densities decay more slowly than predicted by the inverse-square dependence ${\hat R}_P^{-2}$.  From the separation between the distributions for different distances one can infer not only the best fit to the value of $\alpha$ at each polar angle (figures~\ref{F25}, \ref{F29} and \ref{F33}) but also an estimate of the slow dependence of the value of $\alpha$ on distance (see figure~\ref{F24}).

The angular distributions in figure~\ref{F27} differ from those shown in figures~\ref{F22} and \ref{F31} because an additional mechanism of focusing comes into play when the observation point is closer to the equatorial plane than half the width of the source distribution normal to this plane.  In that case the space-time distance between the observation point and the source points is stationary also with respect to the axial coordinate $z$ of any volume element of the source distribution that lies at the same distance from the equatorial plane as the observation point.  This gives rise to an intense narrow beam propagating along the equatorial plane whose angular width decreases as $R_P^{-1}$ with distance (figure~\ref{F28}).  The flux density of this narrowing beam decays faster than that of the non-spherically decaying radiation outside the equatorial plane: it decays with a value of $\alpha$ that nearly equals $2$ (figure~\ref{F29}).  So, the power that propagates into the solid angle subtended by this equatorial beam decreases as $R_P^{1-\alpha}\simeq R_P^{-1}$ with distance. 

Even in the case of the emission depicted in figure~\ref{F28}, the decreasing power carried by the equatorial beam is not sufficient to compensate for the change ${\hat R}_P^{2-\alpha}$ with distance of the power carried by the radiation beam that decays non-spherically.  The way the present radiation meets the requirements of the conservation of energy is through having an energy density whose derivative with respect to time at points inside the non-spherically decaying beam is negative when time-averaged (instead of being zero as in a conventional radiation).  In the equation of continuity stating the conservation of energy [(\ref{C1}) or its time-averaged free-space version (\ref{C32})], the positive flux of energy out of a closed surface is thus balanced by the negative temporal rate of change of the energy contained in the volume bounded by that surface (appendix~\ref{AppC}).  

This is confirmed by the fact that, for the non-spherically decaying radiation beam, time-averaged value of the temporal rate of change of the electromagnetic energy density decays as $R_P^{-\beta}$ with an exponent $\beta$ whose value and angular distribution are related to those of the exponent $\alpha$.  That the change per unit time in the amount of electromagnetic energy contained inside a shell bounded by two spheres centred on the source compensates for the difference in the fluxes of power across these spheres is corroborated by the numerical results presented in figures~\ref{F25} and \ref{fC2}.  

The above two related features of the present radiation (its non-spherical decay with distance and the decrease in its energy density with time) which distinguish it from any other known radiation, stem from the transient nature of the process by which it is emitted.  Because of the nonlinearity of the relationship between the retarded time $t$ and the observation time $t_P$ (figures~\ref{F4} and \ref{F36}), the retarded distribution of the present source bears no resemblance to its actual distribution shown in figure~\ref{F1}.  In the case of the example in figure~\ref{F36}, its retarded distribution consists of several disjoint parts that continually change shape in the course of a rotation period.  Even though the retarded distribution of the source has the same shape at the beginning and the end of each rotation period, the rate at which it changes shape with time is not the same in any two rotation periods (see the final paragraphs of appendix~\ref{AppC}).  At points where they approach the observer with the speed of light along the radiation direction, the boundaries of the retarded distribution of the present source change with time at a rate that depends on the time elapsed since the source was switched on monotonically.  The fact that the present radiation never attains a steady state in which the time-averaged value of the temporal rate of change of its energy density vanishes can thus be traced back to a corresponding transient feature of its source: to the monotonically varying rate at which the topology of the retarded distribution of the source changes with time.  The slower rate of decay of the flux density of this radiation with distance is, in turn, required by the conservation of energy wherever the time-averaged value of the temporal rate of change of its energy density is negative.

As explained in \S\S~\ref{sec:source} and \ref{subsec:constraint}, the source I have analysed can be identified with a single Fourier component of any charge-current whose distribution pattern rotates rigidly.  The description of this source in (\ref{E1}) entails two frequencies: the rotation frequency of the distribution pattern of the source, $\omega$, and the frequency of the radiation generated by the source $m\omega$, where the harmonic number $m$ can be arbitrarily large.  Each value of $m$ designates a given Fourier component both of a member of the set of source distributions in question and of its radiation.  The emphasis in this paper has been on establishing the existence of a new class of solutions of Maxwell's equations by analysing a simple prototype of its required source in detail.  The choice of the values of the parameters of the specific examples of this prototype (including the choice $m=10$) for which I have numerically evaluated the characteristics of the emission has likewise been made to emphasize the feasibility of experimentally realizing such sources in the laboratory \citep[see also][]{ArdavanA:Exponr}.  The effects illustrated by these examples are not only expected to be generic but also to be much stronger when the integer $m$ is large (\S~\ref{subsec:AsymptoticForMinusCin}). 

The results reported in this paper are therefore relevant not only to long-range transmitters in communications technology but also to astrophysical objects containing rapidly rotating neutron stars (such as pulsars) for which the value of $m$ exceeds $10^8$.  Numerical computations based on the force-free, MHD and particle-in-cell formalisms have now firmly established~\citep[see, e.g.,][]{SpitkovskyA:Oblique,Contopoulos:2012} that not only does the distribution pattern of the charge-current in the magnetospheres of such objects rotate rigidly with a superluminally moving outer part, thus belonging to the same class of source distributions as the one I have analysed, but in addition it entails current sheets and so the superposition of a large number of monochromatic source distributions of the type considered in this paper.  \citet{Tchekhovskoy:etal} have concocted an analytic expression for the fields in the magnetosphere of an oblique rotator that fits the results of these numerical simulations very well.  The distribution of the charge-current associated with the magnetospheric current sheet that is formed outside the light cylinder is described according to their analytic expression by a Dirac delta function.  This current sheet would not of course have a vanishing width once the dissipation processes that take place within it are taken into account more accurately than can be accounted for by the force-free or MHD approximations.  Nevertheless, the fact that the Dirac delta function has an infinite number of matching Fourier components makes it clear that the parameter $m$ does indeed have a wide range of values for the distribution of the plasma that constitutes the magnetosphere of any rapidly rotating non-aligned neutron star.  

The thickness of the magnetospheric current sheet in such objects (which sets a lower limit on the wavelength of the radiation this source can emit by the present emission mechanism) is dictated by microphysical processes that are not well understood.  The standard Harris solution of the Vlasov--Maxwell equations that is commonly used in analysing a current sheet is not applicable in the present case because the current sheet in a pulsar magnetosphere moves faster than light and so has no rest frame.  The large value of the harmonic number $m$ associated with a thin current sheet together with the power-law dependence of the Poynting flux of the present radiation on $m$ (Sec.~\ref{sec:total}) suggest, however, that the frequency of the radiation that is generated in the magnetosphere of a rapidly rotating non-aligned neutron star can encompass a broad spectrum ranging from radio waves to gamma rays.  This multi-wavelength radiation escapes the dense plasma constituting the neutron star's magnetosphere in the same way that the radiation generated by the accelerating charged particles invoked in most current attempts at modelling the pulsar radiation does.

It is often presumed that the plasma equations used in the numerical computations of  the magnetospheric structure of an oblique rotator should, at the same time, predict any radiation that the resulting structure would be capable of emitting~\citep{SpitkovskyA:Oblique,Contopoulos:2012}.  Irrespective of the formalism on which they are based (whether MHD, force-free or particle-in-cell), the plasma equations used in these computations are formulated in terms of the electric and magnetic fields (as opposed to potentials).  I have already shown in \S~\ref{sec:potential}, however, that the gauge freedom offered by the solution of Maxwell's equations in terms of {\it potentials} plays an indispensable role in the prediction of the present radiation.  The absence of high-frequency radiation (and, specifically, the type of radiation I have described) is in fact hardwired into the numerical computations that have been performed to determine the magnetospheric structure of an oblique rotator by the imposition of the standard boundary conditions on the fields in the far zone (see \S~\ref{sec:potential}).  The observed fact, too, that the spin-down power in young pulsars is much greater than the electromagnetic power they emit indicates that the physical principles underlying the mechanism of radiation in these objects are independent of those dictating their magnetospheric structure.

That the magnetospheric current sheet may be responsible for the observed radiation from rapidly rotating neutron stars has already been put forward in the literature but with an emphasis on the microscale structure of the sheet and magnetic reconnection~\citep[][and the references therein]{Uzdensky, Philippov2019}.  According to the results obtained here, in contrast, it is the accelerated motion at a superluminal speed of the sharply localized {\it macroscopic} distribution pattern of this current sheet that underlies its candidacy as a possible source of the radiation received from such objects.  Microphysical processes play no role in determining the topology and motion of the distribution pattern of the current sheet, i.e., the features that dictate the distinctive characteristics of the radiation it would generate by the present emission mechanism.  Determination of the thickness of the current sheet does go beyond the approximations used in the numerical computations~\citep{Uzdensky} and would require a consideration of these processes on plasma scales but it would be possible to incorporate this thickness in the description of the macroscopic charges and currents that are associated with the current sheet {\it a posteriori}.  This thickness can be incorporated, for example, in the formulation of the semi-analytic expressions that are provided by~\citet{Tchekhovskoy:etal}.      
  
Before the results of this paper can be applied to the astrophysical objects that have originally motivated the present study  \citep{ArdavanH:Nature}, however, it would be necessary not only to explore a very different region of the parameter space but also to replace the simple monochromatic source distribution of figure~\ref{F1} (for which the radiation is azimuthally symmetric) by one describing the magnetospheric structure of an oblique rotator \citep[such as that reported in][]{SpitkovskyA:Oblique,Contopoulos:2012,Tchekhovskoy:etal}.  Moreover, an exploration of the parameter space of even the simple prototype of superluminally rotating sources described in \S~\ref{sec:source}, which would be needed for adapting its design to its various applications in technology~\citep{ArdavanA:Patent} remains to be done. 

These notwithstanding, the mere fact that the inhomogeneous Maxwell equations possess solutions corresponding to physically tenable sources that describe the emission of non-spherically decaying radiation has far-reaching implications:  not only for communications technology and the radiation mechanism of astrophysical objects containing rapidly rotating neutron stars (such as pulsars) but also for the interpretation of other observed phenomena in astrophysics.  It has implications, for example, for the interpretation of the energetics of the multi-wavelength emissions (such as radio and gamma-ray bursts) whose sources lie at cosmological distances ($\sim10^{28}$
 cm).  It is widely accepted that some of these objects release as much energy as $10^{54}$ ergs (i.e., an energy comparable to that which would be released by the annihilation of the Sun) over a short time interval of the order of a second~\citep{FRB, Piron}.  The unquestioned assumption on which this consensus is based is that the radiation fields of all sources necessarily decay as predicted by the inverse-square law.  This assumption is brought into question by the results of the present analysis, however.  Given that the emission from such objects could in principle be decaying non-spherically with distance, an alternative interpretation of the same observational data based on the findings of the present paper would yield much lower, physically more realistic, estimates of the energy released by these objects.  

\bigskip
\begin{flushleft}
${\rm{\bf ACKNOWLEDGEMENTS}}$
\end{flushleft}
\smallskip

The apparatus described in \S~\ref{sec:source} was invented by Arzhang Ardavan.  I am indebted to him for the contribution his physical insight has made to this work and to Alex Schekochihin for discussions that have notably improved this paper.  I am grateful also to Clay Thompson for verifying my numerical results by repeating the computations described in this paper with MATLAB and to CommScope Inc for financially supporting his work.  My own computations were performed with Mathematica and were partly carried out at the Advanced Research Computing facilities of Oxford University.
 
\appendix
\section{Hadamard's finite part of a divergent integral: an illustrative example} 
\label{appA}

The need for introducing Hadamard's regularization in the theory of generalized functions, in which the order of integration and differentiation can be interchanged, is illustrated by considering the derivative of the function represented by the following double integral:
\begin{equation}
F(z)=\int_z^\infty{\textrm d}x\int_0^\infty {\textrm d}y\,f(x)\delta(y^3-x+z),\quad z\geq0,
\label{A1}
\end{equation}
where the function $f(x)$ and its derivative $f^\prime(x)$ are continuous with finite supports and $\delta$ is the Dirac delta function.  Performing the integration with respect to $x$ prior to differentiating this integral, we obtain 
\begin{equation}
F(z)= \int_0^\infty {\textrm d}y\,f(y^3+z),
\label{A2}
\end{equation}
and hence
\begin{equation}
F^\prime(z)= \int_0^\infty {\textrm d}y\,f^\prime(y^3+z),
\label{A3}
\end{equation}
where a prime denotes differentiation with respect to the argument of the function.  The right-hand sides of (\ref{A2}) and (\ref{A3}) are both well defined and finite. 

On the other hand, if we interchange the order of integration and differentiation \citep[disregarding limits of integration as in][]{HadamardJ:lecCau}, we obtain
\begin{equation}
F^\prime(z)=\int_z^\infty{\textrm d}x\int_0^\infty {\textrm d}y\,f(x)\delta^\prime(y^3-x+z).
\label{A4}
\end{equation}
Evaluation of the $x$ integral in this expression reproduces (\ref{A3}), i.e., yields an expression with a finite value for $F^\prime(z)$.  However, the evaluation of the $y$ integral results in the following alternative expression for the same function
\begin{equation}
F^\prime(z)=\frac{2}{9}\int_z^\infty{\rm d}x\,\frac{f(x)}{(x-z)^{5/3}},
\label{A5}
\end{equation}
which is divergent. 
 
The paradox is resolved once one interprets the divergent integral in (\ref{A5}) as a generalized function and equates it to its Hadamard's finite part~\citep[see, e.g.,][]{HoskinsRF:GenFun}.   Integrating the right-hand side of (\ref{A5}) by parts to obtain
\begin{equation}
F^\prime(z)=-\frac{f(x)}{3(x-z)^{2/3}}\Bigg\vert_z^\infty+\frac{1}{3}\int_z^\infty {\textrm d}x\,\frac{f^\prime(x)}{(x-z)^{2/3}}
\label{A6}
\end{equation}
and discarding the divergent (integrated) term in (\ref{A6}), we find that the following expression for the Hadamard finite part of $F^\prime(z)$
\begin{equation}
{\rm Fp}\{F^\prime(z)\}=\int_z^\infty {\textrm d}x\,\frac{f^\prime(x)}{3(x-z)^{2/3}}=\int_0^\infty {\textrm d}\eta\,f^\prime(\eta^3+z)
\label{A7}
\end{equation}
has the same value as that found in (\ref{A3}).  In other words, there is no discrepancy between the values of the two single integrals in (\ref{A3}) and (\ref{A5}) if these integrals are interpreted as generalized rather than classical functions.

An alternative way of calculating the Hadamard finite part of the divergent integral in (\ref{A5}) is (i) to subtract from the function $f(x)$ as many terms of its Taylor expansion about the singular point $x=z$ as are needed to render the singularity of the integrand integrable, (ii) to add to the integrand what has thus been subtracted from it and (iii) to integrate the added terms discarding all divergent contributions, i.e., to let
\begin{eqnarray}
{\rm Fp} \{F^\prime(z)\}&=&\frac{2}{9}\int_z^\infty{\rm d}x\,\frac{f(x)-f(z)}{(x-z)^{5/3}}-\frac{f(z)}{3(x-z)^{2/3}}\Bigg\vert_{x=\infty}\nonumber\\*
&=&\frac{2}{9}\int_z^\infty{\rm d}x\,\frac{f(x)-f(z)}{(x-z)^{5/3}}
\label{A8}
\end{eqnarray}
in the present case.  That this equals the right-hand side of (\ref{A7}) follows from an integration by parts for which the integrated term now vanishes~\citep{HadamardJ:lecCau,HoskinsRF:GenFun}.

\section{Why a conventional approach to the problem does not work}
\label{appB}
My previous works on the radiation by superluminal sources~\citep{ArdavanH:Genfnd,ArdavanH:JMP99,ArdavanH:Speapc,ArdavanH:Morph,ArdavanH:Funda} have been criticized~\citep{Hewish2,HannayJH:ComIGf,McDonald,HannayJH:JMP,
HannayJH:Speapc,Hannay_Morphology,Hannay:09,Contopoulos:2012} either on the basis of the wave or the plasma equations for the fields [e.g., (\ref{E8})], or on the basis of the following classical form of the retarded potential 
\begin{equation}
A^\mu({\bf x}_P,t_P)=\frac{1}{c}\int{\rm d}^3{\bf x}\frac{j^\mu({\bf x},t_{\rm ret})}{R},
\label{B1}
\end{equation}
with
\begin{equation}
t_{\rm ret}=t_P-R/c,
\label{B2}
\end{equation}
which is obtained by performing the integration with respect to $t$ in (\ref{E14}).  The fact that the customarily used retarded solutions of the wave equations for the fields (as opposed to those for the potentials) do not in the present case satisfy the required boundary conditions at infinity has already been discussed in \S~\ref{sec:potential}.  In this appendix I also explain why a simple-minded approach based on (\ref{B1}) fails to capture the unconventional features of the radiation from an extended source whose distribution pattern rotates superluminally.  Together with \S~\ref{sec:potential}, the analysis that follows supersedes my published replies~\citep{ArdavanH:RepCGf,McDonaldReply,ArdavanH:Speapc1,Ardavan_RepMorph,ArdavanH:RepFund} to the critiques of my earlier works on this problem. 

Let us apply (\ref{B1}) to the experimentally realized source distribution described in \S~\ref{sec:source}, for which the cylindrical components of the current density ${\bf j}=\partial{\bf P}/\partial t$ are described by the real part of
\begin{eqnarray}
j_{r,\varphi,z}&=&{\rm i}m\omega s_{r,\varphi,z}(r,z)\exp[-{\rm i}m(\varphi-\omega t)],\nonumber\\*
&&\qquad\qquad r_L\le r\le r_U,\quad -z_0\le z\le z_0,\quad 0\le\varphi-\omega t<2\pi
\label{B3}
\end{eqnarray}
[see (\ref{E1}), (\ref{E7}), (\ref{E16}) and (\ref{E18})].  The resulting expression for, say, the $z$ component of the vector potential is 
\begin{equation}  
A_z=\frac{{\rm i}m\omega}{c}\int_{r_L}^{r_U} r{\rm d}r\int_{-z_0}^{z_0}{\rm d}z\, s_z\int_{{\cal R}_\varphi}{\rm d}\varphi\frac{\exp(-{\rm i}m{\hat\varphi}_{\rm ret})}{R},
\label{B4}
\end{equation}
where
\begin{eqnarray}
{\hat\varphi}_{\rm ret}&=&\varphi-\omega t_{\rm ret}\nonumber\\*
&=&\varphi+[({\hat z}-{\hat z}_P)^2+{{\hat r}_P}^2+{\hat r}^2-2{\hat r}_P{\hat r}\cos(\varphi-\varphi_P)]^{1/2}-\omega t_P,
\label{B5}
\end{eqnarray}
with ${\hat r}=r\omega/c$, ${\hat z}=z\omega/c$, etc., and ${\cal R}_\varphi$ is the range of $\varphi$ over which the constraint $0\le{\hat\varphi}_{\rm ret}<2\pi$ is satisfied, i.e., is the support of the retarded distribution of the source.  (For a discussion of the significance and indispensability of this constraint see \S~\ref{subsec:constraint}.) 

\begin{figure}
\centerline{\includegraphics[width=12cm]{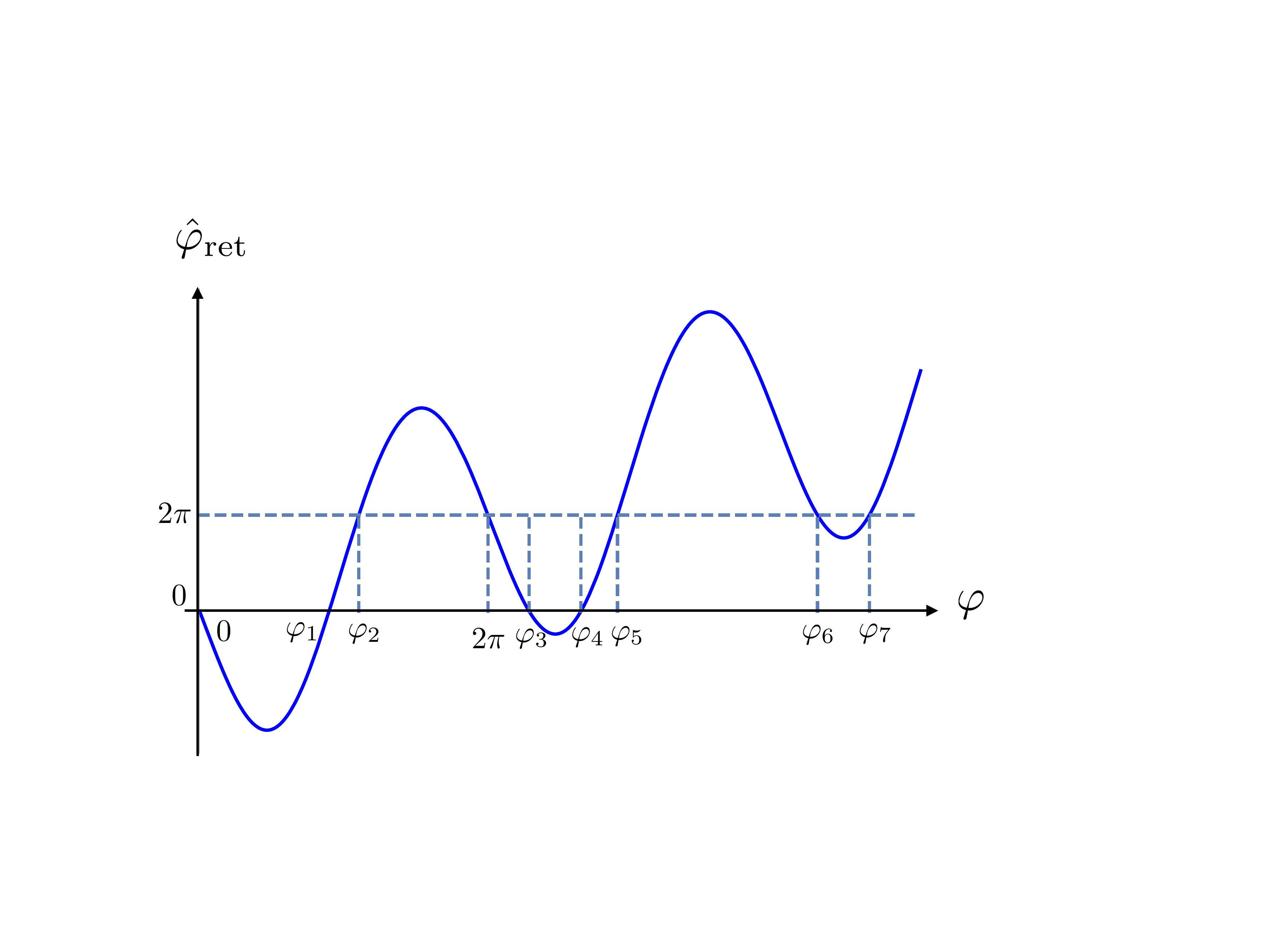}}
\caption{The function ${\hat\varphi}_{\rm ret}$ versus $\varphi$ for the following fixed set of values of $(r,z;r_P,\varphi_P,z_P,t_P)$ at which $\Delta$ is positive: ${\hat r}=10$, ${\hat z}=0$, ${\hat r}_P=89.13$, $\varphi_P=\pi/2$, ${\hat z}_P=45.34$, $t_P= 98.92\omega^{-1}$.  In this example the time and location of the observer is such that the detected field receives simultaneous contributions from the first three rotation cycles of the source point with the initial ($t=0$) position $\varphi=0$, i.e., from $0\le\varphi\le6\pi$.  The range ${\cal R}_\varphi$ of $\varphi$ for which ${\hat\varphi}_{\rm ret}$ falls between $0$ and $2\pi$ consists of the four disjoint intervals $\varphi_1\le\varphi<\varphi_2$, $2\pi\le\varphi<\varphi_3$, $\varphi_4\le\varphi<\varphi_5$ and $\varphi_6<\varphi<\varphi_7$ representing the azimuthal extent of the retarded distribution of the source.  The points ($0,\varphi_1$) and ($\varphi_2,2\pi,\varphi_5$) of the intersections of the above curve with the the lines ${\hat\varphi}_{\rm ret}=0$ and ${\hat\varphi}_{\rm ret}=2\pi$ coalesce onto inflection points when the coordinates $(r,z;r_P,z_P)$ assume values for which $\Delta$ vanishes and the source point lies on the cusp locus $C$ of the bifurcation surface.}
\label{F36}
\end{figure}

Figure~\ref{F36} shows the dependence of the function ${\hat\varphi}_{\rm ret}$ on the coordinate $\varphi$ for a fixed set of values of $(r,z;r_P,\varphi_P,z_P,t_P)$ at which the discriminant $\Delta$ defined in (\ref{E33}) is positive.  [${\hat\varphi}_{\rm ret}$ differs from the function $g$ defined in (\ref{E23}) and plotted in figure~\ref{F4} only by ${\hat\varphi}_P$ which is constant for fixed space-time coordinates of the observation point.]  Note that the $(r,z)$-coordinates of the source point are here kept fixed and the coordinate $\varphi={\hat\varphi}+\omega t$ marks the continually increasing azimuthal position of the source element which was located at $\varphi={\hat\varphi}$ at the time $t=0$ on the circle $r=$ const, $z=$ const.  Given that ${\hat\varphi}$ thus labels each volume element of the rotating source by its azimuthal position at $t=0$, the vertical and horizontal axes in figure~\ref{F36} respectively show which source elements on the circle $r=$ const, $z=$ const, make a contribution toward the radiation received at $(r_P,\varphi_P,z_P; t_P)$ and during which rotation period, i.e., over which $\varphi$-interval.   

The source density in (\ref{B1}) is evaluated on the collapsing sphere $\vert{\bf x}-{\bf x}_P\vert=c(t_P-t)$ in the space of source points whose centre lies on the observation point ${\bf x}_P$ and whose radius shrinks to zero at the observation time $t_P$.  If the speed of the source is sufficiently higher than $c$ so that the separation between the neighbouring extrema of the curve shown in figure~\ref{F36} is greater than $2\pi$ (as in figure~\ref{F36}), then this sphere could be intersected by the rotating source element several times ($3$, $5$, $7$, $\cdots$ times) as it collapses.  In other words, there could then be an odd number of simultaneously received contributions that are made by the same source element over a retarded time interval exceeding one rotation period~\citep{BolotovskiiBM:Radbcm}.  
  
The illustrative example depicted in figure~\ref{F36} shows that, during the first three rotation cycles $0\le\varphi<6\pi$, certain source elements (the ones labelled by values of ${\hat\varphi}_{\rm ret}$ close to $2\pi$) make their contributions toward the field observed at $(r_P,\varphi_P,z_P; t_P)$ at five retarded times (i.e., when passing through five distinct azimuthal positions), while each of the other source elements makes its contribution at three retarded times. This figure also shows that there are intervals of $\varphi$ within the cycles $0\le\varphi<6\pi$ from which no contribution reaches the field observed at $(r_P,\varphi_P,z_P;t_P)$.  According to figure~\ref{F36}, the source elements whose ${\hat\varphi}$ labels satisfy the constraint $0\le{\hat\varphi}_{\rm ret}<2\pi$ are those whose retarded positions lie in the intervals $\varphi_1\le\varphi<\varphi_2$, $2\pi\le\varphi<\varphi_3$, $\varphi_4\le\varphi<\varphi_5$ and $\varphi_6<\varphi<\varphi_7$.  Thus in contrast to the retarded distribution of a stationary or subluminally moving source which occupies an azimuthal interval of length $2\pi$ at most, the volume over which the integration in (\ref{B1}) extends for $\Delta>0$ is so stretched around the rotation axis and perforated as to occupy an azimuthal interval of length $6\pi$.  For $\Delta<0$, on the other hand, ${\hat\varphi}_{\rm ret}$ is a monotonic function of $\varphi$ (see figure~\ref{F4}) and so the range ${\cal R}_\varphi$ consists of the single cycle $0\le\varphi<2\pi$.

Hence, for the example shown in figure~\ref{F36}, the volume integral in (\ref{B4}) assumes the form  
\begin{eqnarray}
A_z&=&\frac{{\rm i}m\omega}{c}\int_{r_L}^{r_U} r{\rm d}r\int_{-z_0}^{z_0}{\rm d}z\, s_z(r,z)\Big[{\rm H}(\Delta)\Bigg(\int_{\varphi_1}^{\varphi_2}+\int_{2\pi}^{\varphi_3}+\int_{\varphi_4}^{\varphi_5}+\int_{\varphi_6}^{\varphi_7}\Bigg){\rm d}\varphi\,\frac{\exp(-{\rm i}m{\hat\varphi}_{\rm ret})}{R}\nonumber\\*
&&+{\rm H}(-\Delta)\int_0^{2\pi}{\rm d}\varphi\,\frac{\exp(-{\rm i}m{\hat\varphi}_{\rm ret})}{R}\Big],
\label{B6}
\end{eqnarray}
in which the Heaviside step functions take account of the fact that the contributing source distribution at the retarded time consists, in general, of both volume elements that approach the observer along the radiation direction with a speed exceeding $c$, for which $\Delta>0$, and elements that approach the observer with a speed lower than $c$, for which $\Delta<0$ (see \S~\ref{subsec:Cusp}).  The limits $\varphi_j$ ($j=1,2,\cdots,7$) of the $\varphi$-integrations are given, as functions of $(r,z,r_P,z_P;t_P)$, by the solutions of the transcendental equations ${\hat\varphi}_{\rm ret}=0$ and ${\hat\varphi}_{\rm ret}=2\pi$. 

Because the limits of integration in this alternative formulation of the retarded potential depend on the space-time coordinates of the observation point, calculation of the field entails the use of Leibniz's formula for the differentiation of a definite integral whose derivative receives contributions also from the variations of its limits
\begin{equation}
\frac{\rm d}{{\rm d}x}\int_{\beta(x)}^{\alpha(x)}f(x,\xi)\,{\rm d}\xi=f(x,\alpha)\frac{{\rm d}\alpha}{{\rm d}x}-f(x,\beta)\frac{{\rm d}\beta}{{\rm d}x}+\int_{\beta(x)}^{\alpha(x)}\frac{\partial f}{\partial x}\,{\rm d}\xi
\label{B7}
\end{equation}
\citep[see, e.g.,][]{Courant1967}.  If we differentiate $A_z$ with respect to $t_P$, for example, the resulting expression would contain terms that involve the derivatives $\partial\varphi_j/\partial t_P$ of the limits of integration $\varphi_j$.  Differentiating the transcendental equation ${\hat\varphi}_{\rm ret}=0$ with respect to $t_P$, we find that 
\begin{equation}
\frac{\partial\varphi_j}{\partial t_P}=\frac{\omega}{1+{\hat r}{\hat r}_P\sin(\varphi-\varphi_P)/{\hat R}}\Big\vert_{\varphi=\varphi_j}=\frac{\omega}{\partial g/\partial\varphi}\Big\vert_{\varphi=\varphi_j},
\label{B8}
\end{equation}
in which the function $g$ is that defined in (\ref{E23}) [see (\ref{E35})]. 

At the points of intersection of the cusp locus $C$ [described by (\ref{E39})] with the source distribution, where the roots ($0,\varphi_1$) and ($\varphi_2,2\pi,\varphi_5$) of ${\hat\varphi}_{\rm ret}=0$ and ${\hat\varphi}_{\rm ret}=2\pi$ coalesce onto inflection points (see figure~\ref{F4}), not only $\partial g/\partial\varphi\vert_{\varphi=\varphi_j}$ but also $\partial^2 g/\partial\varphi^2\vert_{\varphi=\varphi_j}$ vanishes (see \S~\ref{subsec:Green's function}).  At such points, the terms in the derivative of the potential that arise from the differentiation of the limits of integration in (\ref{B6}) contain divergent factors as demonstrated by (\ref{B8}).  The divergence of the derivatives of the limits of integration contravenes the conditions for the differentiability of the $\varphi$-integrals in (\ref{B6}) as classical functions, i.e., contravenes the validity of Leibniz's formula~\citep{Courant1967}.   

The fact that the source cannot be infinitely long lived, i.e., that its trajectory has to have a boundary, is essential to the validity of the above result.  Because the integrand in (\ref{B4}) is a periodic function of $\varphi$, the contributions toward the value of $A_z$ from the $\varphi$ intervals in a given rotation cycle over which the curve ${\hat\varphi}_{\rm ret}(\varphi)$ falls outside the strip $0\le{\hat\varphi}_{\rm ret}<2\pi$ are compensated, in the case of an infinitely long-lived source, by the contributions from the $\varphi$ intervals in other rotation cycles over which this curve lies inside the strip. In other words, the sum of all contributions for a source whose trajectory extends over $-\infty<\varphi<\infty$ amounts to the contribution that would have been attributed to a single cycle had the constraint $0\le{\hat\varphi}_{\rm ret}<2\pi$ been overlooked.  However, this does not hold true in the case of the source element depicted in figure~\ref{F36} (which is turned on at $t=0$ when it is at $\varphi=0$) because its trajectory only extends over $0<\varphi<\infty$. The contributions towards the value of $A_z$ from the three intervals $2\pi\le\varphi\le\varphi_3$, $\varphi_4\le\varphi\le\varphi_5$ and $\varphi_6\le\varphi\le\varphi_7$ jointly compensate for the missing contribution from the interval $0\le\varphi\le\varphi_1$ of the first rotation cycle but the contribution from the missing interval $\varphi_2\le\varphi\le2\pi$ of this cycle remains un-compensated.  Consequently, the divergent contribution from the derivative of $\varphi_2$ towards the value of the field is not cancelled out by any other contribution in this case.

Thus the alternative formulation (\ref{B1})  of the retarded potential merely replaces the singularity of the integrand in (\ref{E27}), i.e., the singularity of the derivative of the Green's function for the problem, by the singularity of the derivatives of the limits of integration.  In contrast to the singularity of the derivative of the Green's function which can be rigorously handled by Hadamard's regularization technique~\citep{HadamardJ:lecCau}, however, the singularity encountered in this appendix vitiates the applicability of (\ref{B1}) to sources whose radiation field has to be found by differentiating the expression for their retarded potential (see \S~\ref{sec:potential}).

\section{How the requirements of the conservation of energy are met by the radiation described in this paper}
\label{AppC}

In this appendix I show explicitly that, notwithstanding the non-spherical decay of their amplitudes, the radiation fields ${\bf E}$ and ${\bf B}$ that are derived in the present paper do comply with the statement of conservation of energy embodied in the Poynting theorem
\begin{equation}
\int_{\cal D}{\rm d}^3{\bf x}_P\frac{\partial}{\partial t_P}\left(\frac{{\bf E}^2+{\bf B}^2}{8\pi}\right)+\int_{\partial{\cal D}}{\rm d}^2{\bf x}_P\cdot\left(\frac{c}{4\pi}{\bf E\times B}\right)
=-\int_{\cal D}{\rm d}^3{\bf x}_P\,{\bf j}\cdot{\bf E}
\label{C1}
\end{equation}
\citep[see][]{JacksonJD:Classical}.  Here $\partial{\cal D}$ stands for the closed surface bounding the volume ${\cal D}$.  In the case of a conventional radiation, for which the phase difference between ${\bf E}$ and $\partial{\bf E}/\partial t_P$ and between ${\bf B}$ and $\partial{\bf B}/\partial t_P$ is $\pi/2$, time-averaged value of the first term in (\ref{C1}) vanishes so that in free space where ${\bf j}=0$ the flux of energy into any closed region (e.g., into the volume bounded by two spheres centred on the source) equals the flux of energy out of it.  In the present case, on the other hand, time-averaged rate of change of the energy density of the non-spherically decaying radiation contained within a closed region of space is as shown in this appendix negative, so that the flux of energy into that region can be smaller than the flux of energy out of it.

The dependence of the radiation field described by (\ref{E27}) on the observation time $t_P$ arises through the variable $\phi$ in the expression for the Green's function $G_{nj}$ in (\ref{E32}).   Differentiating (\ref{E27}) with respect to $t_P$ and noting that $\partial\delta(g-\phi)/\partial t_P=\omega\partial\delta(g-\phi)/\partial{\hat\varphi}$ according to (\ref{E24}), we obtain
\begin{equation}
\left[\matrix{\partial{\bf E}/\partial t_P\cr\partial{\bf B}/\partial t_P\cr}\right]=-\sum_{n=1}^2\sum_{j=1}^3\int_{\mathcal S}{\hat r}{\textrm d}{\hat r}\,{\textrm d}{\hat\varphi}\,{\textrm d}{\hat z}\,\frac{\partial^2 G_{nj}}{\partial{\hat\varphi}^2}\left[\matrix{{\bf u}_{nj}\cr{\bf v}_{nj}\cr}\right].
\label{C2}
\end{equation}
[Note that, according to (\ref{B7}), the dependence on ${\hat\varphi}$ of the limits of integration in (\ref{E32}) does not contribute toward the values of the derivatives of $G_{nj}$ with respect to ${\hat\varphi}$.]  The ${\hat\varphi}$-integral in this expression can be evaluated in exactly the same way as in (\ref{E71}) (see \S~\ref{subsec:Hadamard}).  Breaking up the volume of integration in the expression for the derivative of one of the radiation fields, e.g., $\partial{\bf E}/\partial t_P$, into the domains of validity of $G_{nj}^{\rm in}$, $G_{nj}^{\rm out}$ and $G_{nj}^{\rm sub}$, we can write the ${\hat\varphi}$-integral over ${\bf u}_{nj}$ in (\ref{C2}) as
\begin{eqnarray}
{\mathbf I}_{{\hat\varphi}{\hat\varphi}}&\equiv&\int_0^{2\pi}{\textrm d}{\hat\varphi}\,{\mathbf u}_{nj}\frac{\partial^2 G_{nj}}{\partial{\hat\varphi}^2}\nonumber\\*
&=&{\rm H}(\Delta)\left[\left(\int_0^{{\hat\varphi}_-}+\int_{{\hat\varphi}_+}^{2\pi}\right){\textrm d}{\hat\varphi}\,{\mathbf u}_{nj}\frac{\partial^2 G_{nj}^{\rm out}}{\partial{\hat\varphi}^2}+\int_{{\hat\varphi}_-}^{{\hat\varphi}_+}{\textrm d}{\hat\varphi}\,{\mathbf u}_{nj}\frac{\partial^2 G_{nj}^{\rm in}}{\partial{\hat\varphi}^2}\right]\nonumber\\*
&&+{\rm H}(-\Delta)\int_0^{2\pi}{\textrm d}{\hat\varphi}\,{\mathbf u}_{nj}\frac{\partial^2 G_{nj}^{\rm sub}}{\partial{\hat\varphi}^2}.
\label{C3}
\end{eqnarray}
If we now integrate every term of the above expression by parts, recall that ${\hat\varphi}=0$ labels the same source point as does ${\hat\varphi}=2\pi$, and use the fact that the exact version of $G_{nj}$ given in (\ref{E32}) is periodic in ${\hat\varphi}$ as well as in $\varphi$ (with the same period $2\pi$), we arrive at
\begin{eqnarray}
{\mathbf I}_{{\hat\varphi}{\hat\varphi}}&=&{\rm H}(\Delta)\Bigg\{\left[{\mathbf u}_{nj}\left(\frac{\partial G_{nj}^{\rm in}}{\partial{\hat\varphi}}-\frac{\partial G_{nj}^{\rm out}}{\partial{\hat\varphi}}\right)\right]_{{\hat\varphi}={\hat\varphi}_-}^{{\hat\varphi}={\hat\varphi_+}}-\left(\int_0^{{\hat\varphi}_-}+\int_{{\hat\varphi}_+}^{2\pi}\right){\textrm d}{\hat\varphi}\,\frac{\partial{\mathbf u}_{nj}}{\partial{\hat\varphi}}\frac{\partial G_{nj}^{\rm out}}{\partial{\hat\varphi}}\nonumber\\*
&&-\int_{{\hat\varphi}_-}^{{\hat\varphi}_+}{\textrm d}{\hat\varphi}\,\frac{\partial{\mathbf u}_{nj}}{\partial{\hat\varphi}}\frac{\partial G_{nj}^{\rm in}}{\partial{\hat\varphi}}\Bigg\}-{\rm H}(-\Delta)\int_0^{2\pi}{\textrm d}{\hat\varphi}\,\frac{\partial{\mathbf u}_{nj}}{\partial{\hat\varphi}}\frac{\partial G_{nj}^{\rm sub}}{\partial{\hat\varphi}},
\label{C4}
\end{eqnarray}
an expression that reduces to 
\begin{equation}
{\mathbf I}_{{\hat\varphi}{\hat\varphi}}={\rm H}(\Delta)\left[{\mathbf u}_{nj}\left(\frac{\partial G_{nj}^{\rm in}}{\partial{\hat\varphi}}-\frac{\partial G_{nj}^{\rm out}}{\partial{\hat\varphi}}\right)\right]_{{\hat\varphi}={\hat\varphi}_-}^{{\hat\varphi}={\hat\varphi_+}}-\int_0^{2\pi}{\textrm d}{\hat\varphi}\,\frac{\partial{\mathbf u}_{nj}}{\partial{\hat\varphi}}\frac{\partial G_{nj}}{\partial{\hat\varphi}},
\label{C5}
\end{equation}
once the integrals over $\partial G_{nj}^{\rm in}/\partial{\hat\varphi}$, $\partial G_{nj}^{\rm out}/\partial{\hat\varphi}$ and $\partial G_{nj}^{\rm sub}/\partial{\hat\varphi}$ are combined in the light of (\ref{E69}). 

The integral in (\ref{C5}) differs from the integral ${\bf I}_{\hat\varphi}$ which was evaluated in (\ref{E71})-(\ref{E73}) only in that $\partial{\bf u}_{nj}/\partial{\hat\varphi}$ here replaces ${\bf u}_{nj}$ in ${\bf I}_{\hat\varphi}$.  Performing another integration by parts, as in the evaluation of ${\bf I}_{\hat\varphi}$, we obtain
\begin{equation} 
{\mathbf I}_{{\hat\varphi}{\hat\varphi}}={\rm H}(\Delta)\Bigg[{\mathbf u}_{nj}\left(\frac{\partial G_{nj}^{\rm in}}{\partial{\hat\varphi}}-\frac{\partial G_{nj}^{\rm out}}{\partial{\hat\varphi}}\right)-\frac{\partial{\bf u}_{nj}}{\partial{\hat\varphi}}\left(G^{\rm in}_{nj}-G^{\rm out}_{nj}\right)\Bigg]_{{\hat\varphi}={\hat\varphi}_-}^{{\hat\varphi}={\hat\varphi_+}}+\int_0^{2\pi}{\textrm d}{\hat\varphi}\,\frac{\partial^2{\mathbf u}_{nj}}{\partial{\hat\varphi}^2} G_{nj}
\label{C6}
\end{equation}
from (\ref{C5}).  It can be seen from the last paragraph of \S~\ref{subsec:Expansion} that $G_{nj}^{\rm in}$ and $\partial G_{nj}^{\rm in}/\partial{\hat\varphi}$ both diverge at ${\hat\varphi}={\hat\varphi}_\pm$ (figures~\ref{F9} and~\ref{F10}).  The physically relevant part of ${\mathbf I}_{{\hat\varphi}{\hat\varphi}}$ is given by the right-hand side of (\ref{C6}) without the divergent terms involving $G_{nj}^{\rm in}\vert_{{\hat\varphi}={\hat\varphi}_\pm}$ and $\partial G_{nj}^{\rm in}/\partial{\hat\varphi}\vert_{{\hat\varphi}={\hat\varphi}_\pm}$, 
\begin{equation}
{\rm  Fp}\{{\mathbf I}_{{\hat\varphi}{\hat\varphi}}\}={\rm H}(\Delta)\left[G_{nj}^{\rm out}\frac{\partial{\mathbf u}_{nj}}{\partial{\hat\varphi}}-{\bf u}_{nj}\frac{\partial G^{\rm out}_{nj}}{\partial{\hat\varphi}}\right]_{{\hat\varphi}={\hat\varphi}_-}^{{\hat\varphi}={\hat\varphi_+}}+\int_0^{2\pi}{\textrm d}{\hat\varphi}\,\frac{\partial^2{\mathbf u}_{nj}}{\partial{\hat\varphi}^2}G_{nj},
\label{C7}
\end{equation}
where ${\rm Fp}\{{\mathbf I}_{{\hat\varphi}{\hat\varphi}}\}$ denotes the Hadamard finite part of the divergent integral ${\mathbf I}_{{\hat\varphi}{\hat\varphi}}$ \citep[see][]{HadamardJ:lecCau,HoskinsRF:GenFun}.  This procedure applies also to the expression for $\partial{\bf B}/\partial t_P$ in (\ref{C2}) except that ${\bf u}_{nj}$ in (\ref{C3})-(\ref{C7}) is everywhere replaced by ${\bf v}_{nj}$.  Hence, 
\begin{eqnarray}
\left[\matrix{\partial{\bf E}/\partial t_P\cr\partial{\bf B}/\partial t_P\cr}\right]&=&-\sum_{n=1}^2\sum_{j=1}^3\Bigg\{\int_{\cal S}{\hat r}{\rm d}{\hat r}\,{\rm d}{\hat\varphi}\,{\textrm d}{\hat z}\,G_{nj}\frac{\partial^2}{\partial{\hat\varphi}^2}\left[\matrix{{\mathbf u}_{nj}\cr{\mathbf v}_{nj}\cr}\right]\nonumber\\*
&&+\int_{\mathcal S^\prime}{\hat r}{\textrm d}{\hat r}\,{\textrm d}{\hat z}\,{\rm H}(\Delta)\,\Bigg[G^{\rm out}_{nj}\frac{\partial}{\partial{\hat\varphi}}\left[\matrix{{\mathbf u}_{nj}\cr{\mathbf v}_{nj}\cr}\right]-\left[\matrix{{\mathbf u}_{nj}\cr{\mathbf v}_{nj}\cr}\right]\frac{\partial G_{nj}^{\rm out}}{\partial{\hat\varphi}}\Bigg]_{{\hat\varphi}={\hat\varphi}_-}^{{\hat\varphi}={\hat\varphi}_+}\Bigg\}
\label{C8}
\end{eqnarray}
according to (\ref{C2}), (\ref{C3}) and (\ref{C7}). 

In the case of the charge and current densities associated with the polarization distribution (\ref{E1}) for which the source term $[{\bf u}_{nj}\,\,\,{\bf v}_{nj}]$ assumes the form given in (\ref{E78}), the above expression becomes 
\begin{eqnarray}
\left[\matrix{\partial{\bf E}/\partial t_P\cr\partial{\bf B}/\partial t_P\cr}\right]&=&{\rm i}m\omega\sum_{n=1}^2\sum_{j=1}^3\Bigg\{m^2\int_{\cal S}{\hat r}{\textrm d}{\hat r}\,{\textrm d}{\hat\varphi}\,{\textrm d}{\hat z}\,\exp(-{\rm i}m{\hat\varphi}) G_{nj}\left[\matrix{{\tilde{\bf u}}_{nj}\cr{\tilde{\bf v}}_{nj}\cr}\right]\nonumber\\*
&&+\int_{\mathcal S^\prime}{\hat r}{\textrm d}{\hat r}\,{\textrm d}{\hat z}\,{\rm H}(\Delta)\left[\matrix{{\tilde{\bf u}}_{nj}\cr{\tilde{\bf v}}_{nj}\cr}\right]\left[\exp(-{\rm i}m{\hat\varphi})\left({\rm i}mG^{\rm out}_{nj}+\frac{\partial G_{nj}^{\rm out}}{\partial{\hat\varphi}}\right)\right]_{{\hat\varphi}={\hat\varphi}_-}^{{\hat\varphi}={\hat\varphi}_+}\Bigg\}.\qquad
\label{C9}
\end{eqnarray}
This can in turn be written as
\begin{equation}
\left[\matrix{\partial{\bf E}/\partial t_P\cr\partial{\bf B}/\partial t_P\cr}\right]={\rm i}m\omega\Bigg\{\left[\matrix{{\bf E}\cr{\bf B}\cr}\right]+\sum_{n=1}^2\sum_{j=1}^3\int_{\mathcal S^\prime}{\hat r}{\textrm d}{\hat r}\,{\textrm d}{\hat z}\,{\rm H}(\Delta)\left[\matrix{{\tilde{\bf u}}_{nj}\cr{\tilde{\bf v}}_{nj}\cr}\right]\left[\exp(-{\rm i}m{\hat\varphi})\frac{\partial G_{nj}^{\rm out}}{\partial{\hat\varphi}}\right]_{{\hat\varphi}={\hat\varphi}_-}^{{\hat\varphi}={\hat\varphi}_+}\Bigg\}
\label{C10}
\end{equation}
in the light of (\ref{E75}), (\ref{E84}) and (\ref{E85}).  The first term in (\ref{C10}) arises from the sinusoidal oscillations of the field $\matrix{[{\bf E}&{\bf B}]}$ at the frequency $m\omega$ as in any monochromatic radiation field.  However, the second term which arises from the retardation effects reflected in the discontinuities of the Green's function $G_{nj}$, is not normally encountered in the case of a conventional radiation.

\begin{figure}
\centerline{\includegraphics[width=13cm]{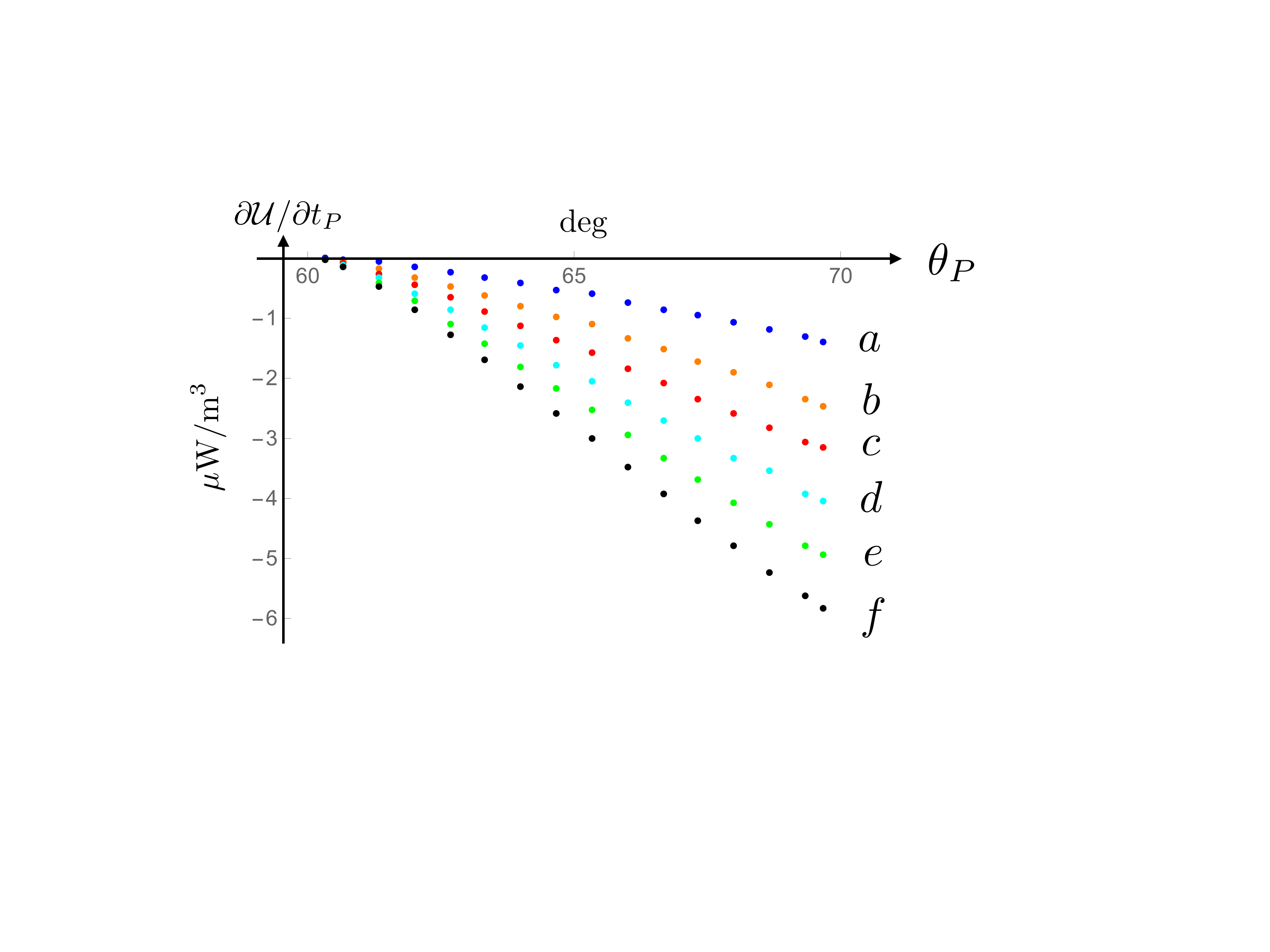}}
\caption{Time-averaged value of the temporal rate of change of the radiation energy density for Case Ia (described in \S~\ref{sec:numericalIa}) at polar angles where the radiation decays non-spherically.  The curves $a$, $b$, $c$, $d$, $e$ and $f$ respectively correspond to the following values of the distance ${\hat R}_P$: $10$ (blue), $10^2$ (orange), $10^3$ (red), $10^4$ (cyan), $10^5$ (green) and $10^6$ (black).  The radiation frequency and the electric current density have the values $\nu=2.5$ GHz and $\vert j_z\vert=0.01$ amp/m${}^2$, respectively, and the ratio of the radiation to rotation frequencies is $m=10$.  To display all six sets of results on the same graph, I have here multiplied the ordinates of the points for ${\hat R}_P=10^2$, $10^3$, $10^4$, $10^5$ and $10^6$ by the factors $10^2$, $10^4$, $10^6$, $10^8$ and $10^{10}$, respectively.}
\label{fC1}
\end{figure}

The Green's function $G_{nj}$ depends on ${\hat\varphi}$ both through the limits of integration in (\ref{E32}) and through the variable $\phi$ which appears in the argument of the Dirac delta function in this equation [see (\ref{E24})].  However, since the integrand in (\ref{E32}) has the same value at both limits of integration and the derivatives of the limits of integration both equal unity, derivative of $G_{nj}$ with respect to ${\hat\varphi}$ receives a non-zero contribution only from the dependence of the delta function on ${\hat\varphi}$ [see (\ref{B7})],
\begin{equation}
\frac{\partial G_{nj}}{\partial{\hat\varphi}}=-\sum_{k=1}^\infty\int_{{\hat\varphi}+2(k-1)\pi}^{{\hat\varphi}+2k\pi} {\rm d}\varphi\,h_{nj}\delta^\prime(g-\phi),
\label{C11}
\end{equation}
where $\delta^\prime$ stands for the derivative of the delta function with respect to its argument and
\begin{equation}
\left[\matrix{h_{n1}\cr h_{n2}\cr h_{n3}\cr}\right]={1\over {\hat R}^n}\left[\matrix{\cos(\varphi-\varphi_P)\cr \sin(\varphi-\varphi_P)\cr 1\cr}\right].
\label{C12}
\end{equation}
Integrating the right-hand side of (\ref{C11}) by parts, we obtain
\begin{equation}
\frac{\partial G_{nj}}{\partial{\hat\varphi}}=\sum_{k=1}^\infty\int_{{\hat\varphi}+2(k-1)\pi}^{{\hat\varphi}+2k\pi} {\rm d}\varphi\,\frac{\partial}{\partial\varphi}\left(\frac{h_{nj}}{\partial g/\partial\varphi}\right)\delta(g-\phi)
\label{C13}
\end{equation}
where the Jacobian $\vert\partial g/\partial\varphi\vert$ stems from the fact that $\partial\delta(g-\phi)/\partial\varphi=\delta^\prime(g-\phi)\partial g/\partial\varphi$.  As in \S~\ref{subsec:Expansion}, a uniform asymptotic approximation to this integral, for small $c_1$, can be found by the method of \citet{ChesterC:Extstd} in the time domain \citep{BurridgeR:Asyeir}.

We have seen that, where it is analytic (i.e., for all ${\bf x}\ne{\bf x}_P$), the function $g(\varphi)$ can be transformed into the cubic function defined in (\ref{E40}).  Inserting (\ref{E40}) and its derivative
\begin{equation}
\frac{\partial g}{\partial\varphi}=\frac{\nu^2-c_1^2}{{\rm d}\varphi/{\rm d}\nu}
\label{C14}
\end{equation}
in (\ref{C13}), we find that
\begin{equation}
\frac{\partial G_{nj}}{\partial{\hat\varphi}}=\sum_{k=1}^\infty {\mathcal H}\int_{-\infty}^\infty{\textrm d}\nu\,\left[-\frac{F_{nj}}{(\nu^2-c_1^2)^2}+\frac{F^\prime_{nj}}{\nu^2-c_1^2}\right]\delta(\textstyle{\frac{1}{3}}\nu^3-{c_1}^2\nu+c_2-\phi),
\label{C15}
\end{equation}
where
\begin{equation}
F_{nj}=\left(\frac{{\rm d}\varphi}{{\rm d}\nu}\right)^3\frac{\partial^2 g}{\partial\varphi^2}\,h_{nj},
\label{C16}
\end{equation}
\begin{equation}
F^\prime_{nj}=\left(\frac{{\rm d}\varphi}{{\rm d}\nu}\right)^2\frac{\partial h_{nj}}{\partial\varphi},
\label{C17}
\end{equation}
and ${\mathcal H}$ is the step function defined in (\ref{E44}).

\begin{figure}
\centerline{\includegraphics[width=11cm]{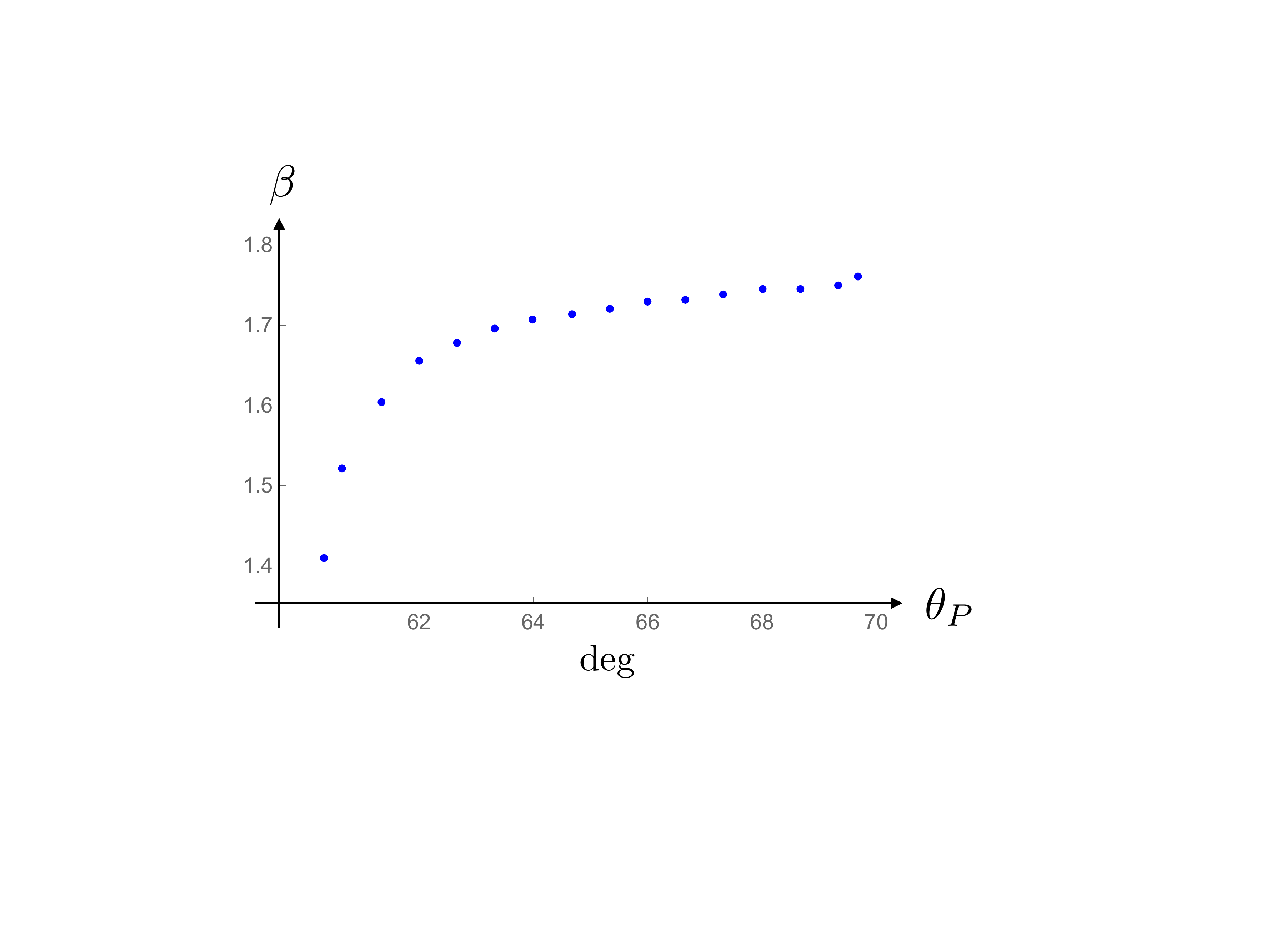}}
\caption{The exponent $\beta$ in the dependence ${\hat R}_P^{-\beta}$ of $\partial{\cal U}/\partial t_P$ (shown in figure~\ref{fC1}) on distance at polar angles $\theta_P$ where the radiation decays non-spherically.}  
\label{fC2}
\end{figure}

The leading term in the asymptotic expansion of the integral in (\ref{C15}) receives contributions only from the first term in the integrand of this integral: the factor $\vert\nu^2-c_1^2\vert$ in the ratio of the two terms relegates the contribution from $F^\prime_{nj}$ to the higher-order terms of the expansion \citep[see][]{ChesterC:Extstd}. Replacing the integral in (\ref{C15}) by the leading term in its asymptotic expansion for small $c_1$, we obtain
\begin{equation}
\frac{\partial G_{nj}}{\partial{\hat\varphi}}\simeq-\sum_{k=1}^\infty {\mathcal H}\int_{-\infty}^\infty{\textrm d}\nu\,\frac{P_{nj}+Q_{nj}\nu}{(\nu^2-c_1^2)^2}\delta(\textstyle{\frac{1}{3}}\nu^3-{c_1}^2\nu+c_2-\phi),\qquad c_1\ll1,
\label{C18}
\end{equation}
where
\begin{equation}
P_{nj}=\textstyle{\frac{1}{2}}(F_{nj}\vert_{\varphi=\varphi_-}+F_{nj}\vert_{\varphi=\varphi_+}),
\label{C19}
\end{equation}
and
\begin{equation}
Q_{nj}=\textstyle{\frac{1}{2}}{c_1}^{-1}(F_{nj}\vert_{\varphi=\varphi_-}-F_{nj}\vert_{\varphi=\varphi_+}).
\label{C20}
\end{equation}
According to (\ref{E36}) and (\ref{E65}),
\begin{equation}
P_{nj}=2c_1^2q_{nj},\quad{\rm and}\quad Q_{nj}=2p_{nj},
\label{C21}
\end{equation}
where the functions $p_{nj}$ and $q_{nj}$, which were encountered in the asymptotic expansion of $G_{nj}$ itself in (\ref{E46}), have the values given by (\ref{E67}) and (\ref{E68}). 

For the purposes of calculating $\partial{\bf E}/\partial t_P$ and $\partial{\bf B}/\partial t_P$ by means of the expression in (\ref{C10}), we need to evaluate $\partial G_{nj}/\partial{\hat\varphi}$ only outside the bifurcation surface, i.e., for $\vert\chi\vert>1$ [see (\ref{E52}) and (\ref{E59})].  In this region, the argument of the delta function in (\ref{C18}) has a single zero at $\nu=\nu_{\rm out}$ given in (\ref{E51}).  The integration with respect to $\nu$ in (\ref{C18}) therefore results in
\begin{eqnarray}
\frac{\partial G^{\rm out}_{nj}}{\partial{\hat\varphi}}&=&-\sum_{k=1}^\infty {\mathcal H}\frac{P_{nj}+Q_{nj}\nu}{\vert\nu^2-c_1^2\vert^3}\bigg\vert_{\nu=\nu_{\rm out}}\nonumber\\*
&=&-\sum_{k=1}^\infty {\mathcal H}\frac{2\sinh^3\left({\textstyle\frac{1}{3}}{\rm arccosh}\vert\chi\vert\right)}{c_1^5\vert \chi^2-1\vert^{3/2}}\left[c_1q_{nj}+2p_{nj}{\rm sgn}(\chi)\cosh\left({\textstyle\frac{1}{3}}{\rm arccosh}\vert\chi\vert\right)\right],\qquad
\label{C22}
\end{eqnarray}
and hence
\begin{equation}
\frac{\partial G^{\rm out}_{nj}}{\partial{\hat\varphi}}\Bigg\vert_{\chi=\pm1}=-\frac{2}{27 c_1^5}[c_1 q_{nj}\pm2p_{nj}],\qquad c_1\ll1.
\label{C23}
\end{equation}
Note that the summation over $k$ drops out of (\ref{C23}) because its summand depends on $k$ only through ${\cal H}\vert_{{\hat\varphi}={\hat\varphi}_\pm}$ and the sum $\sum_{k=1}^\infty{\mathcal H}\vert_{{\hat\varphi}={\hat\varphi}_\pm}$ equals unity. 

Equation~(\ref{C23}) now yields the following expression for the factor that contains $\partial G^{\rm out}_{nj}/\partial{\hat\varphi}\vert_{{\hat\varphi}={\hat\varphi}_\pm}$ in (\ref{C10}):
\begin{eqnarray}
\bigg[\exp(-{\rm i}m{\hat\varphi})\frac{\partial G_{nj}^{\rm out}}{\partial{\hat\varphi}}\bigg]_{{\hat\varphi}={\hat\varphi}_-}^{{\hat\varphi}={\hat\varphi}_+}&=&-\left({\textstyle\frac{2}{3}}\right)^3\frac{p_{nj}\cos\left({\textstyle\frac{2}{3}}mc_1^3\right)-{\textstyle\frac{1}{2}}{\rm i}c_1q_{nj}\sin\left({\textstyle\frac{2}{3}}mc_1^3\right)}{c_1^5}\nonumber\\*
&&\times\exp[-{\rm i}m(c_2+{\hat\varphi}_P)] ,
\label{C24}
\end{eqnarray}
where (\ref{E24}) and (\ref{E41}) have been used to express ${\hat\varphi}_-$ and ${\hat\varphi}_+$ in terms of $c_1$ and $c_2$.  When the cusp curve of the bifurcation surface of the observation point intersects the source distribution (i.e., in the case relevant to the present discussion in which the Poynting flux decays non-spherically), the ${\hat r}$-integration in (\ref{C10}) extends over the interval ${\hat r}_C\le{\hat r}\le{\hat r}_U$ (see figure~\ref{F11}).  Inserting (\ref{C24}) in (\ref{C10}) and writing the integral over ${\cal S}^\prime$ as a double integral, we obtain the following expression:
\begin{eqnarray}
\left[\matrix{\partial{\bf E}/\partial t_P\cr\partial{\bf B}/\partial t_P\cr}\right]&=&{\rm i}m\omega\Bigg\{\left[\matrix{{\bf E}\cr{\bf B}\cr}\right]+\left({\textstyle\frac{2}{3}}\right)^3\exp(-{\rm i}m{\hat\varphi}_P)\sum_{n=1}^2\sum_{j=1}^3\int_{-{\hat z}_0}^{{\hat z}_0}{\textrm d}{\hat z}\int_{{\hat r}_C}^{{\hat r}_U}{\hat r}{\textrm d}{\hat r}\,\left[\matrix{{\tilde{\bf u}}_{nj}\cr{\tilde{\bf v}}_{nj}\cr}\right]\nonumber\\*
&&\times\exp(-{\rm i}m c_2)c_1^{-5}\left[-p_{nj}\cos\left({\textstyle\frac{2}{3}}mc_1^3\right)\right.\left.+{\textstyle\frac{1}{2}}{\rm i}c_1q_{nj}\sin\left({\textstyle\frac{2}{3}}mc_1^3\right)\right]\Bigg\},\nonumber\\*
&&\nonumber\\*
&&\qquad\qquad m\gg1,\quad\theta_L\le\theta_P\le\theta_U,\quad\pi-\theta_U\le\theta_P\le\pi-\theta_L,
\label{C25}
\end{eqnarray}
where $\theta_L$ and $\theta_U$ are the polar angles defined in (\ref{E114}) and (\ref{E115}).  Both terms of the integrand in this equation are singular at the boundary ${\hat r}={\hat r}_C$ of the domain of integration where $c_1$ vanishes [see (\ref{E146})].  While the singularity of the term involving $\sin\left({\textstyle\frac{2}{3}}mc_1^3\right)$ is like that of $({\hat r}-{\hat r}_C)^{-1/2}$ and so is integrable, the singularity of the term involving $\cos\left({\textstyle\frac{2}{3}}mc_1^3\right)$ which is like that of $({\hat r}-{\hat r}_C)^{-5/2}$ needs to be handled by means of the Hadamard regularization technique.  

If we denote the divergent integral over ${\hat r}$ in (\ref{C25}) by 
\begin{equation}
\left[\matrix{{\bf I}\cr {\bf J}\cr}\right]=\sum_{n=1}^2\sum_{j=1}^3\int_{{\hat r}_C}^{{\hat r}_U}{\hat r}{\textrm d}{\hat r}\,\exp(-{\rm i}mc_2)\left[\matrix{{\tilde{\bf u}}_{nj}\cr{\tilde{\bf v}}_{nj}\cr}\right] p_{nj}c_1^{-5}\cos\left({\textstyle\frac{2}{3}}mc_1^3\right),
\label{C26}
\end{equation}
then the first step in finding its Hadamard's finite part is to cast it into the following canonical form by simultaneously multiplying and dividing its integrand by $({\hat r}-{\hat r}_C)^{5/2}$,
\begin{equation}
\left[\matrix{{\bf I}\cr {\bf J}\cr}\right]=\int_{{\hat r}_C}^{{\hat r}_U}{\textrm d}{\hat r}\,\left[\matrix{{\tilde{\bf U}}\cr{\tilde{\bf V}}\cr}\right]({\hat r}-{\hat r}_C)^{-5/2},
\label{C27}
\end{equation}
in which
\begin{equation}
\left[\matrix{{\tilde{\bf U}}\cr{\tilde{\bf V}}\cr}\right]=\sum_{n=1}^2\sum_{j=1}^3{\hat r}\exp(-{\rm i}mc_2)p_{nj}c_1^{-5}\cos\left({\textstyle\frac{2}{3}}mc_1^3\right)({\hat r}-{\hat r}_C)^{5/2}\left[\matrix{{\tilde{\bf u}}_{nj}\cr{\tilde{\bf v}}_{nj}\cr}\right].
\label{C28}
\end{equation}
This form of the integrand consists of two factors: the factor $[\matrix{{\tilde{\bf U}}&{\tilde{\bf V}}}]$ which is a regular function of $({\hat r}-{\hat r}_C)^{1/2}$ throughout the integration domain [see (\ref{E146})] and the factor $({\hat r}-{\hat r}_C)^{-5/2}$ which explicitly specifies the order of the singularity.  Hadamard's finite part of the integral in (\ref{C27}) can be found by expressing its integrand in terms of $\xi=({\hat r}-{\hat r}_C)^{1/2}$, performing four consecutive integrations by parts and discarding the integrated terms that diverge at $\xi=0$.  Since the integrated terms at ${\hat r}={\hat r}_U$ vanish for any current density that smoothly vanishes at this boundary of the source distribution, this procedure results in 
\begin{eqnarray}
{\rm Fp}\left\{\left[\matrix{{\bf I}\cr {\bf J}\cr}\right]\right\}&=&{\rm Fp}\left\{2\int_0^{({\hat r}_U-{\hat r}_C)^{1/2}}{\rm d}\xi\, \xi^{-4}\left[\matrix{{\tilde{\bf U}}\cr{\tilde{\bf V}}\cr}\right]\right\}\nonumber\\*
&=&-\frac{1}{3}\int_0^{({\hat r}_U-{\hat r}_C)^{1/2}}{\rm d}\xi\,\ln(\xi)\frac{\partial^4}{\partial\xi^4}\left[\matrix{{\tilde{\bf U}}\cr{\tilde{\bf V}}\cr}\right]
\label{C29}
\end{eqnarray}
\citep[see][and appendix~\ref{appA}]{HadamardJ:lecCau,HoskinsRF:GenFun}.  To evaluate the above expression numerically, it is of course necessary to remove the indeterminacy of $\matrix{[{\tilde{\bf U}} & {\tilde{\bf V}}]}$ at $\xi=0$ before performing the differentiations by replacing the numerator and the denominator in (\ref{C28}) each by its individual Taylor expansion in a small neighbourhood ($\xi\leq{\hat R}_P^{-2}$) of this point. 

Replacing the divergent integral in (\ref{C25}) by its Hadamard finite part, given by (\ref{C29}), we arrive at
\begin{eqnarray}
\left[\matrix{\partial{\bf E}/\partial t_P\cr\partial{\bf B}/\partial t_P\cr}\right]&=&{\rm i}m\omega\Bigg\{\left[\matrix{{\bf E}\cr{\bf B}\cr}\right]+\left({\textstyle\frac{2}{3}}\right)^3\exp(-{\rm i}m{\hat\varphi}_P)\int_{-{\hat z}_0}^{{\hat z}_0}{\textrm d}{\hat z}\Bigg[-{\rm Fp}\left\{\left[\matrix{{\bf I}\cr {\bf J}\cr}\right]\right\}\nonumber\\*
&&+{\textstyle\frac{1}{2}}{\rm i}\sum_{n=1}^2\sum_{j=1}^3\int_{{\hat r}_C}^{{\hat r}_U}{\textrm d}{\hat r}\,{\hat r}q_{nj}\exp(-{\rm i}mc_2)\left[\matrix{{\tilde{\bf u}}_{nj}\cr{\tilde{\bf v}}_{nj}\cr}\right] c_1^{-4}\sin\left({\textstyle\frac{2}{3}}mc_1^3\right)\Bigg]\Bigg\}\quad
\label{C30}
\end{eqnarray}
for $m\gg1$ and $\theta_L\le\theta_P\le\theta_U$ or $\pi-\theta_U\le\theta_P\le\pi-\theta_L$, i.e., for polar angles at which the cusp curve of the bifurcation surface intersects the source distribution across its entire ${\hat z}$-extent (see figure~\ref{F11}).  As in (\ref{E173}), the right-hand side of the above equation depends on the observation time $t_P$ through the oscillating factor $\exp(-{\rm i}m{\hat\varphi}_P)$ which multiplies all its terms.  Hence, if we denote the time-averaged rate of change of the energy density in the non-spherically decaying part of the radiation field by $\partial{\cal U}/\partial t_P$, then
\begin{equation}
\frac{\partial{\cal U}}{\partial t_P}=\frac{1}{4\pi}\bigg\langle\Re({\bf E}){\bf\cdot}\Re\left(\frac{\partial{\bf E}}{\partial t_P}\right)+\Re({\bf B}){\bf\cdot}\Re\left(\frac{\partial{\bf B}}{\partial t_P}\right)\bigg\rangle=\frac{1}{8\pi}\Re\left({\bf E}^*{\bf\cdot}\frac{\partial{\bf E}}{\partial t_P}+{\bf B}^*{\bf\cdot}\frac{\partial{\bf B}}{\partial t_P}\right),
\label{C31}
\end{equation}
where the angular brackets denote averaging with respect to $t_P$ over an integral multiple of the oscillation period $2\pi/(m\omega)$.  (Note that ${\bf E}^*\cdot\partial{\bf E}/\partial t_P+{\bf B}^*\cdot\partial{\bf B}/\partial t_P$ is not necessarily real in the present case.)  This shows that the first term in (\ref{C30}), which arises from the sinusoidal oscillations of the field $\matrix{[{\bf E}&{\bf B}]}$ at the frequency $m\omega$, makes no contribution towards the value of the time-averaged quantity $\partial{\cal U}/\partial t_P$ because the factor ${\rm i}$ in this term renders its oscillations out of phase with those of $\matrix{[{\bf E}&{\bf B}]}$ by $\pi/2$.  The second term in (\ref{C30}), which is particular to the present radiation process, on the other hand, results in a value for $\partial{\cal U}/\partial t_P$ that is clearly non-zero.

To confirm that, as expected on physical grounds, the non-zero value of $\partial{\cal U}/\partial t_P$ predicted by (\ref{C30}) and (\ref{E173}) is in fact negative, I have evaluated this quantity for the parameters of Case Ia described in \S~\ref{sec:numericalIa} with $\vert j_z\vert=0.01$ amp/m$^2$.  The result is shown in figure~\ref{fC1} at several distances (${\hat R}_P=10$, $10^2$, $10^3$, $10^4$, $10^5$ and $10^6$) within the angular interval ($60^\circ<\theta_P<70^\circ$) where the Poynting vector decays non-spherically.  To make the figure more transparent, I have shifted the results for ${\hat R}_P=10^2$, $10^3$, $10^4$, $10^5$ and $10^6$ relative to that for ${\hat R}_P=10$ by multiplying them by $10^2$, $10^4$, $10^6$, $10^8$ and $10^{10}$, respectively.  Figure~\ref{fC1} shows not only that $\partial{\cal U}/\partial t_P$ is negative wherever the radial component of the Poynting vector decays non-spherically (see figures~\ref{F22} and \ref{F25}), but also that its absolute value diminishes with distance like the value of the radial component of the Poynting vector: as ${\hat R}_P^{-\beta}$ with $1<\beta<2$ . 

I have employed the same procedure as that illustrated in figure~\ref{E24} to find the exponent $\beta$ in the power law ${\hat R}_P^{-\beta}$ that best fits the dependence of $\partial{\cal U}/\partial t_P$ on distance at various values of $\theta_P$.  The result, which is shown in figure~\ref{fC2}, is consistent with the angular dependence of $\alpha$ depicted in figure~\ref{F25}.   

According to (\ref{C31}), the time-averaged version of the Poynting theorem (\ref{C1}) in free space (where ${\bf j}=0$) has the form: 
\begin{equation}
\int_{\cal D}{\rm d}^3{\bf x}_P\frac{\partial{\cal U}}{\partial t_P}+\int_{\partial{\cal D}}{\rm d}^2{\bf x}_P\cdot{\bf S}=0,
\label{C32}
\end{equation}
in which ${\bf S}$ is the time-averaged Poynting vector defined in (\ref{E187}).   Because $\partial{\cal U}/\partial t_P$ is negative throughout any volume ${\cal D}$ that contains the non-spherically decaying radiation field, this equation can only be satisfied by a positive value of the Poynting flux across a closed surface $\partial{\cal D}$ enclosing ${\cal D}$.  Consider two concentric spheres centred on the source both of which intersect the volume occupied by the propagating radiation at a given observation time $t_P$.  A positive value of the Poynting flux across the closed surface consisting of these two spheres means that the total energy that leaves the outer sphere per unit time is greater than the total energy that enters the inner sphere per unit time.  This, on the other hand, is possible only if the magnitude of the time-averaged Poynting vector ${\bf S}$ diminishes with the distance $R_P$ from the source more slowly than $R_P^{-2}$.  The non-spherical decay of the radiation discussed in this paper is thus required by the conservation of energy given that the time-averaged rate of change of the energy density of this radiation is negative.

The fact that the present radiation never attains a steady state can be traced back to the following transient feature of the retarded distribution of its source.  The retarded distribution of the polarization described by (\ref{E1}) is given by 
\begin{equation}
P_{r,\varphi,z}(r,\varphi,z,t_{\rm ret})=s_{r,\varphi,z}(r,z)\cos(m{\hat\varphi}_{\rm ret}), \qquad0\le{\hat\varphi}_{\rm ret}<2\pi,
\label{C33}
\end{equation}
where $t_{\rm ret}$ and ${\hat\varphi}_{\rm ret}$ are defined in (\ref{B2}) and (\ref{B5}) (see \S~\ref{subsec:constraint} and appendix~\ref{appB}).  Because of the nonlinearity of the relationship between the retarded time $t$ and the observation time $t_P$, this retarded distribution bears no resemblance to the actual distribution shown in figure~\ref{F1}.  In the case of the example plotted in figure~\ref{F36}, the above equation describes a retarded distribution of the source whose azimuthal extent consists of the four disjoint intervals $\varphi_1\le\varphi\le\varphi_2$, $2\pi\le\varphi\le\varphi_3$, $\varphi_4\le\varphi\le\varphi_5$ and $\varphi_6\le\varphi\le\varphi_7$.  For fixed values of $({\hat r},{\hat z},{\hat r}_P,{\hat z}_P)$, the curve shown in figure~\ref{F36} 
is lowered by $2\pi$ as the observation time $t_P$ advances by $2\pi/\omega$ without changing shape, so that, in general, the retarded distribution of the source at a given observation point returns to its original shape after a rotation period.  However, as we shall see below the changes that the shape of the retarded distribution of this (or any other superluminally rotating) source undergoes from one period to another occur with different rates during different periods \citep[see also the retarded distribution of the example analysed in][]{ArdavanH:RepFund}.

The temporal rate of change $\partial{\varphi}_j/\partial t_P$ of the position $\varphi_j({\hat r},{\hat z},{\hat r}_P,{\hat z}_P,t_P)$ of each point on a boundary of the azimuthal support of the retarded distribution of the source described by (\ref{E1}) is given by (\ref{B8}).  For an observation point $({\hat r}_P,{\hat z}_P)$ inside the envelope of wave fronts emanating from the source element at $({\hat r},\varphi_j,{\hat z})$ near either the sheet $\phi=\phi_-$ or the sheet $\phi=\phi_+$ of this envelope, the value of $\varphi_j$ is close to that of either $\varphi_-$ or $\varphi_+$, respectively: recall that the integer $k$ in the expressions for these angles in (\ref{E34}) is selected to correspond to the rotation period whose contribution reaches the observation point $({\hat r}_P,\varphi_P,{\hat z}_P)$ at the observation time $t_P$.  The value of $\partial\varphi_j/\partial t_P$ for such an observation point can therefore be obtained by expanding the denominator in (\ref{B8}) in a Taylor series in powers of $\varphi_j-\varphi_+$ or $\varphi_j-\varphi_-$.  The dominant term of the resulting series for $\varphi_j\simeq\varphi_\pm$ is
\begin{equation}
\frac{\partial\varphi_j}{\partial t_P}\simeq\mp\frac{\omega {\hat R}_\pm}{\Delta^{1/2}(\varphi_j-\varphi_\pm)},\qquad\vert\varphi_j-\varphi_\pm\vert\ll1,
\label{C34}
\end{equation}
as can be readily seen from the values of the derivative of $\partial g/\partial\varphi$ at $\varphi=\varphi_\pm$ in (\ref{E36}).  

The right-hand side of (\ref{C34}) is infinitely large on either sheet of the envelope in question and changes sign from one sheet to another.  It also depends on the integer $k$ enumerating successive rotations, which appears in the expressions for $\varphi_\pm$ in (\ref{E34}), monotonically.  At a fixed observation point close to one of the sheets $\phi=\phi_\pm$ of the envelope of wave fronts emanating from the volume element of the source at $({\hat r},\varphi_j,{\hat z})$, the rate $\partial{\varphi}_j/\partial t_P$ monotonically increases or decreases (depending on its sign) as the number of rotations $k$ executed by the source since $t=0$ increases.  On the cusp locus of the envelope where $\Delta=0$, this rate is infinitely large.  Thus the rate at which the boundaries of the azimuthal support of the retarded distribution of the source change with time depends on the time elapsed since the source was switched on monotonically.

\bibliographystyle{jpp}

\bibliography{Ardavan_JPP}

\begin{thebibliography}{44}
\expandafter\ifx\csname natexlab\endcsname\relax\def\natexlab#1{#1}\fi
\def\au#1{#1} \def\ed#1{#1} \def\yr#1{#1}\def\at#1{#1}\def\jt#1{\textit{#1}}
  \def\bt#1{#1}\def\bvol#1{\textbf{#1}} \def\vol#1{#1} \def\pg#1{#1}
  \def\publ#1{#1}\def\arxiv#1{#1}\def\org#1{#1}\def\st#1{\textit{#1}}

\bibitem[Ardavan \& Ardavan(2010)]{ArdavanA:Patent}
{\sc \au{Ardavan, A.} \& \au{Ardavan, H.}} \yr{2010} Apparatus for generating
  focused electromagnetic radiation. European patent EP1112578.

\bibitem[Ardavan {\em et~al.\/}(2004{\natexlab{{\em a\/}}})Ardavan, Ardavan \&
  Singleton]{McDonaldReply}
{\sc \au{Ardavan, A.}, \au{Ardavan, H.} \& \au{Singleton, J.}}
  \yr{2004{\natexlab{{\em a\/}}}}  \at{Synchrotron-\v{C}erenkov radiation}.
  \jt{Science}  \bvol{303}~(5656),  \pg{311}.

\bibitem[Ardavan {\em et~al.\/}(2004{\natexlab{{\em b\/}}})Ardavan, Hayes,
  Singleton, Ardavan, Fopma \& Halliday]{ArdavanA:Exponr}
{\sc \au{Ardavan, A.}, \au{Hayes, W.}, \au{Singleton, J.}, \au{Ardavan, H.},
  \au{Fopma, J.} \& \au{Halliday, D.}} \yr{2004{\natexlab{{\em b\/}}}}
  \at{Experimental observation of nonspherically-decaying radiation from a
  rotating superluminal source}.  \jt{J. Appl. Phys.}  \bvol{96},
  \pg{7760--7777(E)}.

\bibitem[Ardavan(1981)]{ArdavanH:Nature}
{\sc \au{Ardavan, H.}} \yr{1981}  \at{Is the light cylinder the site of
  emission in pulsars?}  \jt{Nature}  \bvol{289},  \pg{44--45}.

\bibitem[Ardavan(1998)]{ArdavanH:Genfnd}
{\sc \au{Ardavan, H.}} \yr{1998}  \at{Generation of focused, nonspherically
  decaying pulses of electromagnetic radiation}.  \jt{Phys. Rev. E}  \bvol{58},
   \pg{6659--6684}.

\bibitem[Ardavan(1999)]{ArdavanH:JMP99}
{\sc \au{Ardavan, H.}} \yr{1999}  \at{Method of handling the divergences in the
  radiation theory of sources that move faster than their waves}.  \jt{J. Math.
  Phys.}  \bvol{40},  \pg{4331--4336}.

\bibitem[Ardavan(2000)]{ArdavanH:RepCGf}
{\sc \au{Ardavan, H.}} \yr{2000}  \at{{R}eply to {C}omments on {G}eneration of
  focused, nonspherically decaying pulses of electromagnetic radiation}.
  \jt{Phys. Rev. E}  \bvol{62}~(2),  \pg{3010--3013}.

\bibitem[Ardavan {\em et~al.\/}(2004{\natexlab{{\em c\/}}})Ardavan, Ardavan \&
  Singleton]{ArdavanH:Speapc}
{\sc \au{Ardavan, H.}, \au{Ardavan, A.} \& \au{Singleton, J.}}
  \yr{2004{\natexlab{{\em c\/}}}}  \at{Spectral and polarization
  characteristics of the nonspherically decaying radiation generated by
  polarization currents with superluminally rotating distribution patterns}.
  \jt{J. Opt. Soc. Am. A}  \bvol{21},  \pg{858--872}.

\bibitem[Ardavan {\em et~al.\/}(2006)Ardavan, Ardavan \&
  Singleton]{ArdavanH:Speapc1}
{\sc \au{Ardavan, H.}, \au{Ardavan, A.} \& \au{Singleton, J.}} \yr{2006}
  \at{Spectral and polarization characteristics of the nonspherically decaying
  radiation generated by polarization currents with superluminally rotating
  distribution patterns: reply to comment}.  \jt{J. Opt. Soc. Am. A}
  \bvol{23}~(6),  \pg{1535--1539}.

\bibitem[Ardavan {\em et~al.\/}(2007)Ardavan, Ardavan, Singleton, Fasel \&
  Schmidt]{ArdavanH:Morph}
{\sc \au{Ardavan, H.}, \au{Ardavan, A.}, \au{Singleton, J.}, \au{Fasel, J.} \&
  \au{Schmidt, A.}} \yr{2007}  \at{Morphology of the nonspherically decaying
  radiation beam generated by a rotating superluminal source}.  \jt{J. Opt.
  Soc. Am. A}  \bvol{24},  \pg{2443--2456}.

\bibitem[Ardavan {\em et~al.\/}(2008{\natexlab{{\em a\/}}})Ardavan, Ardavan,
  Singleton, Fasel \& Schmidt]{ArdavanH:Funda}
{\sc \au{Ardavan, H.}, \au{Ardavan, A.}, \au{Singleton, J.}, \au{Fasel, J.} \&
  \au{Schmidt, A.}} \yr{2008{\natexlab{{\em a\/}}}}  \at{Fundamental role of
  the retarded potential in the electrodynamics of superluminal sources}.
  \jt{J. Opt. Soc. Am. A}  \bvol{25},  \pg{543--557}.

\bibitem[Ardavan {\em et~al.\/}(2008{\natexlab{{\em b\/}}})Ardavan, Ardavan,
  Singleton, Fasel \& Schmidt]{Ardavan_RepMorph}
{\sc \au{Ardavan, H.}, \au{Ardavan, A.}, \au{Singleton, J.}, \au{Fasel, J.} \&
  \au{Schmidt, A.}} \yr{2008{\natexlab{{\em b\/}}}}  \at{Morphology of the
  nonspherically decaying radiation beam generated by a rotating superluminal
  source: reply to comment}.  \jt{J. Opt. Soc. Am. A}  \bvol{25}~(9),
  \pg{2167--2169}.

\bibitem[Ardavan {\em et~al.\/}(2009{\natexlab{{\em a\/}}})Ardavan, Ardavan,
  Singleton, Fasel \& Schmidt]{ArdavanH:RepFund}
{\sc \au{Ardavan, H.}, \au{Ardavan, A.}, \au{Singleton, J.}, \au{Fasel, J.} \&
  \au{Schmidt, A.}} \yr{2009{\natexlab{{\em a\/}}}}  \at{Fundamental role of
  the retarded potential in the electrodynamics of superluminal sources: reply
  to comment}.  \jt{J. Opt. Soc. Am. A}  \bvol{26}~(10),  \pg{2109--2113}.

\bibitem[Ardavan {\em et~al.\/}(2009{\natexlab{{\em b\/}}})Ardavan, Ardavan,
  Singleton, Fasel \& Schmidt]{ArdavanH:Inad}
{\sc \au{Ardavan, H.}, \au{Ardavan, A.}, \au{Singleton, J.}, \au{Fasel, J.} \&
  \au{Schmidt, A.}} \yr{2009{\natexlab{{\em b\/}}}}  \at{Inadequacies in the
  conventional treatment of the radiation field of moving sources}.  \jt{J.
  Math. Phys.}  \bvol{50},  \pg{103510(1)--103510(12)}.

\bibitem[Ardavan {\em et~al.\/}(2008{\natexlab{{\em c\/}}})Ardavan, Ardavan,
  Singleton \& Perez]{ArdavanH:Pul}
{\sc \au{Ardavan, H.}, \au{Ardavan, A.}, \au{Singleton, J.} \& \au{Perez,
  M.~R.}} \yr{2008{\natexlab{{\em c\/}}}}  \at{Mechanism of generation of the
  emission bands in the dynamic spectrum of the {C}rab pulsar}.  \jt{Mon. Not.
  R. Astron. Soc.}  \bvol{388},  \pg{873--883}.

\bibitem[Bender \& Orszag(1999)]{BenderOrszag}
{\sc \au{Bender, C.~M.} \& \au{Orszag, S.~A.}} \yr{1999} {\em Advanced
  mathematical methods for scientists and engineers I: asymptotic methods and
  perturbation theory\/}.  \publ{London: Springer}.

\bibitem[Bolotovskii \& Bykov(1990)]{BolotovskiiBM:Radbcm}
{\sc \au{Bolotovskii, B.~M.} \& \au{Bykov, V.~P.}} \yr{1990}  \at{Radiation by
  charges moving faster than light}.  \jt{Sov. Phys-Usp.}  \bvol{33},
  \pg{477--487}.

\bibitem[Bolotovskii \& Ginzburg(1972)]{BolotovskiiBM:VaveaD}
{\sc \au{Bolotovskii, B.~M.} \& \au{Ginzburg, V.~L.}} \yr{1972}  \at{The
  {V}avilov-\v{C}erenkov effect and the doppler effect in the motion of sources
  with superluminal velocity in vacuum}.  \jt{Sov. Phys-Usp.}  \bvol{15},
  \pg{184--192}.

\bibitem[Bolotovskii \& Serov(2005)]{BolotovskiiBM:Radsse}
{\sc \au{Bolotovskii, B.~M.} \& \au{Serov, A.~V.}} \yr{2005}  \at{Radiation of
  superhumanal sources in empty space}.  \jt{Sov. Phys-Usp.}  \bvol{48},
  \pg{903--915}.

\bibitem[Burridge(1995)]{BurridgeR:Asyeir}
{\sc \au{Burridge, R.}} \yr{1995}  \at{Asymptotic evaluation of integrals
  related to time-dependent fields near caustics}.  \jt{SIAM J. Appl. Math.}
  \bvol{55},  \pg{390--409}.

\bibitem[Chatterjee {\em et~al.\/}(2017)Chatterjee, Law, Wharton,
  Burke-Spolaor, Hessels, Bower, Cordes, Tendulkar, Bassa, Demorest, Butler,
  Seymour, Scholz, Abruzzo, Bogdanov, Kaspi, Keimpema, Lazio, Marcote \&
  McLaughlin]{FRB}
{\sc \au{Chatterjee, S.}, \au{Law, C.~J.}, \au{Wharton, R.~S.},
  \au{Burke-Spolaor, M.}, \au{Hessels, J. W.~T.}, \au{Bower, G.~C.},
  \au{Cordes, J.~M.}, \au{Tendulkar, S.~P.}, \au{Bassa, C.~G.}, \au{Demorest,
  P.}, \au{Butler, B.~J.}, \au{Seymour, A.}, \au{Scholz, P.}, \au{Abruzzo,
  M.~W.}, \au{Bogdanov, S.}, \au{Kaspi, V.~M.}, \au{Keimpema, A.}, \au{Lazio,
  T. J.~W.}, \au{Marcote, B.} \& \au{McLaughlin, M.}} \yr{2017}  \at{A direct
  localization of a fast radio burst and its host}.  \jt{Nature}  \bvol{541},
  \pg{58 -- 61}.

\bibitem[Chester {\em et~al.\/}(1957)Chester, Friedman \&
  Ursell]{ChesterC:Extstd}
{\sc \au{Chester, C.}, \au{Friedman, B.} \& \au{Ursell, F.}} \yr{1957}  \at{An
  extension of the method of steepest descent}.  \jt{Proc. Cambridge Philos.
  Soc.}  \bvol{53},  \pg{599--611}.

\bibitem[Courant(1967)]{Courant1967}
{\sc \au{Courant, R.}} \yr{1967} {\em Differential and Integral Calculus\/}, ,
  \vol{vol. 2, Chap. 4}.  \publ{Blackie}.

\bibitem[Ginzburg(1972)]{GinzburgVL:vaveaa}
{\sc \au{Ginzburg, V.~L.}} \yr{1972}  \at{Vavilov-\v{C}erenkov effect and
  anomalous doppler effect in a medium in which the wave phase velocity exceeds
  the velocity of light in vacuum}.  \jt{Sov. Phys-JETP}  \bvol{35},
  \pg{92--93}.

\bibitem[Gradshteyn \& Ryzhik(1980)]{Gradshteyn}
{\sc \au{Gradshteyn, I.~S.} \& \au{Ryzhik, I.~M.}} \yr{1980} {\em Table of
  Integrals,Series and Products\/}.  \publ{Academic Press}.

\bibitem[Hadamard(2003)]{HadamardJ:lecCau}
{\sc \au{Hadamard, J.}} \yr{2003} {\em Lectures on Cauchy's problem in linear
  partial differental equations\/}.  \publ{New York: Dover}.

\bibitem[Hannay(2000)]{HannayJH:ComIGf}
{\sc \au{Hannay, J.~H.}} \yr{2000}  \at{Comment {II} on `{G}eneration of
  focused, nonspherically decaying pulses of electromagnetic radiation'}.
  \jt{Phys. Rev. E}  \bvol{62}~(2),  \pg{3008--3009}.

\bibitem[Hannay(2001)]{HannayJH:JMP}
{\sc \au{Hannay, J.~H.}} \yr{2001}  \at{Comment on `{M}ethod of handling the
  divergences in the radiation theory of sources that move faster than their
  waves'}.  \jt{J. Math. Phys.}  \bvol{42},  \pg{3973--3974}.

\bibitem[Hannay(2006)]{HannayJH:Speapc}
{\sc \au{Hannay, J.~H.}} \yr{2006}  \at{Spectral and polarization
  characteristics of the nonspherically decaying radiation generated by
  polarization currents with superluminally rotating distribution patterns:
  comment}.  \jt{J. Opt. Soc. Am. A}  \bvol{23}~(6),  \pg{1084--7529}.

\bibitem[Hannay(2008)]{Hannay_Morphology}
{\sc \au{Hannay, J.~H.}} \yr{2008}  \at{Morphology of the nonspherically
  decaying radiation generated by a rotating superluminal source: comment}.
  \jt{J. Opt. Soc. Am. A}  \bvol{25},  \pg{2165--2166}.

\bibitem[Hannay(2009)]{Hannay:09}
{\sc \au{Hannay, J.~H.}} \yr{2009}  \at{Fundamental role of the retarded
  potential in the electrodynamics of superluminal sources: comment.}  \jt{J.
  Opt. Soc. Am. A}  \bvol{26},  \pg{2107--2109}.

\bibitem[Hewish(2000)]{Hewish2}
{\sc \au{Hewish, A.}} \yr{2000}  \at{Comment {I} on `{G}eneration of focused,
  nonspherically decaying pulses of electromagnetic radiation'}.  \jt{Phys.
  Rev. E}  \bvol{62},  \pg{3007}.

\bibitem[Hoskins(2009)]{HoskinsRF:GenFun}
{\sc \au{Hoskins, R.~F.}} \yr{2009} {\em Delta functions: an introduction to
  generalised functions\/}, 2nd edn.  \publ{Oxford: Woodhead}.

\bibitem[Jackson(1999)]{JacksonJD:Classical}
{\sc \au{Jackson, J.~D.}} \yr{1999} {\em Classical electrodynamics\/}, 3rd edn.
   \publ{New York: Wiley}.

\bibitem[Jay(1984)]{IEEE}
{\sc \au{Jay, F.}}, ed. \yr{1984} {\em IEEE Standard dictionary of electrical
  and electronics terms\/}, 3rd edn.  \publ{New York, NY: Institute of
  Electrical and Electronics Engineers}.

\bibitem[Kalapotharakos {\em et~al.\/}(2012)Kalapotharakos, Contopoulos \&
  Kazanas]{Contopoulos:2012}
{\sc \au{Kalapotharakos, C.}, \au{Contopoulos, I.} \& \au{Kazanas, D.}}
  \yr{2012}  \at{The extended pulsar magnetosphere}.  \jt{Mon. Not. Astron.
  Soc.}  \bvol{420},  \pg{2793--2798}.

\bibitem[McDonald(2004)]{McDonald}
{\sc \au{McDonald, K.}} \yr{2004}  \at{Synchrotron-\v{C}erenkov radiation}.
  \jt{Science}  \bvol{303}~(5656),  \pg{310}.

\bibitem[Morse \& Feshbach(1953)]{MorsePM:Methods1}
{\sc \au{Morse, P.~M.} \& \au{Feshbach, H.}} \yr{1953} {\em Methods of
  Theoretical Physics\/}, ,  \vol{vol.~1}.  \publ{New York: McGraw-Hill}.

\bibitem[Olver {\em et~al.\/}(2010)Olver, Lozier, Boisvert \& Clark]{Olver}
{\sc \au{Olver, F. W.~J.}, \au{Lozier, D.~W.}, \au{Boisvert, R.~F.} \&
  \au{Clark, C.~W.}}, ed. \yr{2010} {\em NIST Handbook of Mathematical
  Functions\/}.  \publ{Cambridge University Press}.

\bibitem[Philippov {\em et~al.\/}(2019)Philippov, Uzdensky, Spitkovsky \&
  Cerutti]{Philippov2019}
{\sc \au{Philippov, A.}, \au{Uzdensky, D.~A.}, \au{Spitkovsky, A.} \&
  \au{Cerutti, B.}} \yr{2019} Pulsar radio emission mechanism: radio nanoshots
  as a low frequency afterglow of relativistic magnetic reconnection.
  ArXiv:1902.07730[astro-ph.HE].

\bibitem[Piron(2016)]{Piron}
{\sc \au{Piron, F.}} \yr{2016}  \at{Gamma-ray bursts at high and very high
  energies}.  \jt{Comptes Rendus Physique}  \bvol{17},  \pg{617--631}.

\bibitem[Spitkovsky(2006)]{SpitkovskyA:Oblique}
{\sc \au{Spitkovsky, A.}} \yr{2006}  \at{Time-dependent force-free pulsar
  magnetospheres: axisymmetric and oblique rotators}.  \jt{Astrophys. J.}
  \bvol{648},  \pg{L51--L54}.

\bibitem[Tchekhovskoy {\em et~al.\/}(2016)Tchekhovskoy, Philippov \&
  Spitkovsky]{Tchekhovskoy:etal}
{\sc \au{Tchekhovskoy, A.}, \au{Philippov, A.} \& \au{Spitkovsky, A.}}
  \yr{2016}  \at{Three-dimensional analytical description of magnetized winds
  from oblique pulsars}.  \jt{Mon. Not. R. Astron. Soc.}  \bvol{457},
  \pg{3384--3395}.

\bibitem[Uzdensky \& Spitkovsky(2014)]{Uzdensky}
{\sc \au{Uzdensky, D.~A.} \& \au{Spitkovsky, A.}} \yr{2014}  \at{Physical
  conditions in the reconnection layer in pulsar magnetospheres}.
  \jt{Astrophys. J.}  \bvol{780},  \pg{3(1)--3(7)}.

\end{thebibliography}

\end{document}